\documentclass[a4paper,fleqn,usenatbib]{mnras}

\usepackage{newtxtext,newtxmath}
\usepackage[T1]{fontenc}
\usepackage{ae,aecompl}

\usepackage{graphicx}	% Including figure files
\usepackage{amsmath}	% Advanced maths commands
\usepackage{amssymb}	% Extra maths symbols

\usepackage{color}
\usepackage{fancybox}
\usepackage{subfigure}
\usepackage{rotating}
\usepackage{import}
\usepackage{pdfpages}
\usepackage{xcolor}
\usepackage{hyperref}
\usepackage{verbatim} 
\usepackage{CJK}
\usepackage{etoolbox}
\makeatletter
\patchcmd\@combinedblfloats{\box\@outputbox}{\unvbox\@outputbox}{}{%
   \errmessage{\noexpand\@combinedblfloats could not be patched}%
}%
 \makeatother
\usepackage{adjustbox}

\newcommand{\angstrom}{{\rm \AA}}

\newcommand{\kms}{$\rm km~s^{-1}$}

\newcommand{\HeIIa}{He{\sevenrm\,II}\,$\lambda$4686}
\newcommand{\hb}{H{$\beta$}}
\newcommand{\ha}{H{$\alpha$}}

\newcommand{\FeII}{Fe{\sevenrm\,II}}

\newcommand{\MgII}{Mg{\sevenrm\,II}}
\newcommand{\MgIIa}{Mg{\sevenrm\,II}\,$\lambda$2800}

\newcommand{\OIII}{[O{\sevenrm\,III}]}

\newcommand{\OIIIa}{[O{\sevenrm\,III}]\,$\lambda$4959}
\newcommand{\OIIIb}{[O{\sevenrm\,III}]\,$\lambda$5007}
\newcommand{\OIIIc}{[O{\sevenrm\,III}]\,$\lambda\lambda$4959,5007}
\newcommand{\OIa}{[O{\sevenrm\,I}]\,$\lambda$6300}
\newcommand{\NII}{[N{\sevenrm\,II}]}

\newcommand{\SII}{[S{\sevenrm\,II}]}

\font\sevenrm=cmr7 scaled 1000

\begin{document}

\title[Radial Velocity Test for Sub-pc BSBHs in Quasars]{Constraining Sub-Parsec Binary Supermassive Black Holes in Quasars with Multi-Epoch Spectroscopy. III. Candidates from Continued Radial Velocity Tests}

% The list of authors, and the short list which is used in the headers.
% If you need two or more lines of authors, add an extra line using \newauthor

\author[H. Guo et al.]{
Hengxiao Guo,$^{1,2}$\thanks{E-mail: hengxiao@illinois.edu (HG), xinliuxl@illinois.edu (XL)}
Xin Liu,$^{1,2}$
Yue Shen,$^{1,2,7}$
Abraham Loeb,$^{3,4}$
TalaWanda Monroe$^{5}$ and
\newauthor
Jason Xavier Prochaska$^{6}$
\\
% List of institutions
$^{1}$Department of Astronomy, University of Illinois at Urbana-Champaign, Urbana, IL 61801, USA\\
$^{2}$National Center for Supercomputing Applications, University of Illinois at Urbana-Champaign, 605 East Springfield Avenue, \\
Champaign, IL 61820, USA\\
$^{3}$Harvard-Smithsonian Center for Astrophysics, 60 Garden Street, Cambridge, MA 02138, USA\\
$^{4}$Institute for Theory and Computation, Harvard University, 60 Garden Street, Cambridge, MA 02138, USA\\
$^{5}$Space Telescope Science Institute, 3700 San Martin Drive, Baltimore, MD 21218, USA\\
$^{6}$University of California Observatories-Lick Observatory, University of California, 1156 High Street, Santa Cruz, CA 95064, USA\\
$^{7}$Alfred P. Sloan Research Fellow\\
}

% These dates will be filled out by the publisher
\date{Accepted 2018 October 23. Received 2018 October 16; in original form 2018 August 31}

% Enter the current year, for the copyright statements etc.
\pubyear{2018}

% Don't change these lines

\label{firstpage}
\pagerange{\pageref{firstpage}--\pageref{lastpage}}
\maketitle

\begin{abstract}
Quasars whose broad emission lines show temporal, bulk radial velocity (RV) shifts have been proposed as candidate sub-parsec (sub-pc), binary supermassive black holes (BSBHs). We identified a sample of 16 BSBH candidates based on two-epoch spectroscopy among 52 quasars with significant RV shifts over a few rest-frame years. The candidates showed consistent velocity shifts independently measured from two broad lines (H$\beta$ and H$\alpha$ or Mg${\rm \,II}$) without significant changes in the broad-line profiles. Here in the third paper of the series, we present further third- and fourth-epoch spectroscopy for 12 of the 16 candidates for continued RV tests, spanning $\sim$5--15 yr in the quasars' rest frames. Cross-correlation analysis of the broad H$\beta$ calibrated against [O${\rm\,III}]\,\lambda 5007$ suggests that 5 of the 12 quasars remain valid as BSBH candidates. They show broad H$\beta$ RV curves that are consistent with binary orbital motion without significant changes in the broad line profiles. Their broad H$\alpha$ (or Mg${\rm \,II}$) lines display RV shifts that are either consistent with or smaller than those seen in broad H$\beta$. The RV shifts can be explained by a $\sim$0.05--0.1 pc BSBH with an orbital period of $\sim$40--130 yr, assuming a mass ratio of 0.5--2 and a circular orbit. However, the parameters are not well constrained given the few epochs that sample only a small portion of the hypothesized binary orbital cycle. The apparent occurrence rate of sub-pc BSBHs is $\lesssim$13$\pm$5\% among all SDSS quasars, with no significant difference in the subsets with and without single-epoch broad line velocity offsets. Dedicated long-term spectroscopic monitoring is still needed to further confirm or reject these BSBH candidates. 
\end{abstract}


%%%%%%%%%%%%%%%%%%%%%%%%%%%%%%%%%%%%%%%%%%%%%%%%%%

%%%%%%%%%%%%%%%%% BODY OF PAPER %%%%%%%%%%%%%%%%%%
%%%%%%%%%%%%%%%%%%%%%%%%%%%%%%
\section{Introduction}\label{sec:intro}

LIGO has detected gravitational waves (GWs) from stellar-mass binary black hole mergers \citep{LIGO2016}. GW sources should exist outside the LIGO frequency \citep[e.g.,][]{eLISAConsortium2013,Colpi2017,Schutz2018}, and this series of papers aims at identifying candidate binary supermassive black holes (BSBHs). A BSBH consists of two black holes, each with a mass of $\sim10^6$--$10^9$ M$_{\odot}$. BSBHs are expected from galaxy mergers \citep{begelman80,ebisuzaki91,Quinlan1996,haehnelt02,volonteri03}, since most massive galaxies harbor supermassive black holes \citep[SMBHs;][]{kormendy95,ff05}. The final coalescences would produce the loudest GW signals \citep{thorne76,haehnelt94,vecchio97,Jaffe2003}. The more massive BSBHs are being constrained with the upper limits from pulsar-timing arrays \citep[e.g.,][]{Arzoumanian2014,Zhu2014,Huerta2015,Sesana2015,Sesana2018,Shannon2015,Arzoumanian2016,Babak2016,Ellis2016,Middleton2016,Middleton2018,Rosado2016,Simon2016,Taylor2016,Kelley2017a,Mingarelli2017,Arzoumanian2018,Holgado2018,Tiburzi2018}, whereas the less massive BSBHs are among the primary science targets for the planned space-based GW observatories such as LISA \citep[e.g.,][]{Sesana2004,Klein2016,Amaro-Seoane2017,Danzmann2017}. They are laboratories to directly test general relativity in the strong field regime and to study the cosmic evolution of galaxies and cosmology \citep[e.g.,][]{Baumgarte2003,holz05,valtonen08,hughes09,Centrella2010,Babak2011,Amaro-Seoane2013,Arun2013,DEGN,Colpi2014,Berti2015}.

The orbital decay of BSBHs may slow down or stall at $\sim$pc scales \citep[e.g.,][]{begelman80,milosavljevic01,zier01,yu02,Vasiliev2013,Dvorkin2017,Tamburello2017}, or the barrier may be overcome in gaseous environments \citep[e.g.,][]{gould00,escala04,hayasaki07,Hayasaki2009,Cuadra2009,Lodato2009,Chapon2013,Rafikov2012,delValle2015}, in triaxial or axisymmetric galaxies \citep[e.g.,][]{yu02,berczik06,Preto2011,Khan2013,Khan2016,Vasiliev2015,Gualandris2017,Kelley2017}, and/or by interacting with a third SMBH in hierarchical mergers \citep[e.g.,][]{valtonen96,blaes02,hoffman07,Kulkarni2012,Tanikawa2014,Bonetti2018}. The accretion of gas and the dynamical evolution of BSBHs are likely to be coupled \citep{Ivanov1999,Armitage2002,Bode2010,Bode2012,haiman09,Farris2010,Farris2011,Farris2014,Farris2015,Kocsis2012a,Shi2012,DOrazio2013,Shapiro2013} such that the occurrence rate of BSBHs depends on the initial conditions and gaseous environments at earlier phases \citep[e.g., thermodynamics of the host galaxy interstellar medium;][]{Dotti2007,dotti09,Dotti2012,Fiacconi2013,Mayer2013,Tremmel2018}.  Quantifying the occurrence rate of BSBHs at various merger phases is therefore important for understanding the associated gas and stellar dynamical processes. This is a challenging problem for three main reasons. First, BSBHs are expected to be rare \citep[e.g.,][]{Foreman2009,Volonteri2009}, and only a fraction of them accrete enough gas to be ``seen''. Second, the physical separations of BSBHs that are gravitationally bound to each other ($\lesssim$ a few pc) are too small for direct imaging. Even VLBI cannot resolve BSBHs except for in the local universe \citep{burke11}. CSO 0402+379 (discovered by VLBI as a double flat-spectrum radio source separated by 7 pc) remains the only secure case known \citep[][see \citealt{Kharb2017}, however, for a possible 0.35-pc BSBH candidate in NGC 7674]{rodriguez06,Bansal2017}. Third, various astrophysical processes complicate their identification such as bright hot spots in radio jets \citep[e.g.,][]{Wrobel2014}.  Until recently, only a handful cases of dual active galactic nuclei (AGNs) -- galactic-scale progenitors of BSBHs -- were known \citep{owen85,junkkarinen01,komossa03,ballo04,hudson06,max07,bianchi08,guidetti08}. While great strides have been made in identifying dual AGNs at kpc scales \citep[e.g.,][]{gerke07,comerford08,comerford11b,Comerford2015,Liu2010a,Liu2013,Liu2018b,green10,fabbiano11,Fu2011a,Fu2012,Fu2015,Fu2015a,Koss2011,koss12,Koss2016,rosario11,teng12,Woo2014,Wrobel2014a,McGurk2015,Muller-Sanchez2015,Shangguan2016,Ellison2017,Satyapal2017}, there is no confirmed BSBH at sub-pc scales \citep[for recent reviews, see e.g., ][]{popovic11,Burke-Spolaor2013,Bogdanovic2015,Komossa2015a}. 

\begin{figure}
  \centering
    \includegraphics[width=80mm]{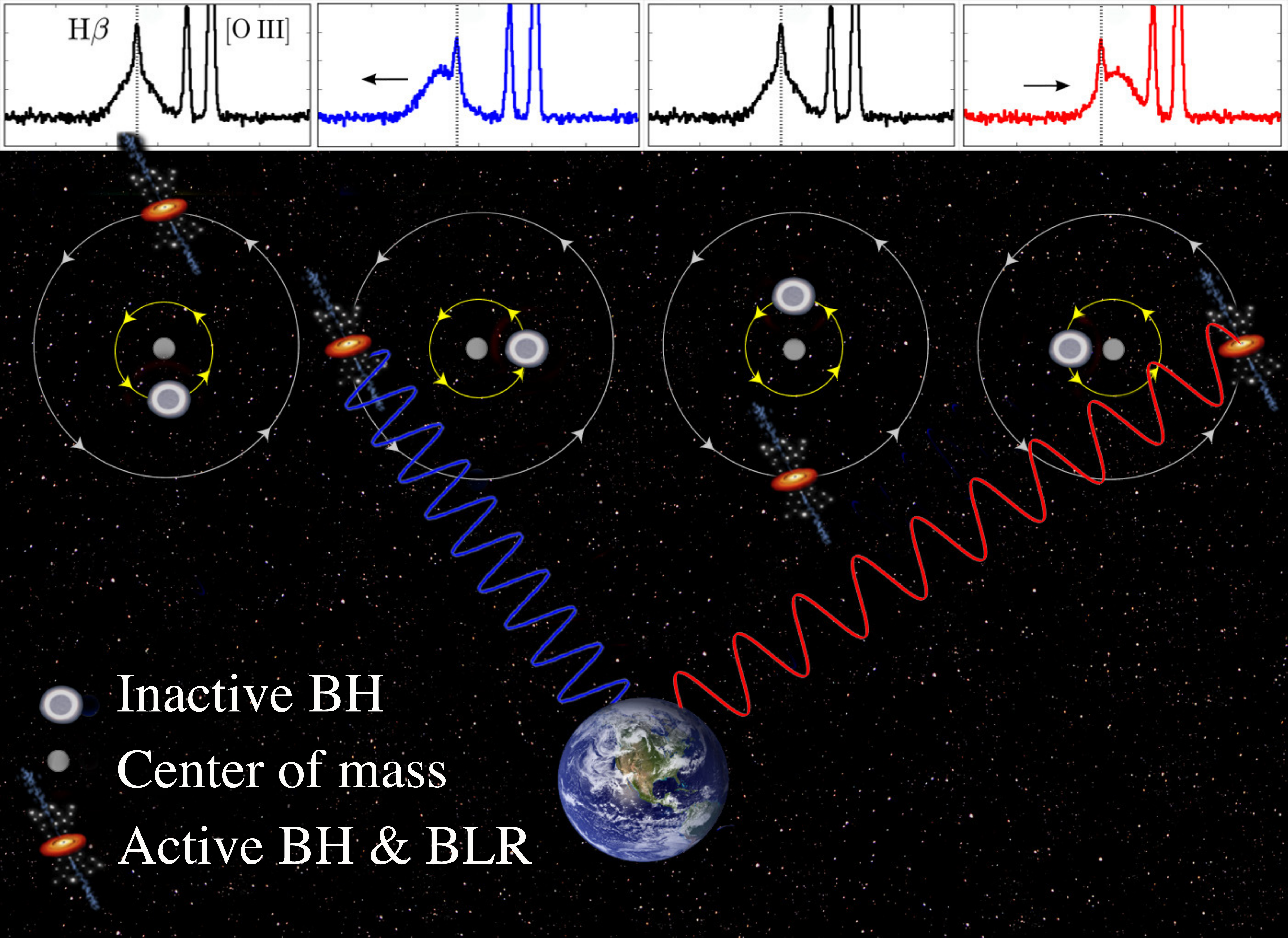}
   \caption{Cartoon illustration of the RV method for identifying sub-pc BSBHs. Here we assume that the smaller BH is active taking its accretion disk and BLR orbiting the common center of mass. The center of mass of the BSBH is assumed to be at rest with the host galaxy, anchored by the narrow emission lines.}
    \label{fig:cartoon}
\end{figure}

Alternatively, BSBH candidates may be identified by measuring the bulk radial velocity (RV) drifts as a function of time in quasar broad emission lines \citep[e.g.,][]{gaskell83,Bogdanovic2008,boroson09,gaskell10,shen10,popovic11,Bon2012,eracleous11,Decarli2013,McKernan2015,Nguyen2016,Simic2016,Pflueger2018}, in analogy to RV searches for exoplanets (Figure \ref{fig:cartoon}). Only one of the two BHs in a BSBH is assumed to be active, powering its own broad-line region (BLR). The binary separation needs to be sufficiently large compared to the BLR size such that the broad-line velocity traces the binary motion, yet small enough that the acceleration is detectable over the time baseline of typical observations \citep[e.g.,][]{eracleous11,Ju2013,shen13,Liu2014}. However, most of previous work has focused on a small population of low-redshift quasars and Seyfert galaxies that show double peaks with extreme velocity offsets or double shoulders \citep[e.g.,][]{gaskell96a,eracleous94,eracleous97,eracleous03,boroson09,lauer09,tsalmantza11,Bon2012,Decarli2013,Li2016}. These extreme, kinematically offset quasars, originally proposed as due to BSBHs where both members are active \citep[e.g.,][]{gaskell83,peterson87,gaskell96a}, are most likely due to rotation and relativistic effects in the accretion disks around single BHs rather than BSBHs \citep[e.g., so-called ``disk emitters'',][]{capriotti79,halpern88,Chen1989,Chen1989a,Laor1991,Popovic1995,eracleous95,eracleous97,eracleous99,strateva03,gezari07,chornock10,lewis10,Liuj2016}. 

Unlike previous work, we focus on the general quasar population \citep[][hereafter Paper I; see also \citealt{Ju2013,Wang2017}]{shen13} and those with single-peaked offset broad emission lines \citep[][hereafter Paper II; see also \citealt{tsalmantza11,eracleous11,Decarli2013,Runnoe2017}]{Liu2014}.  We have studied the temporal broad-line velocity shifts using the largest sample of quasars with multi-epoch spectroscopy (Papers I \& II) based on the SDSS DR7 spectroscopic quasar catalog \citep{schneider10,shen11}. They include data both from repeated SDSS observations for the general quasar population (Paper I) and from combining our follow-up observations for the sample of quasars with kinematically offset broad emission lines (Paper II). The general quasar sample includes $\sim$2000 pairs of observations in total of which $\sim$700 pairs have good measurements (1 $\sigma$ error $\sim$40 km s$^{-1}$) of the velocity shifts between two epochs (Paper I). These pilot studies allow us to: (i) tentatively constrain the abundance of sub-pc BSBHs in the general and offset quasar populations, with caveats on the assumed models for the accretion flow and geometry of the BLR gas \citep{Cuadra2009,Montuori2011}, and (ii) yield 16 BSBH candidates for further tests. The 16 BSBH candidates show significant RV shifts in the broad H$\beta$ lines (corroborated by either broad H$\alpha$ or \MgII ) over a few yrs (rest frame), yet with no significant changes in the emission-line profile (i.e., the shifts represent a change in {\it bulk} velocity rather than variation in the broad-line profiles, which is more likely due to BLR kinematics around single BHs rather than BSBHs). The existing two-epoch spectroscopy represents a first step toward confirming sub-pc BSBHs and in sorting out the origins for the broad-line velocity shifts.

We have been conducting third- and more-epoch spectroscopy to further test the binary hypothesis for the 16 BSBH candidates. As the third paper in this series, our primary goal is to identify strong cases in 12 of the 16 BSBH candidates by continued RV tests. With a constant acceleration under the binary hypothesis, the velocity shifts are expected to be a few hundred km s$^{-1}$ in a few yrs with no significant changes in the broad emission line profile \citep{Runnoe2017,Wang2017}. On the other hand, objects with stochastic accelerations and/or changes in the broad emission line profile will be likely due to alternative scenarios such as structural changes in the BLR on the dynamical time scale, often observed in accretion disk emitters, and/or asymmetric reverberation in the BLRs of single BHs \citep{Barth2015}. 

The rest of the paper is organized as follows.  \S \ref{sec:sample} presents our sample selection and identification of the BSBH candidates. We describe our follow-up spectroscopy, data reduction, and data analysis in \S \ref{sec:obs}. We present our results in \S \ref{sec:result}, discuss their uncertainties and implications in \S \ref{sec:discuss}, and conclude in \S \ref{sec:conclude}.

Throughout this paper, we assume a concordance cosmology with $\Omega_m = 0.3$, $\Omega_{\Lambda} = 0.7$, and $H_{0}=70$ km s$^{-1}$ Mpc$^{-1}$, and use the AB magnitude system \citep{oke74}. Following Papers I \& II, we adopt ``offset'' to refer to the velocity difference between two lines in single-epoch spectra, and ``shift'' to denote changes in the line velocity between two epochs. We quote velocity offset relative to observers, i.e., negative values mean blueshifts. All time intervals are in the quasar rest frames by default, unless noted otherwise.

% table 1

\begin{table*}

	\centering
	
	\caption{New Follow-up Spectroscopy of SDSS Quasars Hosting Candidate Sub-pc BSBHs.}
	\label{table:obs}
	\begin{adjustbox}{width=\textwidth}
	\begin{tabular}{lccccccccc} % 
		\hline
		\hline
		       &                                &                           & $r$     & $V_{{\rm off}}$ &          &          & $t_{{\rm exp}}$ & S/N                 & \\
No.~~~~~~~~~~~~~~~~~~~~ & SDSS Designation  & z$_{{\rm sys}}$ & (mag) & (km s$^{-1}$)   & Spec & MJD & (s)                      & (pixel$^{-1}$) & Ref. \\
                  (1) &                           (2) &                     (3) & (4)      & (5)                    & (6)     & (7)    & (8)                      & (9)                  & (10) \\ 
		\hline
01\dotfill & SDSS J032213.89$+$005513.4  & 0.1854 & 16.70 & ~187$\pm$30 &  B\&C & 57252 & 1800 & 41 & [1] \\
02\dotfill & SDSS J082930.60$+$272822.7  & 0.3211 &18.10 & 1487$\pm$61\, &  GMOS-N & 57463 & 3180 & 48 & [2]   \\
03\dotfill & SDSS J084716.04$+$373218.1  & 0.4534 &18.45 &  ~433$\pm$44 &  BOSS & 57452 & 3600 & 53  & [2]  \\
     &                                         &             &           &                         &  GMOS-N & 57463 & 3580 & 95  &  \\
04\dotfill & SDSS J085237.02$+$200411.0  & 0.4615 &18.10  &   ~700$\pm$67 &  BOSS  & 55955 & 13512 & 36  & [2] \\
     &                                         &             &           &                          &  GMOS-N  &  57461 & 4865 & 44  & \\
05\dotfill & SDSS J092837.98$+$602521.0  & 0.2959 &17.01 &  $-$759$\pm$149 &  GMOS-N & 57461 & 1364 & 102 & [2]    \\
06\dotfill & SDSS J103059.09$+$310255.8  & 0.1781 &16.77 & ~~\,642$\pm$119 &  GMOS-N & 57464 & 964 & 86  & [2]    \\
07\dotfill & SDSS J110051.02$+$170934.3 & 0.3476 &18.48 &  1502$\pm$33\, &  GMOS-N  & 57464 & 3288 & 45  & [2]    \\
08\dotfill & SDSS J111230.90$+$181311.4  & 0.1952 &18.13 & ~1016$\pm$270 &  GMOS-N & 57464 & 2488 & 70 & [2]   \\
09\dotfill & SDSS J141020.57$+$364322.7  & 0.4495 &18.20 & $-$292$\pm$330 &  GMOS-N & 57437 & 3292 & 83 & [1] \\ 
10\dotfill & SDSS J153705.95$+$005522.8 & 0.1365 &17.10  & $-$110$\pm$60~ &  B\&C         & 57252 & 1800 & 72 & [1] \\
     &                                        &             &           &                              &   GMOS-N & 57437 & 964 &  29 &      \\
11\dotfill & SDSS J155053.16$+$052112.1 & 0.1104 &16.30  & ~~\,487$\pm$150 &  B\&C        & 57252 & 1800    & 75 & [1] \\ 
     &                                        &             &           &                              &  GMOS-N  & 57437 & 564 & 35 &       \\ 
12\dotfill & SDSS J234932.77$-$003645.8  & 0.2798 &17.20 & $-$172$\pm$30~ &  BOSS & 56932 & 4500 & 32   & [1] \\
     &                                        &             &          &                              &  B\&C  & 57251 & 1800 & 29 &       \\
		\hline
\multicolumn{10}{l}{Column 2: SDSS names with J2000 coordinates given in the form of ``hhmmss.ss+ddmmss.s''}\\
\multicolumn{10}{l}{Column 3: systemic redshift from Paper I\&II}\\
\multicolumn{10}{l}{Column 4: SDSS $r$-band PSF magnitude}\\
\multicolumn{10}{l}{Column 5: broad \hb\ centroid (peak) velocity offset and 1$\sigma$ uncertainty of the first-epoch spectrum reported in Paper II (Paper I) for offset-line (general) quasars}\\
\multicolumn{10}{l}{Column 6: spectrograph used for the follow-up observations}\\
\multicolumn{10}{l}{Column 7: MJD of the follow-up observations}\\
\multicolumn{10}{l}{Column 8: total exposure time of the follow-up observations}\\
\multicolumn{10}{l}{Column 9: median S/N pixel$^{-1}$ around the broad \hb\ region of the follow-up spectra}\\
\multicolumn{10}{l}{Column 10: Original reference that identified the quasar as a sub-pc BSBH candidate. [1]: \citet[][Paper I]{shen13}, [2]: \citet[][Paper II]{Liu2014}}\\
	\end{tabular}
\end{adjustbox}
\end{table*}

%%%%%%%%%%%%%%%%%%%%%%%%%%%%%%%%%%%%%%%%%%
\section{Sample Selection and BSBH Candidate Target Identification}\label{sec:sample}
Our parent sample includes 16 sub-pc BSBH candidates identified from Papers I \& II. It consists of 7 objects selected from the general quasar population (Paper I; \S \ref{subsec:general}) and 9 objects selected from a sample of quasars with kinematically offset broad Balmer emission lines (Paper II; \S \ref{subsec:offset}). Below we provide a summary of the sample selection and target identification. We refer the readers to Papers I \& II for further details. 

\subsection{Candidates from the General Quasar Population}\label{subsec:general}

Paper I presented a systematic search for sub-pc BSBHs in the general broad-line quasar population at $z<0.8$ based on multi-epoch spectroscopy in the SDSS DR7 \citep{SDSSDR7}. The SDSS DR7 quasar catalog consists of 105,783 objects selected to be brighter than $M_{\rm i}$ = $-$22.0 that have at least one broad emission line with the full width at half-maximum (FWHM) larger than 1000 \kms\ or have interesting/complex absorption features \citep{schneider10}. The spectral wavelength coverage is 3800 -- 9200\AA\ with a spectral resolution R $\sim$ 1850 -- 2200. The spectra are stored in vacuum wavelength with a pixel scale of $10^{-4}$ in log-wavelength, corresponding to 69 \kms. All spectra are wavelength calibrated to the heliocentric reference, with an accuracy of better than 5 \kms. \citet{shen11} presented physical properties of the SDSS DR7 quasars including the continuum and emission line measurements, virial black hole mass estimates, and RV offsets of the broad emission lines (such as broad \ha , broad \hb , and broad \MgII ) relative to the systemic redshift from the narrow \OIII\ lines.

Several thousand of the DR7 quasars have multiple spectra taken at different epochs by the SDSS. Among them $\sim$193 pairs of spectra have good enough measurements (with 1$\sigma$ error of $\sim$50 km s$^{-1}$ yr$^{-1}$; the ``superior'' sample of Paper I) of the RV shifts between two epochs separated by up to several years. Out of the $\sim$193 pairs Paper I found 28 objects with significant (99\% confidence) RV shifts in broad \hb .  7 of the 28 have been identified as the best candidates for hosting BSBHs. These candidates show significant RV shifts in the broad H$\beta$ lines in their two-epoch spectra separated over a few yrs, yet with no significant changes in the emission-line profile. Their broad H$\alpha$ or \MgII\ also show velocity shifts consistent with broad \hb . One exception is the case of SDSS J1550+0521, where the velocity shift for \hb\ is larger than that for \ha , which may be explained if the \hb\ BLR is mostly confined to the active BH, while the \ha\ BLR also contains a circumbinary component (which does not accelerate).  \S \ref{sec:obs} presents new third- and fouth-epoch spectroscopy to further test the binary hypothesis for 5 out of the 7 candidates from the general quasar population.

\subsection{Candidates from Quasars with Kinematically Offset Broad Balmer Emission lines}\label{subsec:offset}

Paper II selected a sample of 399 quasars from the SDSS DR7 whose broad \hb\ lines are significantly (99.7\% confidence) offset from the systemic redshift determined from narrow emission lines. The velocity offset has been suggested as evidence for BSBHs, but single-epoch spectra cannot rule out alternative scenarios such as accretion disk emitters around single BHs or recoil BHs (\S \ref{sec:intro}). To test the binary hypothesis, Paper II obtained second-epoch spectroscopy for 50 of the 399 offset-line quasars separated by 5--10 yr from the original SDSS observations. 24 of the 50 show significant (99\% confidence) RV shifts in broad \hb\ with a typical measurement uncertainty of $\sim$10 km s$^{-1}$ yr$^{-1}$. Following the criteria similar as in Paper I, 9 of the 24 with significant RV shifts have been suggested as sub-pc BSBH candidates. The RV shifts for BSBH candidates have been required to be caused by an overall shift in the bulk velocity rather than variation in the broad-line profiles. The RV shifts independently measured from a second broad line (either broad \ha\ or \MgII ) have been required to be consistent with those measured from broad \hb . \S \ref{sec:obs} presents new third- and fouth-epoch spectroscopy to further test the binary hypothesis for 7 out of the 9 candidates from the sample of offset-line quasars.

\section{Observations, Data Reduction, and Data Analysis}\label{sec:obs}

\subsection{Continued Follow-up Spectroscopy}\label{subsec:obs}

\begin{figure*}
  \centering
    \includegraphics[width=180mm]{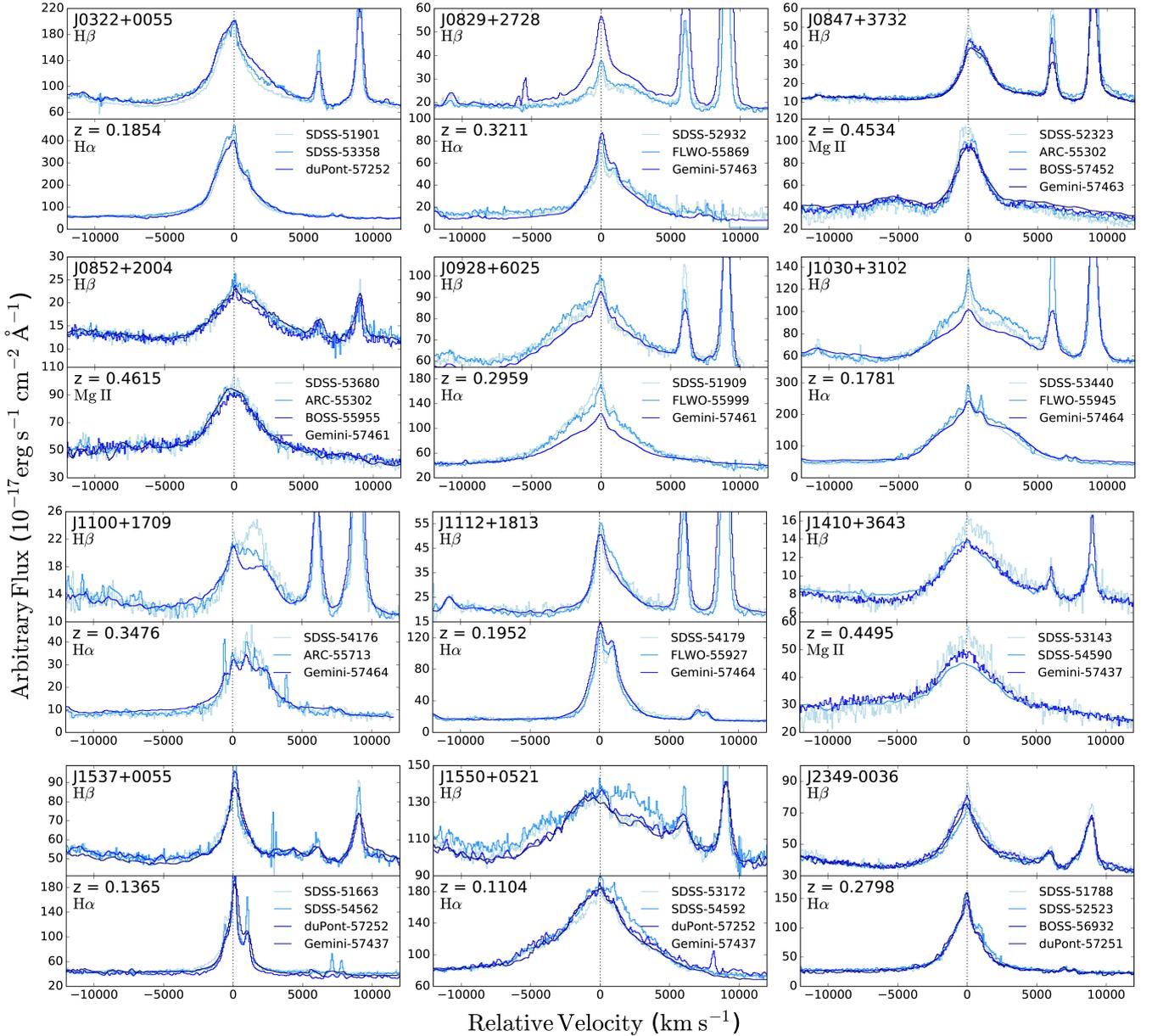}
    \caption{Multi-epoch spectra of the 12 SDSS quasar targets selected as sub-pc BSBH candidates. The spectra have been normalized for display purposes. For each quasar, the top (bottom) panel shows the \hb\ (\ha\ or \MgII ) region centered on the systemic redshift. The previous two-epoch spectra presented in Papers I \& II are shown in lighter shades, whereas the new third- and fouth-epoch spectra, when available, are shown in darker shades. The spectrograph and MJD of the observations are labeled on each panel, along with the systemic redshift and abbreviated name for each quasar.}
    \label{fig:allspec}
\end{figure*}

\subsubsection{Gemini/GMOS-N}

We observed 10 BSBH candidate targets with the Gemini Multi Object Spectrographs (GMOS) on the 8.1\,m Gemini-North Telescope on the summit of Mauna Kea.  Observations were carried out in queue mode over 5 nights on 2016 February 19, and March 13, 14, 16, and 17 UT (Program ID GN-2016A-Q-83; PI Liu). The sky was non-photometric with varied seeing conditions (PSF FWHM $\sim$0\farcs5--1\farcs1). We adopted the GMOS-N longslit with the R150 grating and a 0\farcs5 slit width, which offers a spectral resolution of $R \simeq 630$ ($\sigma_{{\rm inst}}\sim$140 km s$^{-1}$) spanning the wavelength range 400--950 nm with a pixel scale of 1.93 {\AA} pixel$^{-1}$. The slit was oriented at the parallactic angle at the time of observation. Total exposure time ranged from 564s to 13512s for each target, which was divided into four individual exposures dithered at two slightly different central wavelengths to cover CCD gaps and to help reject cosmic rays. Table \ref{table:obs} lists details of the observations for each target.

\subsubsection{du Pont 2.5 m/B\&C}

We observed 4 BSBH candidate targets using the Boller \& Chivens (B\&C) spectrograph on the 2.5\,m Ir$\rm \acute{e}$n$\rm \acute{e}$e du Pont Telescope at the Las Campanas Observatory on the nights of 2015 August 17 and 18. 2 of the 4 targets were also observed by GMOS at similar times to calibrate systematics due to instrumental and observational effects as well as short-term RV variation such as caused by reverberation effects \citep{Barth2015}. The sky was non-photometric with seeing $\sim$1\arcsec . We employed the 300 lines mm$^{-1}$ grating with a 271$''\times$1\farcs 5 slit oriented at the parallactic angle at the time of observation. The spectral coverage was $\sim$6230 {\AA} centered at 6550 {\AA}, with a spectral resolution of $R \simeq 1100$ ($\sigma_{{\rm inst}}\sim$89 \kms) and a pixel scale of 3.0 \angstrom\ pixel$^{-1}$. Total integral exposure time for each object was 1800s (Table \ref{table:obs}). 

\subsubsection{SDSS DR14/BOSS}

3 of the original 16 BSBH candidate targets had later-epoch spectra from the SDSS DR14 \citep{Abolfathi2017}. DR14 is the fourth generation of the SDSS and the first public release of data from the extended Baryon Oscillation Sky Survey \citep{Dawson2016}. It is cumulative, including the most recent reductions and calibrations of all data taken by the SDSS since the first phase began operations in 2000. The cut-off date for DR14 was 2016 July 10 (MJD = 57580).  The 3 targets were observed as part of the Time Domain Spectroscopic Survey \citep{Morganson2015,MacLeod2017}. The BOSS spectra cover the wavelength range of 3650--10400 {\rm \AA} with a spectral resolution of $R\sim$1850--2200 \citep{Dawson2013}, similar to that of the original SDSS spectra which cover the wavelength range of 3800--9200  {\rm \AA} \citep{York2000}.

\begin{figure*}
  \centering
    \includegraphics[width=58mm]{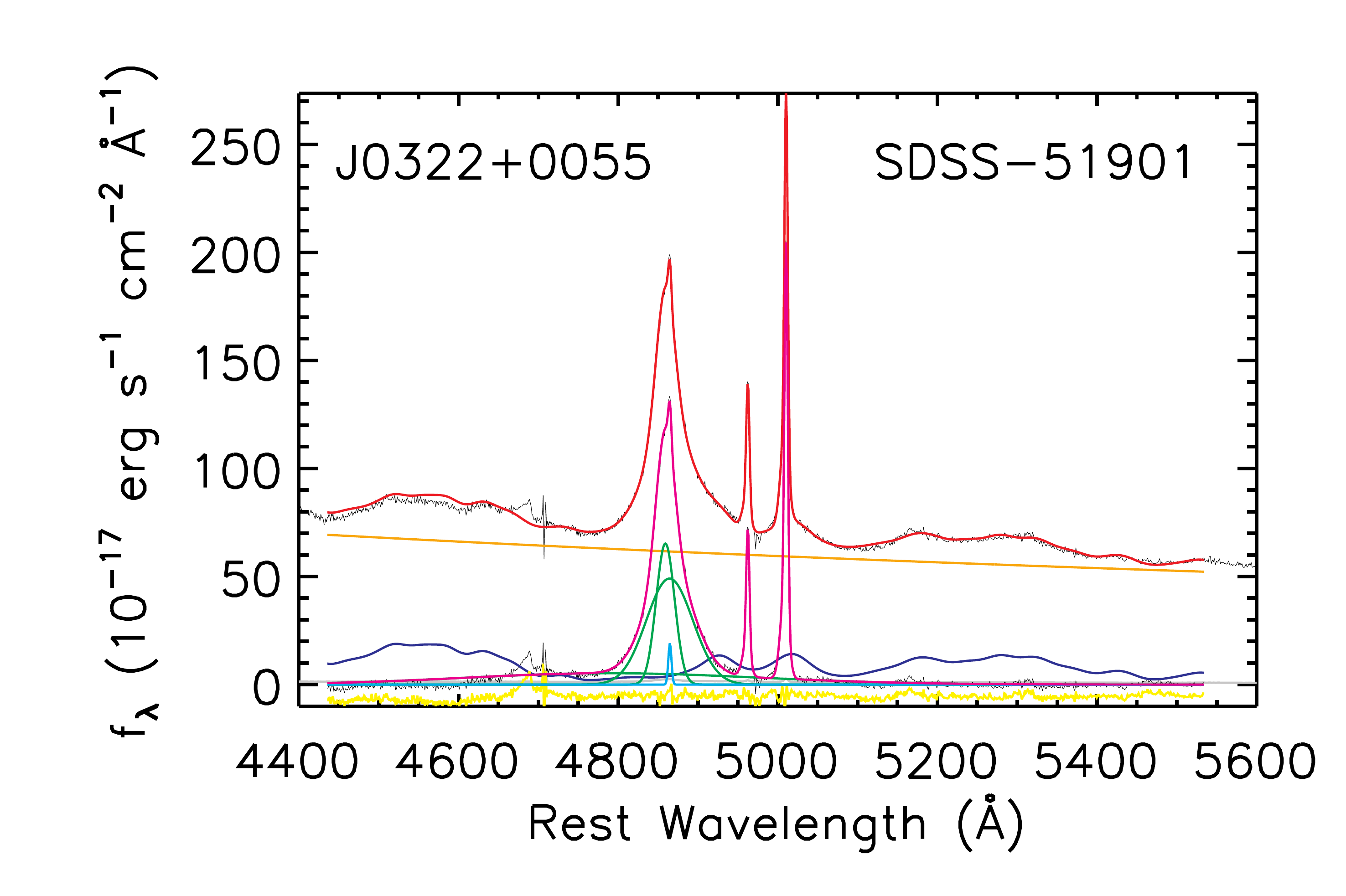}
    \includegraphics[width=58mm]{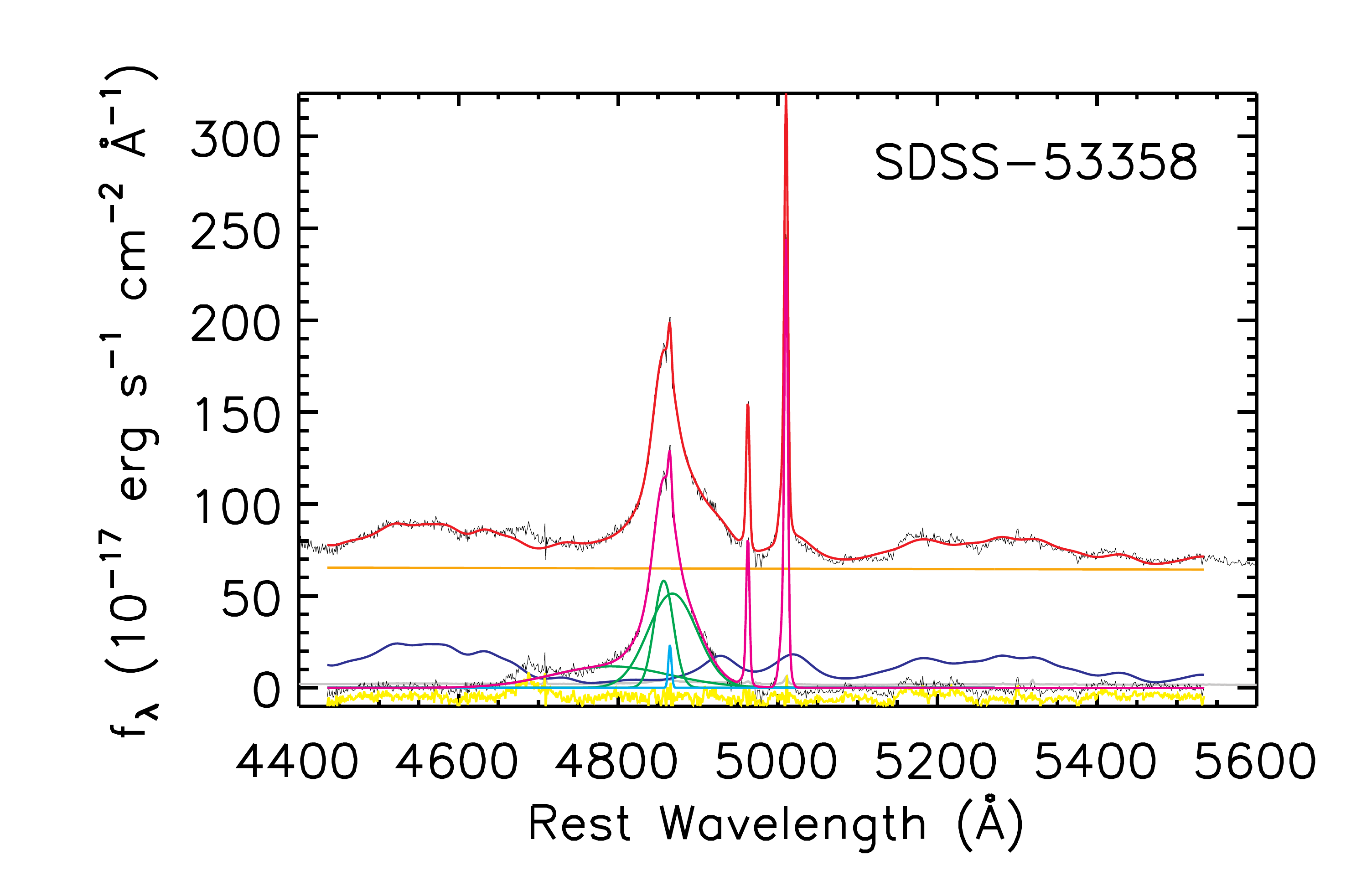}
    \includegraphics[width=58mm]{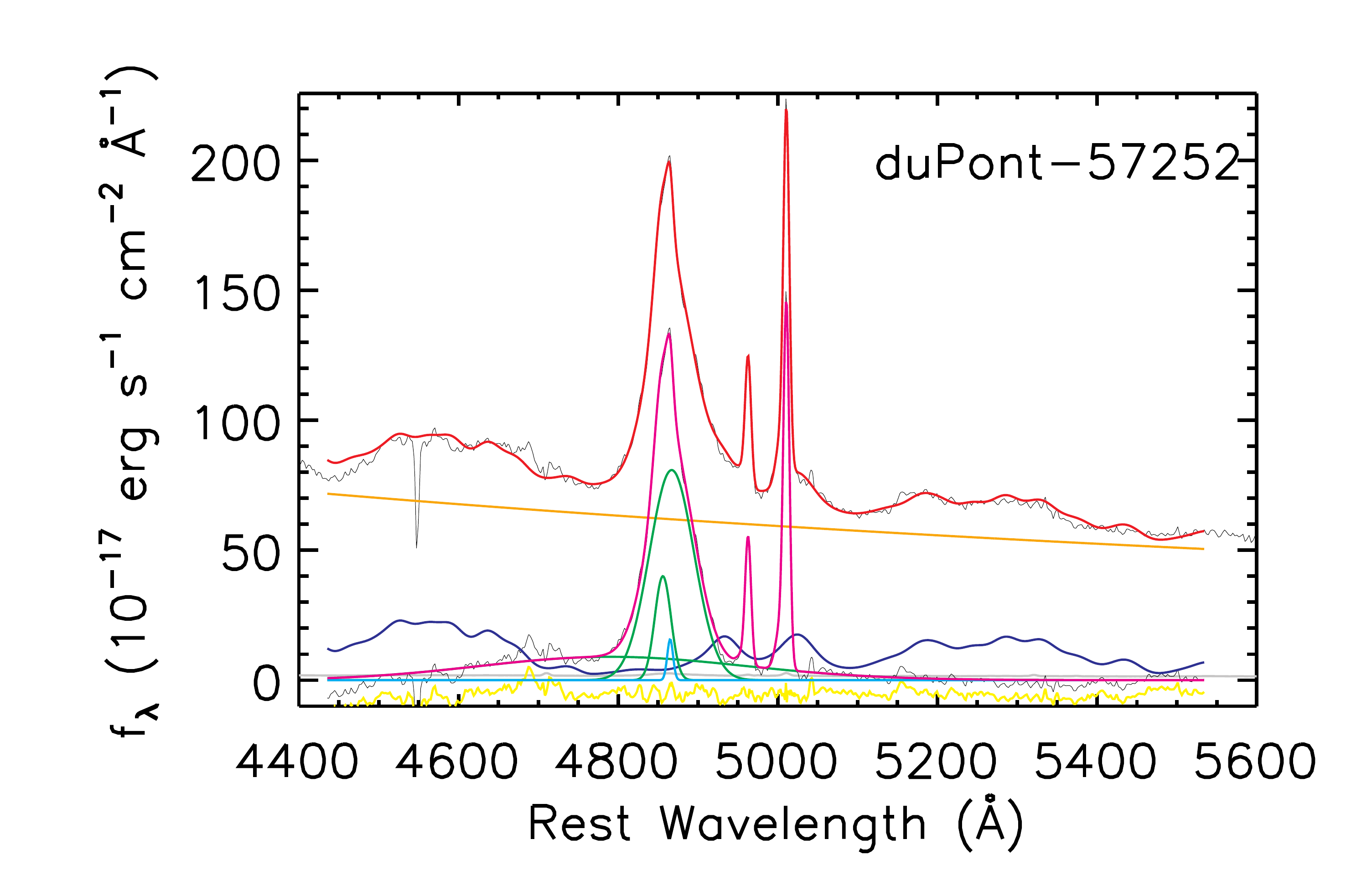}
    \includegraphics[width=58mm]{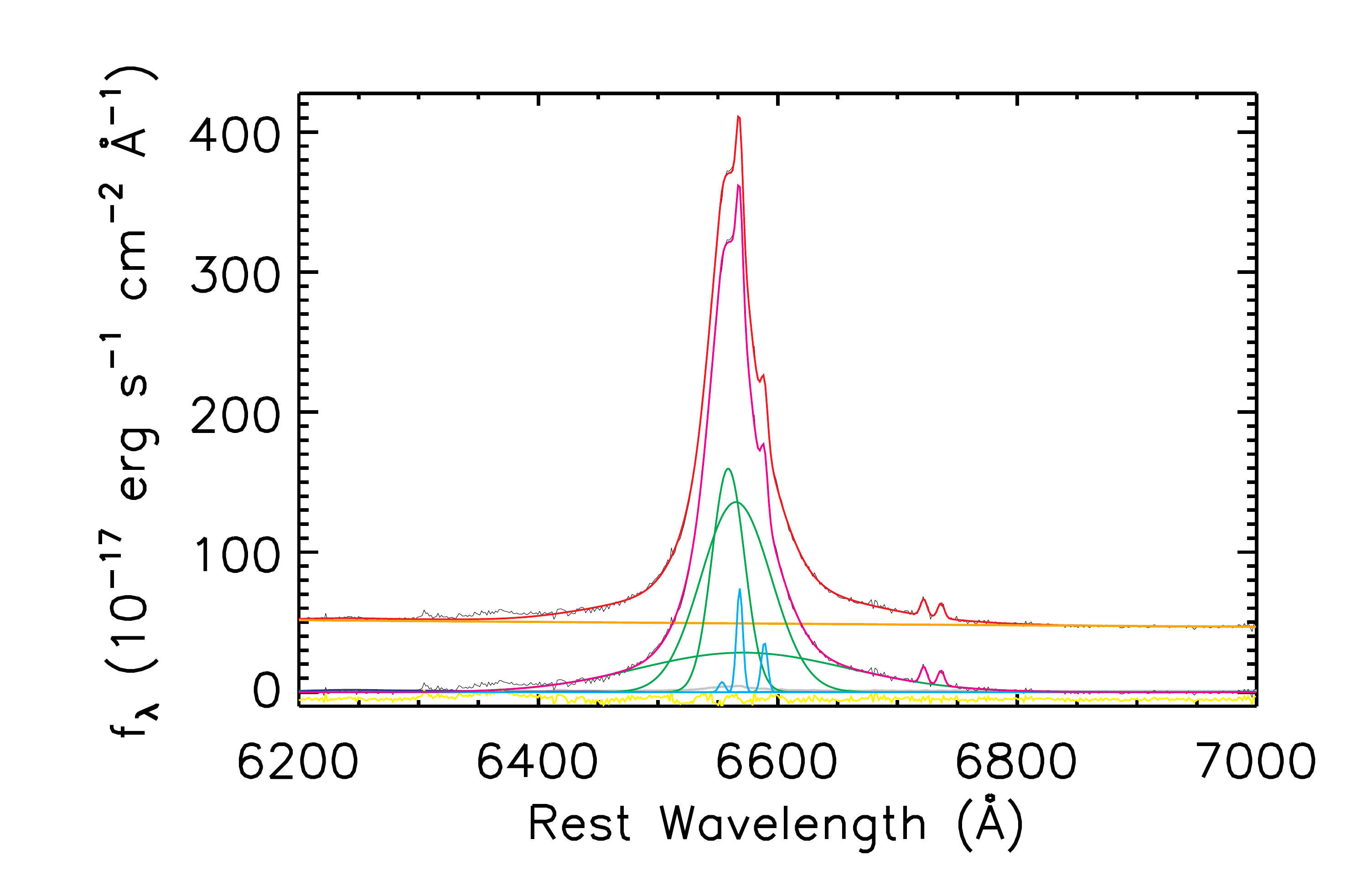}
    \includegraphics[width=58mm]{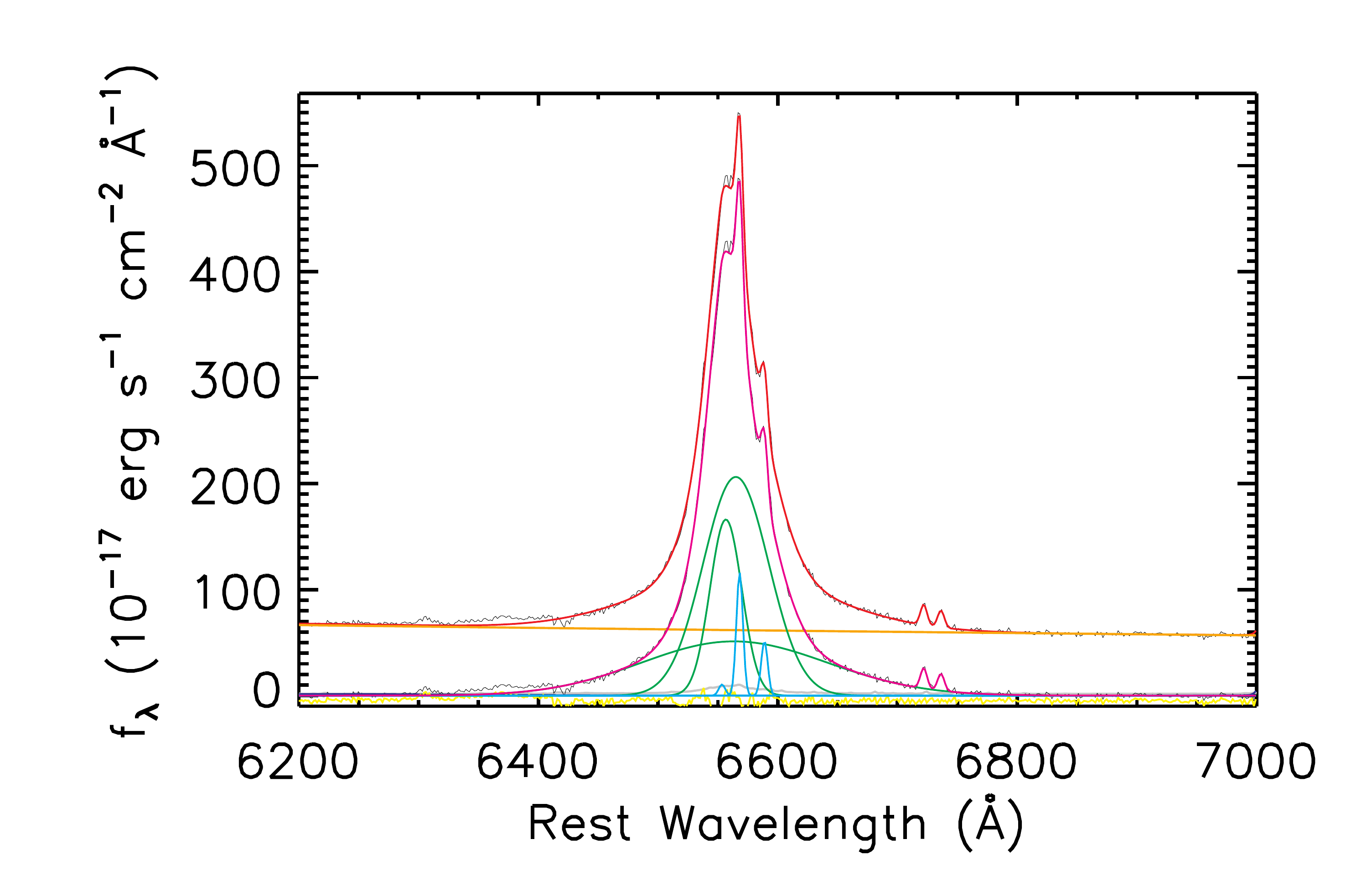}
    \includegraphics[width=58mm]{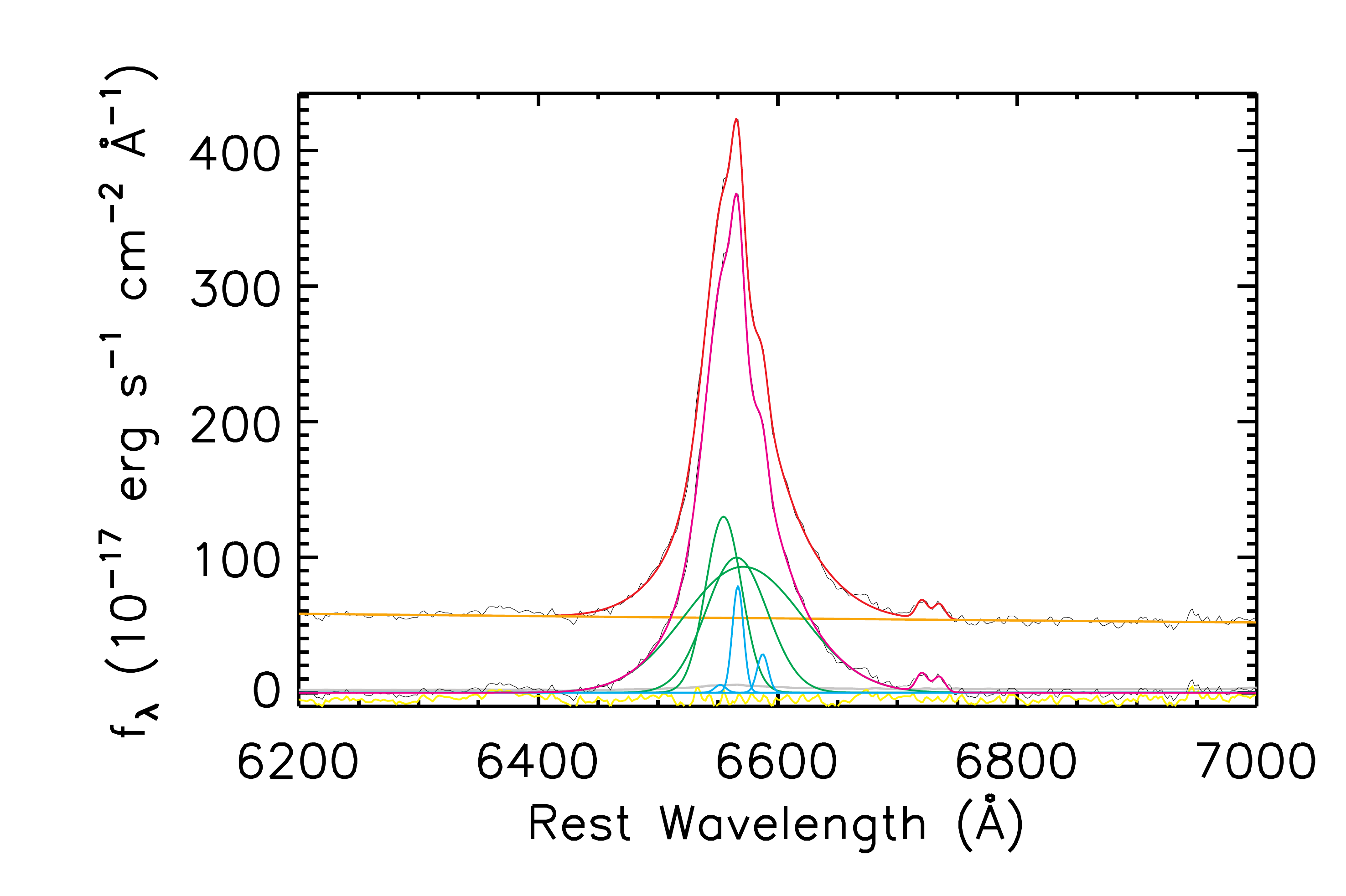}
       \caption{An example of our spectral decomposition modeling. Upper panels show the \hb\ fits whereas lower panels show the \ha\ fits. Three columns represent three different epochs including the first two epochs reported in Papers I \& II and the new third-epoch spectrum presented in this work. In each panel, the upper is the original spectrum whereas the lower is the pseudo-continuum subtracted spectrum, both overplotted with our best-fit models in magenta. Orange denotes our model for the power-law continuum, blue is the \FeII\ template, and cyan and green represents the narrow and broad \hb\ (or \ha ) components, respectively. The \NII$\lambda\lambda$6548,6584 lines are also shown in cyan. Yellow denotes the fitting residual (offset vertically by $-5\times 10^{-17}$ erg s$^{-1}$ cm$^{-2}$ {\rm $\AA^{-1}$}). The \HeIIa\ and \OIa\ lines have been masked out from the fitting. Labeled are the spectrograph and the MJD of each epoch. See \S \ref{subsubsec:specfit} for details.}
    \label{fig:specfit_eg}
\end{figure*}

\subsection{Data Reduction}\label{subsec:reduction}

We reduced our new Gemini\footnote{http://www.gemini.edu/sciops/instruments/gmos/data-format-and-reduction} and du Pont 2.5 m\footnote{http://www.lco.cl/Members/hrojas/website/boller-chivens-spectrograph-manuals/the-boller-and-chivens-spectrograph/?searchterm=6250} follow-up spectra following standard IRAF procedures \citep{tody86}, with particular attention to accurate wavelength calibration. A low-order polynomial wavelength solution was fitted using $\sim$30--90 CuAr (HeNeAr) lamp lines with rms less than 20\% (10\%) for the Gemini (Du Pont 2.5 m) data. One-dimensional spectrum was extracted from each individual frame before flux calibration and telluric correction were applied. The calibrated wavelength arrays were converted from air to vacuum following the SDSS convention and were corrected for heliocentric velocity ($\lesssim$ 30 km s$^{-1}$) following \citet{Piskunov2002}. Finally, we combined all the frames to get a co-added spectrum for each epoch. Table \ref{table:obs} lists the S/N achieved for each follow-up spectroscopic epoch. 

In preparation for cross-correlation analysis, we have re-sampled the Gemini and Du Pont 2.5 m spectra to the same wavelength grids as the SDSS and BOSS spectra, which are linear on a logarithmical scale (i.e., homogeneous in velocity space) with a pixel scale of 10$^{-4}$ in log-wavelength, corresponding to 69 km s$^{-1}$ pixel$^{-1}$. We further correct for any residual absolute wavelength calibration errors when calculating the broad-line RV shifts by setting the zero point according to cross-correlation analysis of the narrow \OIIIb\ emission line (see below \ref{subsubsec:ccf} for details).  Figure \ref{fig:allspec} shows all the new follow-up spectra compared against the previous two-epoch observations before the \OIIIb\ absolute wavelength zero-point correction.

\subsection{Data Analysis}\label{subsec:analysis}

\subsubsection{Spectral Fitting and Decomposition}\label{subsubsec:specfit}

We perform spectral decomposition to separate broad emission lines (\hb , \ha , or \MgII ) from continuum and narrow emission lines using the publicly available code \textsf{PyQSOFit} \citep{Shen2018b,Guo2018}. This is done by a $\chi^2$-based method of fitting spectral models and templates to data \citep[see also][]{shen08,shen11,Guo2014}. Figure \ref{fig:specfit_eg} shows an example of our spectral decomposition modeling of all the three epochs of the quasar SDSS J0322+0055. We provide the spectral fitting results from all epochs for all the other targets in Appendix \ref{appendix:spec}. Below we briefly describe the analysis procedure. 

First, we fit a power-law continuum plus a \FeII\ template \citep{Boroson1992,Vestergaard2001} for the pseudo continuum to a few line-free windows around the broad emission lines (over 4435--4630 {\rm \AA} and 5100--5535 {\rm \AA} for \hb , 6000--6250 {\rm \AA} and  6800--7000 {\rm \AA} for \ha , and 2200--2700 {\rm \AA} and 2900--3090 {\rm \AA} for \MgII ). Second, the pseudo continuum model was subtracted from the data to get the emission-line only spectrum. Third, we fit the continuum-subtracted spectrum using a model with multiple Gaussians for the emission lines. Finally, we subtracted the narrow (broad) lines to get the broad-line-only (narrow-line-only) spectrum for the cross-correlation analysis. For the broad-line component, the multiple Gaussians were only used to reproduce the line profile and bared no physical meaning for individual components. 

More specifically, we modeled the \hb\ emission with one Gaussian for the narrow line component (defined as having a FWHM $<$1200 \kms) and up to three Gaussians for the broad line component (defined as having a FWHM $\geq$ 1200 \kms ). Since blueshifted wings may be present in the narrow \OIIIc\ \citep[e.g.,][possibly from galactic-scale outflows in the narrow line regions]{heckman81,Komossa2008d}, we adopted up to two Gaussians for the \OIIIb\ line (and \OIIIa ) to account for the core and the wing components; the narrow \hb\ component velocity and width were tied to the core \OIII\ component in these cases. We fit the \hb --\OIIIc\ complex over the wavelength range of 4750--5100 {\rm \AA}, except in two for which the range was enlarged to 4700--5100 {\rm \AA} to accommodate the broader \hb\ lines. We have tied the \OIIIb /narrow \hb\ intensity ratio to be the same at all epochs for each quasar\footnote{We do this iteratively by: (i) fitting all epochs independently for which the \OIIIb /narrow \hb\ intensity ratio is allowed to vary and (ii), re-fitting all the spectra with the \OIIIb /narrow \hb\ intensity ratio fixed to be the mean value from all epochs in the previous fits.}. This helped break the model degeneracy between the narrow and broad \hb\ components and was necessary for mitigating bias in the broad \hb\ RV shift between different epochs due to residual narrow \hb\ emission. 

For the \MgIIa\ line (covered by the spectra for 3 of the 12 targets at $z>0.4$), we fit the wavelength range 2700--2900 {\rm \AA}. We model the \MgIIa\ line using a combination of up to two Gaussians for the broad component and one Gaussian for the narrow component. 

For the \ha --\NII --\SII\ complex (covered by the spectra of 9 of the 12 targets at $z<0.4$), we fit the wavelength range 6400--6800  {\rm \AA}. We adopt up to three Gaussians for the broad \ha\ and one Gaussian for the narrow \ha . We adopt four additional Gaussians for the \NII$\lambda\lambda$6548,6584 and \SII$\lambda\lambda$6717,6731 lines. We have also tied the \NII$\lambda$6548/narrow \ha\ and the \NII$\lambda$6584/narrow \ha\ intensity ratios to be consistent among all epochs for each quasar to help break model degeneracy in decomposing narrow- and broad-line components.

 \begin{figure}
  \centering
    \hspace*{-0.2in}
    \includegraphics[width=90mm]{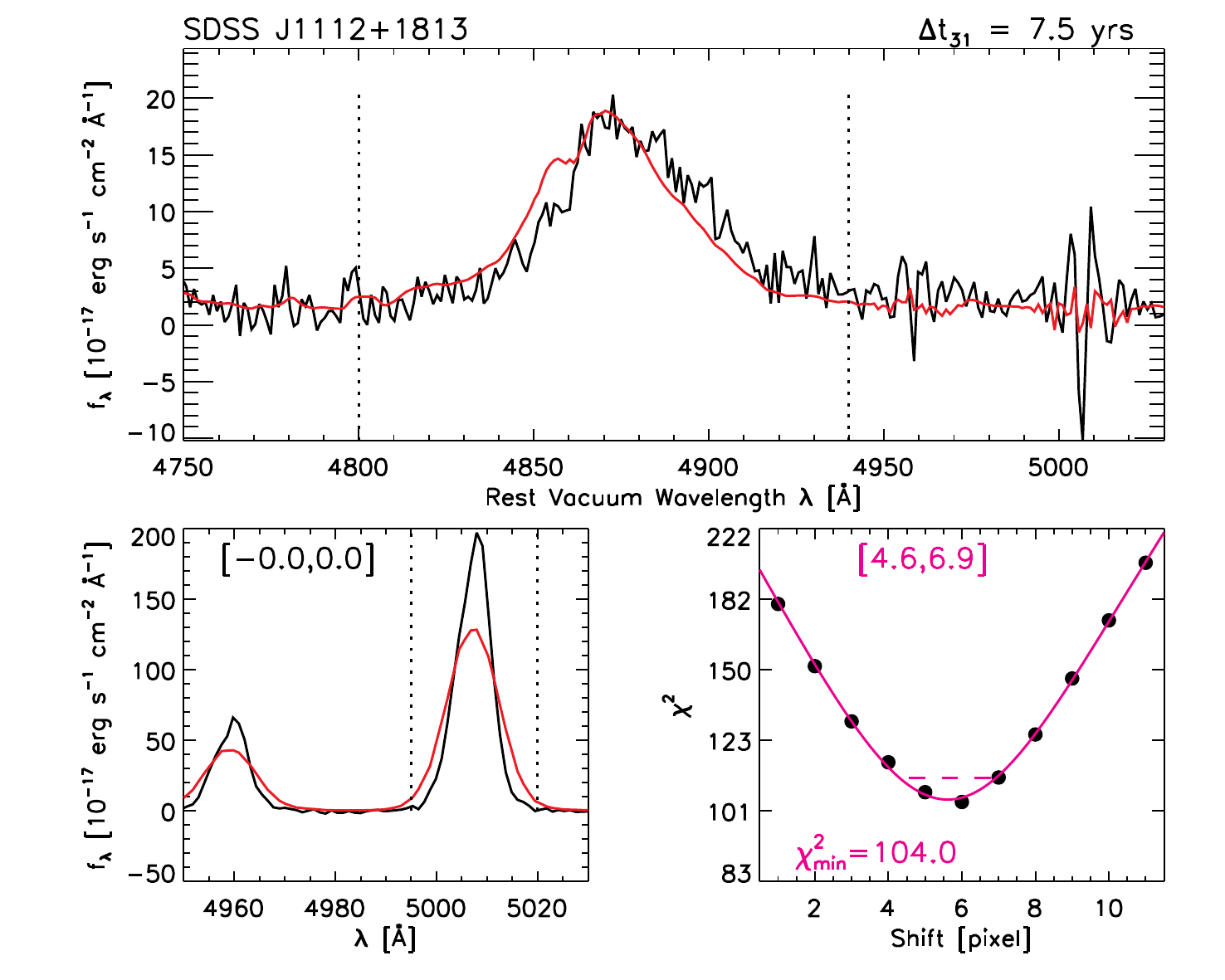}
       \caption{An example of measuring the emission-line RV shift between two epochs of quasar spectra using cross-correlation analysis. Top panel: broad \hb\ spectrum of the first- (black) and third-epoch (red) spectra separated by 7.5 yr for quasar SDSS 1112+1813. The spectra have been normalized by the integrated emission-line flux within the cross-correlation analysis range shown by the dotted lines. Bottom left: same as in the top panel, but for the narrow \OIIIc\ emission lines. Shown in brackets are the 99\% confidence intervals (2.5$\sigma$) in units of pixels of the RV shift for \OIIIb . 1 pixel corresponds to 69 km s$^{-1}$. Here and throughout, positive values mean that the later-epoch spectrum needs to be redshifted to match the emission lines RVs in the first-epoch spectrum (i.e., the emission line in the later-epoch spectrum is blueshifted w.r.t. that in the first-epoch spectrum). Bottom right:  $\chi^2$ curve for the cross-correlation analysis of the broad \hb\ as a function of RV shift in pixels. The solid magenta curve is the sixth-order B-spline fit of the shift grid data points (filled circles). The dashed horizontal bar represents the $\Delta\chi^2=6.63$ (2.5$\sigma$) range, which is shown in the magenta brackets in units of pixels. 
        }
    \label{fig:ccf_eg}
\end{figure}

\subsubsection{Measuring Emission-Line Radial-Velocity Shift with Cross-Correlation Analysis}\label{subsubsec:ccf}

Following Papers I \& II \citep[see also][]{eracleous11,Runnoe2017}, we adopt a $\chi^2$-based cross-correlation analysis (``ccf'' for short) to measure the emission-line RV shift that is expected from the orbital acceleration of a sub-pc BSBH. We focus on the broad-line only spectrum (i.e., \hb , \ha , or \MgII ) because possible changes in the underlying pseudo-continuum (e.g., due to intrinsic quasar variability), if not subtracted properly, could potentially bias the ccf result. The ccf searches for the best-fit RV shift between two epochs by minimizing the $\chi^2$ as a function of the shift:  
\begin{equation}\label{eqn:chi2}
\chi^2 = \sum_i\frac{(f_{1,i}-f_{2,i}^\prime)^{2}}{\sigma_{1,i}^2 + \sigma_{2,i}^{\prime,2}}\ ,
\end{equation}
where $f_{1,i}$ and $f_{2,i}^\prime$ are the flux densities of the $i$th pixel in the Epoch 1 and the shifted Epoch 2 spectra, with $\sigma_{1,i}$ and $\sigma_{2,i}^{\prime}$ being the 1$\sigma$ errors in the flux densities. For multiple epochs, we performed the ccf for all the later epochs against the first epoch spectrum taken by the original SDSS.

For the broad \hb\ (\ha\ or \MgII ) line, the ccf was performed in the wavelength range of 4800--4940 {\rm \AA} (6450--6650 {\rm \AA} for \ha\ or 2750--2850 {\rm \AA} for \MgII ) encompassing most of the broad-line component while excluding extended, noisy wings.  We shifted the later-epoch spectrum by $-30$ to 30 pixels (recall that 1 pixel being 69 km s$^{-1}$) and calculated the $\chi^2$ as a function of the shift. We then fit the $\chi^2$ data points enclosing the minimum value with a sixth-order B-spline function. The minimum $\chi^2$ and the corresponding shift were determined from the model fit, allowing for estimation of sub-pixel shifts. We also quantified the uncertainty of the shift from the best-fit $\chi^2$ model using the intercepts of the B-spline at $\Delta \chi^2 = \chi^2_{{\rm min}}$+6.63, corresponding to 99\% confidence \citep[$\sim$2.5$\sigma$; e.g.,][]{Lampton1976,eracleous11}.

Figure \ref{fig:ccf_eg} shows an example of our ccf where a significant ($>99$\% confidence) RV shift is detected between the third- and first-epoch spectra in the broad \hb\ line without any significant changes in the broad-line profile. We have scaled the later-epoch spectrum by the ratio of the integrated emission line flux of the two epochs over the ccf wavelength range. This was to account for absolute flux variation possibly due to intrinsic quasar variability and/or observational issues (e.g., variable weather conditions and/or difference in slit/fiber coverages).  

To further calibrate the absolute RV zero point, we have also performed the ccf for the \OIIIb\ line in the wavelength range of 4995--5020 {\rm \AA}. In the example shown in Figure \ref{fig:ccf_eg}, the best-fit shift between the two epochs is consistent with being zero for the \OIIIb\ line, serving as a sanity check for our wavelength zero-point calibration. The difference in the apparent \OIII\ line widths between two epochs is caused by the spectral resolution mismatch of our follow-up observations (\S \ref{subsec:obs}) as compared against the first-epoch SDSS spectrum, which does not affect the line centroids (i.e., relevant for RV measurements). In Appendix \ref{appendix:ccf} we provide the ccf results for the \hb\ and \OIIIb\ lines for all targets.

For 8 of the 12 targets, there is a small ($<30$ km s$^{-1}$) but significant, nonzero shift in \OIIIb\ in the follow-up spectra compared against the first spectrum. Assuming these \OIIIb\ shifts were due to residual wavelength calibration errors, we subtract them off from the final broad-line RV shift measurements.

\begin{figure*}
  \centering
    \includegraphics[width=165mm]{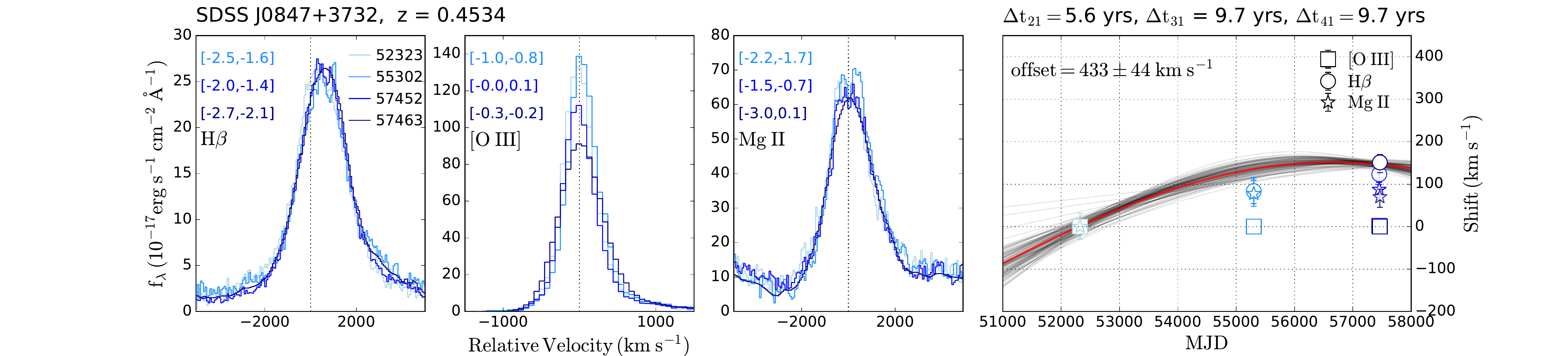}
    \includegraphics[width=165mm]{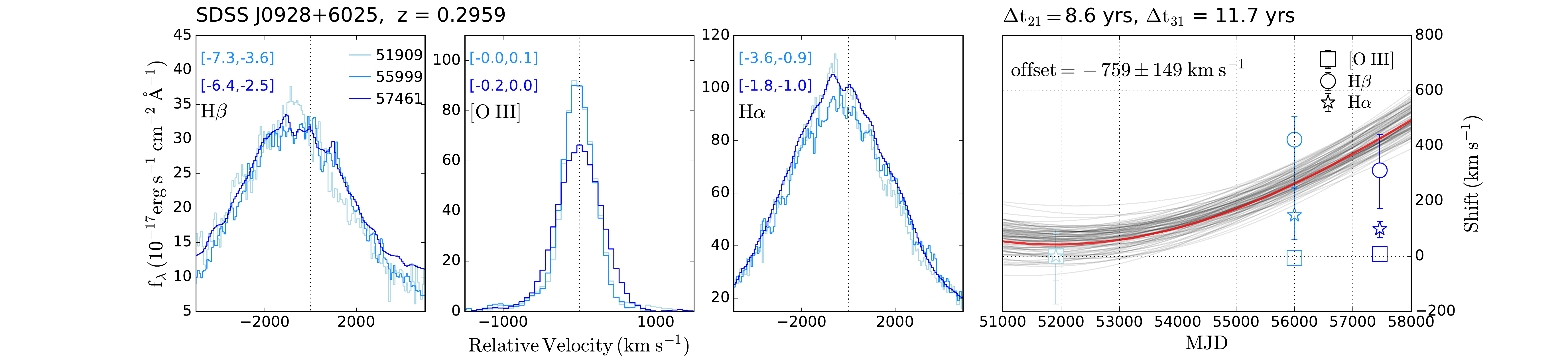}
    \includegraphics[width=165mm]{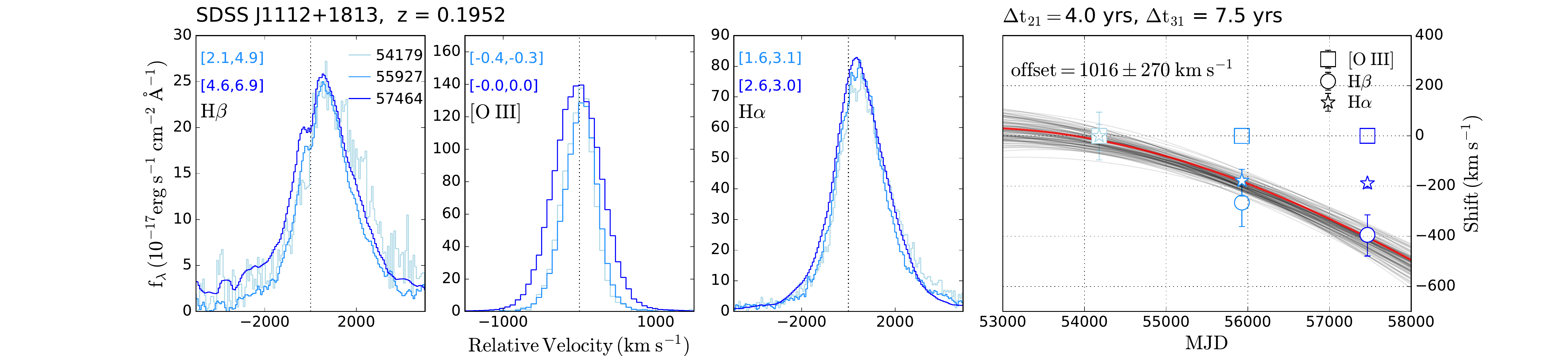}
    \includegraphics[width=165mm]{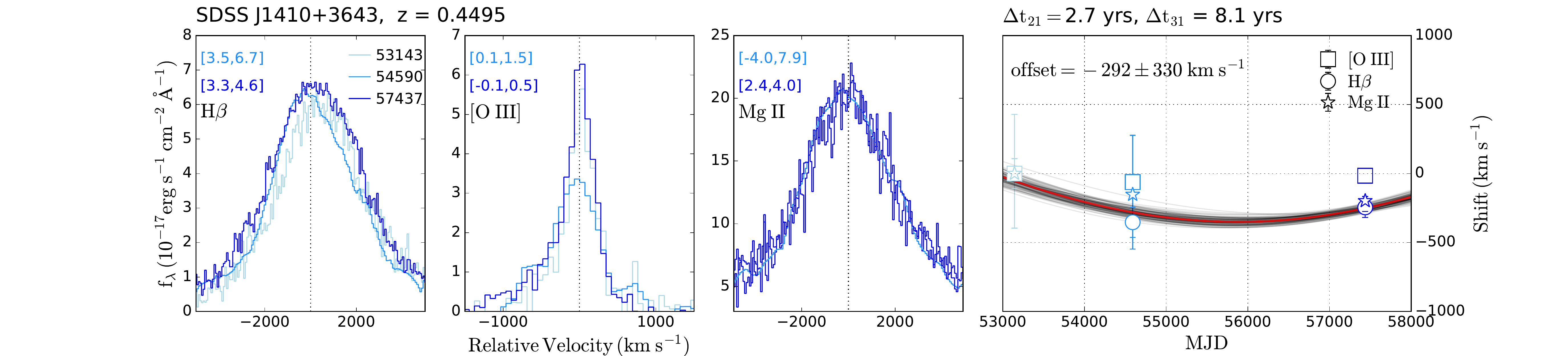}
    \includegraphics[width=165mm]{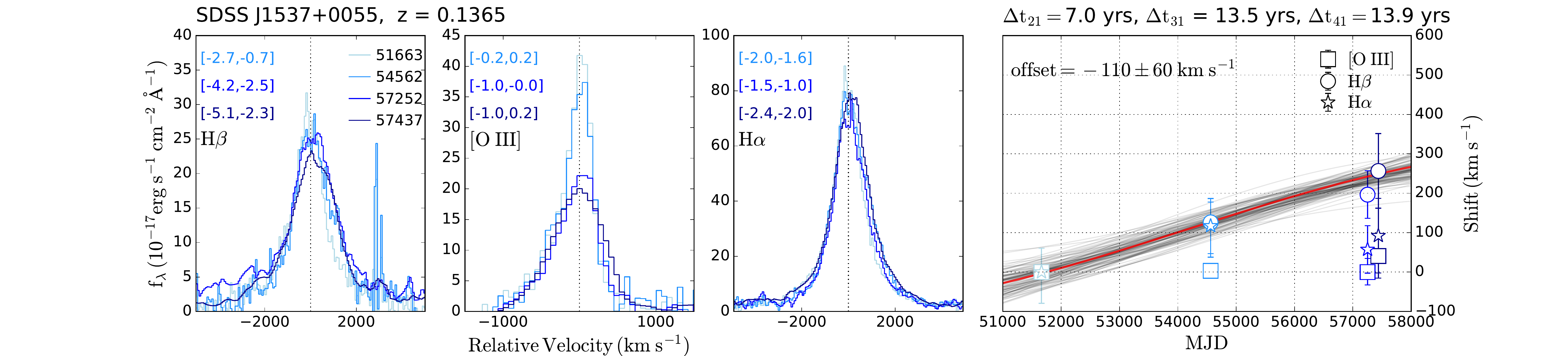}
       \caption{Multi-epoch emission-line RV measurements and modeling for the 5 target quasars as BSBH candidates suggested by continued RV tests. Different rows show different quasars. For each quasar the first three columns show the broad \hb , \OIII, and broad \ha\ (or \MgII ) lines of all epochs. Labeled in brackets are the ccf results  (99\% confidence range in the unit of pixels). Different colors show different epochs with their MJD labeled with darker colors representing later epochs. The last column shows the broad-line RV shifts with the rest-frame time separations labeled on top. The \OIIIb\ RV shift, if nonzero, has been subtracted from the RV shifts of the broad lines. Error bars correspond to 2.5$\sigma$ (99\% confidence). The thick red lines show the best sinusoid fit, whereas the thin gray lines are 100 models randomly drawn within 1$\sigma$ (68\% confidence) from the MCMC analysis. 
     }
    \label{fig:class_bbh}
\end{figure*}

\begin{figure*}
  \centering
    \includegraphics[width=165mm]{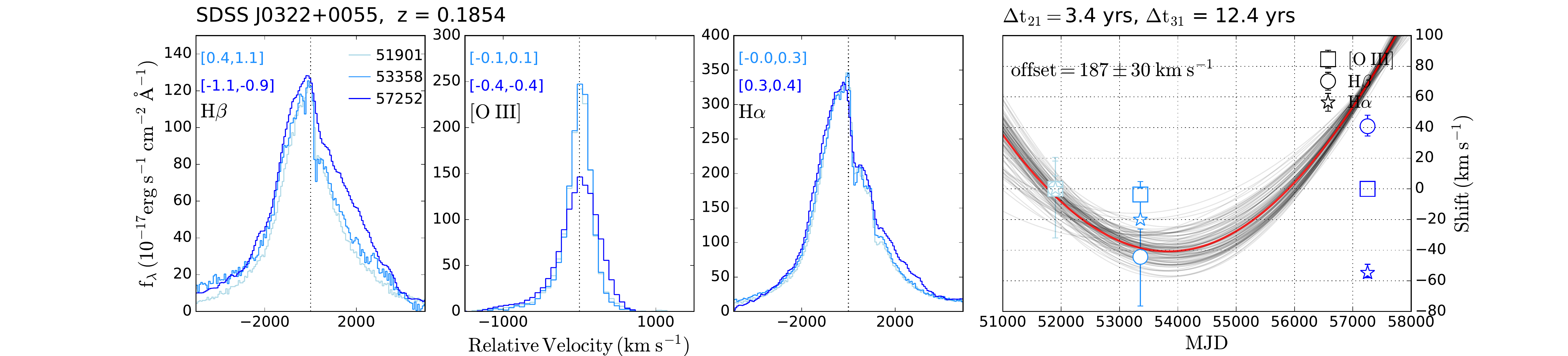}
    \includegraphics[width=165mm]{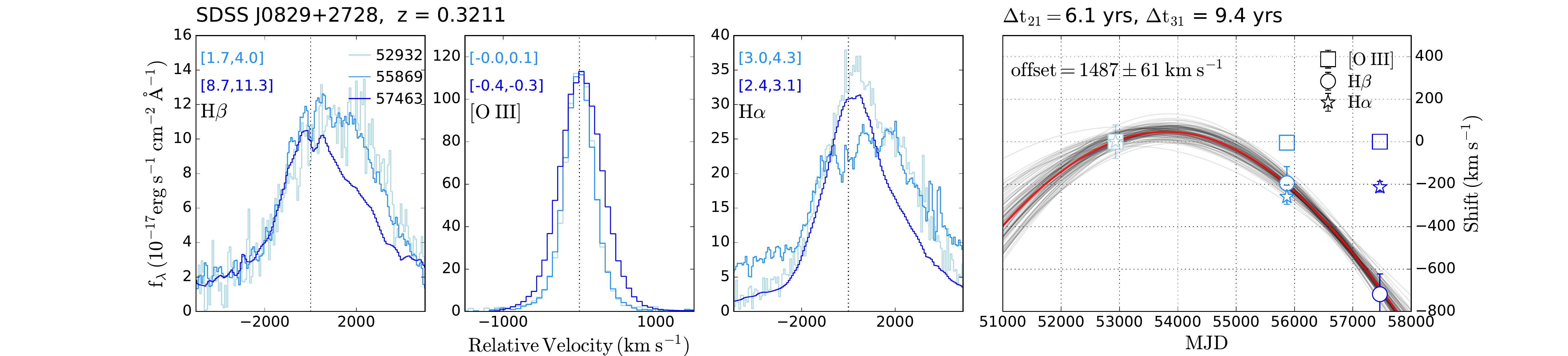}
    \includegraphics[width=165mm]{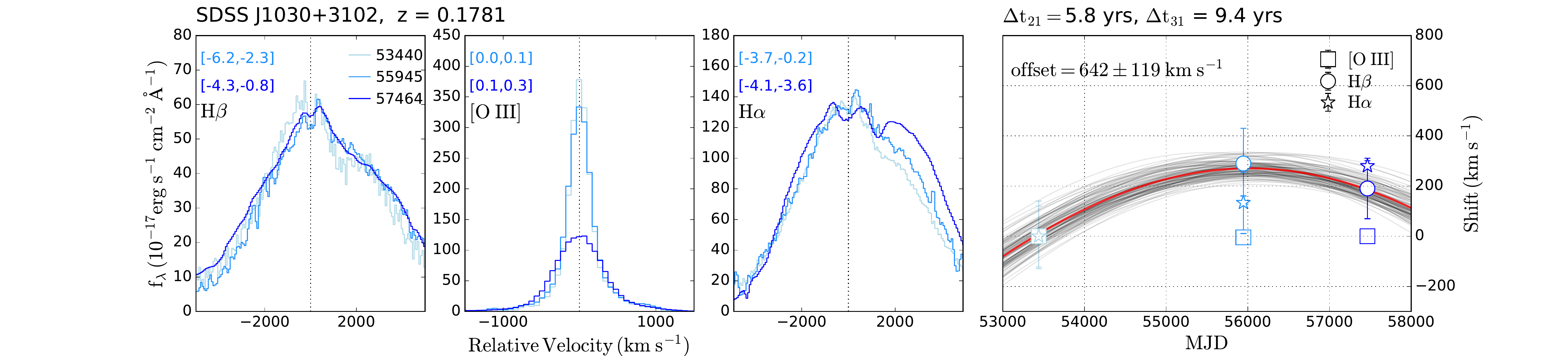}
    \includegraphics[width=165mm]{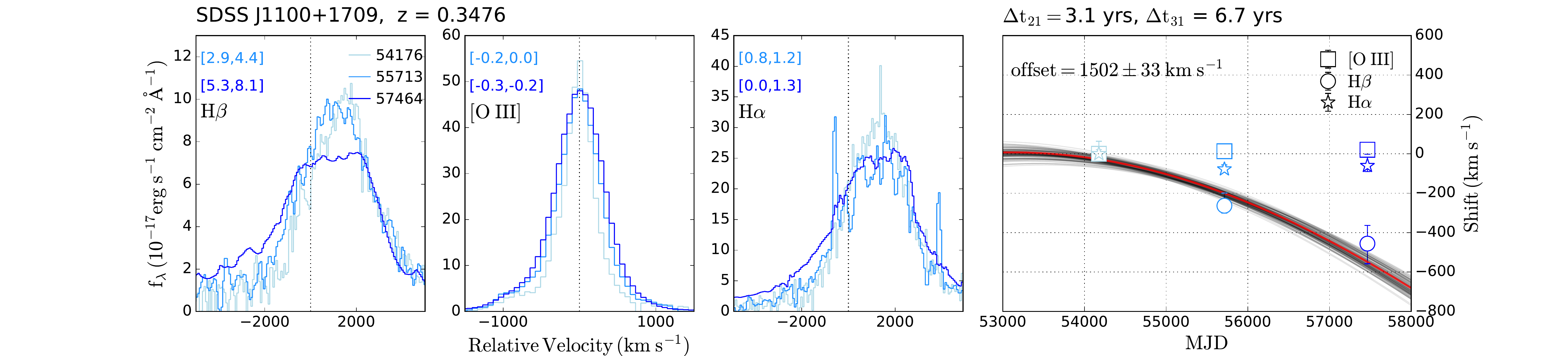}
    \includegraphics[width=165mm]{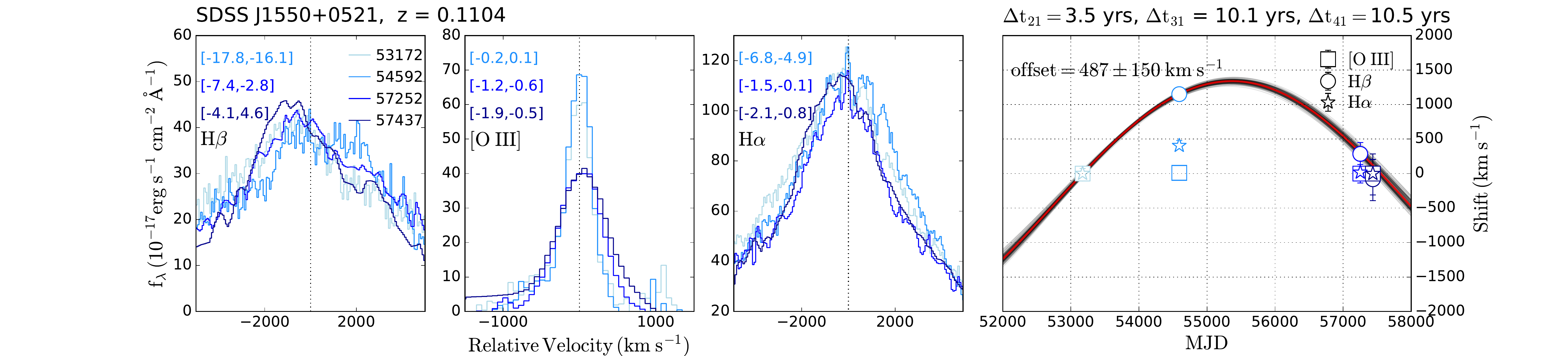}
    \includegraphics[width=165mm]{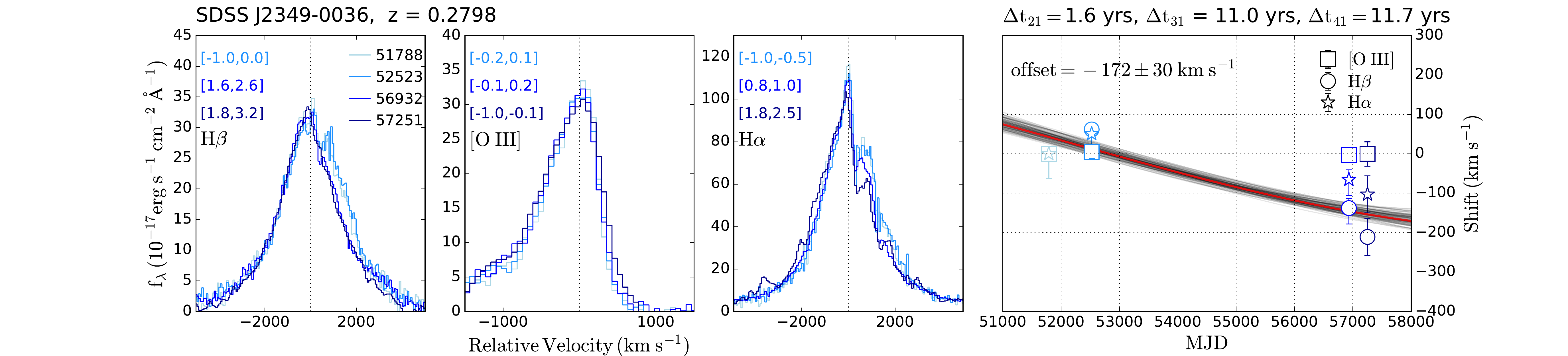}
     \caption{Similar to Figure \ref{fig:class_bbh}, but for our target quasars categorized as BLR variability.}
    \label{fig:class_blr}
\end{figure*}

\begin{figure*}
  \centering
    \includegraphics[width=165mm]{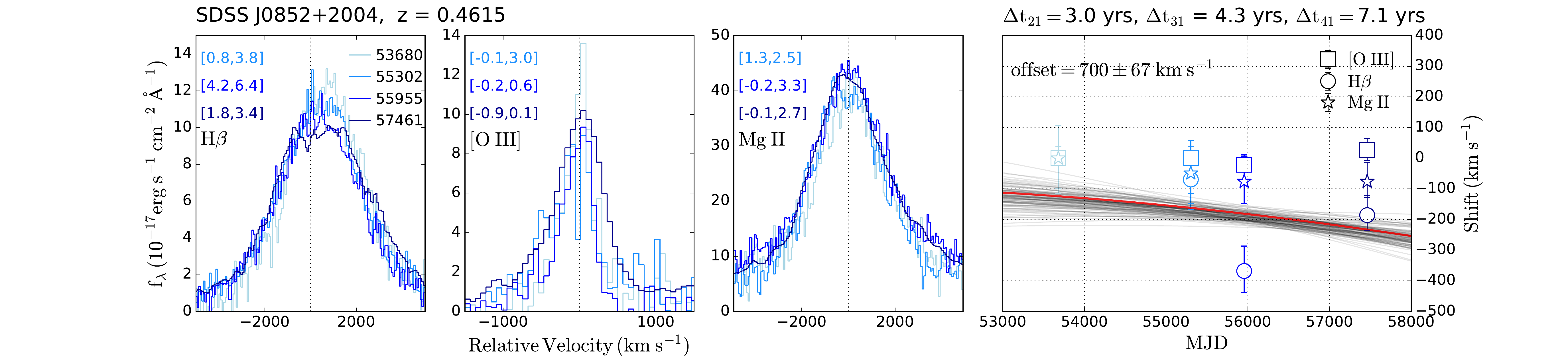}
     \caption{Similar to Figure \ref{fig:class_bbh}, but for a target categorized as ambiguous.}
    \label{fig:class_ambiguous}
\end{figure*}

\section{Results}\label{sec:result}

Figures \ref{fig:class_bbh}--\ref{fig:class_ambiguous} show the ccf results and the inferred broad-line RV curves for all the 12 targets. Table \ref{table:result} lists all RV measurements from the ccf. We detect significant ($>$99\% confidence) RV shifts (i.e., w.r.t. the first-epoch spectrum from the SDSS) for the broad \hb\ line in the new follow-up spectra of all the 12 targets. This is not unexpected since our targets were selected to have significant RV shifts between their previous second- and first-epoch spectra. As discussed, the continued RV shifts may be due to the orbital motion of a sub-pc BSBH and/or BLR variability in single BHs. Below we first classify the targets according to their likely origins of the observed RV shifts in broad emission lines (\S \ref{subsec:classification}). We then present parameter estimation under the BSBH hypothesis to check for self consistency of the models (\S \ref{subsec:bbh_model}).

\subsection{Classification}\label{subsec:classification}

We divide our sample into three categories: (1) BSBH candidates, (2) broad-line variability, and (3) ambiguous cases. These present our best guesses of the ``most likely'' scenarios and are by no means a rigorous classification. Among the 12 targets, we find 5 BSBH candidates, 6 broad-line variability, and 1 ambiguous case as we discuss in detail below. 

\subsubsection{BSBH Candidates}\label{subsubsec:bbh}

We categorize 5 objects as BSBH candidates (Table \ref{table:result}, Category ``1''). Our criteria are defined as: (1) significant ($>99$\% confidence) broad \hb\ velocity shifts are detected between the later-epoch and the first-epoch spectra; (2) the ccf RV shift in broad \hb\ represents an overall bulk velocity shift as verified by visual inspection; there is no significant changes in the broad \hb\ profile as quantified by the line shape parameters (e.g., FWHM, skewness, kurtosis) and verified by visual inspection; (3) the RV shifts independently measured from the broad \ha\ (or \MgII ) are consistent with those of broad \hb\ within uncertainties, or the shift in the broad \ha\ (or \MgII )  is smaller than that of the broad \hb\ (e.g., due to the possibility of an additional circumbinary BLR component with less acceleration; Paper I); and (4) the implied BSBH orbital separation (see \S \ref{subsec:bbh_model} below) is larger than the estimated Roche radius of the BLR so that the hypothesized BSBH model would be self-consistent, although not yet proven. Figures \ref{fig:class_bbh} shows their ccf results and the broad-line RV curves. Below we comment on each case.

{\it SDSS J0847$+$3732.} The quasar was selected by Paper II as a BSBH candidate from the sample of quasars with offset broad \hb\ lines. Continued RV shifts are detected in the broad \hb\ line in both its third- and fourth-epoch spectra with no significant line profile variation.  The third- and fourth-epoch spectra (taken at MJD=57452 by BOSS and 57463 by Gemini, i.e., separated by only 11 days) yield consistent RV acceleration within uncertainties. This verifies that systematic effects are minor for this quasar (e.g., due to instrumental or observational issues and/or short-term variability caused by BLR reverberation). The RV shifts independently measured from the broad \MgII\ are consistent with those of the broad \hb\ within uncertainties. 

{\it SDSS J0928$+$6025.} The quasar was selected by Paper II as a BSBH candidate from the sample of offset-line quasars. Continued RV shift is detected in broad \hb\ in its third-epoch spectra with no significant changes in the line profiles. The broad \ha\ RV shift is also detected but is smaller than that of broad \hb\ in the third-epoch spectrum. 

{\it SDSS J1112$+$1813.} The quasar was selected by Paper II as a BSBH candidate from the sample of offset-line quasars. The detected RV shifts of broad \hb\ monotonically increases with time.  No significant line profile changes are observed. RV shift is also detected in broad \ha\ but is smaller than that of broad \hb\ in the third-epoch spectrum. 

{\it SDSS J1410$+$3643.} The quasar was selected by Paper I as a BSBH candidate from the general quasar population. RV shift is detected in broad \hb\ in its third-epoch spectra with no significant changes in the line profiles, although the acceleration switched signs from the second- to the third-epoch spectra. The RV shift independently measured from the broad \MgII\ is consistent with those of the broad \hb\ within uncertainties in the third-epoch spectrum. 

{\it SDSS J1537$+$0055.} The quasar was selected by Paper I as a BSBH candidate from the general quasar population. Continued broad \hb\ RV shifts are observed in its third- and fourth-epoch spectra with no significant changes in the line profiles. Significant RV shift is also detected in broad \ha\ but is smaller than that of broad \hb\ in the third- and fourth-epoch spectra.

% table 2
\begin{table*}
	\centering
	\caption{Broad \hb\ Radial-Velocity Shifts Measurements from Cross-Correlation Analysis.}
	\label{table:result}
	\begin{adjustbox}{width=\textwidth}
	\begin{tabular}{lcccccrccrccrccc} % 
	        \hline
		\hline
No. & Name & MJD$_1$ & MJD$_2$ & MJD$_3$ & MJD$_4$ & $V^{2,1}_{2}$ & 2.5$\sigma$ & 1$\sigma$ & $V^{3,1}_{2}$ & 2.5$\sigma$ & 1$\sigma$ & $V^{4,1}_{2}$  & 2.5$\sigma$ & 1$\sigma$ & Category  \\
(1)     & (2)  & (3)    & (4)     &   (5)   &       (6)     &    (7)     &    (8)        &     (9)       &      (10)   &    (11)      &   (12)       & (13) & (14) & (15) & (16) \\
\hline
01* & J0322+0055 &  51901 & 53358 & 57252 & ... & $-44$  & $\pm$25 & $\pm$9 & 40 & $\pm$7 & $\pm$3 & ... & ... & ... &  2   \\
02* & J0829+2728 &  51781 & 55869 & 57463 & ... & $-190$& $\pm$79 & $\pm$31 & $-685$& $\pm$89 & $\pm$35 & ... & ... & ... &  2 \\
03* & J0847+3732 &  52323 & 55302 & 57452 & 57463 & 98 & $\pm$31 & $\pm$12 & 123 & $\pm$21 & $\pm$8 & 149& $\pm$22 & $\pm$8 & 1 \\
04  & J0852+2004 & 53680 & 55302 & 55955 & 57461 & $-162$ & $\pm$100 & $\pm$39 & $-368$& $\pm$76 & $\pm$30 & $-186$& $\pm$57 & $\pm$22 &1\\
05 & J0928+6025 & 51909 & 55999 & 57461 & ... & 423 & $\pm$128 & $\pm$51 & 312& $\pm$133 & $\pm$51 & ... & ... & ... & 1 \\
06* & J1030+3102 & 53440 & 55945 & 57464 & ... & 290 & $\pm$135 & $\pm$52 & 195 & $\pm$120 & $\pm$46  & ... & ... & ... & 2 \\
07* & J1100+1709 & 54176 & 55713 & 57464 & ... & $-263$ & $\pm$51 & $\pm$19 & $-451$ & $\pm$97 & $\pm$37 & ... & ... & ... & 2\\
08 & J1112+1813 & 54179 & 55927 & 57464 & ... & $-240$ & $\pm$95 & $\pm$37 & $-394$ & $\pm$82 & $\pm$30 & ... & ... & ... & 1 \\
09 & J1410+3643 & 53143 & 54590 & 57437 & ... & $-353$ & $\pm$109 & $\pm$43 & $-245$ & $\pm$46 & $\pm$19 & ... & ... & ... & 3\\
10* & J1537+0055 & 51663 & 54562 & 57252 & 57437 & 126 & $\pm$70 & $\pm$27 & 201 & $\pm$60 & $\pm$23 & 257 & $\pm$94 & $\pm$37 & 1 \\
11* & J1550+0521 & 53172 & 54592 & 57252 & 57437 & 1154 & $\pm$56 & $\pm$24 & 343 & $\pm$159 & $\pm$62 & $-5$ & $\pm$290 & $\pm$144 & 2 \\
12* & J2349$-$0036 & 51788 & 52523 & 56932 & 57251 & 62 & $\pm$34 & $\pm$13 & $-138$ & $\pm$33 & $\pm$13 & $-209$ & $\pm$47 & $\pm$18 & 2 \\
		\hline
\multicolumn{16}{l}{*: \OIII\ lines showing nonzero velocity shifts, which have been subtracted from the broad-line shift.}\\
\multicolumn{16}{l}{Column 2: abbreviated SDSS name}\\
\multicolumn{16}{l}{Columns 3--6: Modified Julian Dates of all spectroscopic observations}\\
\multicolumn{16}{l}{Columns 7--15: broad \hb\ velocity shift in km s$^{-1}$ measured between the later- and the first-epoch spectra.}\\
\multicolumn{16}{l}{~~~~~~~~~~~~~~~~~~~~~~~Positive (negative) values indicate that the later-epoch spectrum is redshifted (blueshifted) relative to the first-epoch spectrum. }\\
\multicolumn{16}{l}{~~~~~~~~~~~~~~~~~~~~~~~The quoted uncertainties enclose the 2.5$\sigma$ (Columns 8, 11, and 14) and the 1$\sigma$ (Columes 9, 12, and 15) confidence ranges.}\\
\multicolumn{16}{l}{Column 16:  ``1'' for BSBH candidates, ``2'' for broad-line variability, and ``3'' for ambiguous cases. Refer to \S \ref{subsec:classification} for details.}\\
	\end{tabular}
\end{adjustbox}
\end{table*}

\subsubsection{BLR Variability}

It has long been known that variability in the ionizing continuum produces changes in the broad line profiles on the BLR light-travel timescales if the velocity field of the BLR is ordered \citep[e.g.,][]{Blandford1982,Bochkarev1982,Capriotti1982,Peterson1988}. We categorize the 6 quasars shown in Figure \ref{fig:class_blr} as BLR variability (Table \ref{table:result}, Category ``2''). Their third-epoch spectra show significant changes in the broad line profiles of broad \hb\ and/or \ha\ (quantified by changes in the emission-line shape parameters and verified by visual inspection). Our ccf analysis shows that they do have continued RV shifts (Figure \ref{fig:class_blr}). While the line profile change does not necessarily rule out BSBHs \citep[e.g.,][]{shen10,Li2016}, we classify them as BLR variability to be more conservative.
 
{\it SDSS J0322$+$0055.} The quasar was selected by Paper I as a BSBH candidate from the general quasar population. While significant RV shifts are detected in both broad \hb\ and \ha\ in the third-epoch spectra, the broad-line profiles changed significantly, which are most prominently seen in the red wings of the lines.

{\it SDSS J0829$+$2728.} The quasar was selected by Paper II as a BSBH candidate from the sample of offset-line quasars. Monotonic RV shifts are detected in the second- and third-epoch spectra in both broad \hb\ and \ha , but both broad \hb\ and \ha\ of the third-epoch spectra are significantly narrower than those in the first two epochs. This object was also noted by \citet{eracleous11} and by \citet{tsalmantza11} for having significant offset broad lines. \citet{Runnoe2017} also observed substantial profile variability in this quasar.

{\it SDSS J1030$+$3102.} The quasar was selected by Paper II as a BSBH candidate from the sample of offset-line quasars. Continued RV shifts are detected in both broad \hb\ and \ha\ in the third-epoch spectra. While the broad \hb\ profiles are consistent among all three epochs, the broad \ha\ profile changed significantly in the third-epoch spectrum.

{\it SDSS J1100$+$1709.} The quasar was selected by Paper II as a BSBH candidate from the sample of offset-line quasars. Monotonic RV shifts are detected in the second- and third-epoch spectra in broad \hb , whereas no significant RV shift is detected in broad \ha\ in the third-epoch spectrum. The line profiles of both broad \hb\ and \ha\ changed significantly in the third-epoch spectra compared against the previous two epochs.

{\it SDSS J1550$+$0521.} The quasar was selected by Paper I as a BSBH candidate from the general quasar population. Continued RV shifts are detected in the broad \hb\ in its third- and fourth-epoch spectra, but the line profile also changed in both broad \hb\ and \ha . Furthermore, its estimated orbital decay timescale due to gravitational radiation ($\sim$ a few Myr; Table \ref{table:bbh_model}) seems to be too small (i.e., much smaller than the Hubble time) to be compatible with its detection out of a relatively small sample (1 out of 12; see below for details).

{\it SDSS J2349$-$0036.} The quasar was selected by Paper I as a BSBH candidate from the general quasar population. While continued RV shifts are detected in both broad \hb\ and \ha\ in the third- and fourth-epoch spectra, the line profiles have changed significantly compared to previous epochs.

\subsubsection{Ambiguous Cases}\label{subsubsec:ambiguous}

We categorize SDSS J0852$+$2004 (Figure \ref{fig:class_ambiguous}) as ambiguous (Table \ref{table:result}, Category ``3''). It shows continued RV shifts in the broad \hb\ in its second- and fourth-epoch spectra with no significant line profile changes, although the third-vs-first epoch RV acceleration seems to be larger than that of the fourth-vs-first epoch one. This could be due to short-term noise from BLR variability. The broad \MgII\ RV shifts are consistent with those of the broad \hb\ within uncertainties for the second- and fourth-epoch spectra whereas it is smaller for the third-epoch spectrum.

\subsection{Parameters Estimation Under the BSBH Hypothesis with Markov Chain Monte Carlo Analysis}\label{subsec:bbh_model}

Under the BSBH hypothesis, we ask what constraints can be put on the binary orbital parameters given the measured RV shifts, and assess whether they are compatible with the BSBH model assumptions. Rather than providing a proof of the BSBH hypothesis, the test serves as a self-consistency check. This exercise could yield a lower limit on the period and the mass of the BSBH, which could eventually provide a test of the BSBH hypothesis \citep{Runnoe2017}.

We consider a binary on a circular orbit, where BH 2 is active\footnote{This convention is different from Papers I \& II, where we assumed that BH 1 was active. We assume that only the less massive BH 2 is active. We adopt this convention, because simulations have shown that in general the secondary black hole, more appropriated denoted as BH 2, is closer to the gas reservoir and is therefore more likely to be active \citep[e.g.,][]{Cuadra2009,DOrazio2013}.} and powering the observed broad emission lines (Figure \ref{fig:cartoon}; see also \S 2 in Paper I). The orbital period and LOS velocity (relative to systemic velocity) of the active BH at the n-th spectroscopic epoch are
\begin{equation}\label{eq:kepler}
\begin{split}
P & = 2 \pi d^{3/2} (G M_{{\rm tot}})^{-1/2} = 9.4 d^{3/2}_{0.01} M^{-1/2}_{8, {\rm tot}}~{\rm yr}, \\
V^{n}_{2} & = \frac{M_1}{M_{{\rm tot}}} \Bigg(\frac{G M_{{\rm tot}}}{d}\Bigg)^{1/2} \sin I \sin \phi \\
& = 6560 \Bigg(\frac{M_1}{M_{{\rm tot}}} \Bigg) M^{1/2}_{8,{\rm tot}} d^{-1/2}_{0.01} \sin I \sin \phi ~ {\rm km}\,{\rm s}^{-1},
\end{split}
\end{equation}
where subscripts 1 and 2 refer to BH 1 and 2, the superscript $n$ refers to the n-th spectroscopic epoch, $M_{{\rm tot}}\equiv M_1 + M_2$, $I$ is the inclination of the orbit, $d$ is the binary separation, and $\phi = \phi_0 + 2 \pi t/P$ is the orbit phase. We adopt the conventions $M_{8, {\rm tot}} = M_{{\rm tot}}/(10^8~M_{\odot})$ and $d_{0.01} = d/(0.01~{\rm pc})$. We fit the LOS RV shifts (measured at multiple epochs defined as differential RV offsets relative to the first epoch) of the active BH 2 with a sinusoidal model given by: 
\begin{equation}\label{eq:sin}
\begin{split}
V^{n,1}_{2} \equiv V^{n}_2 - V^{1}_{2}  = A~ {\rm sin} \phi - V^{1}_{2}
\end{split}
\end{equation}
where $A\equiv (M_1/M_{{\rm tot}})(GM_{{\rm tot}}/d)^{1/2} \sin I$ is the amplitude and $V^{1}_{2}$ is the LOS velocity of the active BH at the first spectroscopic epoch (measured by $V_{{\rm off}}$ listed in Table \ref{table:obs}). $\phi_0$ is given by $\sin ^{-1} (V^{1}_{2} /A)$ since by definition $V^{1,1}_{2} = 0$. 

We adopt a maximum likelihood approach to estimate the posterior distributions of our model parameters given the RV data and physically motivated priors (see below) under the binary hypothesis. To efficiently draw samples from the posterior probability distributions of the model parameters, we use emcee \citep{Foreman-Mackey2013}, a Python implementation of the affine invariant ensemble sampler for Markov Chain Monte Carlo (MCMC) proposed by \citet{Goodman2010}. The observed $\mathbf{V}^{n,1}_{{\rm obs},2}$ is the observational data to fit. The log-likelihood function is given by
\begin{equation}\label{eq:likelihood}
\begin{split}
{\rm ln}~p(\mathbf{V}^{n,1}_{{\rm obs},2} | \mathbf{k}) & = \\
- \frac{1}{2} \sum_{n=1}^{N} & \bigg\{\frac{\big[V^{n,1}_{{\rm obs},2} - V^{n,1}_{{\rm model},2}(\mathbf{k})\big]^2}{\sigma^2_{{\rm obs},n}} + {\rm ln} \big(\sigma^2_{{\rm obs},n}\big) \bigg\}, 
\end{split}
\end{equation}
where $\mathbf{k} = (A, V^{1}_{2}, P)$ is the vector of free parameters, $N$ the total number of spectroscopic epochs, $\sigma_{{\rm obs},n}$ the 1$\sigma$ error of $V^{n,1}_{{\rm obs},2}$ measured from the ccf analysis, and $V^{n,1}_{{\rm model},2}(\mathbf{k})$ the LOS RV shift calculated from the vector of free parameters $\mathbf{k}$. 

For $A$ we assume a uniform prior, i.e., flat over $[0, |V|_{{\rm max}}]$. We adopt $|V|_{{\rm max}}=4000$ km s$^{-1}$ motivated by the observed distribution of the line-of-sight broad-line velocity offsets in SDSS quasars (e.g., Paper II). For $V^{1}_{2}$ we assume a Gaussian prior with a central value of $V_{{\rm off}}$ and a standard deviation of 1$\sigma$ uncertainty measured from the first-epoch spectrum listed in Table \ref{table:obs}.

% table 3
\begin{table*}
	\centering
	\caption{Binary Black Hole Model Parameters Estimated from Markov Chain Monte Carlo Analysis.}
	\label{table:bbh_model}
	\begin{tabular}{lccccccccccccccc} %
	        \hline
		\hline
 & && & & & & & \multicolumn{3}{c}{$q=0.5$} & & \multicolumn{3}{c}{$q=2$} \\
\cline{9-11}   \cline{13-15} \\
& & log$M_{2}$ & $R_{{\rm BLR}}$ & $a_{{\rm h}}$ & $P_{{\rm min}}$ & $P_{{\rm max}}$ & $P$ & $f_r^{-1} R_{{\rm BLR}}$ & $d$ & $t_{{\rm gr}}$ & & $f_r^{-1} R_{{\rm BLR}}$ & $d$ & $t_{{\rm gr}}$ \\ 
No. & Name & ($M_{\odot}$) & (pc) & (pc) & (yr) & (kyr) & (yr) & (pc) & (pc) & (Gyr) & & (pc) & (pc) & (Gyr)  \\ 
(1) & (2) & (3) & (4) & (5) & (6) & (7) & (8) & (9) & (10) & (11) & & (12) & (13) & (14) \\
\hline   
01 & J0322+0055    & 8.0 & 0.056 & 2.3& 61 & 33 & 71  & 0.18 & 0.056 & 0.90 & & 0.13 & 0.044 & 2.9  \\
02 & J0829+2728    & 8.6 & 0.045 & 5.4& 22 & 33 & 50  & 0.14 & 0.070 & 0.04 & & 0.10 & 0.055 & 0.11 \\
03 & J0847+3732    & 8.1 & 0.051 & 2.7& 47 & 31 & 130 & 0.16  & 0.088 & 2.9 & & 0.12  & 0.070&9.2   \\
04 & J0852+2004    & 8.4 & 0.055 & 4.1 & 37 & 35 & 120 & 0.17  & 0.11 & 0.86 & & 0.13 & 0.087& 2.7 \\
05 & J0928+6025    & 8.9 & 0.068 & 8.2& 29 & 34 &  63  & 0.21  & 0.10 & 0.02 & & 0.15 & 0.081& 0.070 \\
06 & J1030+3102    & 8.7 & 0.043 & 6.2 & 18 & 34 & 47  & 0.13 & 0.072 & 0.02 & & 0.098 & 0.057& 0.070 \\
07 & J1100+1709    & 8.2 & 0.042 & 3.1 & 31 & 33 & 62  & 0.13 & 0.059 & 0.29 & & 0.095 & 0.047& 0.93 \\
08 & J1112+1813    & 7.9 & 0.028 & 2.0 & 24 & 35 & 69  & 0.087 & 0.050 & 1.2 & & 0.064 & 0.040& 3.9 \\
09 & J1410+3643    & 8.4 & 0.044 & 4.1 & 27 & 31 & 38  & 0.14& 0.050 & 0.040 & & 0.10 & 0.039 & 0.17 \\
10 & J1537+0055    & 7.6 & 0.032 & 1.3 & 42 & 33 & 65  & 0.10 & 0.039 & 3.3 & & 0.073 & 0.031& 10 \\
11 & J1550+0521    & 9.0 & 0.036 & 9.4 & 10 & 34 & 26  & 0.11 & 0.061 & 0.0010 & & 0.082 &0.049& 0.0040 \\
12 & J2349-0036     & 8.3 & 0.061 & 3.5 & 49 & 34 & 74  & 0.19 & 0.072 & 0.32 & & 0.14 & 0.057& 1.0 \\
		\hline
\multicolumn{15}{l}{Column 2: abbreviated SDSS name.}\\
\multicolumn{15}{l}{Column 3: virial mass for the active BH from the estimates of \citet{shen11}.}\\
\multicolumn{15}{l}{Column 4: BLR size estimated from the 5100 \angstrom\ continuum luminosity assuming the $R$-$L_{5100}$ relation of \citet{Bentz2009}.}\\
\multicolumn{15}{l}{Column 5: Hard binary separation given by Equation \ref{eq:hardradius}.}\\
\multicolumn{15}{l}{Columns 6 and 7: Lower and upper limits for the adopted prior of $P$ inferred from setting $d$ as $R_{{\rm BLR}}$ and $a_{{\rm h}}$.}\\
\multicolumn{15}{l}{Column 8: Maximum likelihood value of $P$ from the MCMC analysis.}\\
\multicolumn{15}{l}{Columns 9 and 12: Lower limit for the binary separation under the requirement that the BLR size is smaller than the Roche radius. }\\
\multicolumn{15}{l}{Columns 10 and 13: Binary separation inferred using the maximum likelihood value of $P$. }\\
\multicolumn{15}{l}{Columns 11 and 14: Orbital decay timescales due to gravitational radiation. }\\
	\end{tabular}
\end{table*}

For $P$ we adopt a Jeffreys prior (i.e., flat in log$P$, with $\psi \propto 1/P$ over $[P_{{\rm min}}, P_{{\rm max}}]$), with physically motivated lower and upper limits determined as follows. $P_{{\rm min}}$ was estimated according to Equation \ref{eq:kepler} using $d_{{\rm min}} \equiv R_{{\rm BLR}}$, i.e., the separation of the BHs is larger than the radius of the BLR. The typical size of the BLR for \hb\ around a single BH with mass $M_2$ is \citep{shen10}
\begin{equation}\label{eq:blr_size}
\begin{split}
R_{{\rm BLR}} & \sim R_0 (L/L_0)^{1/2} \sim 2.7 \times 10^{-2} \Bigg(\frac{L}{10^{45} \, {\rm erg} \,{\rm s}^{-1}}\Bigg)^{1/2} {\rm pc}, \\
& \sim 3\times 10^{-2} \Bigg(\frac{\lambda_{{\rm Edd}}}{0.1}\Bigg)^{1/2} \Bigg(\frac{M_2}{10^8 M_{\odot}}\Bigg)^{1/2} {\rm pc},
\end{split}
\end{equation}
following the observed $R$--$L$ relation for the reverberation mapping AGN sample at $z<0.4$, with a $\lesssim 40$\% intrinsic scatter in the predicted BLR size\footnote{There is growing evidence \citep[e.g.,][]{Grier2017,LiShen2017} that $z>0.3$ quasars have systematically smaller sizes (inferred from having shorter lags) than the previous $z<0.4$ AGN $R$--$L$ relation due to a combination of selection effects and a physical effect associated with a different BLR size at high luminosities or accretion rates \citep[see also][]{Shen2015a,Shen2016,Du2016}.} \citep{Kaspi2000,Kaspi2005,Bentz2009}. $P_{{\rm max}}$ was estimated using $d_{{\rm max}} \equiv a_{{\rm h}}$, i.e., the separation of the BHs is smaller than the hard binary separation, which is given by \citep[e.g.,][]{DEGN}
\begin{equation}\label{eq:hardradius}
\begin{split}
a_{{\rm h}} \approx 2.7 (1+q)^{-1} \Bigg(\frac{M_2}{10^8 M_{\odot}}\Bigg)\Bigg(\frac{\sigma_{\ast}}{200\, {\rm km}\,{\rm s}^{-1}}\Bigg)^{-2} {\rm pc},
\end{split}
\end{equation}
where $q \equiv M_2 / M_1$ is the binary mass ratio\footnote{The above equation applies to cases where $q<1$.}. $\sigma_{\ast}$ is the stellar velocity dispersion of the quasar host galaxy, which is estimated from $M_{{\rm tot}} = M_2(1+q^{-1})$, assuming that $M_{{\rm tot}}$ follows the $M_{\bullet}$--$\sigma_{\ast}$ relation\footnote{A caveat of this assumption is that the $M_{\bullet}$--$\sigma_{\ast}$ relation may not apply to BSBHs because the binary is disturbing the stellar orbits near the nucleus. This comes down to the question of how quickly the stellar orbits relax after scattering by the BSBH, which is still under debate. Nevertheless, the inferred upper limit in the period prior is $\sim$3 orders of magnitude larger than our best-fit value, and therefore a deviation from the $M_{\bullet}$--$\sigma_{\ast}$ relation still would not affect our results in practice.} \citep{tremaine02,KormendyHo2013,Shen2015b}. Table \ref{table:bbh_model} lists $R_{{\rm BLR}}$ and $a_{{\rm h}}$ as well as the corresponding lower and upper limits on the adopted prior of $P$. 

To explore the parameter space, we used 10 walkers for each set of initial values in a 3D space with each walker corresponding to 50,000 steps. Burn-in phases (2,000 steps) were deleted before connecting 10 chains end to end. We examined each combined chain to ensure that they were likely to be converged. Because our RV measurements only sampled 3 or 4 epochs, the parameter space was not very well constrained. We therefore tried a large range of initial values to make sure that our result was representative of the maximum likelihood from the global posterior distribution\footnote{We looped through different initial values of $P$ spanning the whole range allowed by the prior.  Depending on the initial value, the MCMC chain may be trapped in different local maxima of the loosely constrained parameter space. To avoid running the MCMC chain for too long given our limited computational resources, we first found the local maxima in all the likely converged chains and then chose the global maximum likelihood region in the parameter space according to Equation \ref{eq:likelihood} as our final result.}. Our best-fit models are shown in Figures \ref{fig:class_bbh}--\ref{fig:class_ambiguous} as the red curves, whereas the grey curves show 100 models randomly selected from the 1$\sigma$ range.  Appendix \ref{appendix:mcmc} presents more details on the test of the dependence of our MCMC results on the initial values. Appendix \ref{appendix:noise} discusses the effect of broad-line short-term variability (``jitter'' noise) on our RV result from the ccf analysis.

Table \ref{table:bbh_model} lists the best-fit value for $P$ from the MCMC analysis. We then infer the binary separation $d$ using Equation \ref{eq:kepler} assuming $q=0.5$ or $q=2$. We have assumed $q<1$ so far but below we relax this to account for the more general cases where the more massive BH 1 is active. We compare $f_r d$ against $R_{{\rm BLR}}$ as a self-consistency check of the binary hypothesis. $f_r d$ characterizes the maximum size of the BLR before it is dynamically affected by the companion BH in the system. $f_r d$ can be defined as the average radius of the Roche lobe in a circular binary system \citep[e.g.,][]{Paczynski1971}, where:
\begin{equation}
\begin{split}
f_r & = 0.38 - 0.2 \, {\rm log} \,q, \,\,\,0.05 < q < 1.88 \\
& = 0.46224(1+q)^{-1/3}, \,\,\,q > 1.88.
\end{split}
\end{equation}
We categorize systems that satisfy the condition $d > f_r^{-1} R_{{\rm BLR}}$ as ``BSBH candidates'' in addition to passing the first three criteria as discussed in \S \ref{subsubsec:bbh}. All candidates passed the self-consistency check after accounting for systematic uncertainties in the assumed $R$-$L$ relation \citep[$R_{{\rm BLR}}$ can be a factor of $\sim$3 smaller than the assumed baseline value; e.g., see Figure 11 of][]{Grier2017}. Table \ref{table:bbh_model} also lists the orbital decay timescale due to gravitational radiation assuming a circular binary with a mass ratio of $q=0.5$ or $q=2$, which is given by \citep{Peters1964}
\begin{equation}
t_{{\rm gr}} = \frac{5}{256}\frac{c^5}{G^3}\frac{q^2d^4}{(1+q)M_2^3}.
\end{equation}

\section{Discussion}\label{sec:discuss}

\subsection{Uncertainties and Caveats}\label{subsec:uncertainty_caveat}

First, broad emission-line variability around single SMBHs is the primary uncertainty in identifying BSBH candidate from radial velocities. The AGN BLR has long been known to be dynamic \citep[e.g.,][]{cherepashchuk73,osterbrock76,Capriotti1982,Peterson1988}.  Kinematic changes in the broad emission line profiles have generally been attributed to the asymmetric response to the variable continuum \citep[e.g.,][]{Blandford1982,Peterson1988,Barth2015,Sun2018} and/or changes in the kinematic structure of the BLR \citep[e.g.,][]{Marziani1996,Wandel1999,Peterson1999a,Sergeev2007,Bentz2009,Grier2013}. If the BLR is dominated by radial motion \citep[i.e., inflows or outflows; e.g.,][]{Denney2009} and/or the distribution of gas is significantly non-axisymmetric, the transfer function will be strongly asymmetric about the line center, which will lead to one side of the emission line response to the continuum before the other side and produce fake RV shifts in multi-epoch spectra. In general, however, the profile variations in response to a variable ionizing continuum are much smaller and faster than profile variations due to structural changes in the BLR. The relevant time scales for the broad line kinematic profile changes are the light-travel and dynamical times of the BLR as well as the continuum variability time\footnote{The recombination time is generally short compared to all other time scales or changes would have been averaged out otherwise.}. These timescales range from hours to years for SDSS quasars, which are shorter than or comparable to the cadence (days to years) but are shorter than the typical time baseline of existing RV surveys ($\lesssim$20 yr). Changes of the kinematic structure of the BLR are expected to occur on the dynamical time scale $t_{{\rm dyn}}{\sim}24(R_{{\rm BLR,~0.1}}/{\rm FWHM}_{4000})$ yr, which is similar to the time intervals between the observations presented in this program. Independent from variation of the continuum source, broad-line profile variability may result from structural changes in the BLR such as due to redistribution of the BLR gas in position and/or velocity space, resembling a ``see saw'' pattern. To evaluate these effects on the RV test, \citet{Runnoe2017} performed simulations to study ``see saw'' variability of the \hb\ line profile. These authors have demonstrated that broad cuspy or boxy profiles could easily result in apparent RV shift. 

Second, our baseline BSBH model is oversimplified which neglects the possibility of a circumbinary accretion disk \citep[e.g.,][]{Rafikov2012,Farris2014,Nguyen2018}. We have assumed that only one BH is active and carries its own BLR on a circular orbit, whose motion can be traced by the RV shifts in the broad emission lines. This requires that the binary separation is larger than the BLR size at least. To infer the BLR size we had to assume some empirical correlation, such as the adopted $R$--$L$ relation, which however is subject to uncertainties and significant scatters according to reverberation mapping campaigns \citep[e.g.,][]{Kaspi2000,Bentz2009,Grier2017}. 

Furthermore, we have assumed that the separation of the BHs is larger than the radius of the BLR, estimated using the observed $R$--$L$ relation from reverberation mapped AGN. However, the BLR radius obtained from the $R$--$L$ relation does not signify the outer edge of the BLR but a characteristic radius within it; the BLR is likely to be a few times bigger, and therefore our adopted $d_{{\rm min}}$ is likely to be underestimated by a factor of a few.  An additional caveat is that the BLR would be truncated to a size several times smaller than the Roche lobe radius of the accreting BH \citep[e.g.,][]{Runnoe2015} because of the tidal interaction between the two BHs. This effect is well known in the context of interacting binary stars \citep[e.g.,][]{Paczynski1977}. Nevertheless, these effects would not change our results qualitatively considering the substantial systematic uncertainties in the assumed $R$--$L$ relation \citep[$R_{{\rm BLR}}$ can be a factor of $\sim$3 smaller than the assumed baseline value; e.g., see Figure 11 of][]{Grier2017}.

Finally, another possibility to explain the RV offset is the recoil effect on the merger product, which results from the emission of anisotropic gravitational radiation after the coalescence of two SMBHs due to momentum conservation \citep[e.g.,][]{Baker2006,bogdanovic07,bonning07,campanelli07,gonzalez07,civano10,Dotti2010,Blecha2011,Blecha2016}.  While the returning timescales for recoiling BHs may be sensitive to many parameters and may strongly depend on the magnitude of the recoil velocity \citep[e.g.,][]{Choksi2017}, it is typically on the order of $\sim$Myr, which is much longer than the time baselines of our survey. Therefore, we would expect to see no RV variation in the BLR emission of kicked BHs unless it is caused by BLR variability.

\subsection{Detection Rate of Sub-pc BSBH Candidates}

We started off with 52 systems with significant RV shifts measured in two-epoch spectra from the parent sample consisting of 193 ordinary (the ``superior'' sample in Paper I) and 50 offset-line (Paper II) quasars. Among the 52, we identified a sample of 16 BSBH candidates based on two-epoch spectroscopy. Here with continued RV tests for 12 of the 16 candidates, we further suggest that 5 of the 12 remain valid as BSBH candidates. This indicates that our detection rate is 
\begin{equation}
\sim \frac{5 + \frac{5}{12}(16-12)}{52}\sim 13 \pm 5\% ~(1\sigma~{\rm Poisson~error}).
\end{equation}
We find no significant evidence for a different detection rate between the sample of the ordinary quasar population (\S \ref{subsec:general}; Paper I) and those with offset broad emission lines (\S \ref{subsec:general}; Paper II). The apparent detection rate is $\sim20\pm9$\% in the offset quasar population and is $\sim5\pm4$\% in ordinary quasars, which are consistent within uncertainties given our small sample size.

Theory suggests that BSBHs should spend most of their lifetime ($\lesssim$Gyr) at sub-pc scale before entering the GW-dominated regime. Considering typical quasar life times $\sim$10$^{7}$--10$^{8}$ yr \citep[e.g.,][]{Martini2001}, we would expect a $\sim$1--10\% probability at least to observe sub-pc BSBHs assuming that all quasars are triggered by galaxy mergers with two SMBHs. This is consistent with the apparent rate of sub-pc BSBH candidates found by our work, if most of the candidates turn out to be real BSBHs. On the other hand, if the majority of the candidates were caused by BLR variability, the occurrence rate would be much lower than the naive expectation.  Many scenarios may lead to a lower-than-expected BSBH occurrence rate, such as (i) only a small fraction of quasars are triggered by galaxy mergers with two SMBHs, (ii) BSBHs sweep through the sub-pc scale or stall at larger radii \citep[e.g.,][]{Wang2017}, (iii) the BLR region is much bigger than expected from the $R$--$L$ relation (although growing evidence suggests the opposite; \citealt{Du2016,LiShen2017,Grier2017}) and the associated RV variability behavior is more complicated than being assumed here, and (iv) BSBHs become depleted of gas at the sub-pc scale and/or are radiate inefficient. 

The sub-pc BSBH candidates have estimated orbital periods on the order of decades to centuries (Table \ref{table:bbh_model} and Appendix \ref{appendix:mcmc}), whereas the orbital period constrained by PTAs is of order years \citep[e.g.,][]{Holgado2018,Sesana2018}. Further assumption and modeling are needed to evolve these BSBH candidates into the PTA frequency band to directly compare our results with PTA limits.

\subsection{Comparison with Previous Results}

%\subsubsection{Detection Rate}

\citet{Runnoe2017} conducted a spectroscopic monitoring campaign for 88 $z < 0.7$ quasars whose broad \hb\ lines were selected to be significantly offset from the systemic redshifts by a few thousand \kms\ \citep{eracleous11}. These authors found 29 of the 88 quasars displayed no profile shape variability using three or four-epoch spectra covering a time baseline over 12 yr in the observed frame, among which three objects showed systematic and monotonic velocity changes as their best BSBH candidates. In a similar study but based on \MgII , \citet{Wang2017} found no good BSBH candidate in a sample of 21 quasars at $0.36 <z<2$ with three-epoch spectra. These authors also suggested a low binary fraction ($\lesssim$1\%) in the regime of $\sim0.1$ pc separations based on the analysis of \MgII\ using two-epoch spectra of 1438 quasars with eight-year median time baselines.  

While the statistics is still poor, our apparent detection rate is tentatively higher than but is still broadly consistent with the result independently found by \citet{Runnoe2017}. These authors found 3 best candidates out of 88, or $\sim$3$\pm$2\%, but all the 29 with radial velocity curves are still consistent with the binary hypothesis (so the fraction may be as high as $\sim$33$\pm$6\%). There is a general agreement even though our targets are normal quasars (Paper I) and/or quasars with intermediate broad-line velocity offsets (Paper II). \citet{Barth2015} has suggested that selection of quasars with the largest velocity offsets will bias towards the tail of the distribution of reverberation-induced velocity shifts, resulting in major contamination of false positives in candidate BSBHs.  This is in line with our finding of a tentatively higher but still consistent binary fraction in the sample of ordinary and/or intermediate-offset quasars than in those with the largest offsets. However, our result seems to be higher than the low binary occurrence rate of $\lesssim$1\% found by \citet{Wang2017}.  In addition to BLR variability, another factor that may contribute to the apparent discrepancy may be the difference between the broad \hb\ and \MgII\ lines and their RV shifts.  In the 3 of our 12 targets with both broad \hb\ and \MgII\ coverages, the broad-\hb\ RV shifts are always either larger than or consistent with those in \MgII .  While the sample size is still too small to draw any firm conclusion, this may suggest that RV searches based on the \MgII\ line may lead to biases that would underestimate the binary fraction based on \hb\ (e.g., due to the possibility of an additional circumbinary BLR component with less acceleration).

\section{Conclusions and Future Work}\label{sec:conclude}

We have searched for temporal RV shifts of the broad lines in ordinary (Paper I) and intermediate-offset (Paper II) quasars as signposts for the hypothesized orbital motion from sub-pc BSBHs. Among a parent sample of 52 quasars that show significant RV shifts in the first two epochs, we have identified 16 quasars that showed no broad line profile changes in the previous two epochs (6 from Paper I and 9 from Paper II). Using continued spectroscopic monitoring, we have further obtained a third and/or fourth-epoch spectrum for 12 of the 16 quasars from Gemini/GMOS-N, du Pont 2.5 m/B\&C, and/or SDSS-III/IV/BOSS. We summarize our main findings as follows.

\begin{enumerate}

\item We have used a $\chi^2$-based cross-correlation approach to quantify the velocity shifts between the first and later epochs. We have subtracted the pseudo-continua and narrow emission lines before measuring the velocity shifts from the broad emission lines using both broad \hb\ and broad \ha\ (\MgII ). We have calibrated the relative RV zero point using the narrow \OIII\ lines which were simultaneously observed with the broad emission lines to minimize systematic errors from calibration. We have measured significant RV shifts in the later-epoch spectra w.r.t. the first epoch in all our 12 targets.   

\item We have divided the 12 targets into three categories, including 5 ``BSBH candidates'', 6 ``BLR variability'', and 1 ``ambiguous'' case. We have required that the BSBH candidates show broad \hb\ RV shifts consistent with binary orbital motion (using a self-consistency check; \S \ref{subsec:bbh_model}) without any significant changes in the line profiles. Further requirements include that the RV shifts independently measured from the broad \ha\ (or \MgII ) are either consistent with those of broad \hb\ within uncertainties or smaller than that of the broad \hb\ (e.g., due to the possibility of an additional circumbinary BLR component with less acceleration; Paper I). 

\item We have performed a maximum likelihood analysis to estimate the posterior distributions of model parameters under the binary hypothesis as a self-consistency check. The RV data of our BSBH candidates are best explained with a $\sim$0.05--0.1 pc BSBH with an orbital period of $\sim$40--130 yr, assuming a mass ratio of 0.5--2 and a circular orbit, although the parameter space is not well constrained because of the small number of RV measurements (i.e., 3 or 4 epochs). 

\item Our results suggest that the apparent fraction of the sub-pc BSBH candidates is $\sim$13$\pm$5\% (1$\sigma$ Poisson error) among all SDSS quasars without correcting for selection incompleteness (such as due to viewing angles and/or orbital phases). We find no evidence for a significant difference in the detection rate for the subsets with and without single-epoch broad line velocity offsets ($\sim$20$\pm$9\% and $\sim$5$\pm$4\%). This is broadly consistent with the previous result of \citet{Runnoe2017} within uncertainties, which were based on the spectroscopic monitoring of quasars with the largest single-epoch broad-line velocity offsets. Taken at face value, the fraction is higher than the result suggested by \citet{Wang2017} in a similar study but based on the analysis of \MgII , which may be at least partly due to the difference between broad \hb\ and \MgII .

\end{enumerate}

Dedicated, long-term spectroscopic monitoring (with at least two orbital cycles with enough cadence to sample the orbit well) is still required to further confirm or reject the BSBH candidates given the short-term ``jitter'' noise due to BLR variability. In genuine BSBH systems, we expect that the RV curve is a long-term periodic signal overlapped with a relatively short-term red-noise variability \citep[e.g.,][]{Guo2017}. The RV variation should be uncorrelated with the continuum flux variation to rule out asymmetric reverberation \citep{shen10,Barth2015}. Future large spectroscopic synoptic surveys \citep[e.g.,][]{McConnachie2016,Kollmeier2017} could identify BSBHs using the RV method in low-mass systems (i.e., with shorter orbital periods than the candidates identified in this work). Alternative approaches (based on, e.g., spectral energy distribution of the circumbinary accretion disks, gravitational lensing, quasi-periodic light curves, and/or astrometry) are also needed to finally uncover the elusive population of BSBHs at the sub-pc and smaller scales \citep[e.g.,][]{Yu2003,liufk04,Liufk2014,loeb10,Li2012,Lusso2014,Yan2014,Yan2015,DOrazio2015a,DOrazio2017,DOrazio2018,DOrazio2018a,Graham2015a,Graham2015,Liutt2015,Liutt2016,Charisi2016,Charisi2018,Li2016,Li2017a,Zheng2016}. 

\section*{Acknowledgements}

We thank S. Tremaine for his insight and encouragement, J. Runnoe for helpful comments, and our referee, M. Eracleous, for his prompt and constructive report that helped significantly improve the paper. H.G. thanks Z. Cai and M. Sun for valuable discussions on the MCMC analysis and support by the NSFC (grant No. 11873045). X.L. thanks Percy Gomez for assistance with the Gemini observations. Y.S. acknowledges support from the Alfred P. Sloan Foundation and NSF grant AST-1715579. J.X.P. acknowledges support from the NSF grant AST-1412981.

Based on observations obtained at the Gemini Observatory (Program ID GN-2016A-Q-83), which is operated by the Association of Universities for Research in Astronomy, Inc., under a cooperative agreement with the NSF on behalf of the Gemini partnership: the National Science Foundation (United States), the National Research Council (Canada), CONICYT (Chile), Ministerio de Ciencia, Tecnolog\'{i}a e Innovaci\'{o}n Productiva (Argentina), and Minist\'{e}rio da Ci\^{e}ncia, Tecnologia e Inova\c{c}\~{a}o (Brazil).

Funding for the Sloan Digital Sky Survey IV has been provided by the Alfred P. Sloan Foundation, the U.S. Department of Energy Office of Science, and the Participating Institutions. SDSS-IV acknowledges support and resources from the Center for High-Performance Computing at the University of Utah. The SDSS web site is www.sdss.org.

SDSS-IV is managed by the Astrophysical Research Consortium for the Participating Institutions of the SDSS Collaboration including the Brazilian Participation Group, the Carnegie Institution for Science, Carnegie Mellon University, the Chilean Participation Group, the French Participation Group, Harvard-Smithsonian Center for Astrophysics, Instituto de Astrof\'isica de Canarias, The Johns Hopkins University, Kavli Institute for the Physics and Mathematics of the Universe (IPMU) / University of Tokyo, Lawrence Berkeley National Laboratory, Leibniz Institut f\"ur Astrophysik Potsdam (AIP),  Max-Planck-Institut f\"ur Astronomie (MPIA Heidelberg), Max-Planck-Institut f\"ur Astrophysik (MPA Garching), Max-Planck-Institut f\"ur Extraterrestrische Physik (MPE), National Astronomical Observatories of China, New Mexico State University, New York University, University of Notre Dame, Observat\'ario Nacional / MCTI, The Ohio State University, Pennsylvania State University, Shanghai Astronomical Observatory, United Kingdom Participation Group,Universidad Nacional Aut\'onoma de M\'exico, University of Arizona, University of Colorado Boulder, University of Oxford, University of Portsmouth, University of Utah, University of Virginia, University of Washington, University of Wisconsin, Vanderbilt University, and Yale University.

Facilities: Gemini (GMOS-N), du Pont 2.5 m (B\&C), Sloan

%%%%%%%%%%%%%%%%%%%%%%%%%%%%%%%%%%%%%%%%%%%%%%%%%%



\bibliographystyle{mnras}
%\bibliography{/Users/zeus/Documents/References/binaryrefs} % if your bibtex file is called example.bib

%%%%%%%%%%%%%%%%% APPENDICES %%%%%%%%%%%%%%%%%%%%%

\appendix

\section{Spectral fitting and decomposition results}\label{appendix:spec}
All the figures are available online.

%\begin{comment}
\begin{figure*}
  \centering
    \includegraphics[width=85mm]{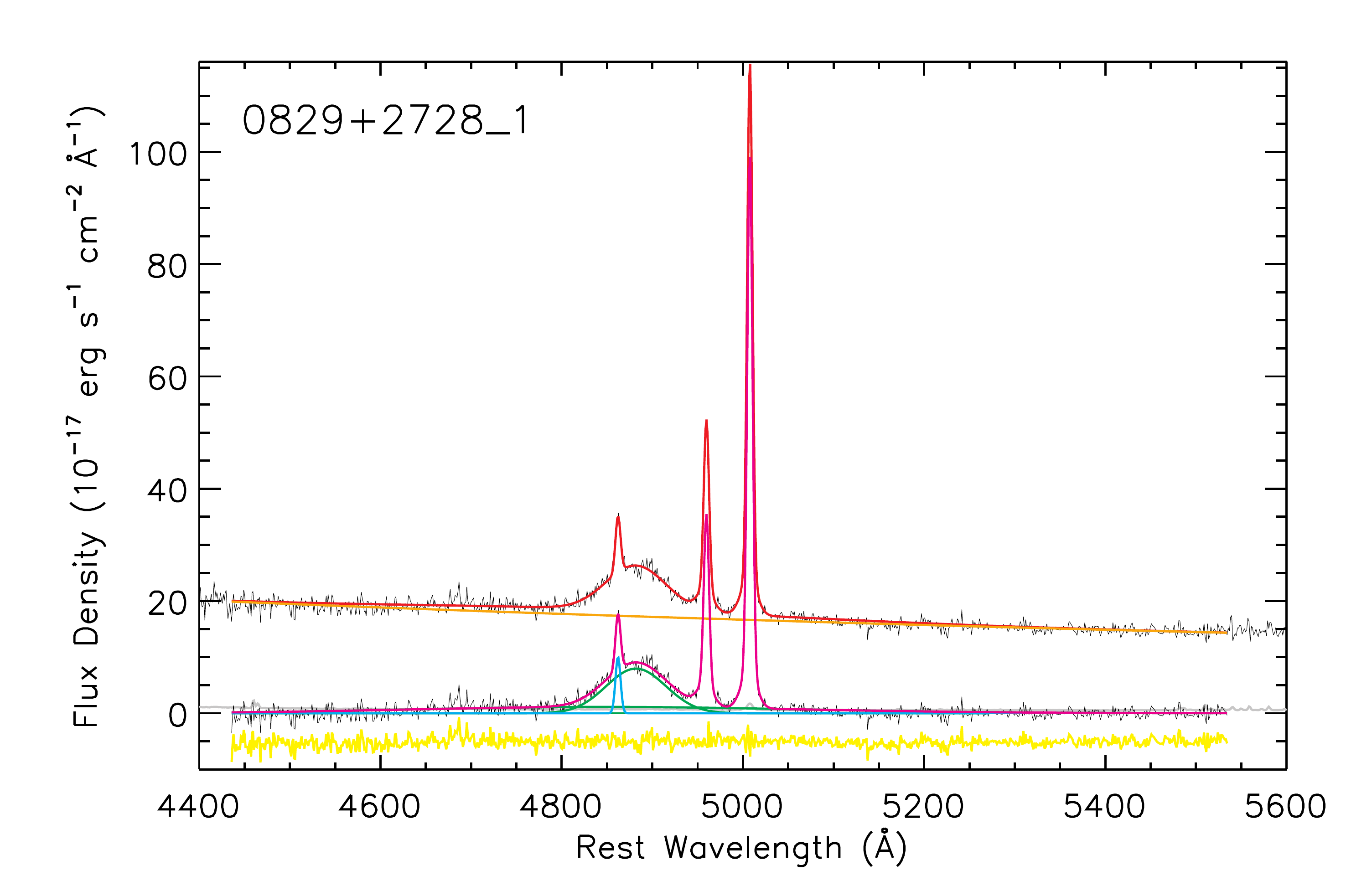}
    \includegraphics[width=85mm]{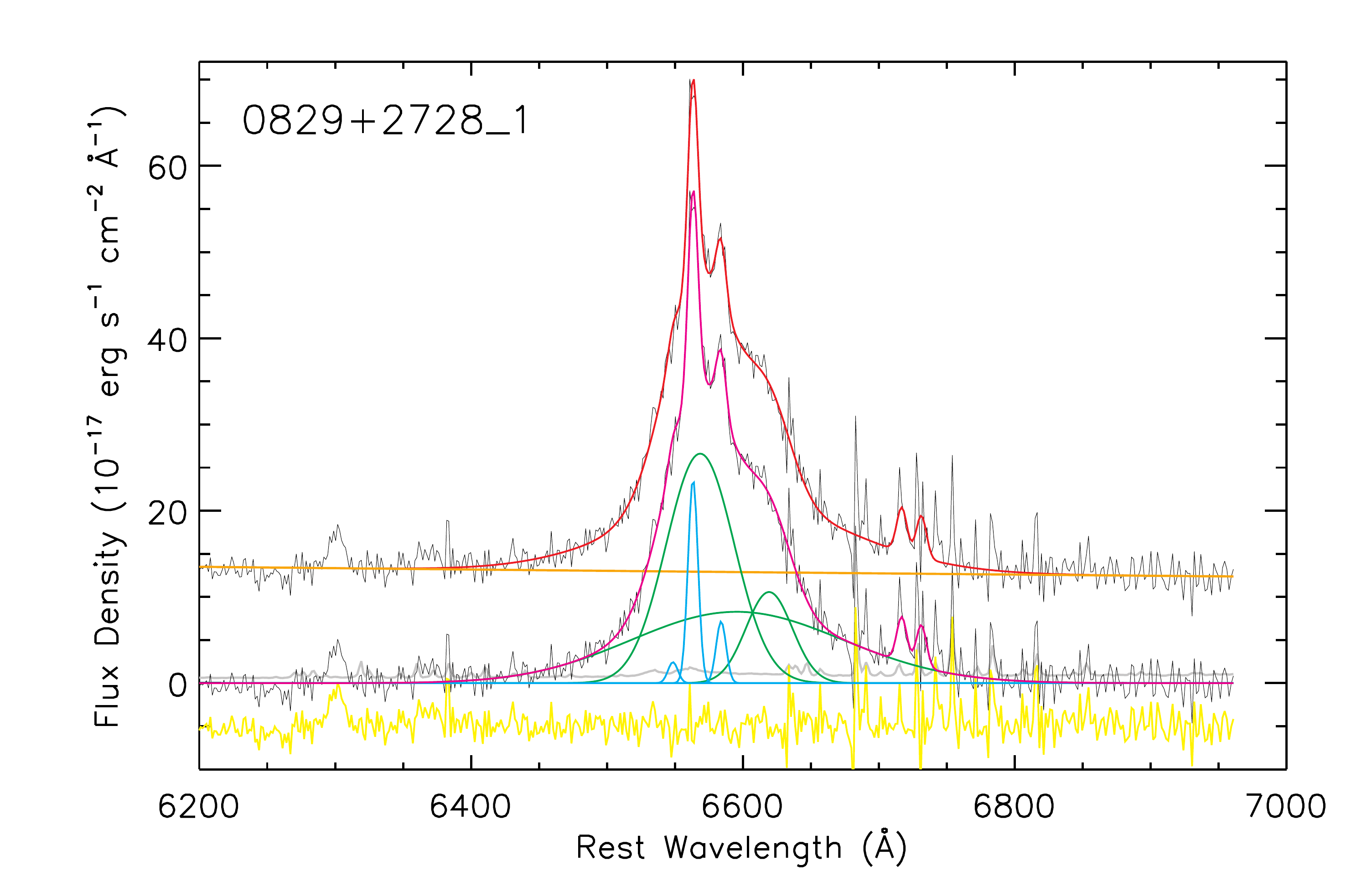}
    \includegraphics[width=85mm]{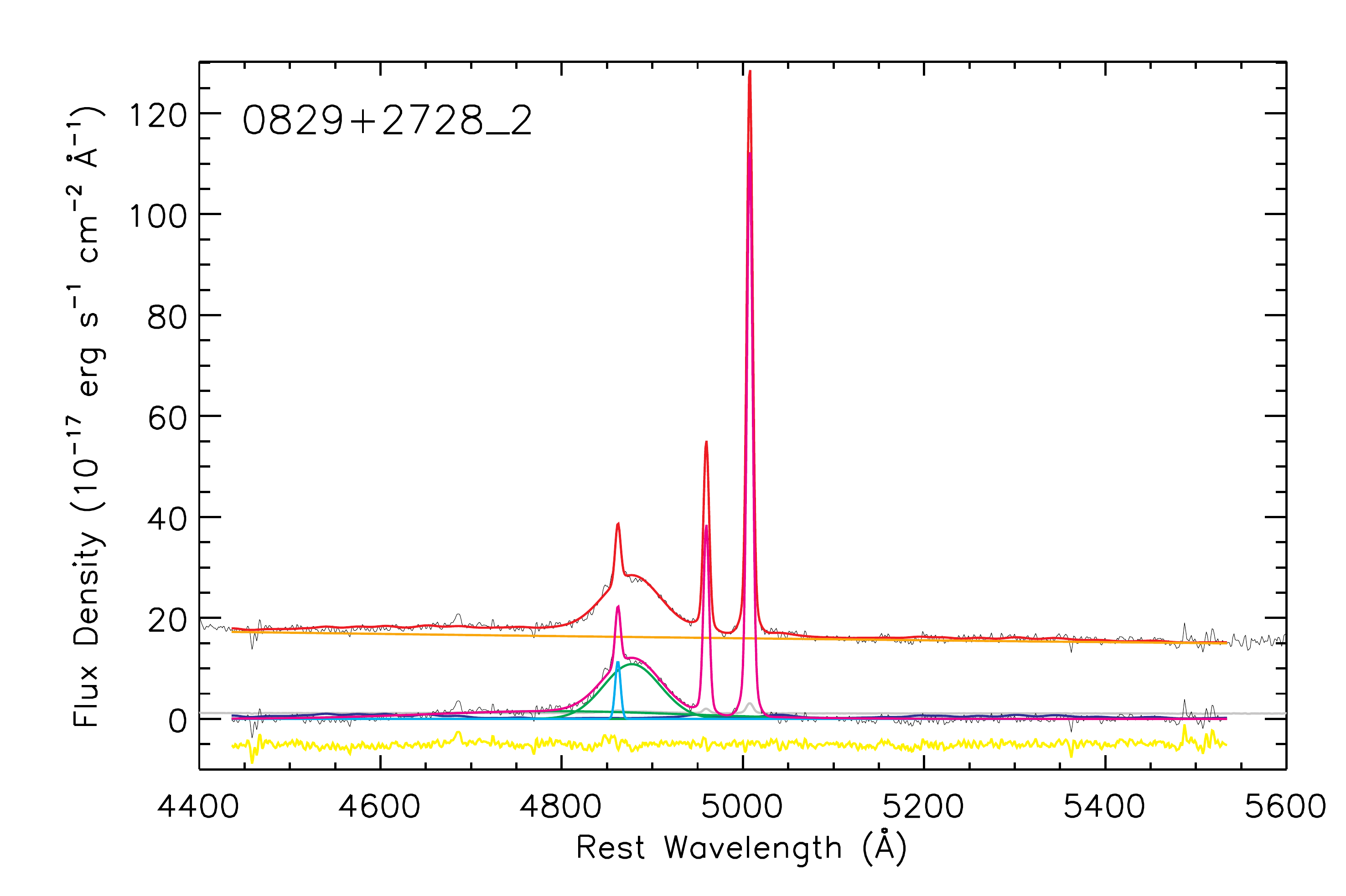}
    \includegraphics[width=85mm]{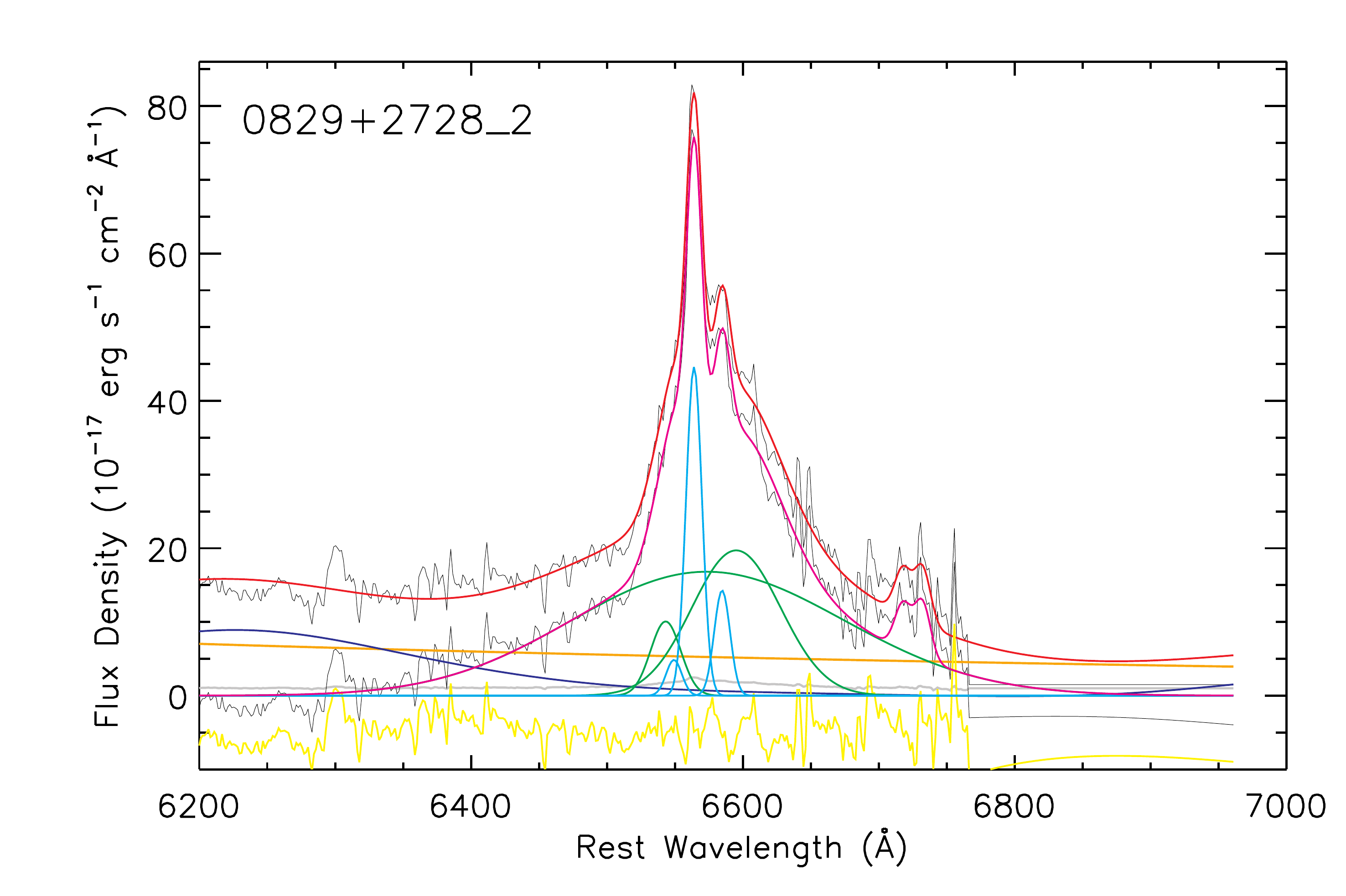}
    \includegraphics[width=85mm]{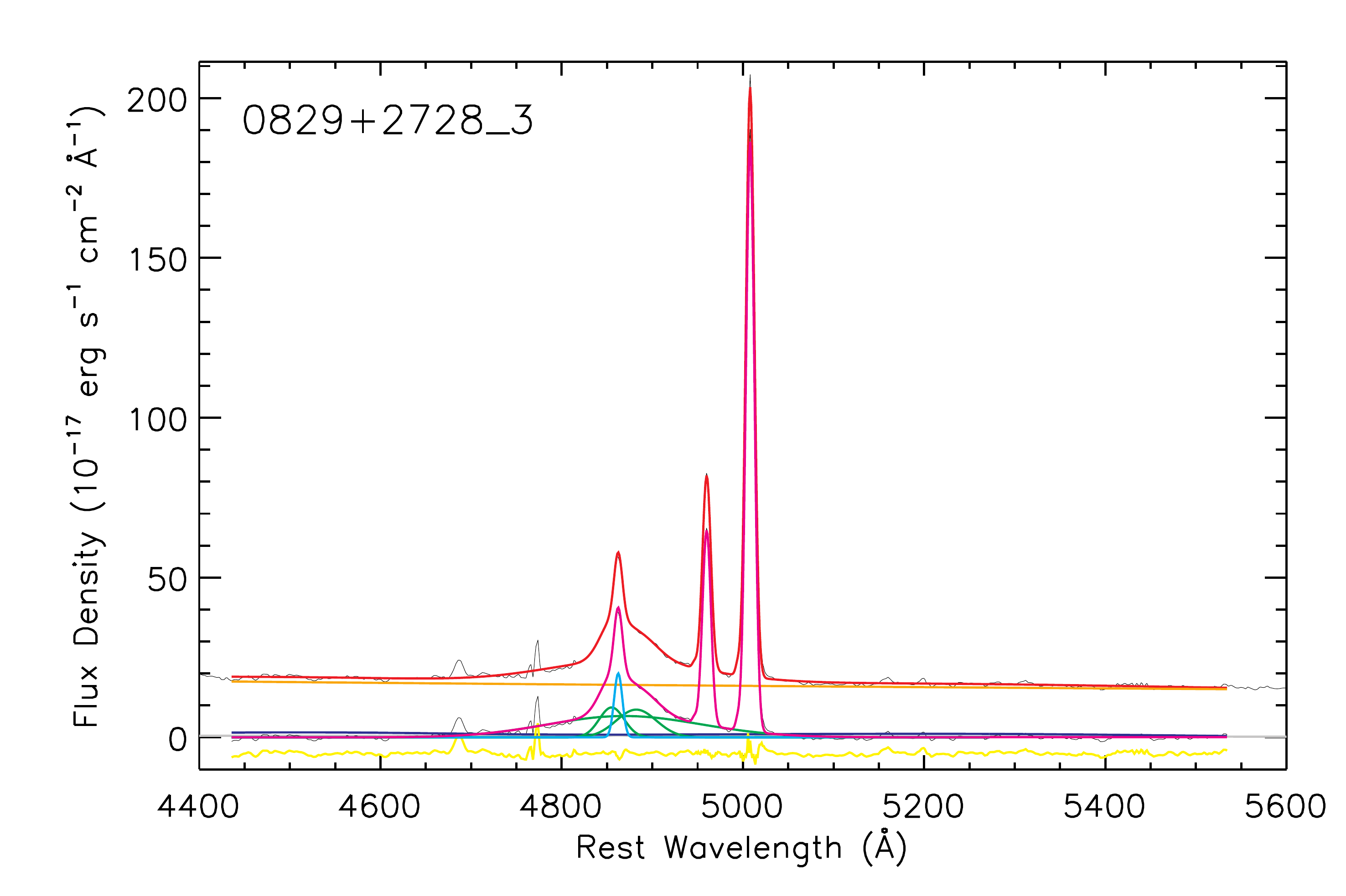}
    \includegraphics[width=85mm]{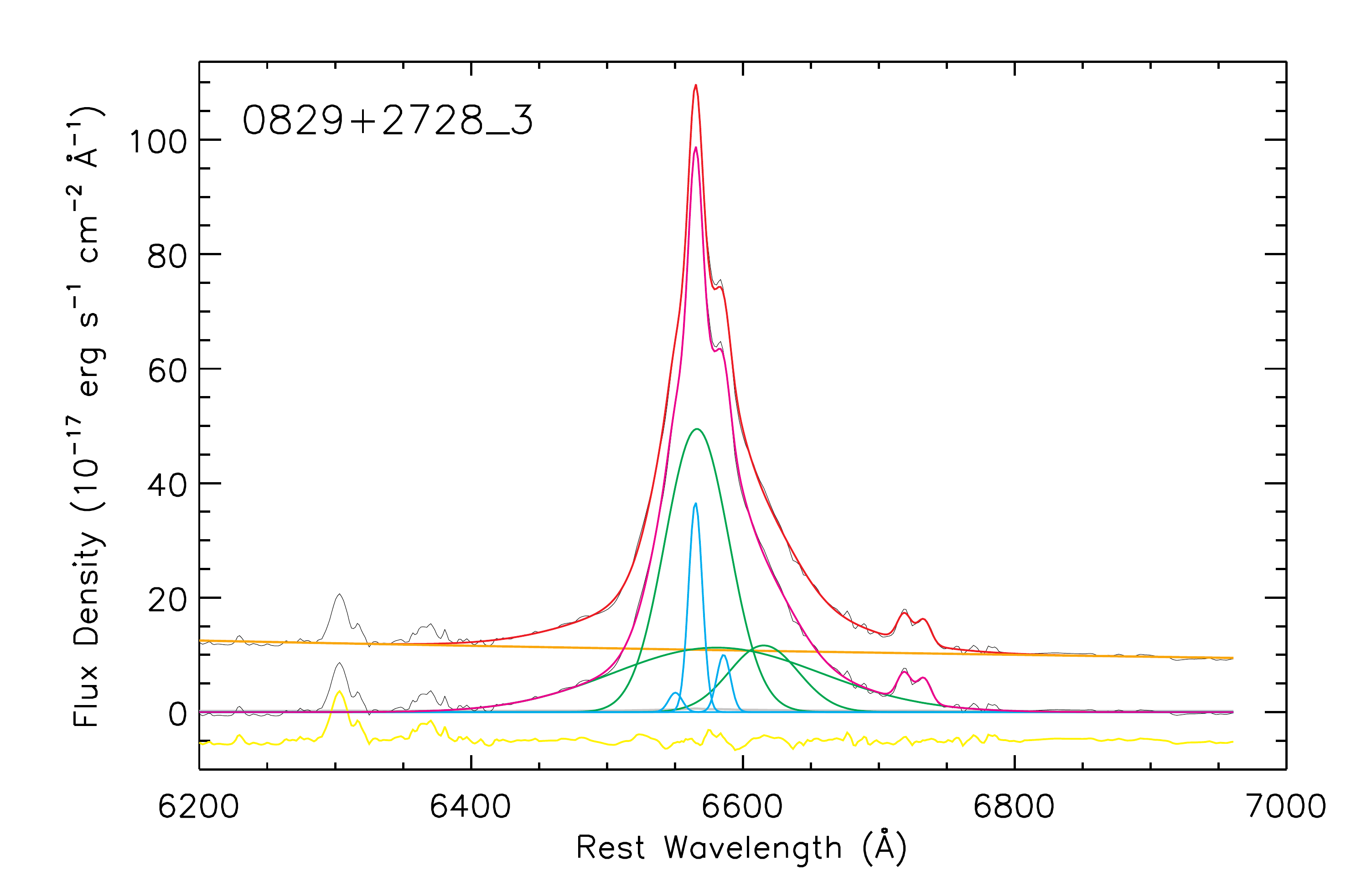}
    \label{fig:allspecfit}
    \caption{Similar to Figure \ref{fig:specfit_eg}, but for the other targets in our sample.}    
\end{figure*}
\clearpage

\begin{figure*}
  \centering
    \includegraphics[width=80mm]{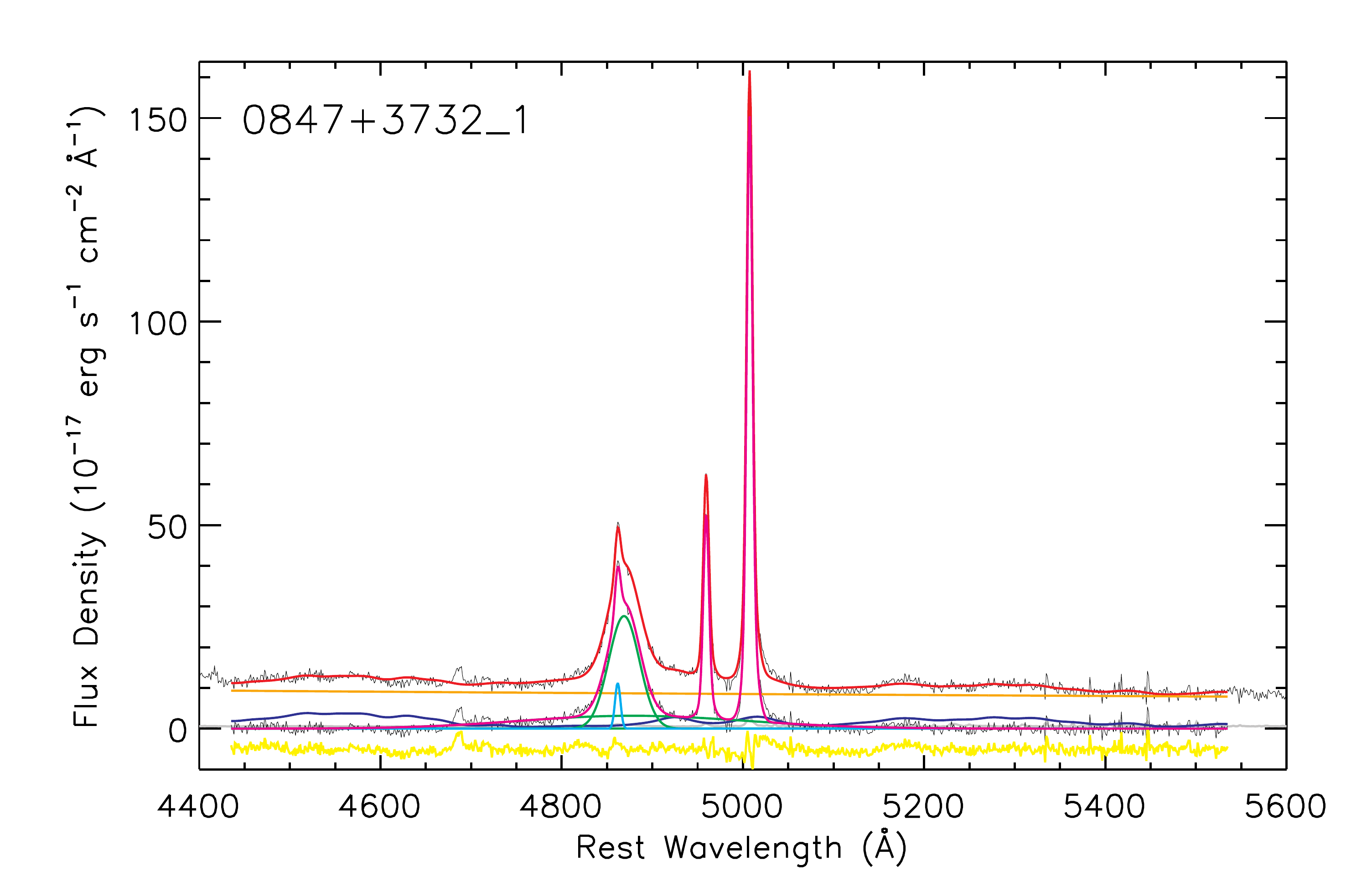}
    \includegraphics[width=80mm]{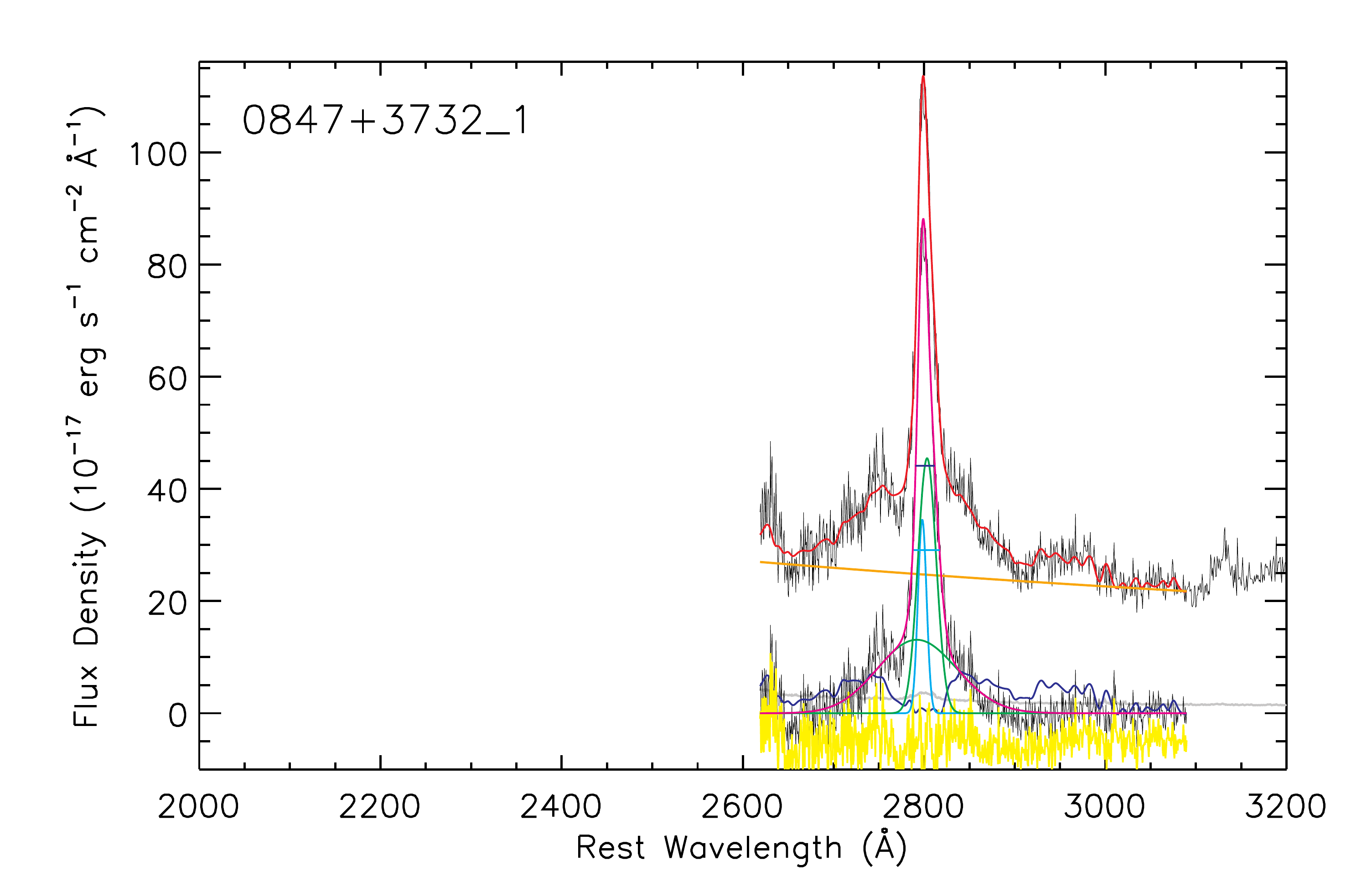}
    \includegraphics[width=80mm]{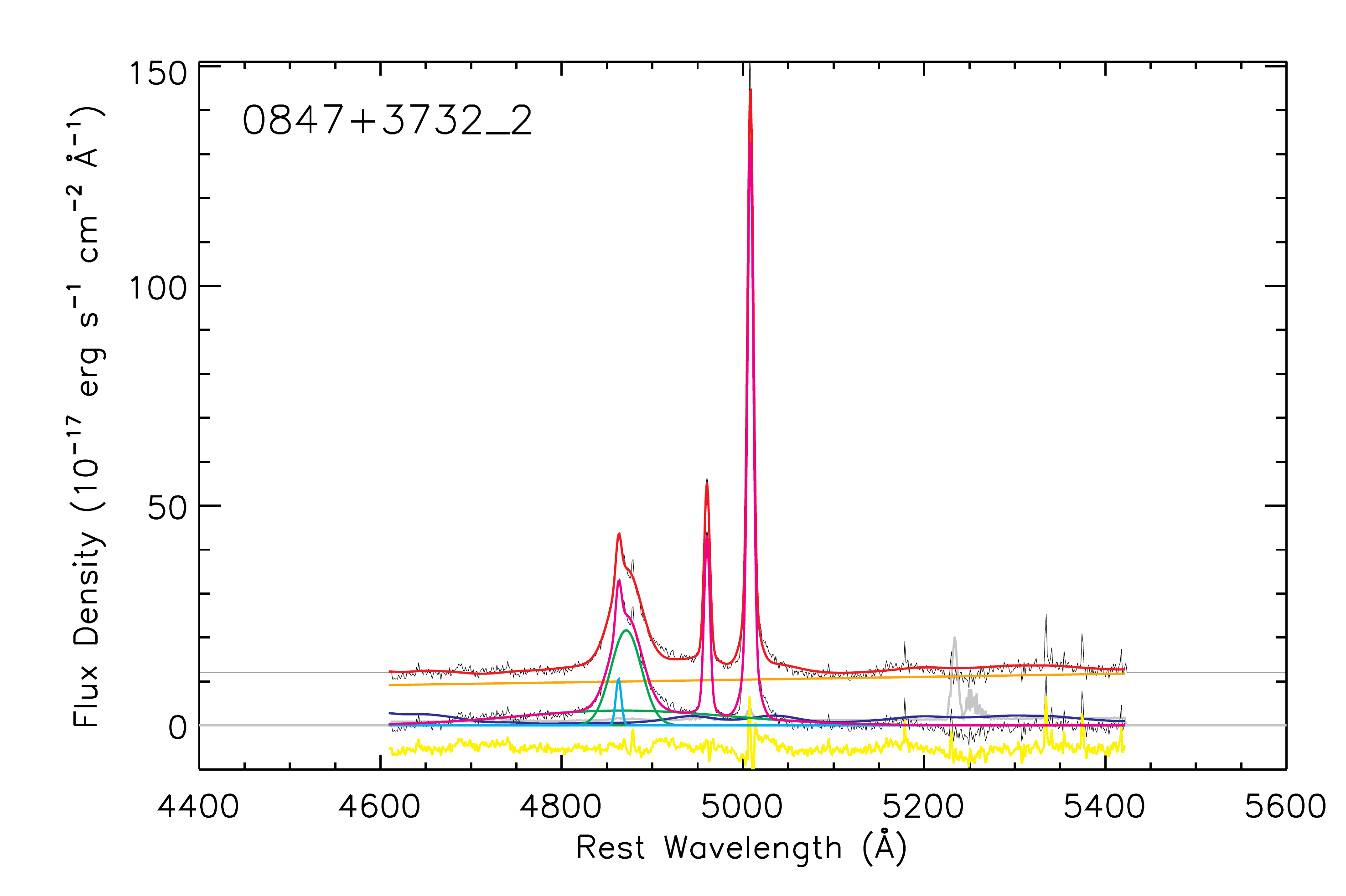}
    \includegraphics[width=80mm]{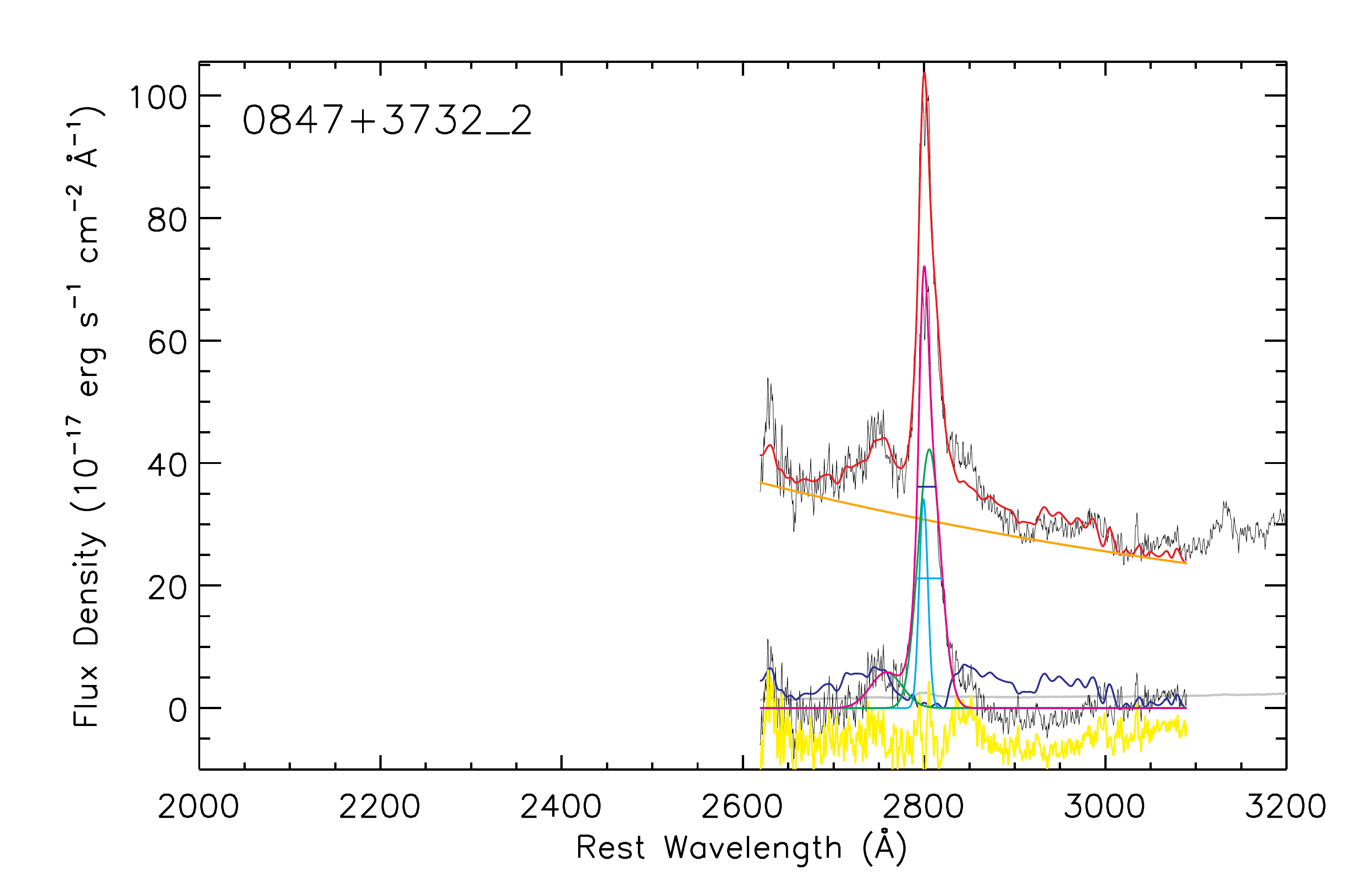}
    \includegraphics[width=80mm]{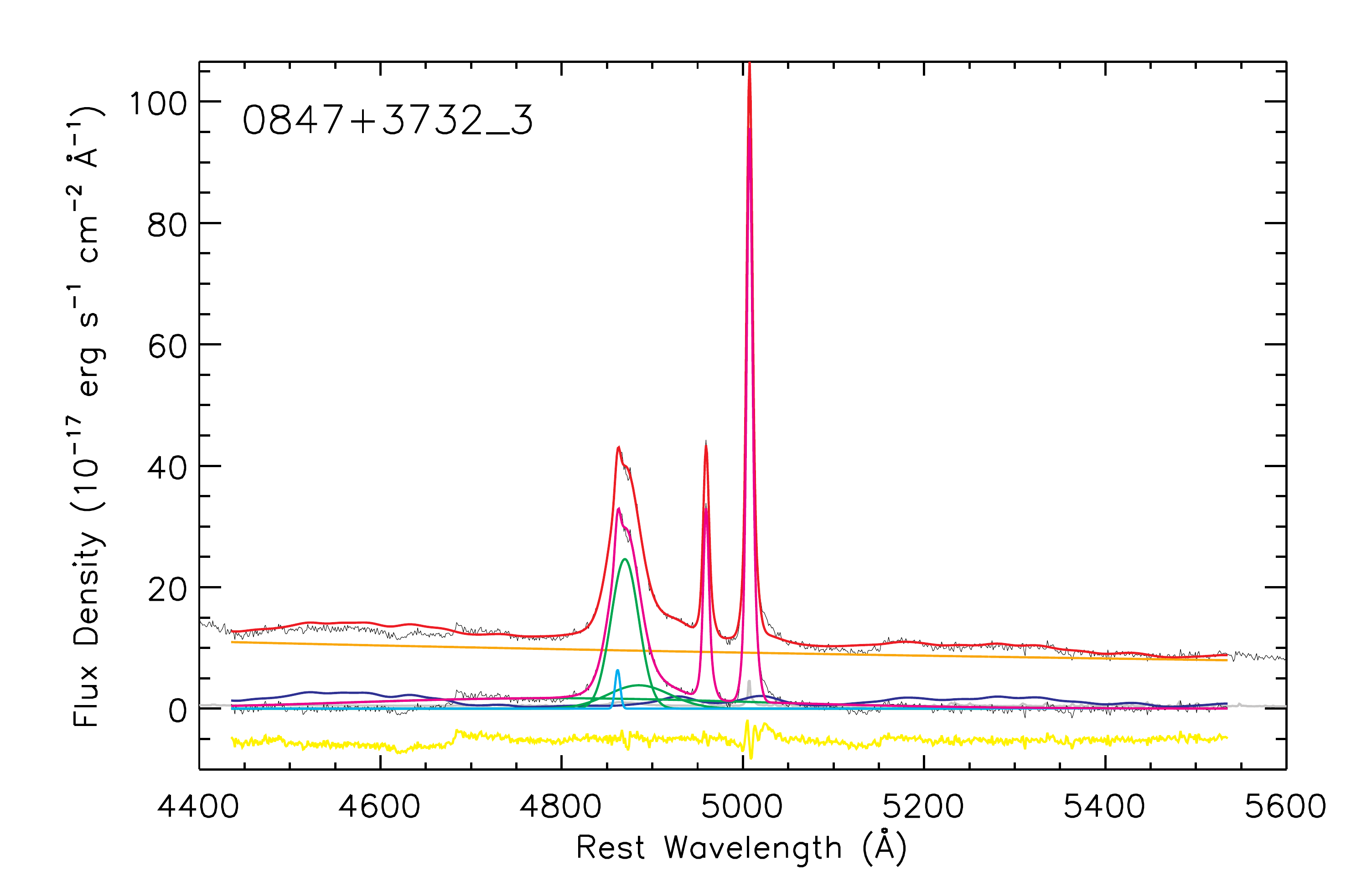}
    \includegraphics[width=80mm]{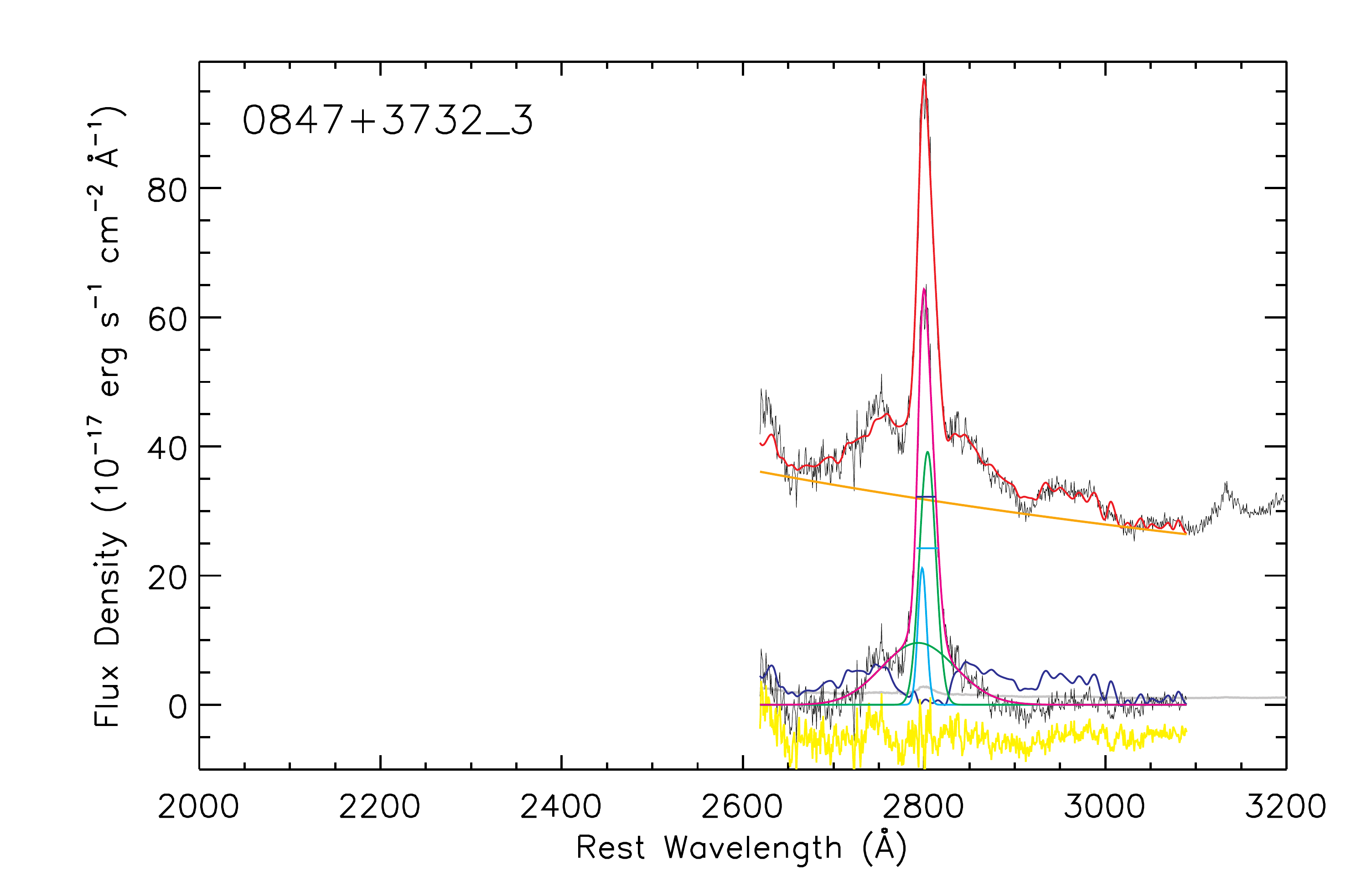}
    \includegraphics[width=80mm]{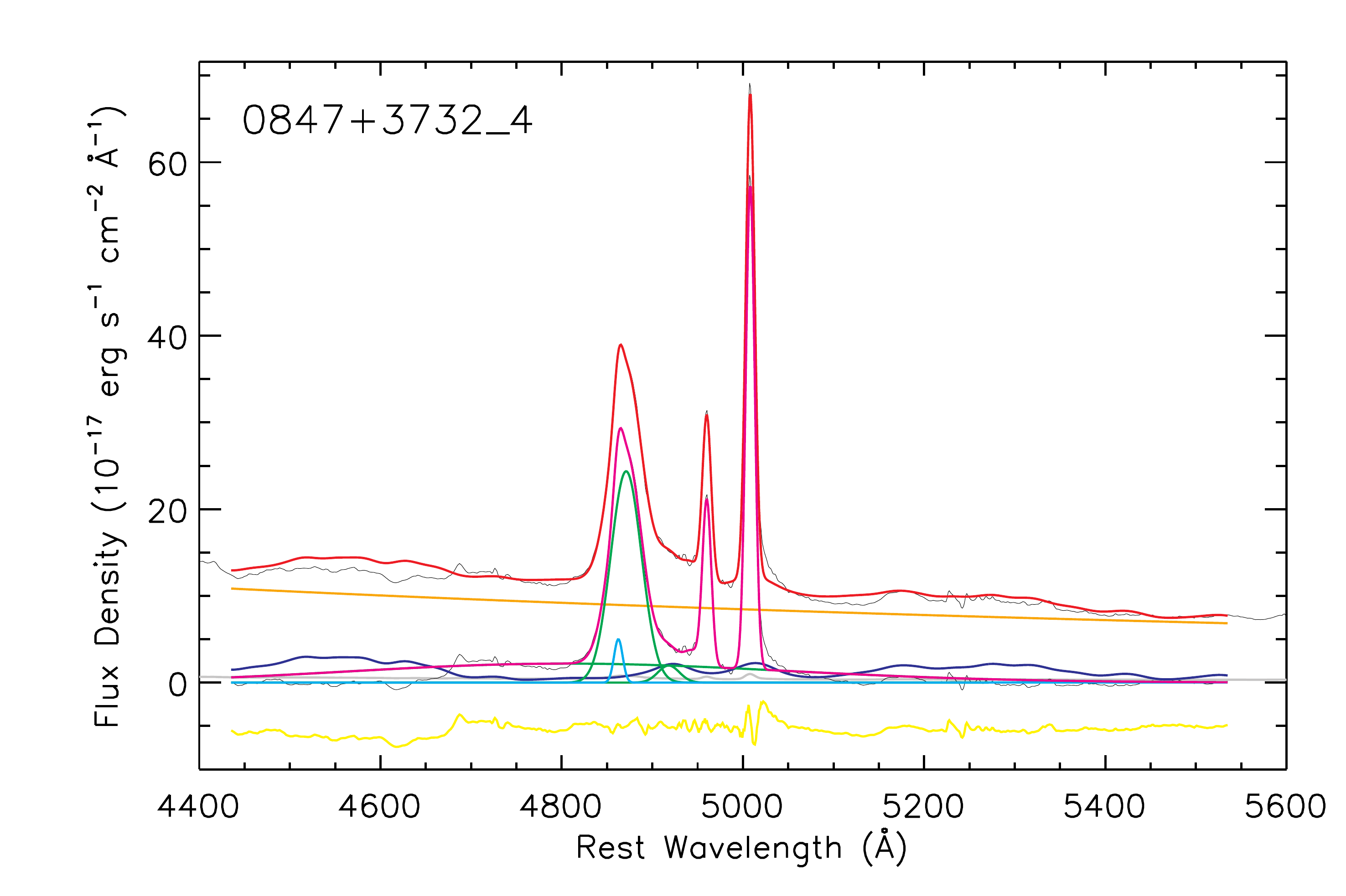}
    \includegraphics[width=80mm]{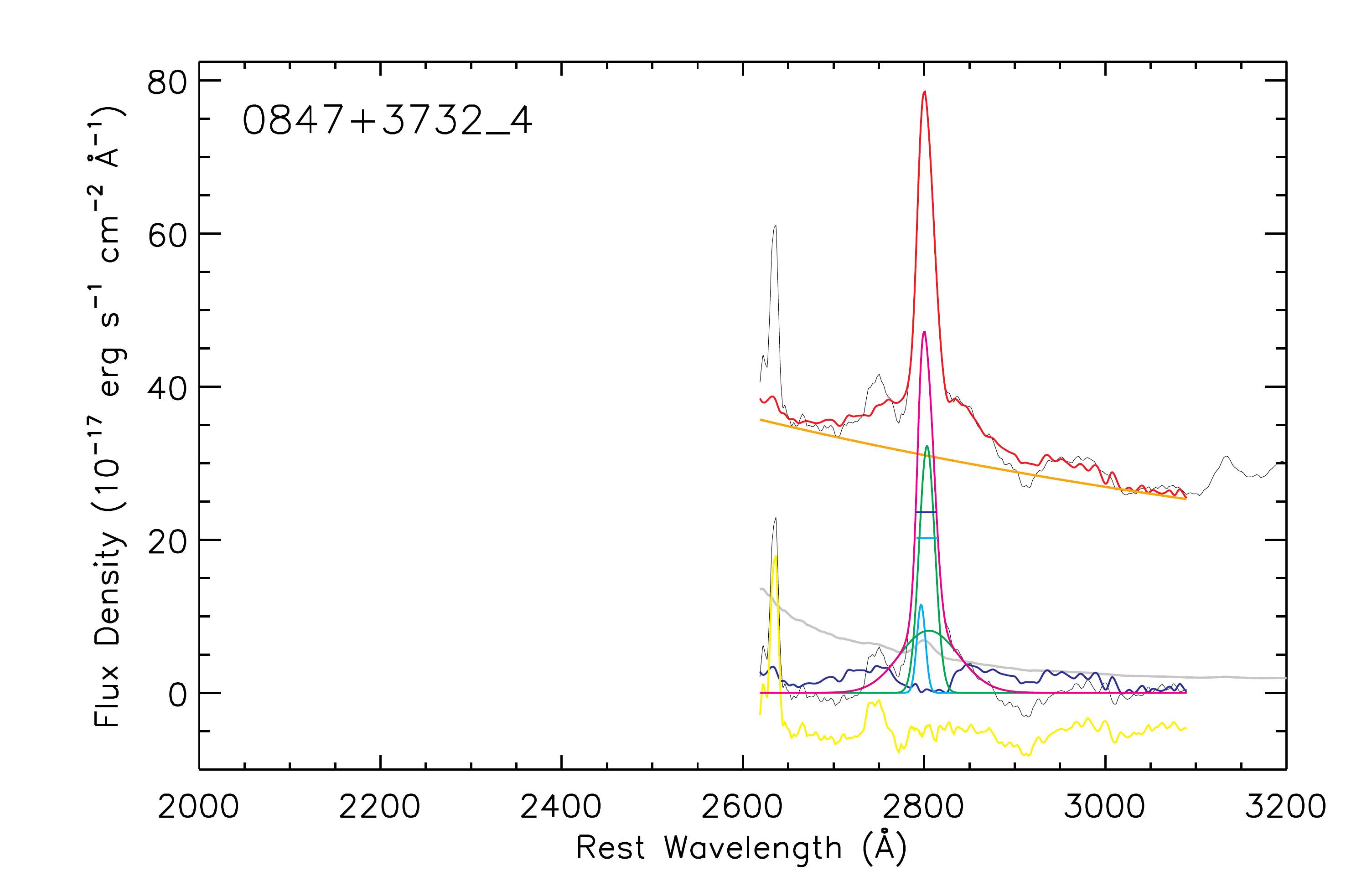}
       \centerline{Figure A1 -- Continued.    }     
\end{figure*}
\clearpage

\begin{figure*}
  \centering
    \includegraphics[width=80mm]{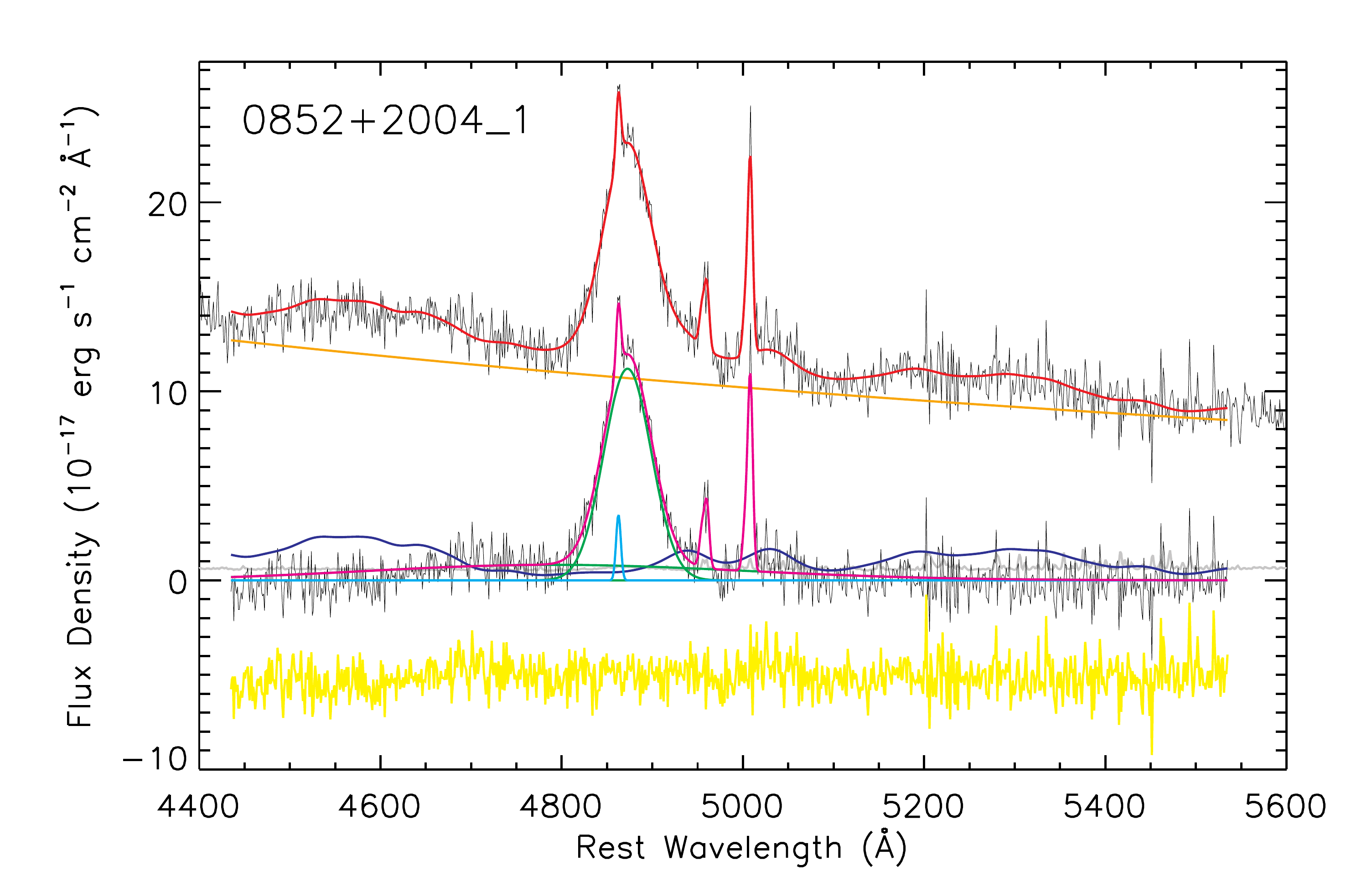}
    \includegraphics[width=80mm]{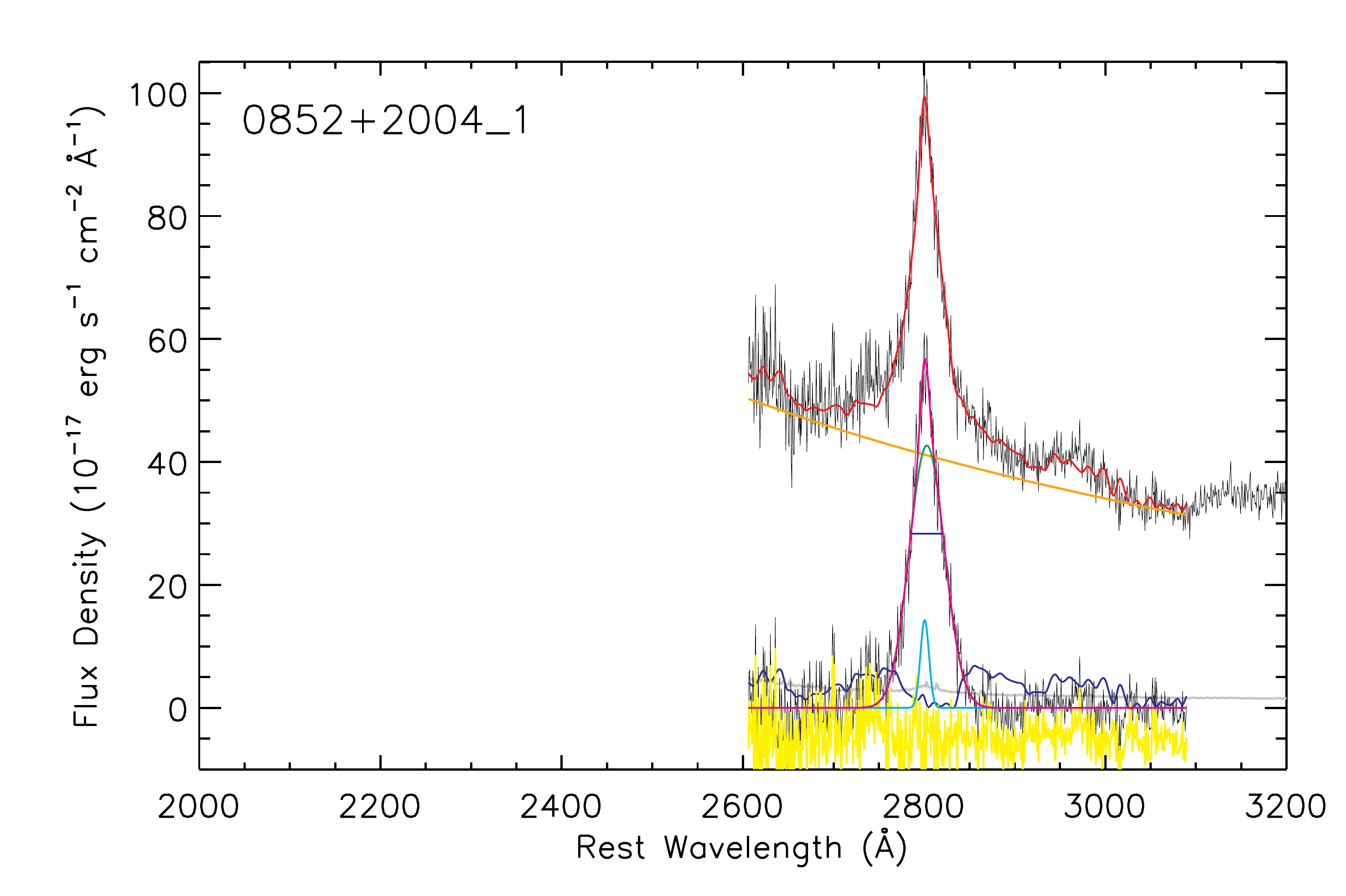}
    \includegraphics[width=80mm]{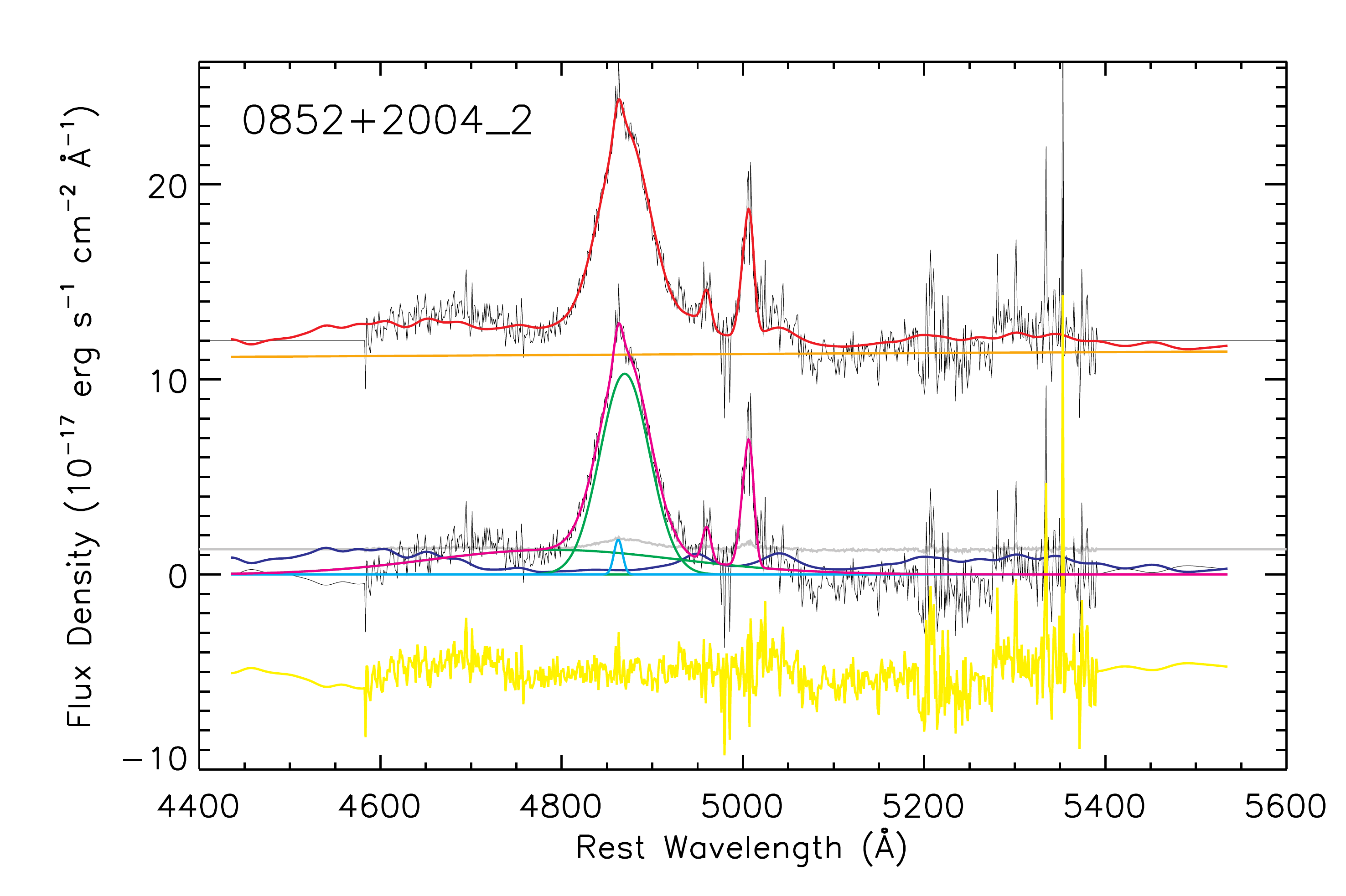}
    \includegraphics[width=80mm]{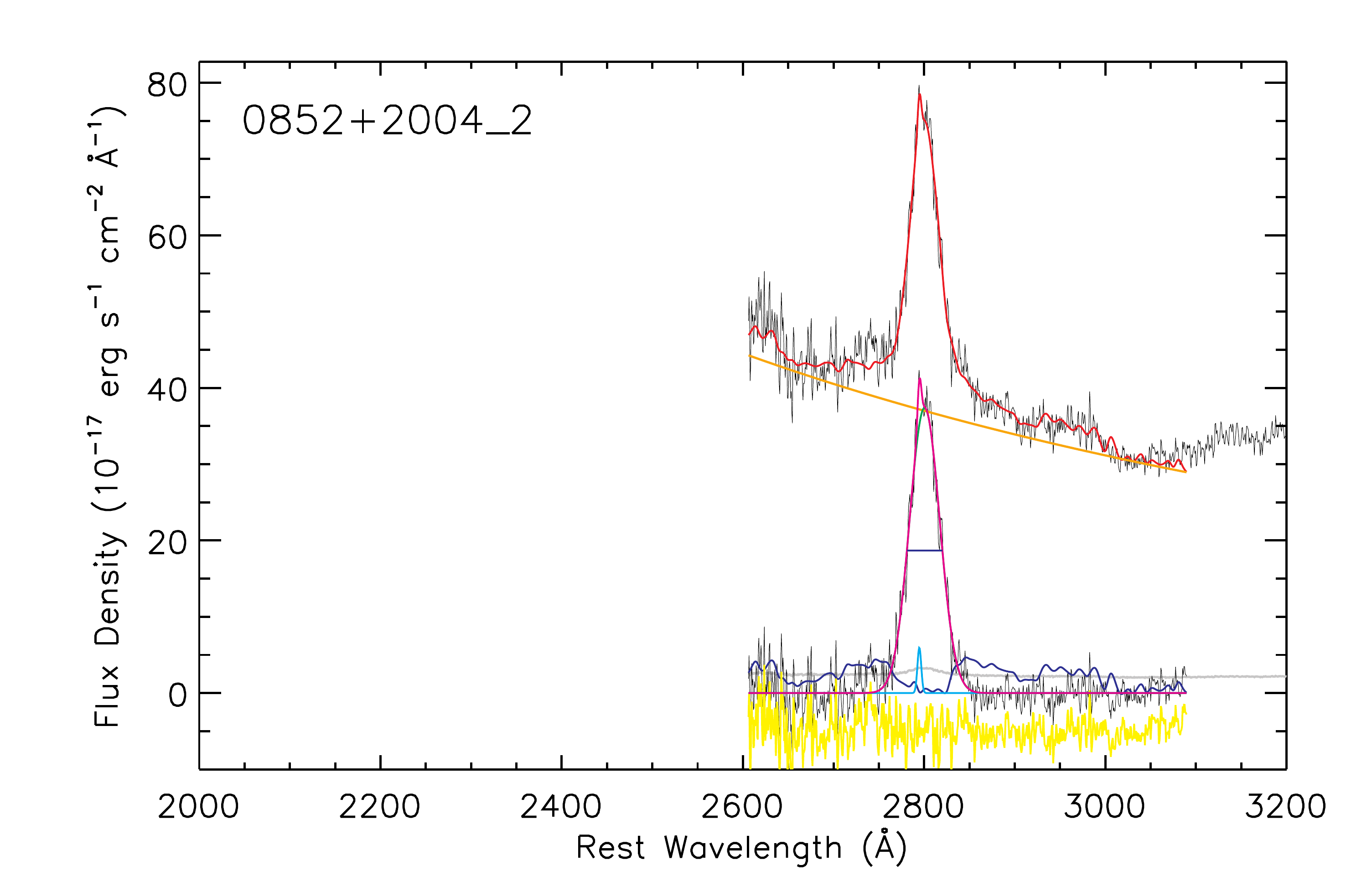}
    \includegraphics[width=80mm]{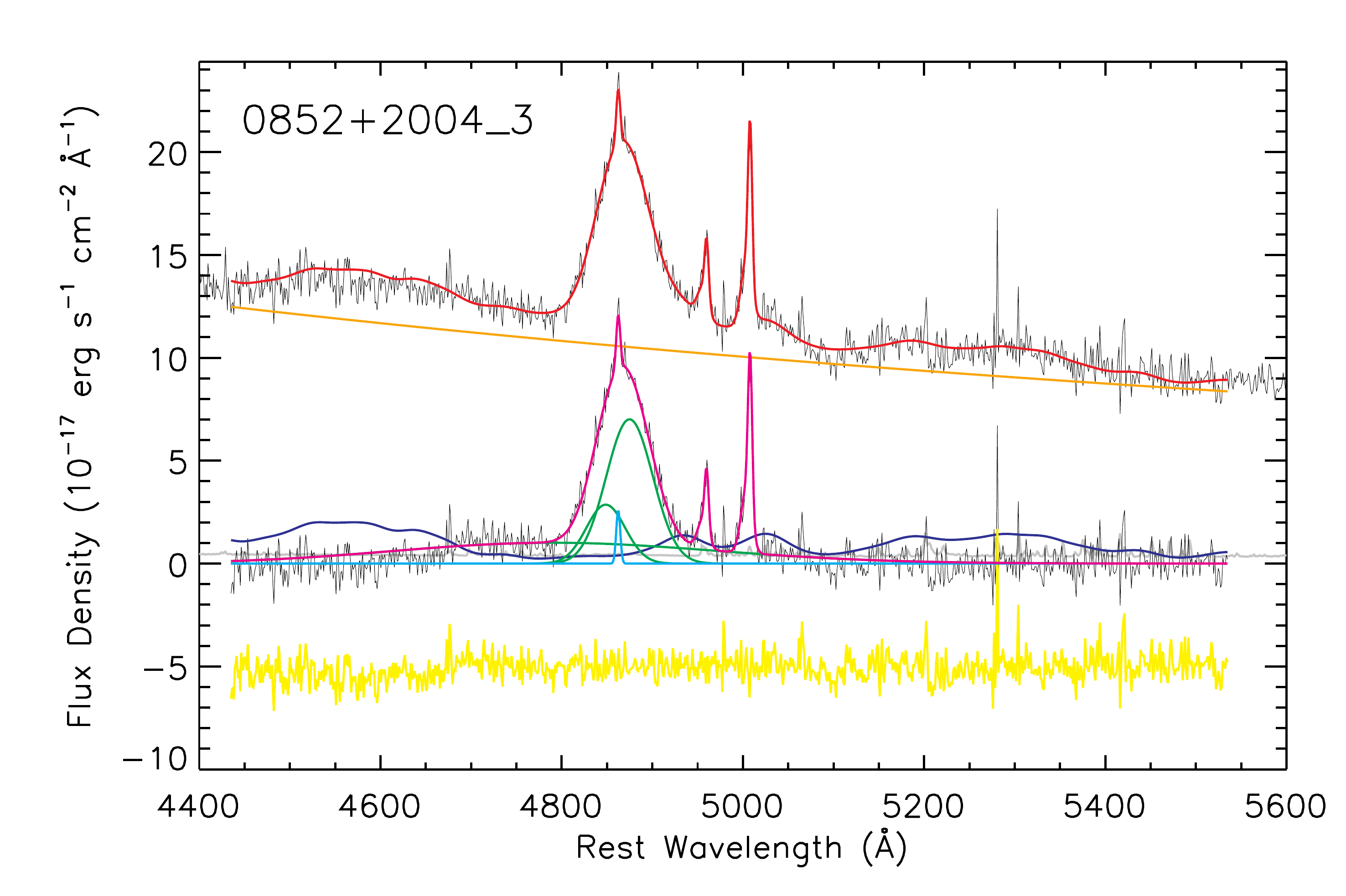}
    \includegraphics[width=80mm]{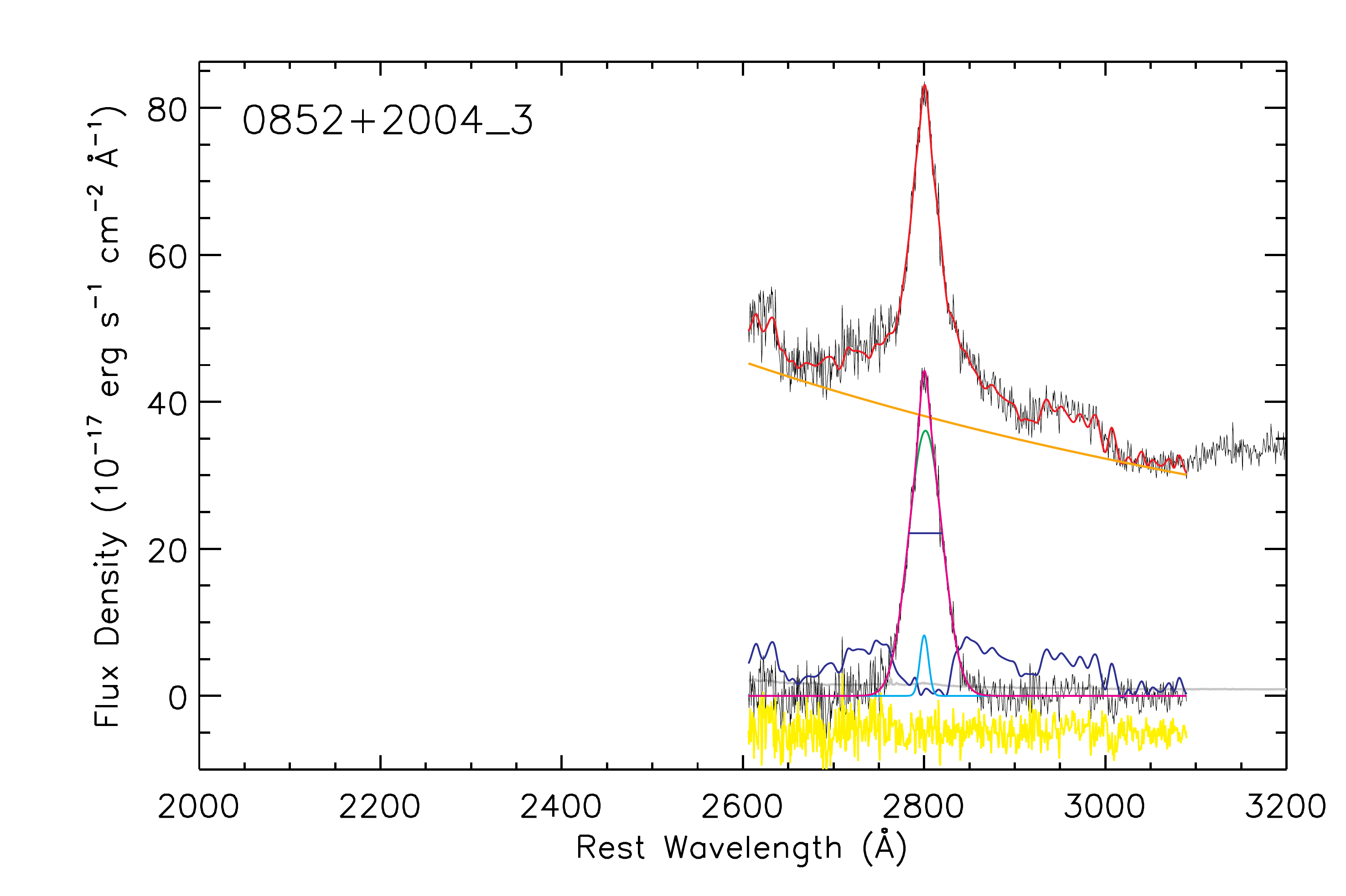}
    \includegraphics[width=80mm]{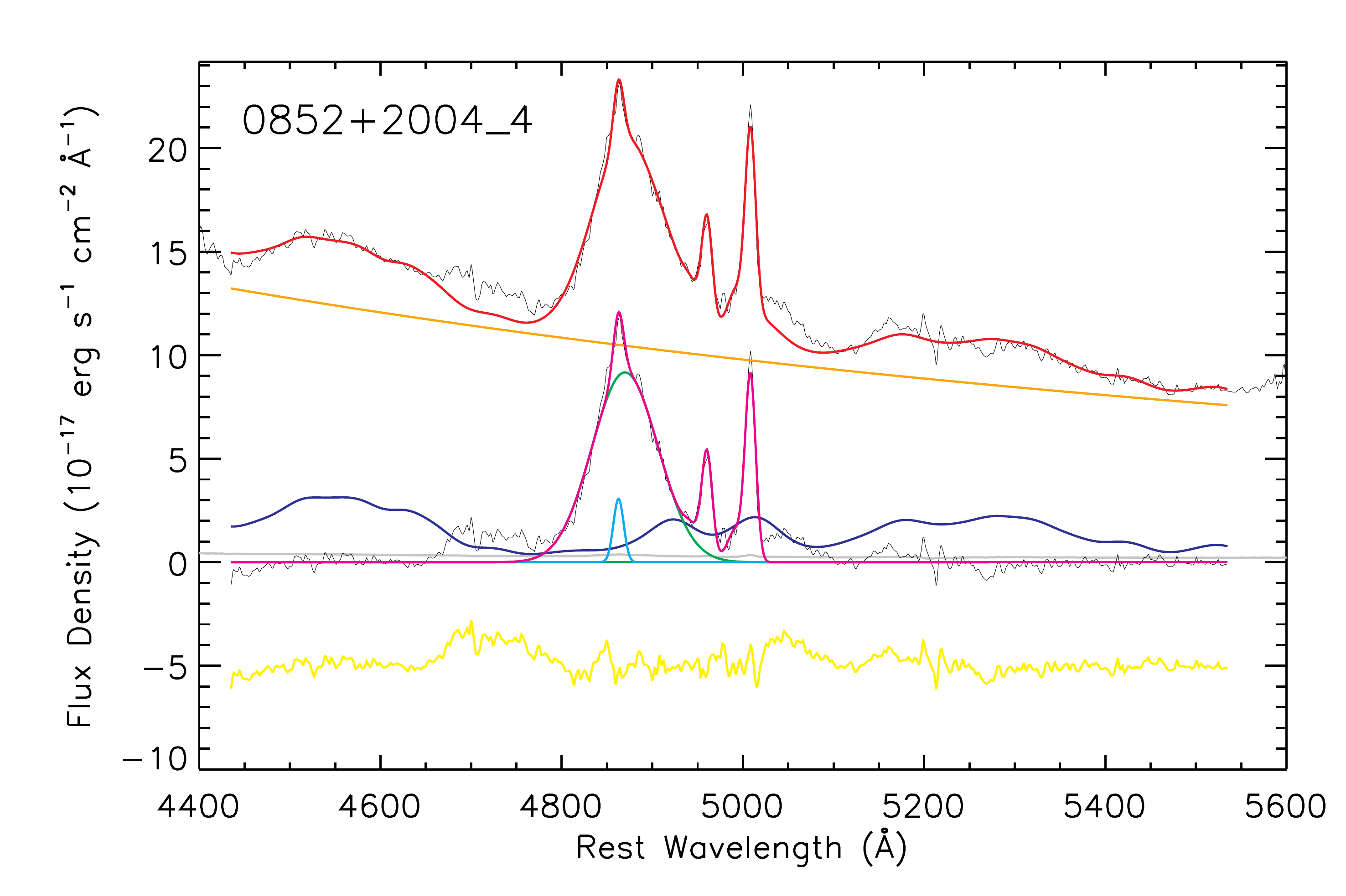}
    \includegraphics[width=80mm]{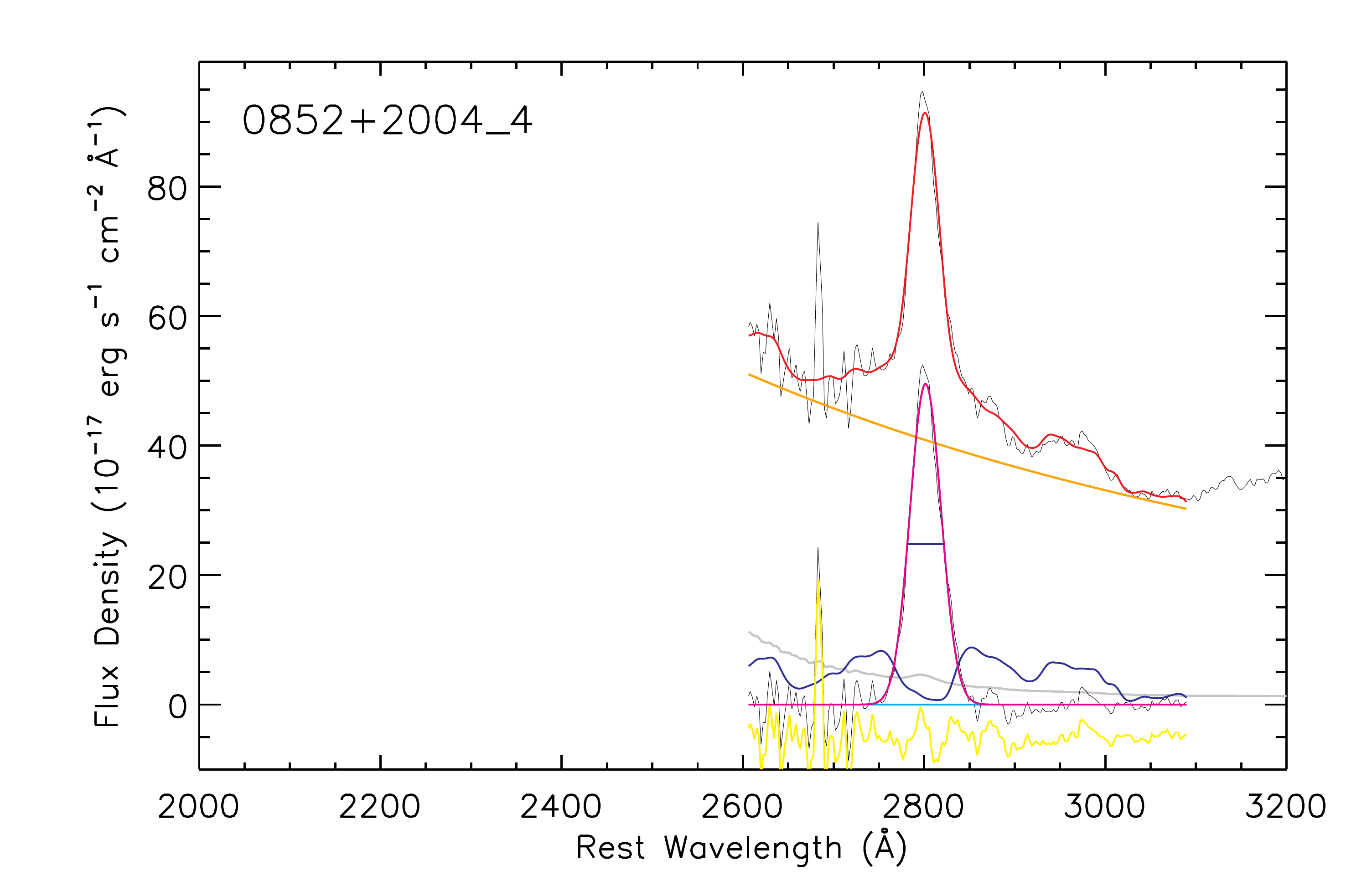}
       \centerline{Figure A1 -- Continued.    }    
\end{figure*}

\begin{figure*}
  \centering
    \includegraphics[width=85mm]{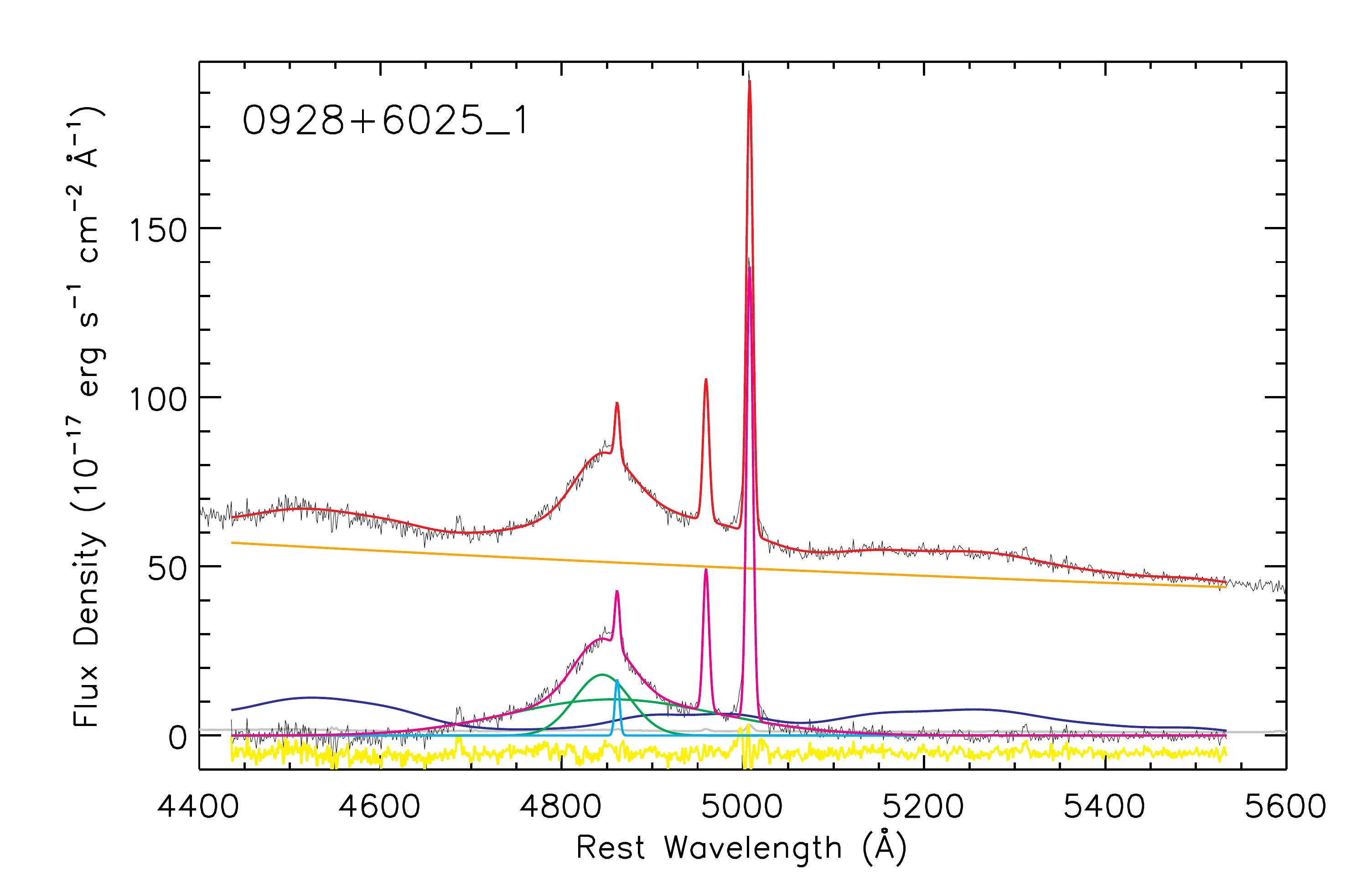}
    \includegraphics[width=85mm]{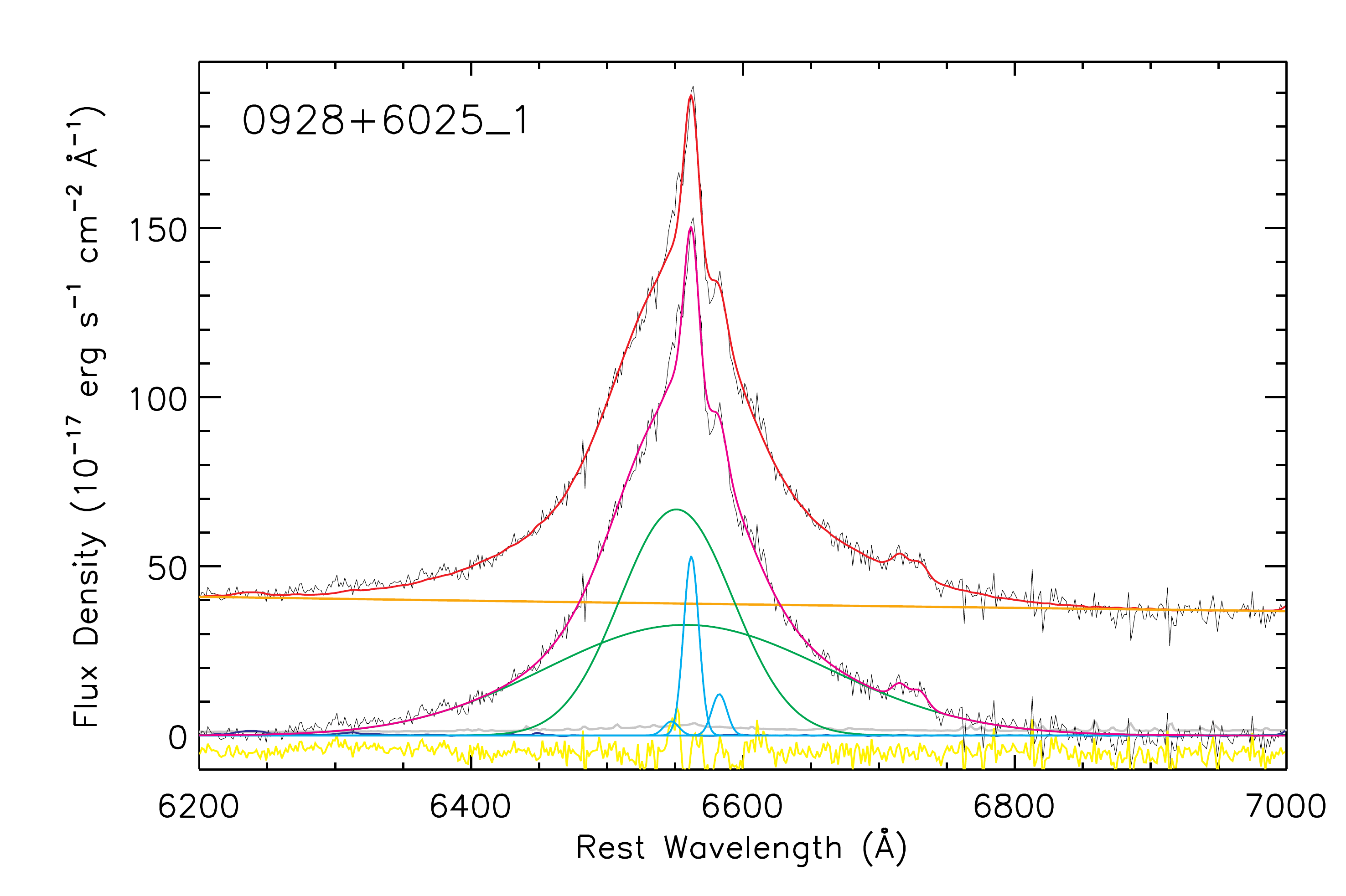}
    \includegraphics[width=85mm]{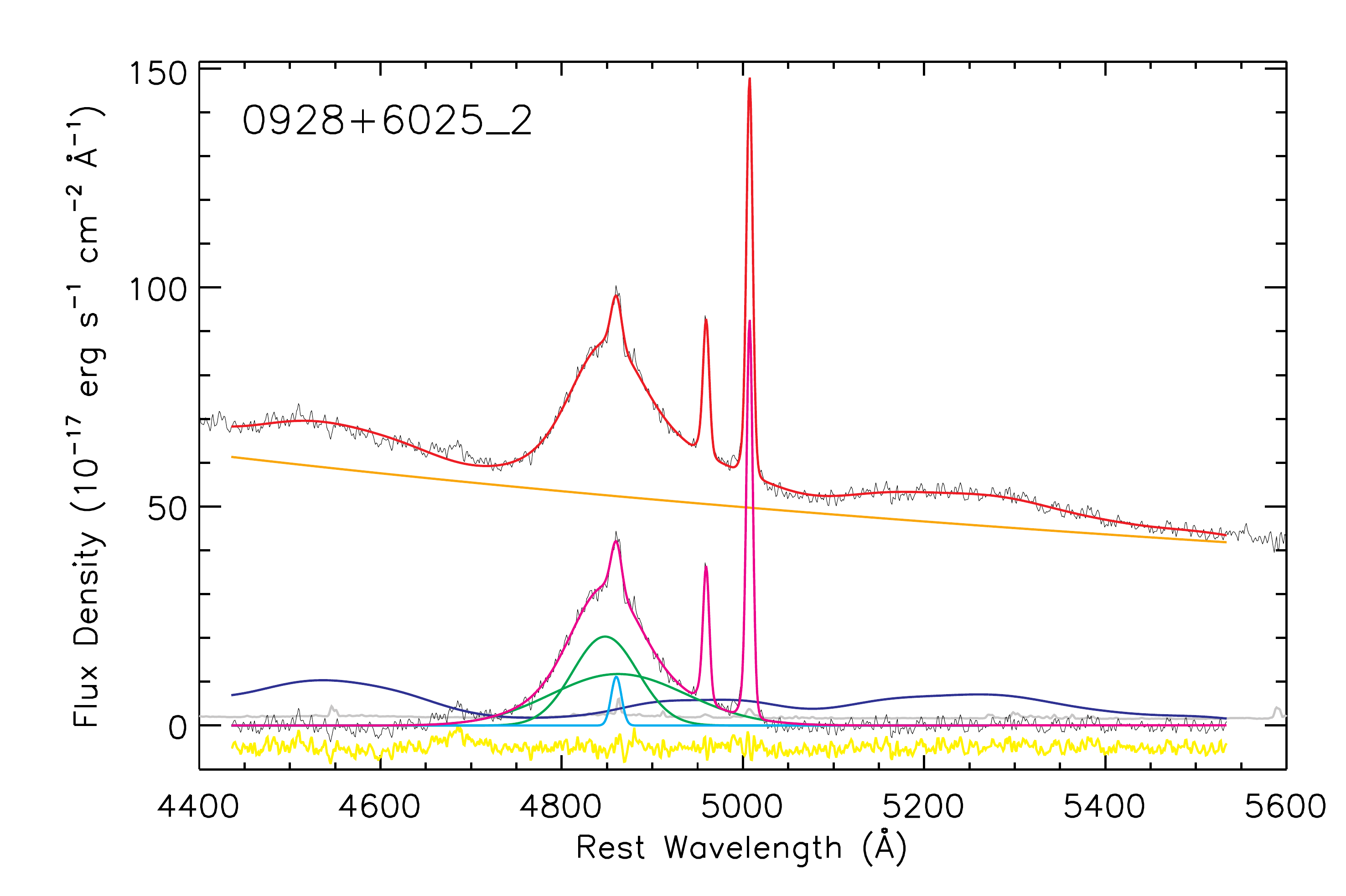}
    \includegraphics[width=85mm]{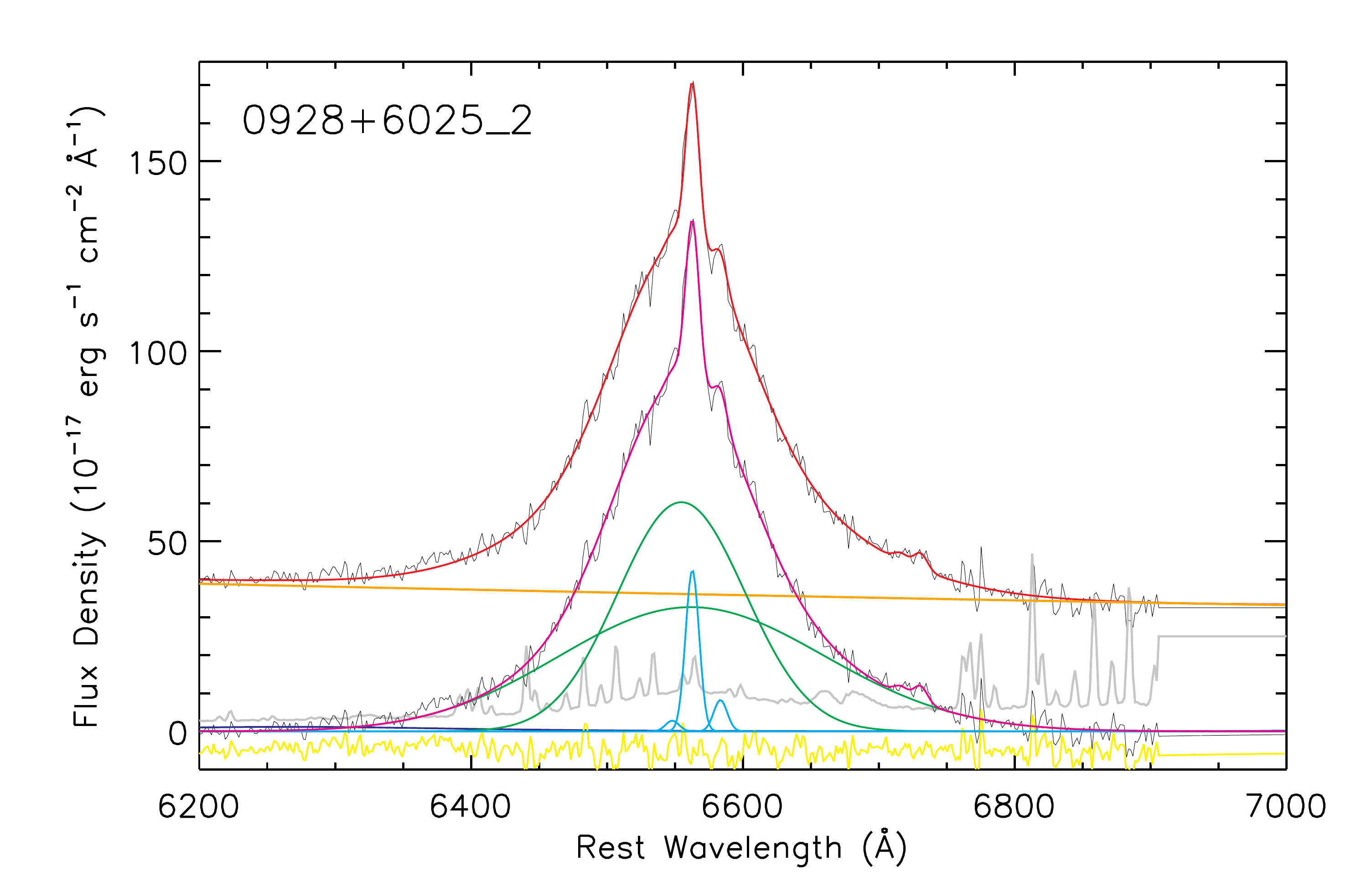}
    \includegraphics[width=85mm]{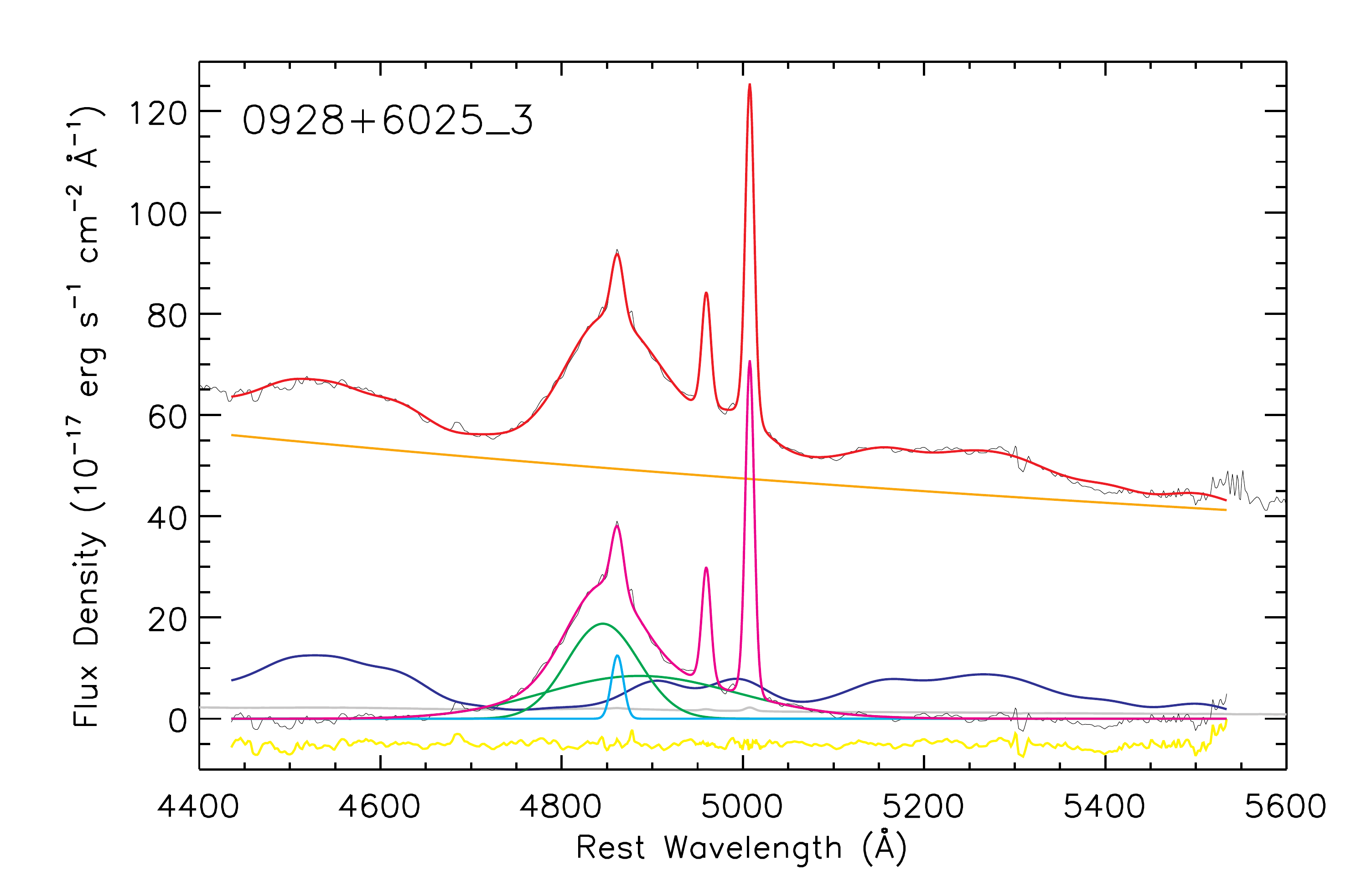}
    \includegraphics[width=85mm]{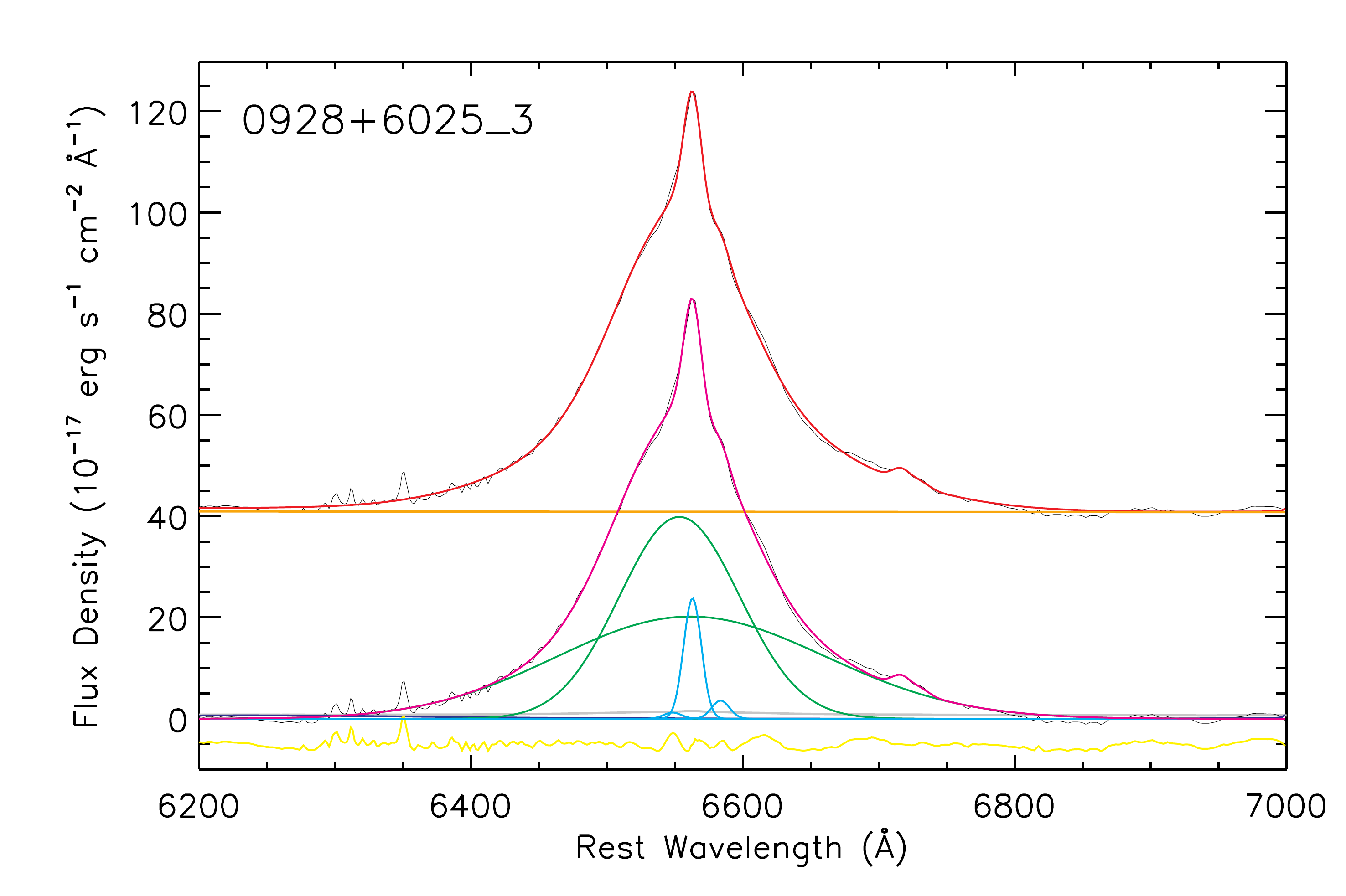}
       \centerline{Figure A1 -- Continued.    }
\end{figure*}
 \clearpage

\begin{figure*}
  \centering 
    \includegraphics[width=85mm]{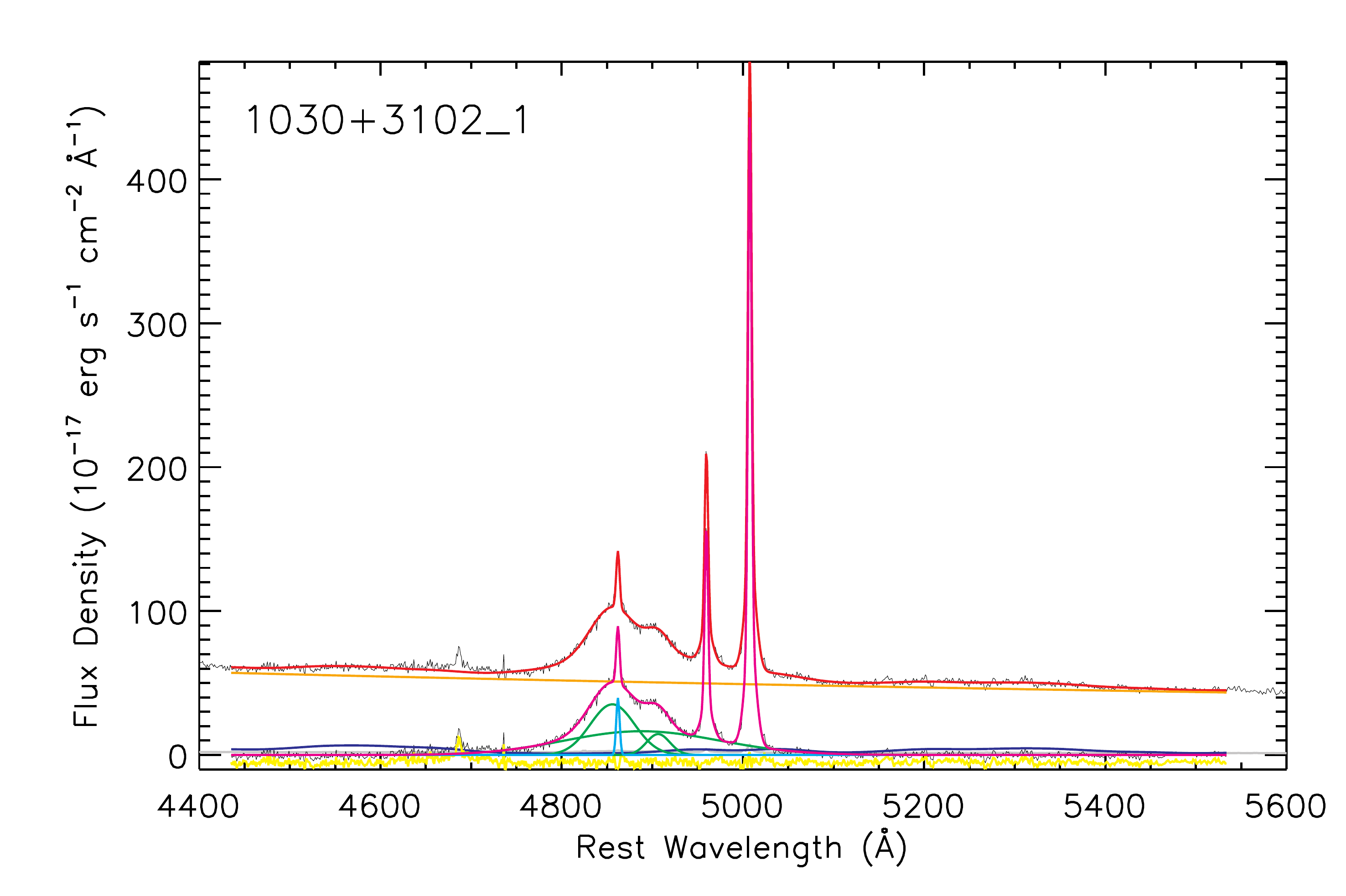}
    \includegraphics[width=85mm]{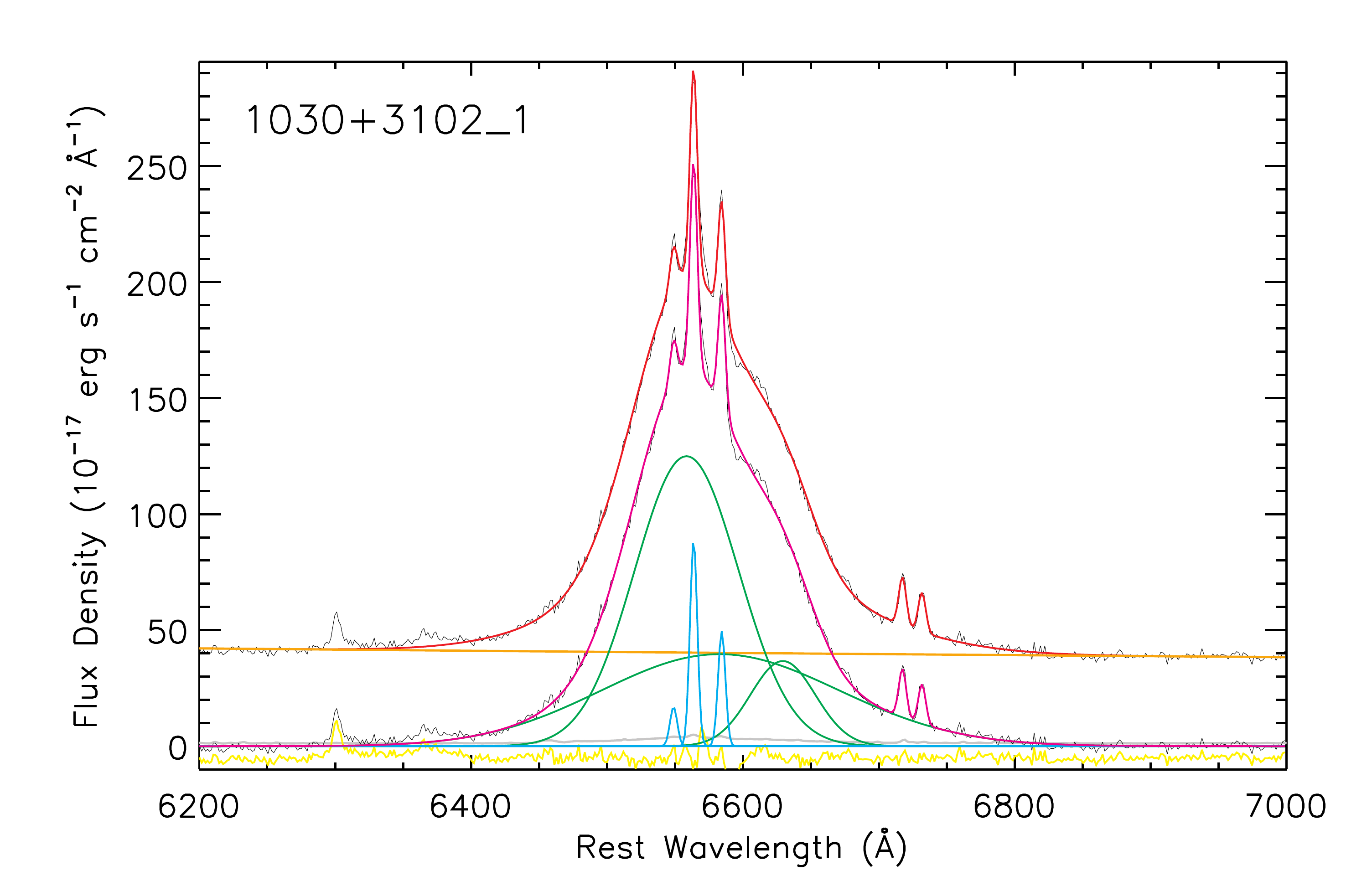}
    \includegraphics[width=85mm]{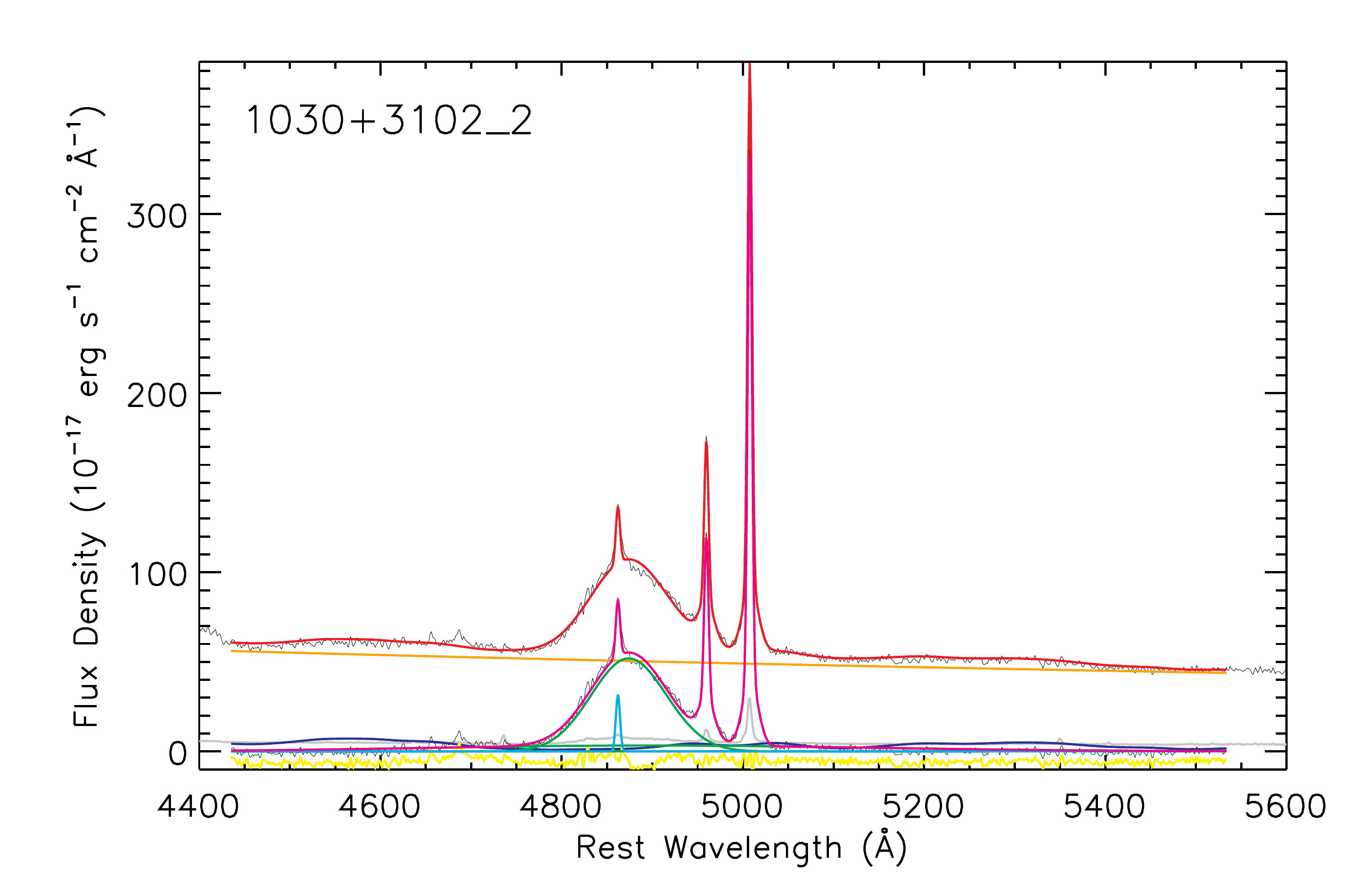}
    \includegraphics[width=85mm]{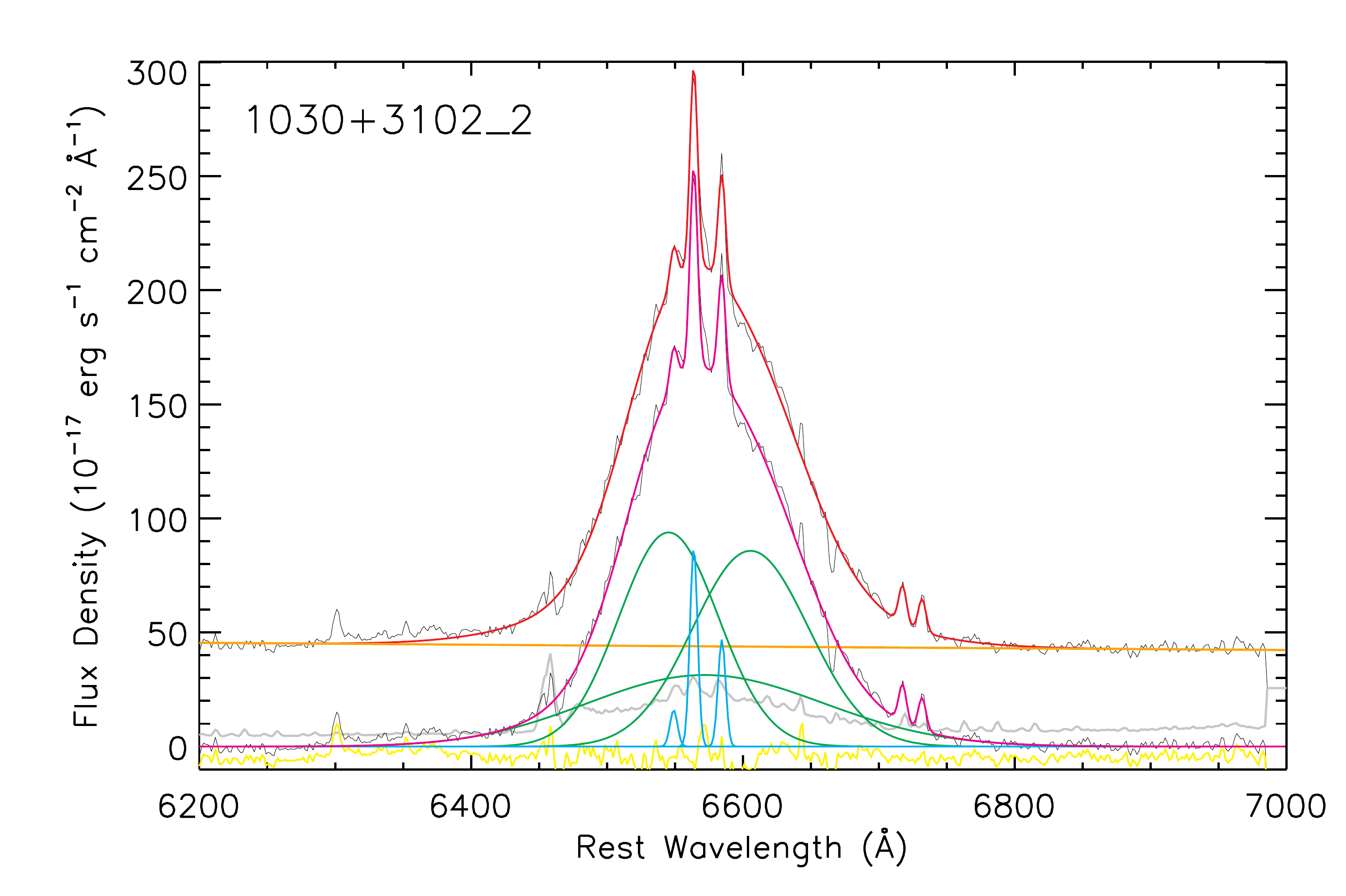}
    \includegraphics[width=85mm]{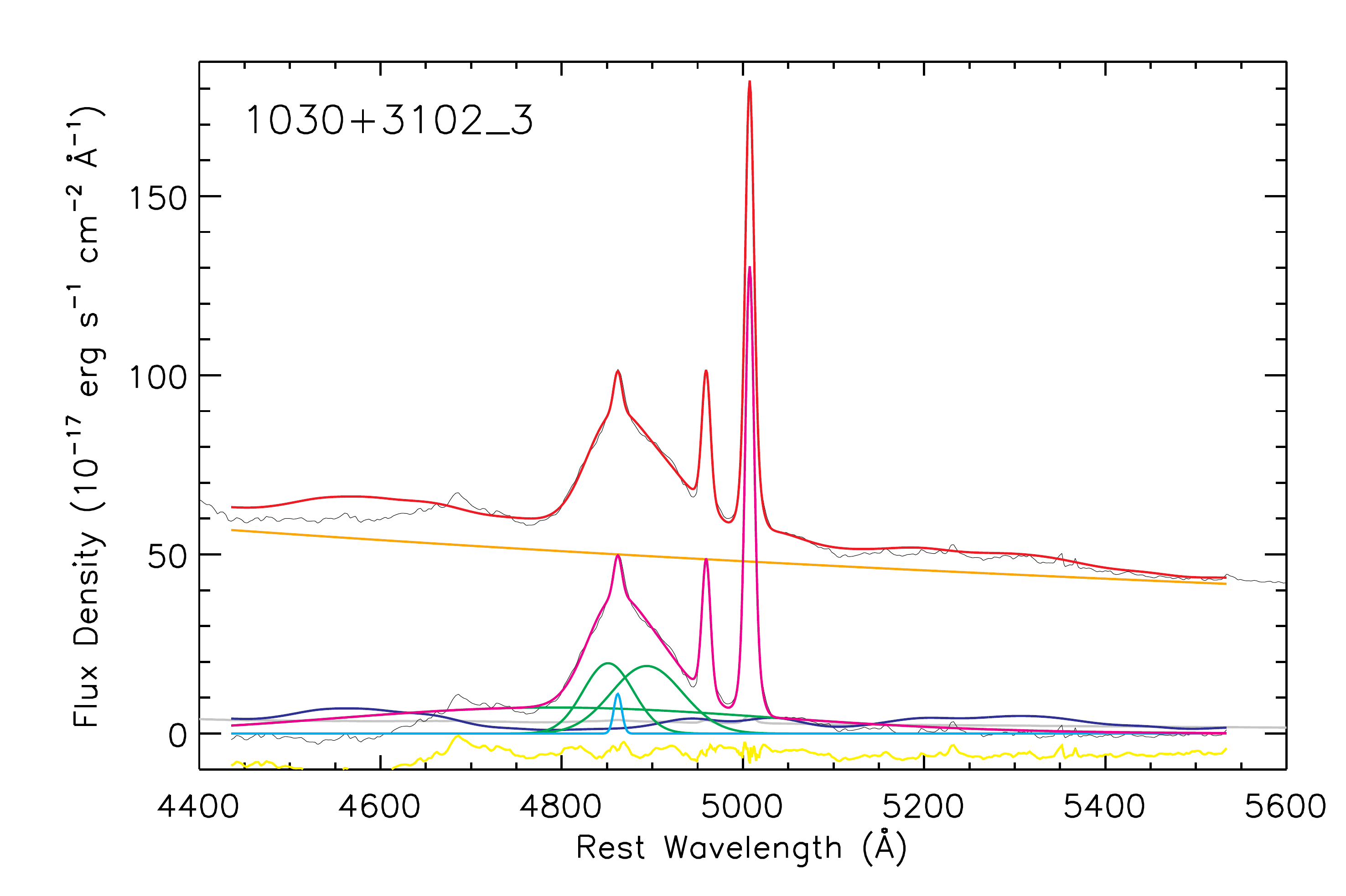}
    \includegraphics[width=85mm]{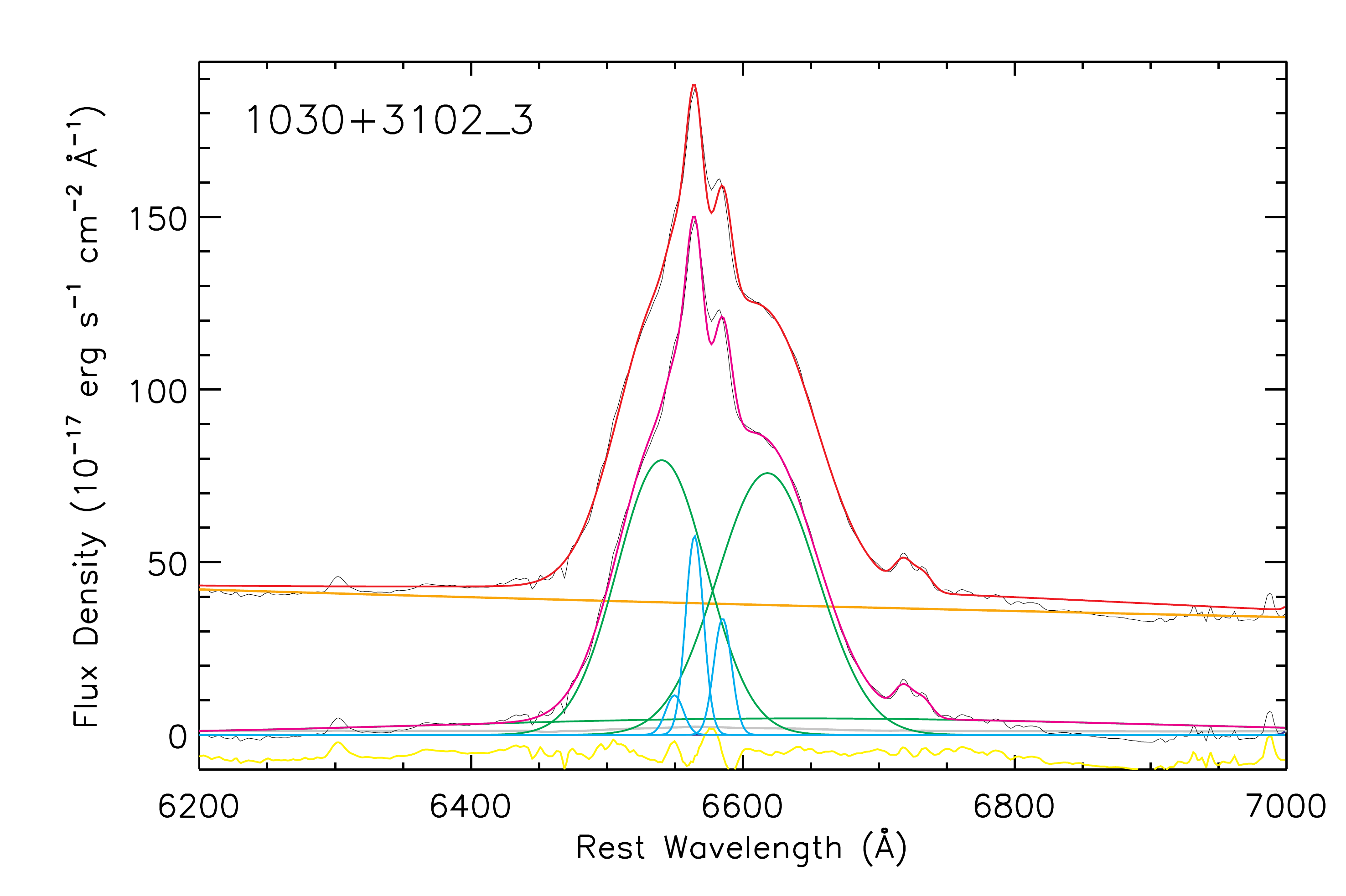}
       \centerline{Figure A1 -- Continued.    } 
\end{figure*}
\clearpage

\begin{figure*}
  \centering
    \includegraphics[width=85mm]{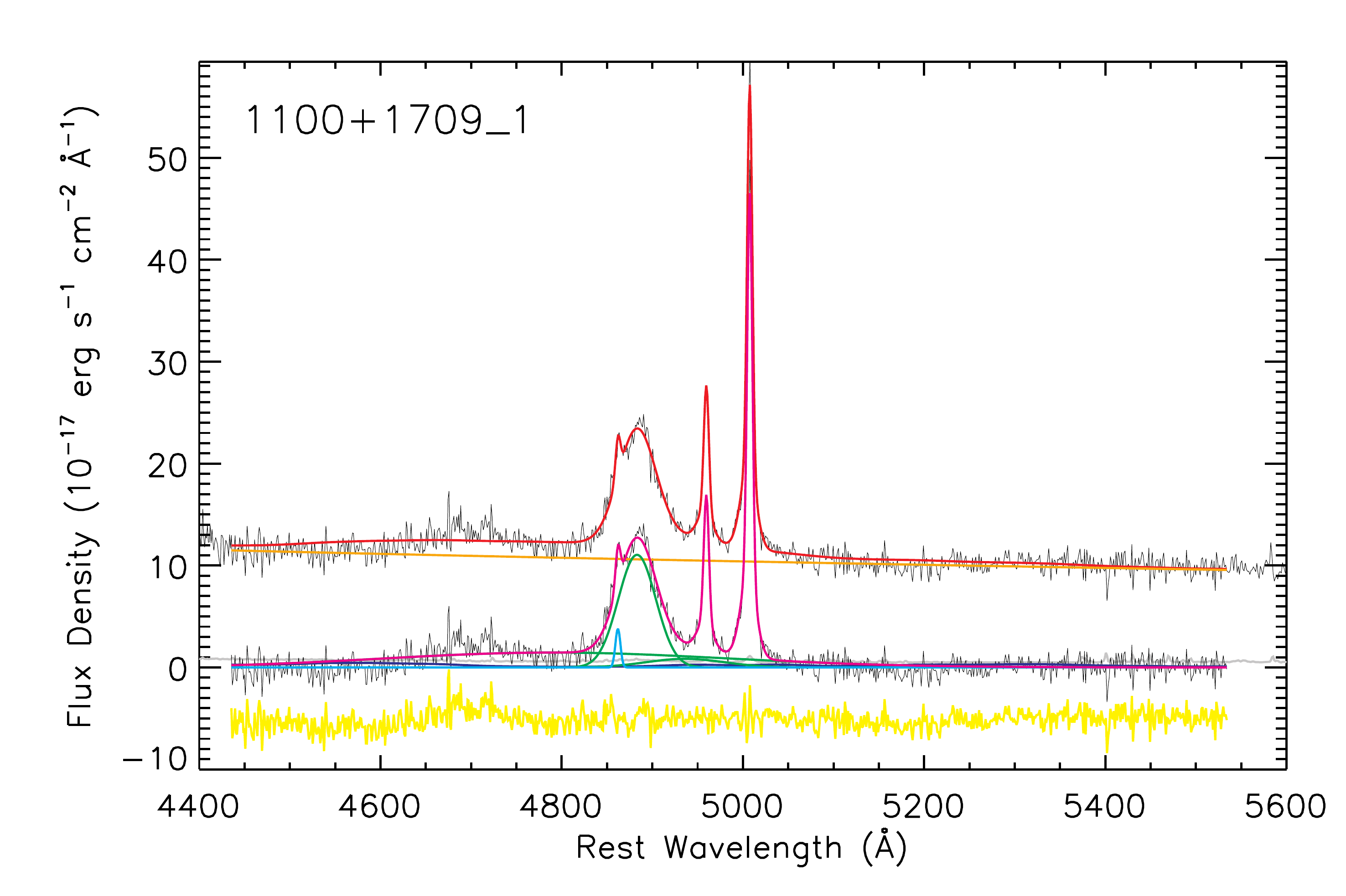}
    \includegraphics[width=85mm]{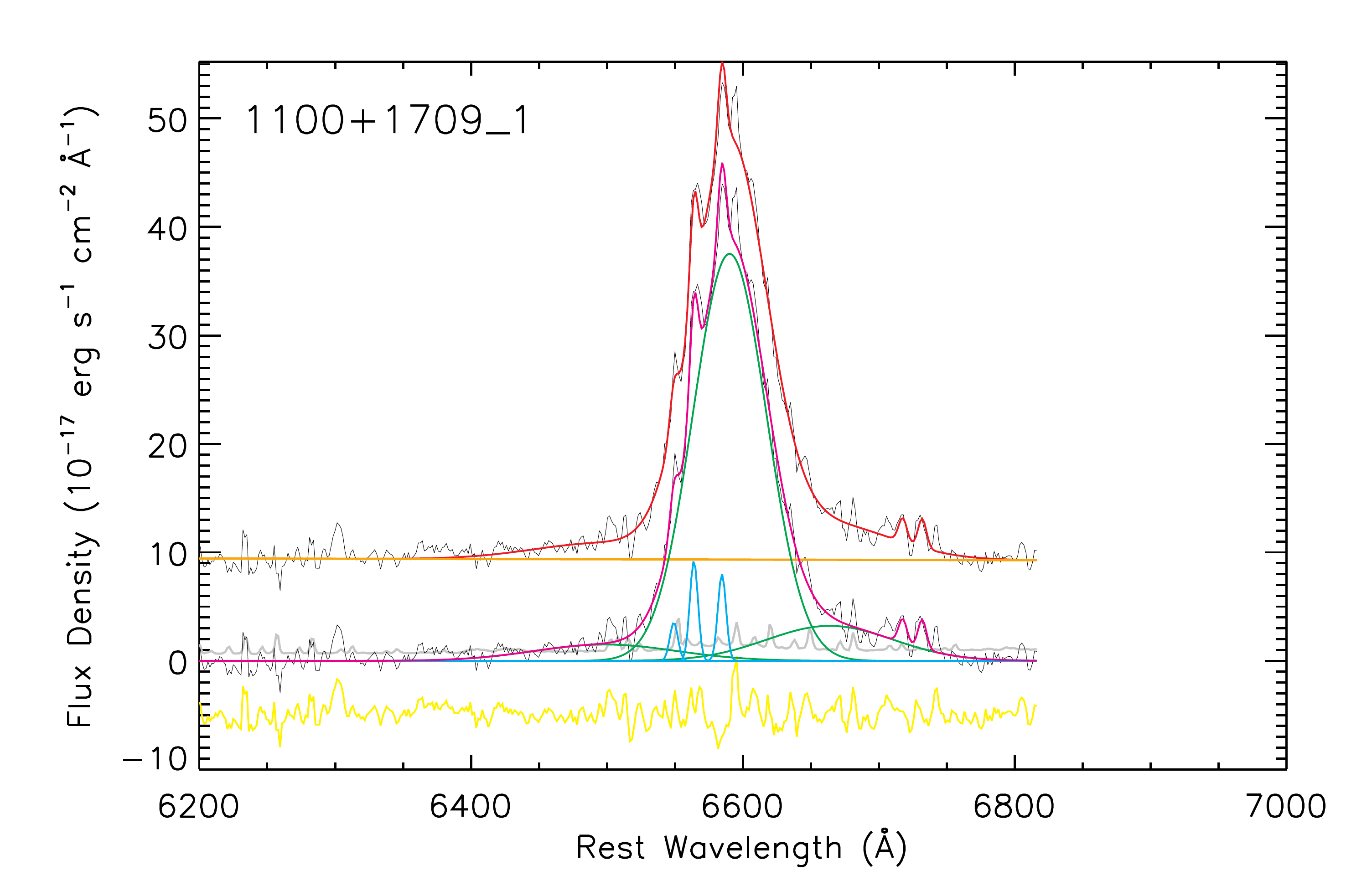}
    \includegraphics[width=85mm]{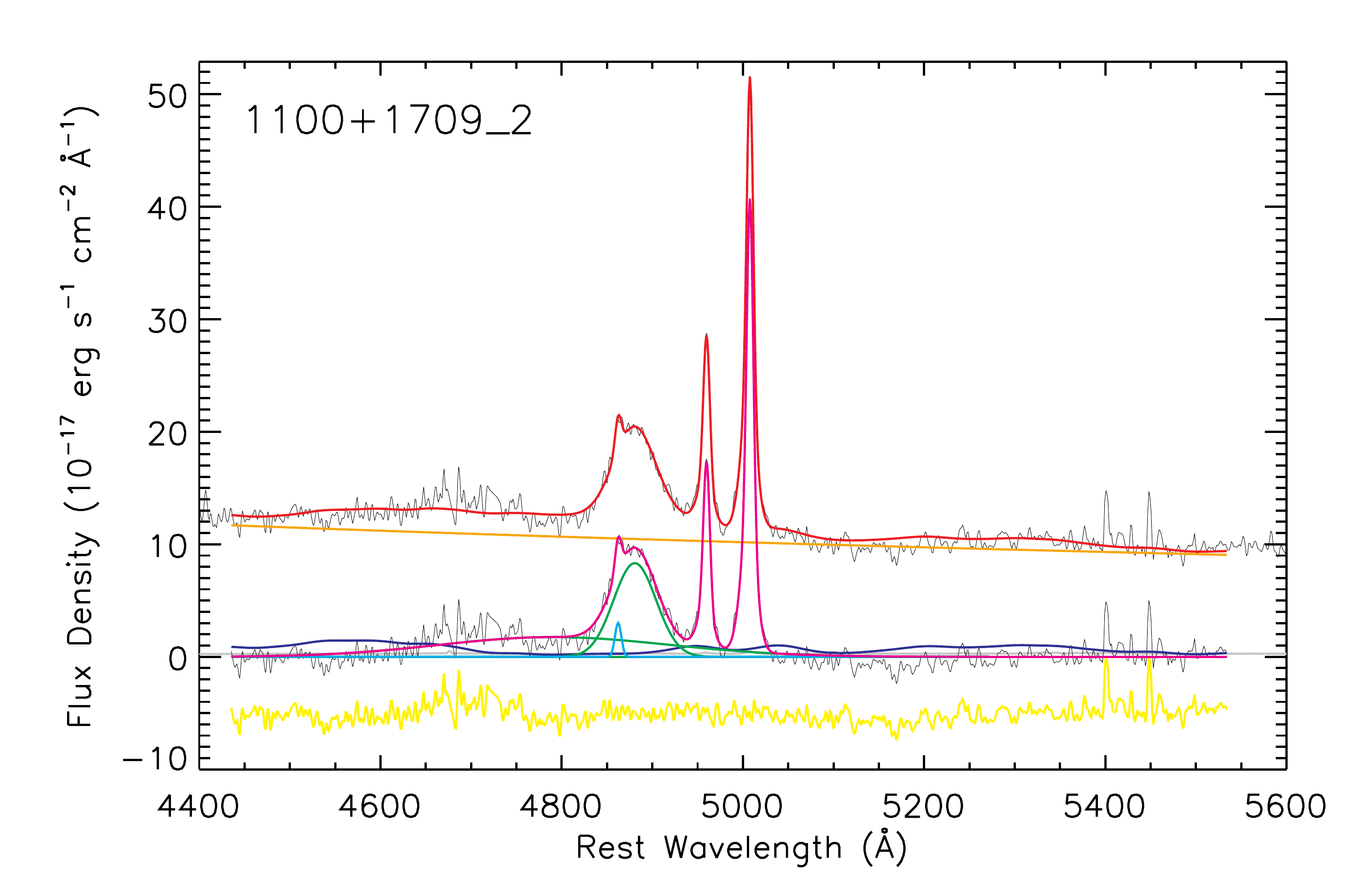}
    \includegraphics[width=85mm]{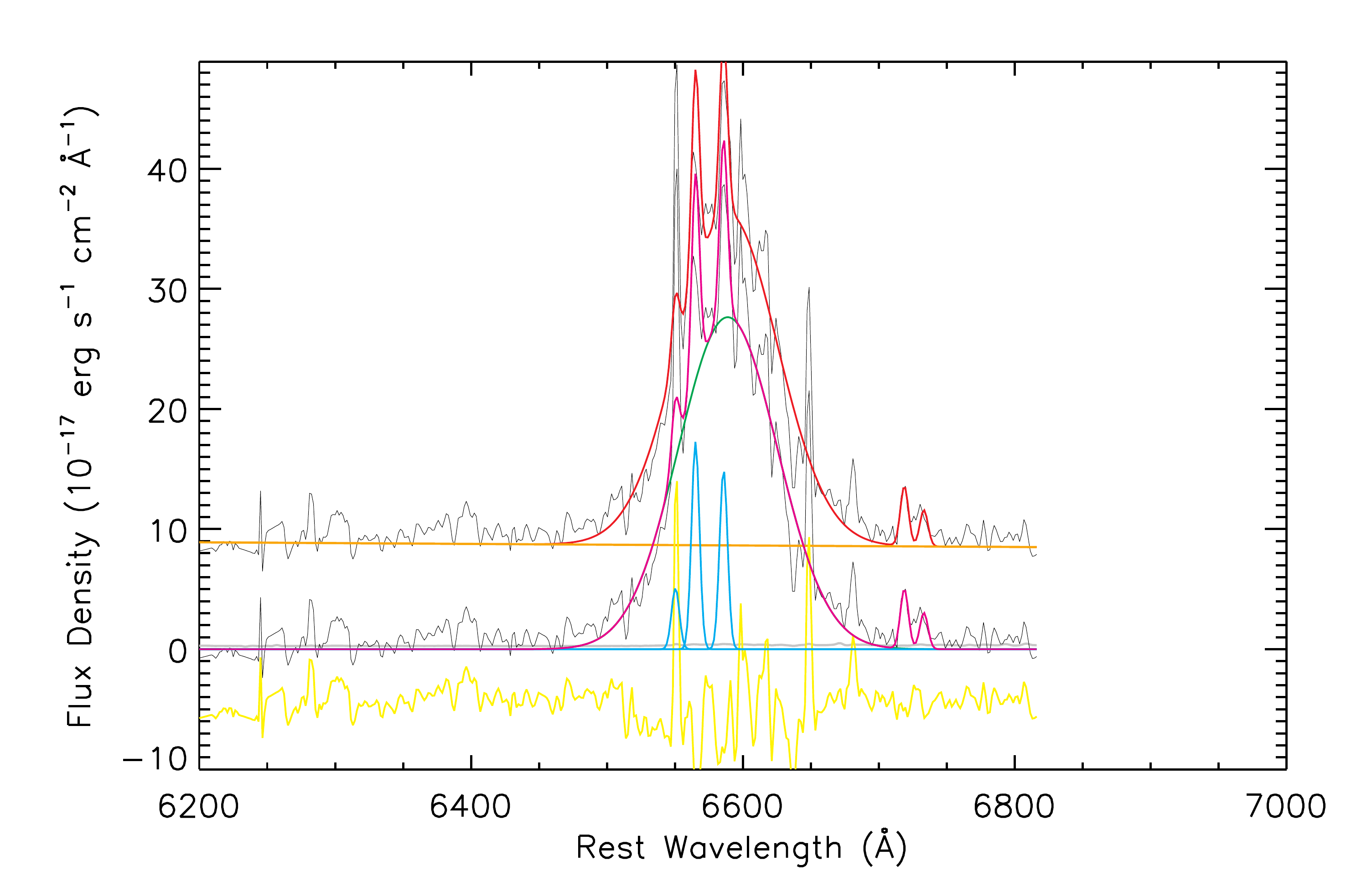}
    \includegraphics[width=85mm]{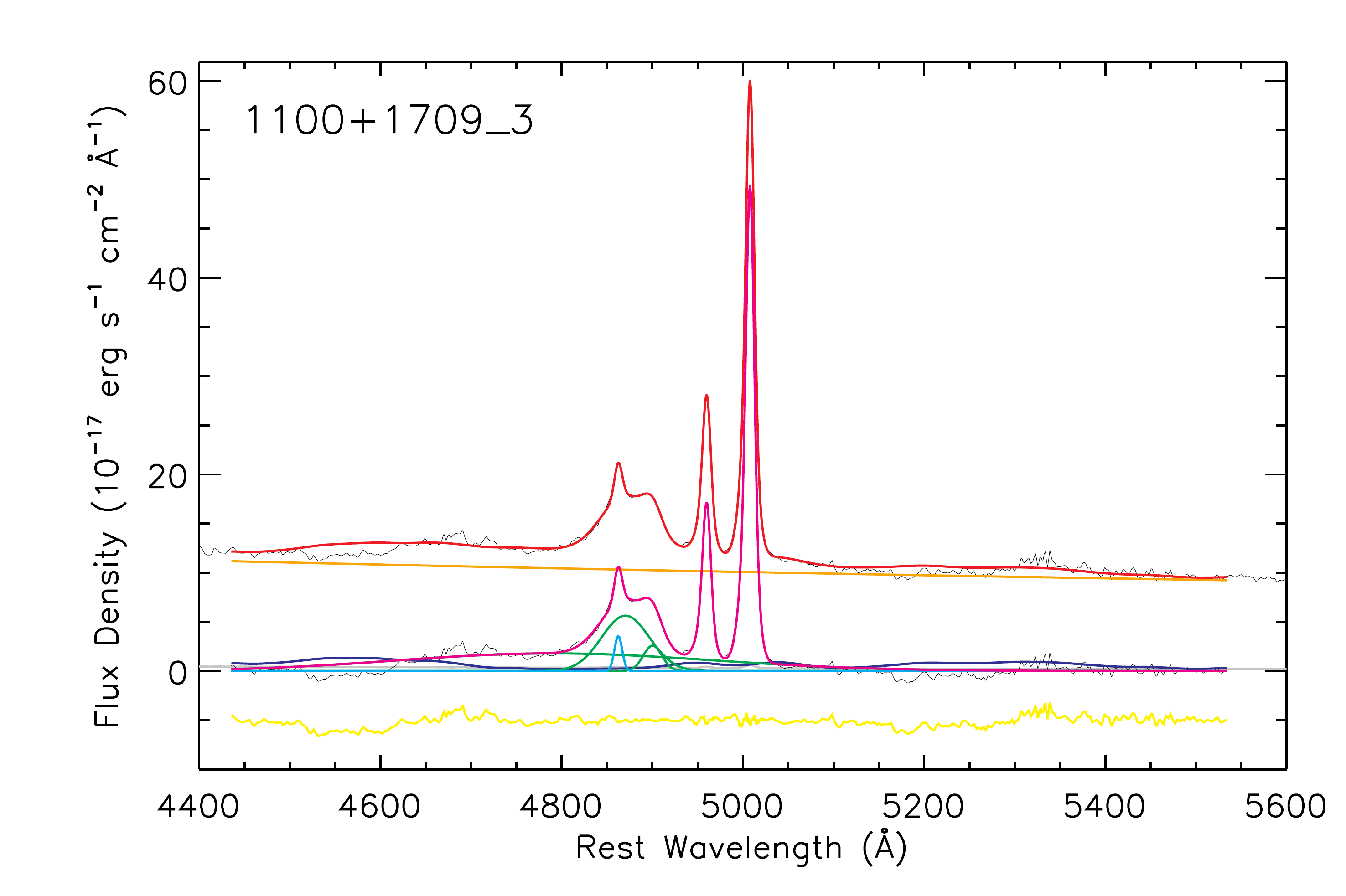}
    \includegraphics[width=85mm]{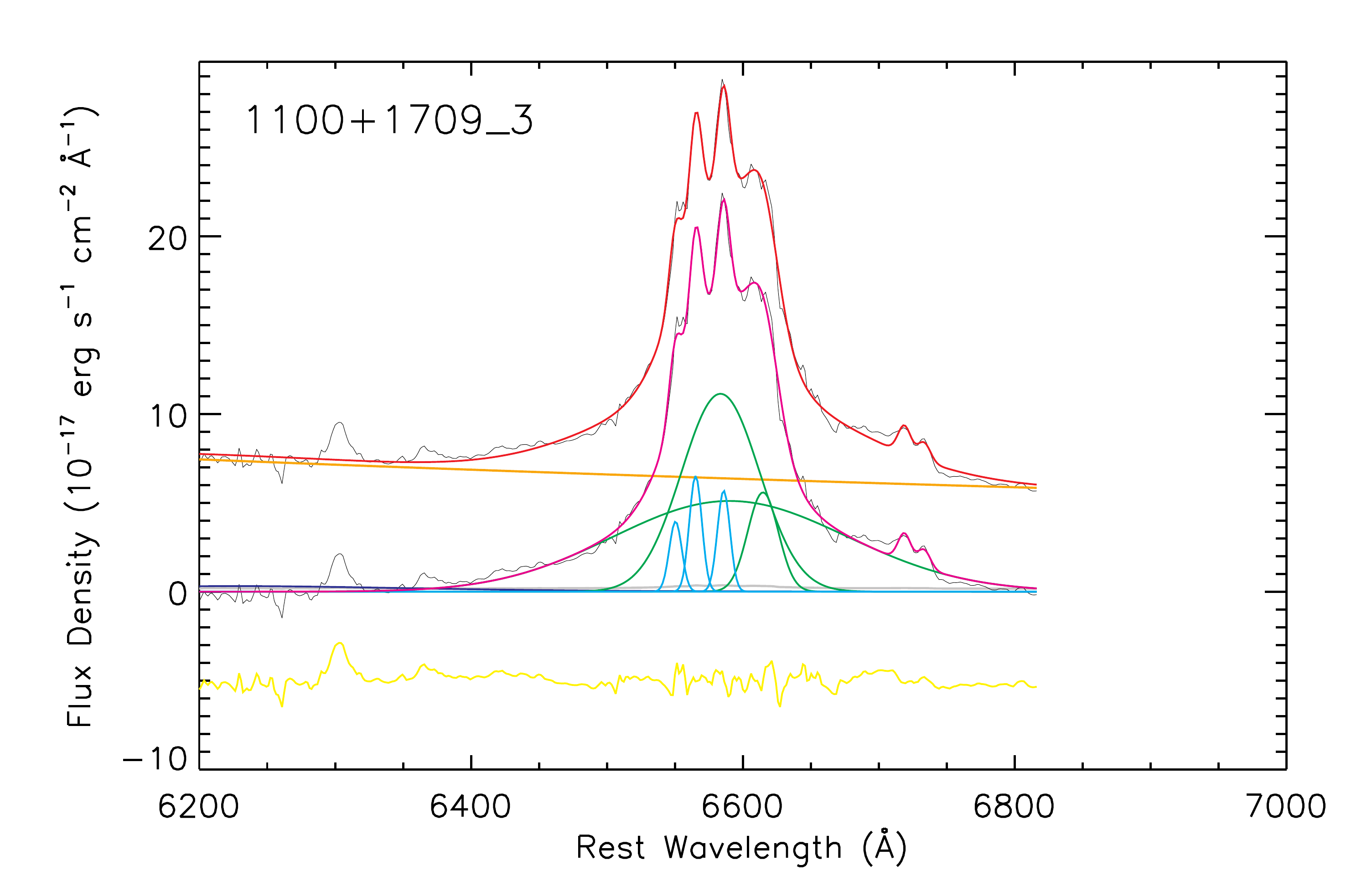}
       \centerline{Figure A1 -- Continued.    }     
\end{figure*}
\clearpage

\begin{figure*}
  \centering
    \includegraphics[width=85mm]{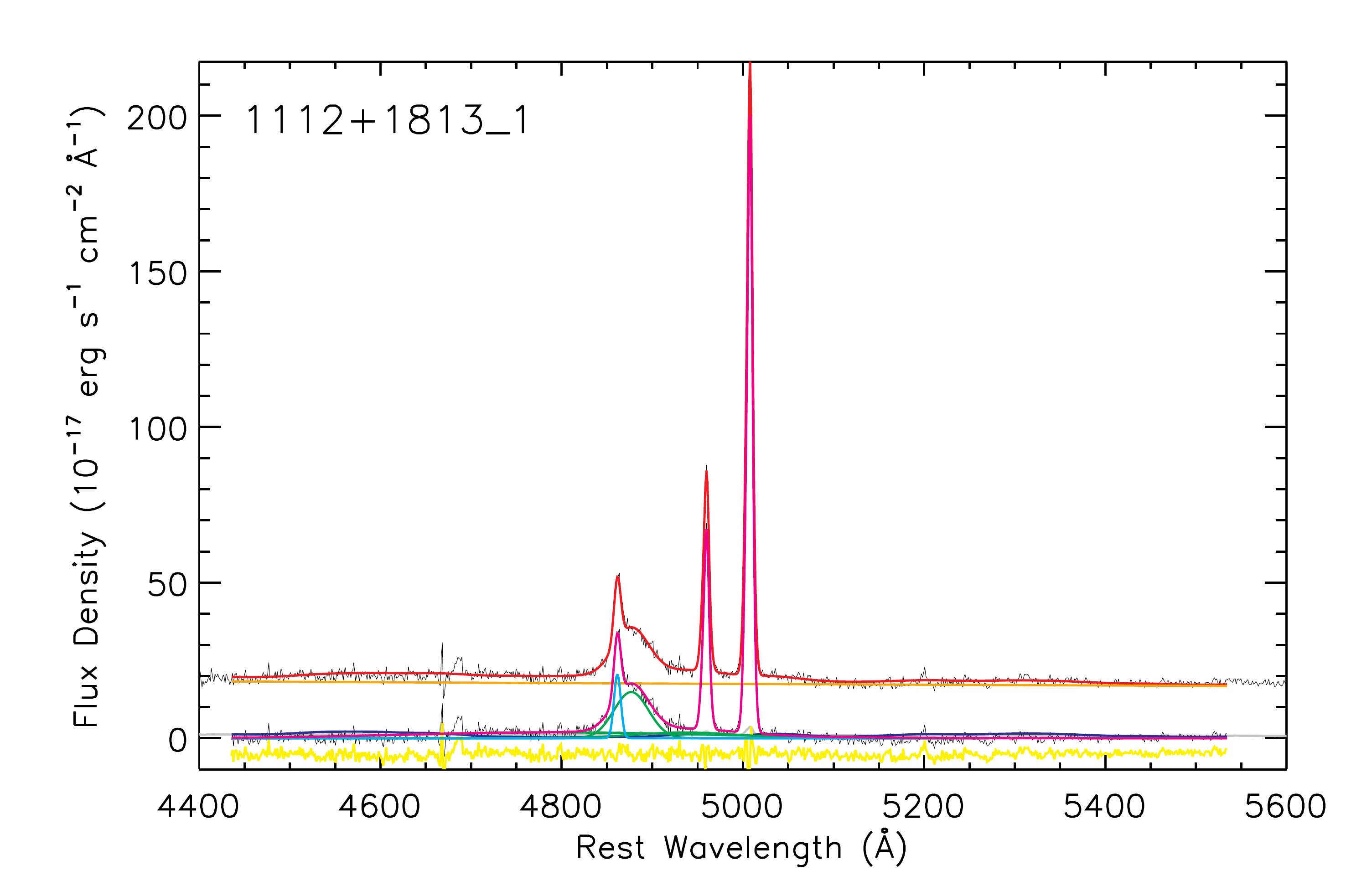}
    \includegraphics[width=85mm]{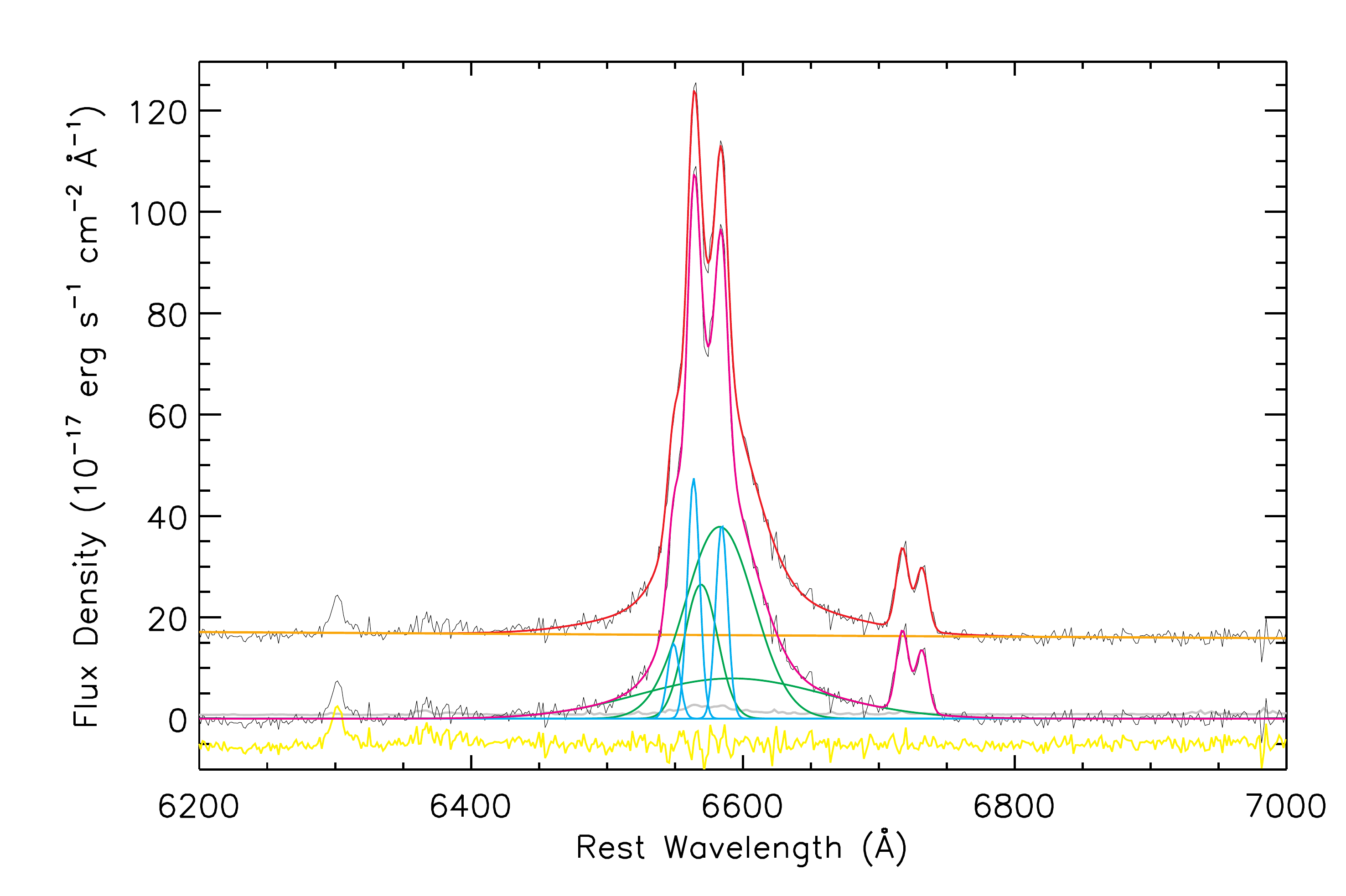}
    \includegraphics[width=85mm]{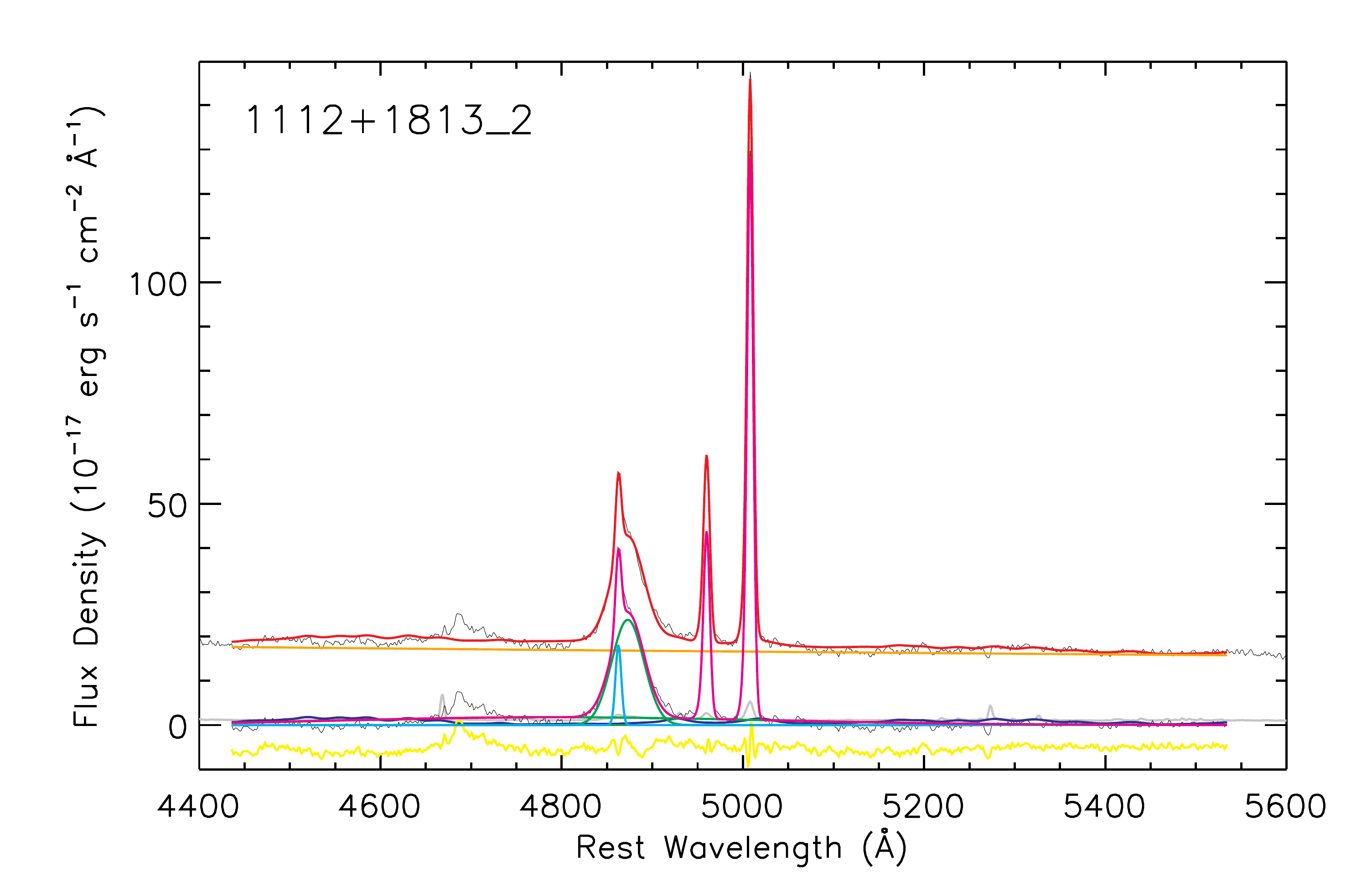}
    \includegraphics[width=85mm]{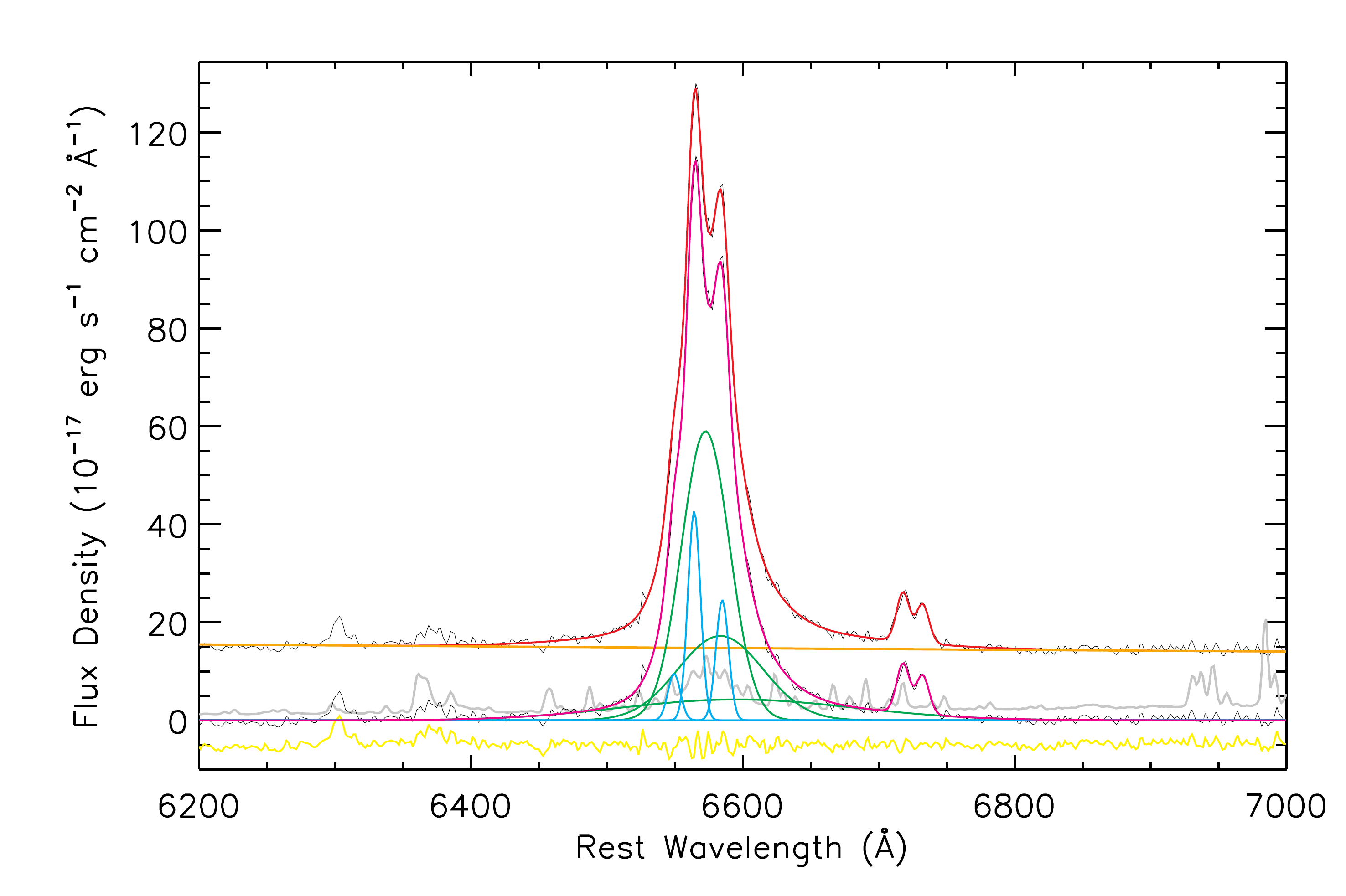}
    \includegraphics[width=85mm]{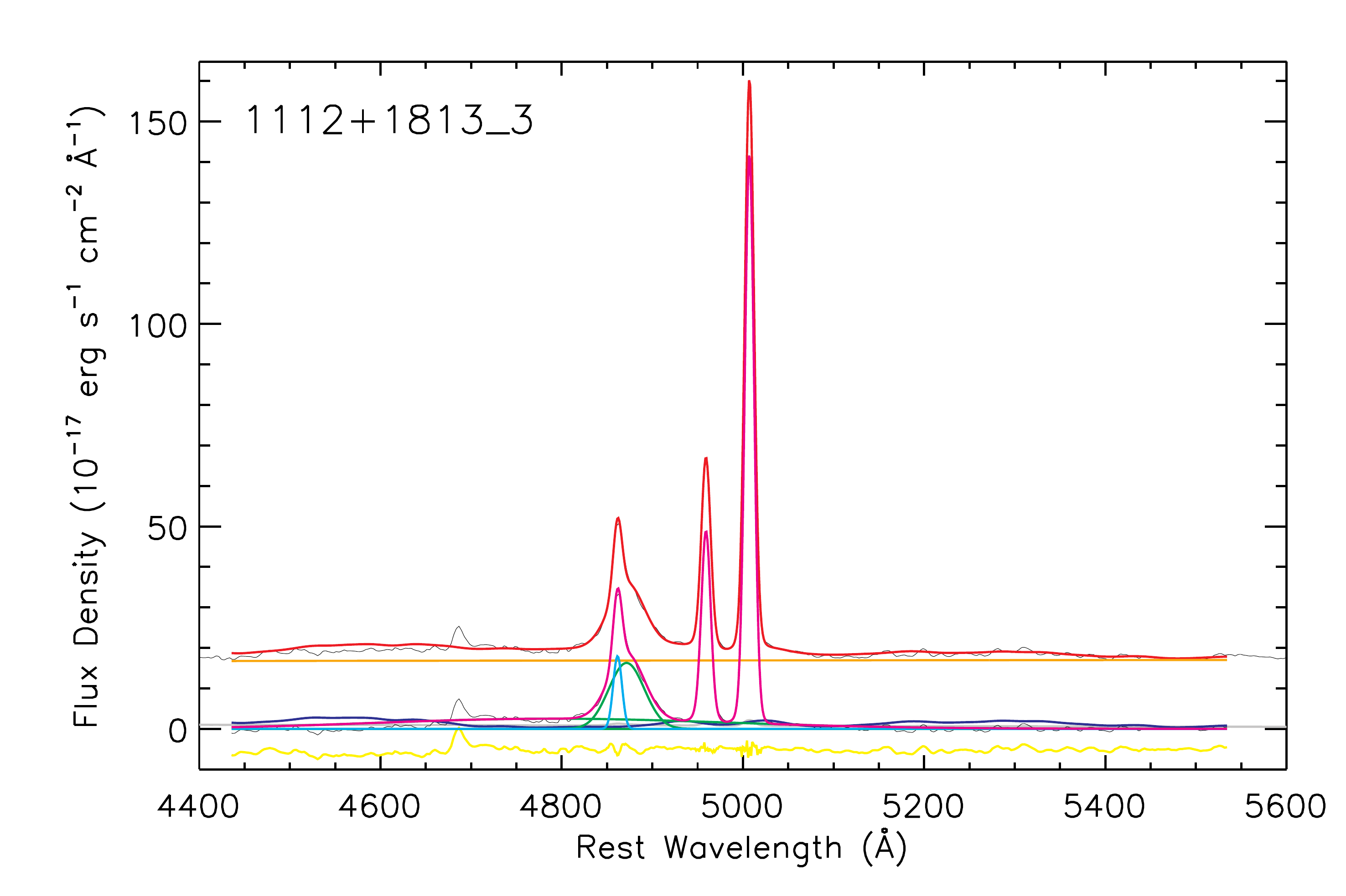}
    \includegraphics[width=85mm]{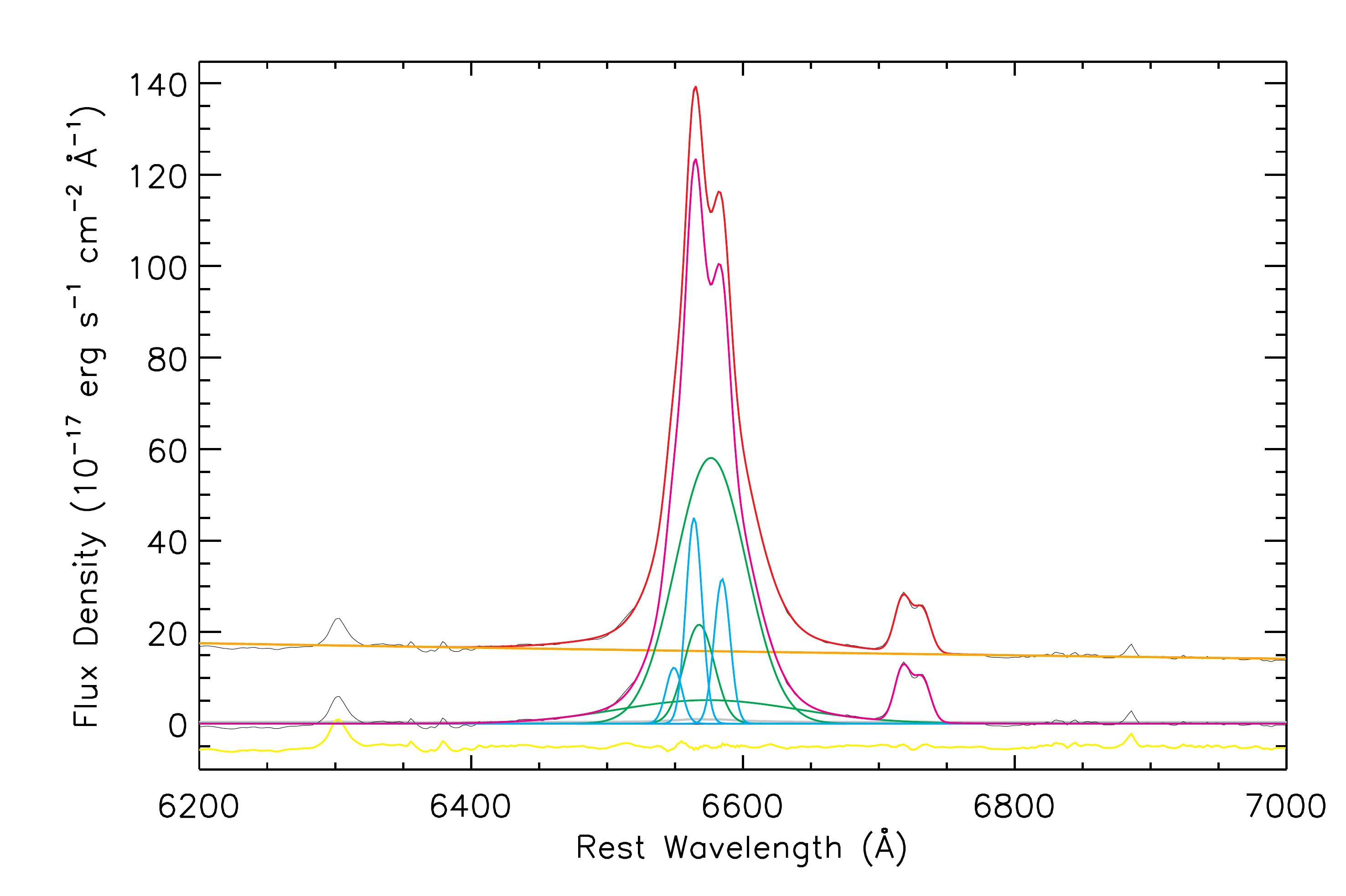}
       \centerline{Figure A1 -- Continued.    }  
\end{figure*}
\clearpage

\begin{figure*}
  \centering
    \includegraphics[width=85mm]{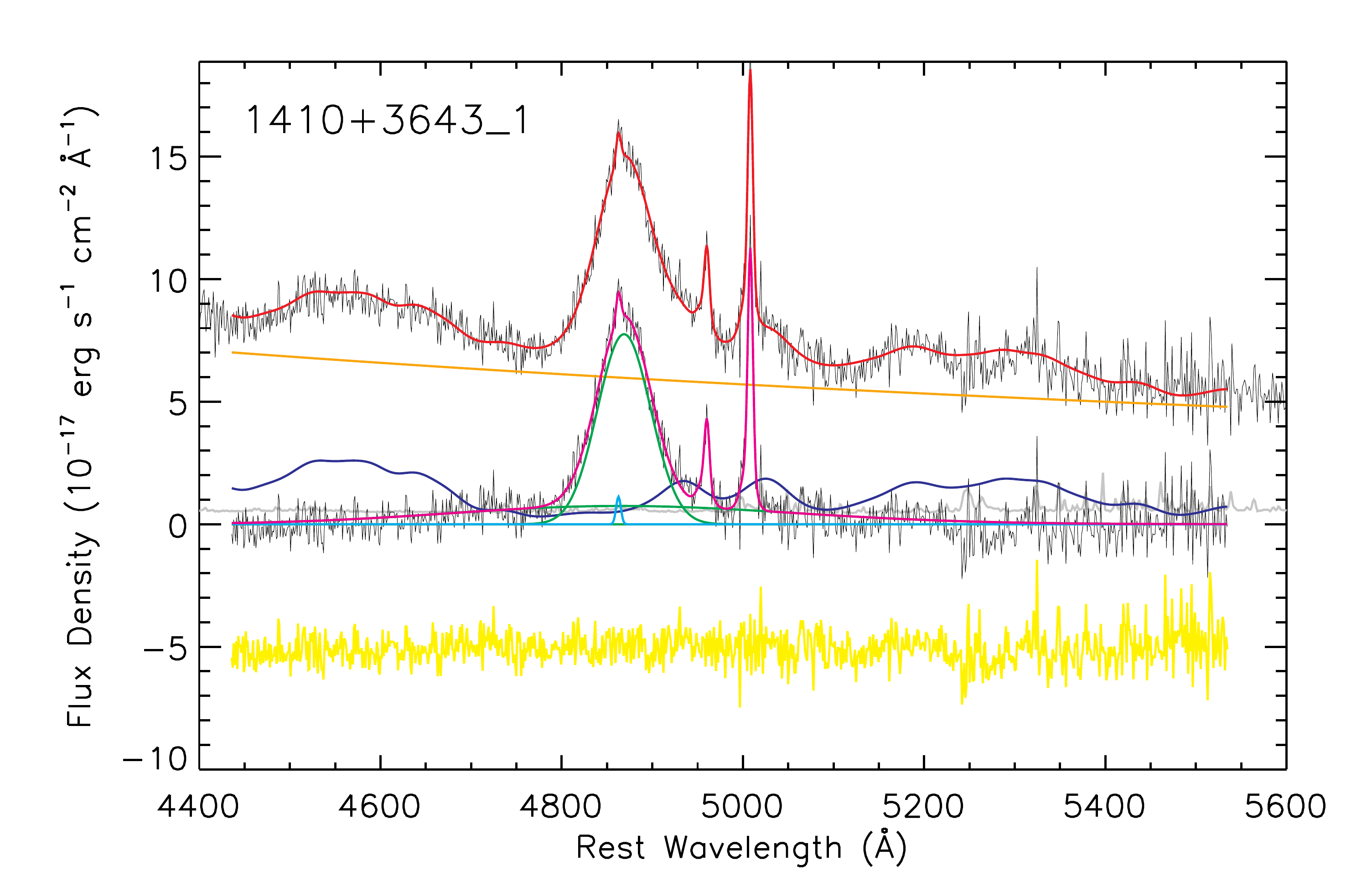}
    \includegraphics[width=85mm]{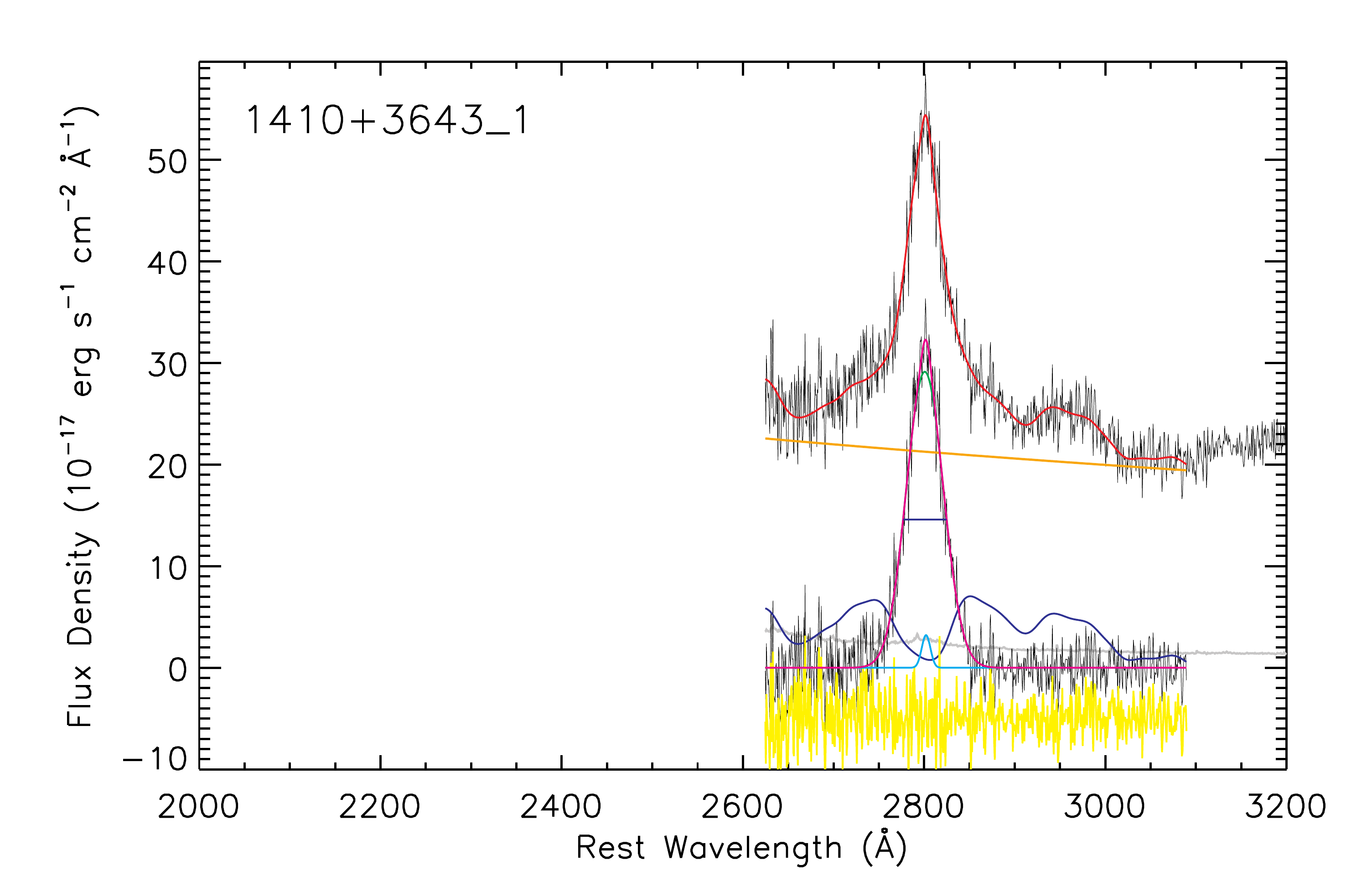}
    \includegraphics[width=85mm]{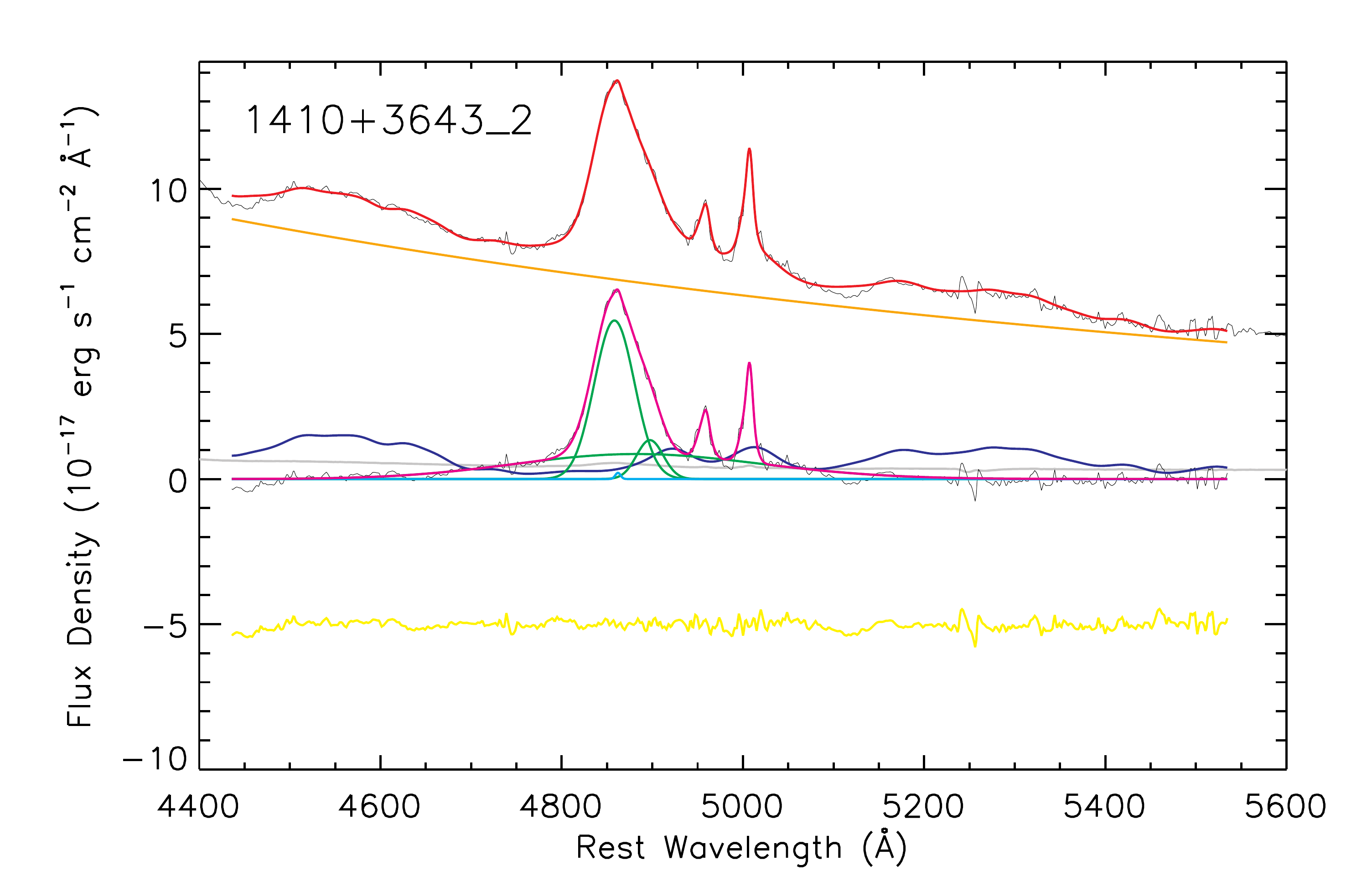}
    \includegraphics[width=85mm]{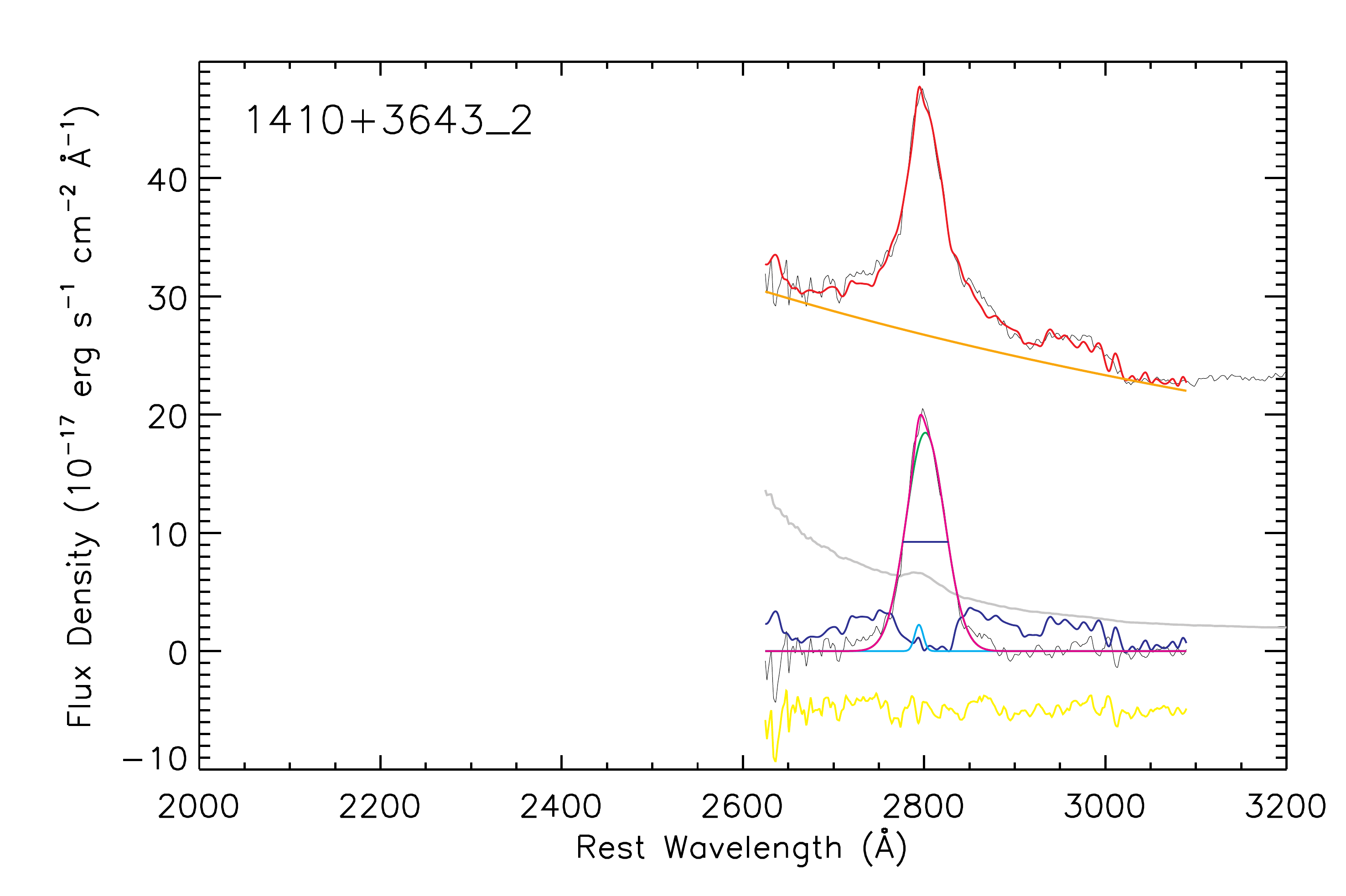}
    \includegraphics[width=85mm]{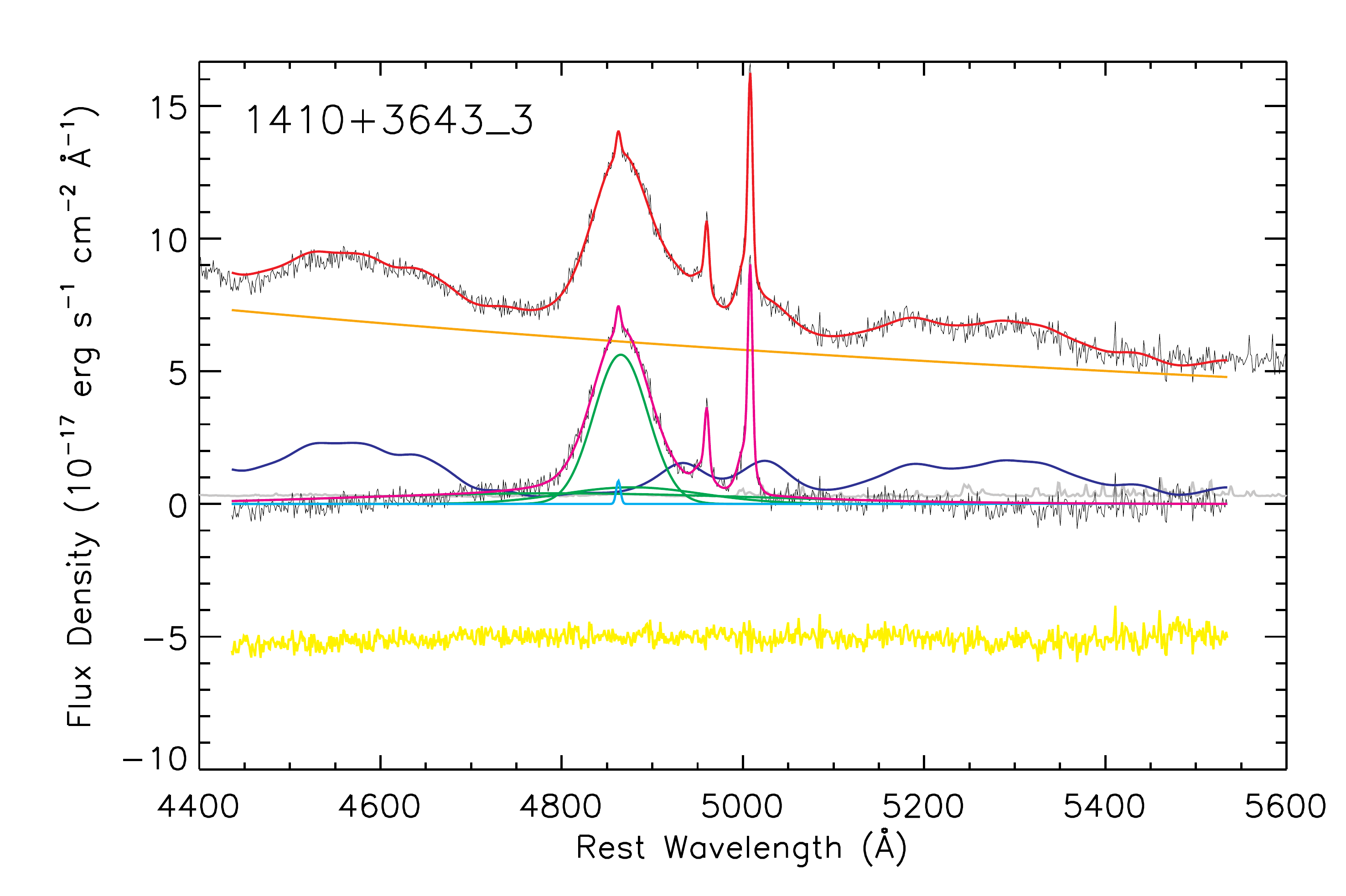}
    \includegraphics[width=85mm]{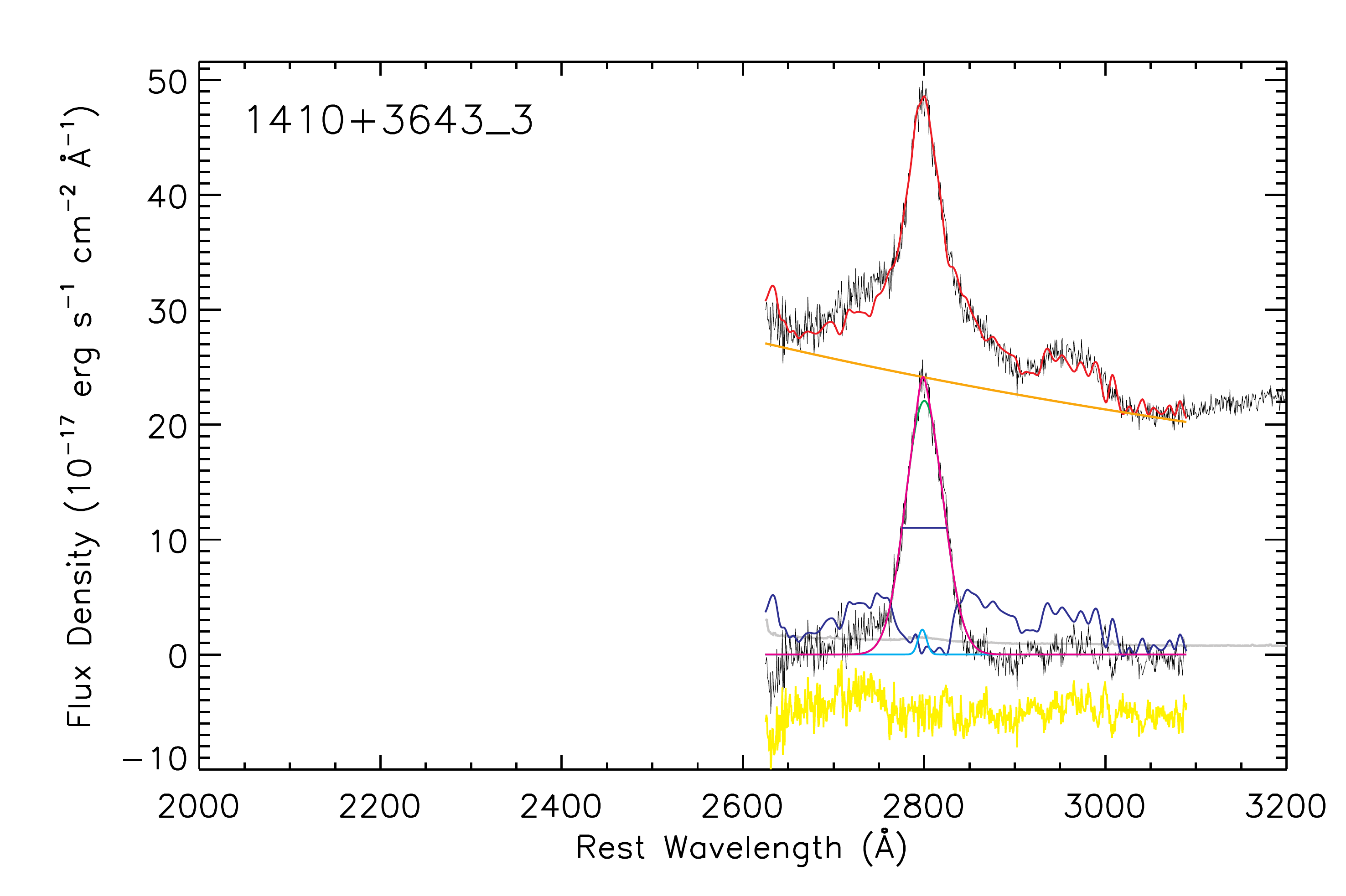}
       \centerline{Figure A1 -- Continued.    }  
\end{figure*}
\clearpage

\begin{figure*}
  \centering
    \includegraphics[width=80mm]{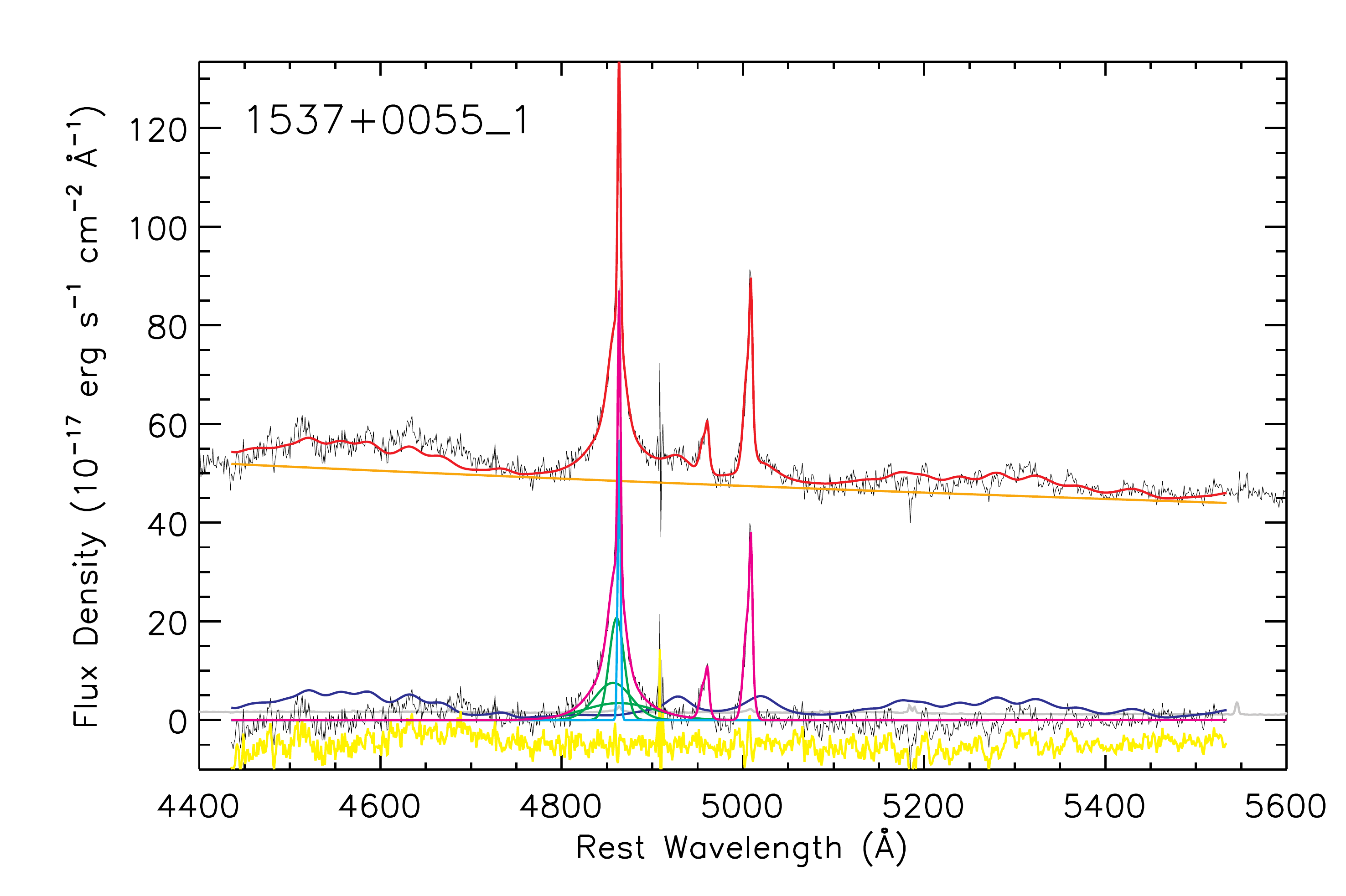}
    \includegraphics[width=80mm]{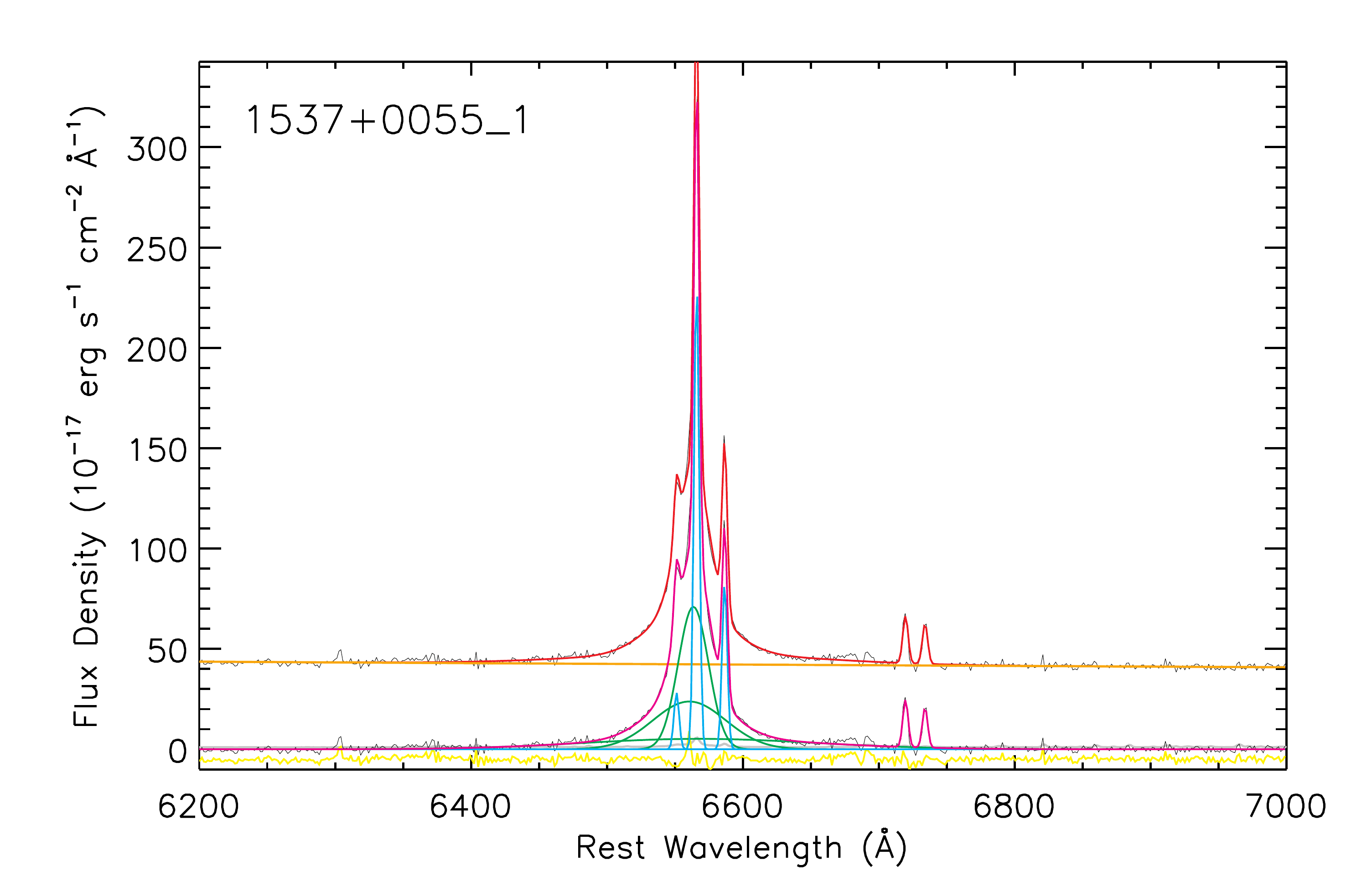}
    \includegraphics[width=80mm]{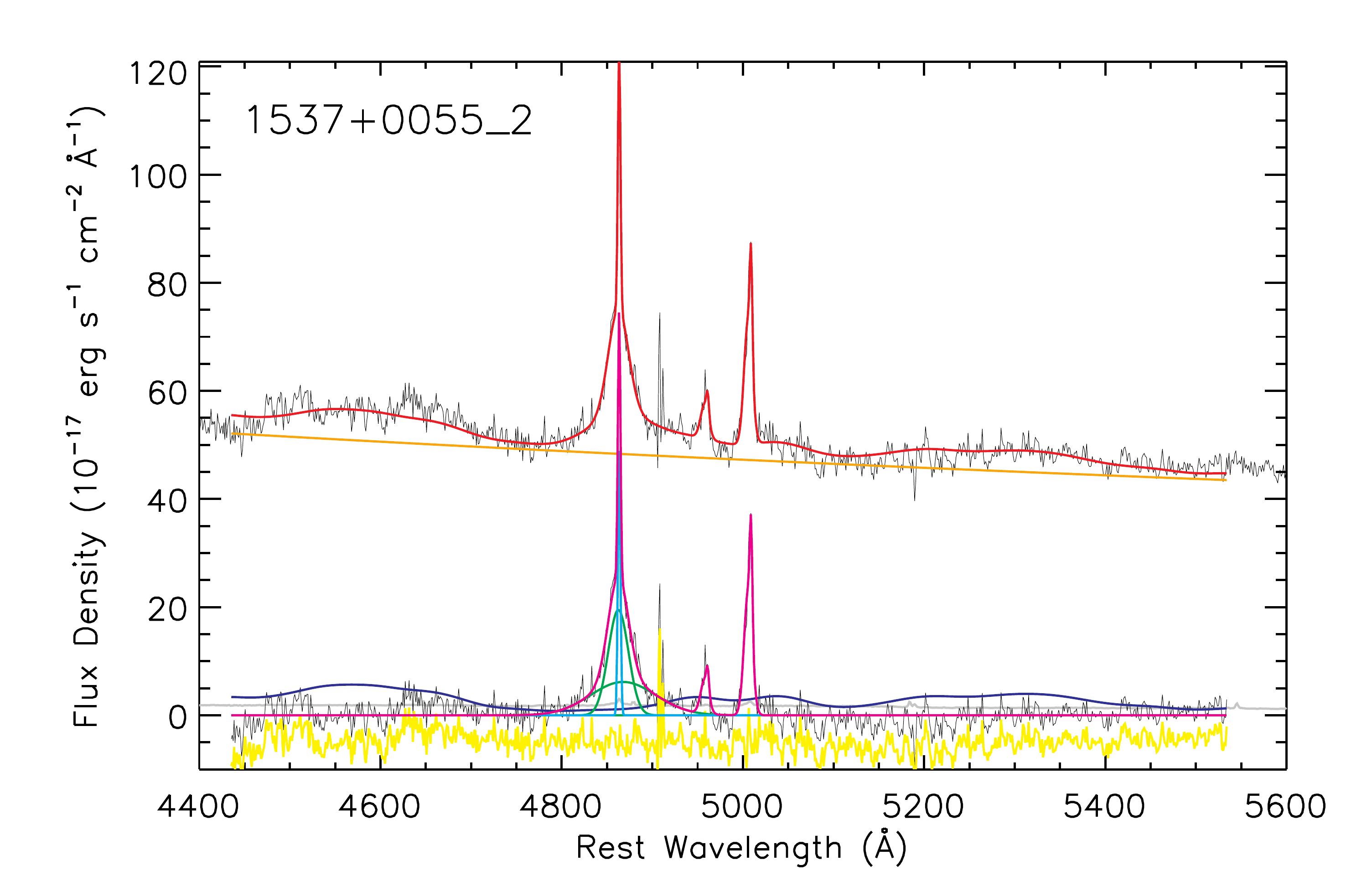}
    \includegraphics[width=80mm]{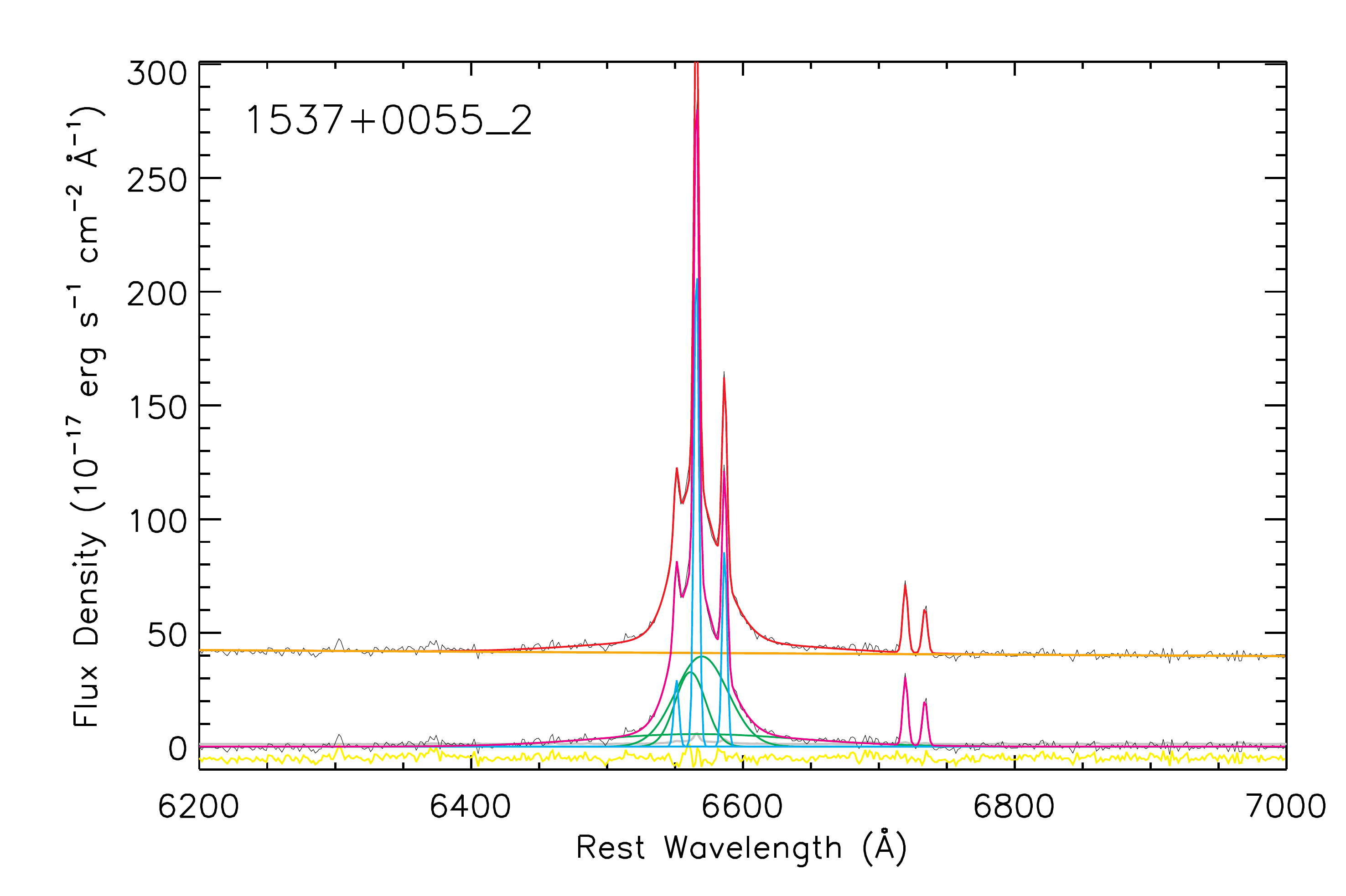}
    \includegraphics[width=80mm]{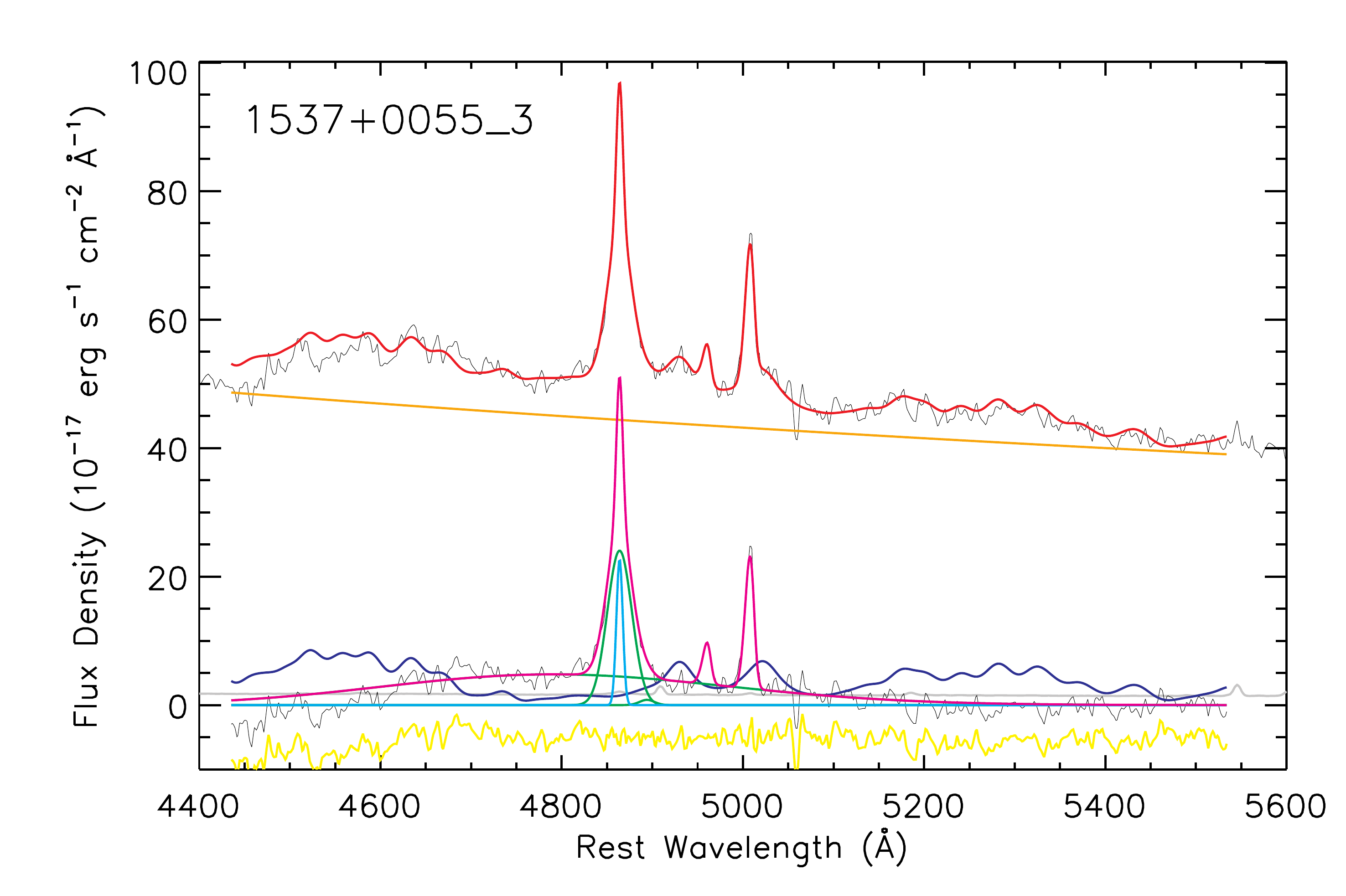}
    \includegraphics[width=80mm]{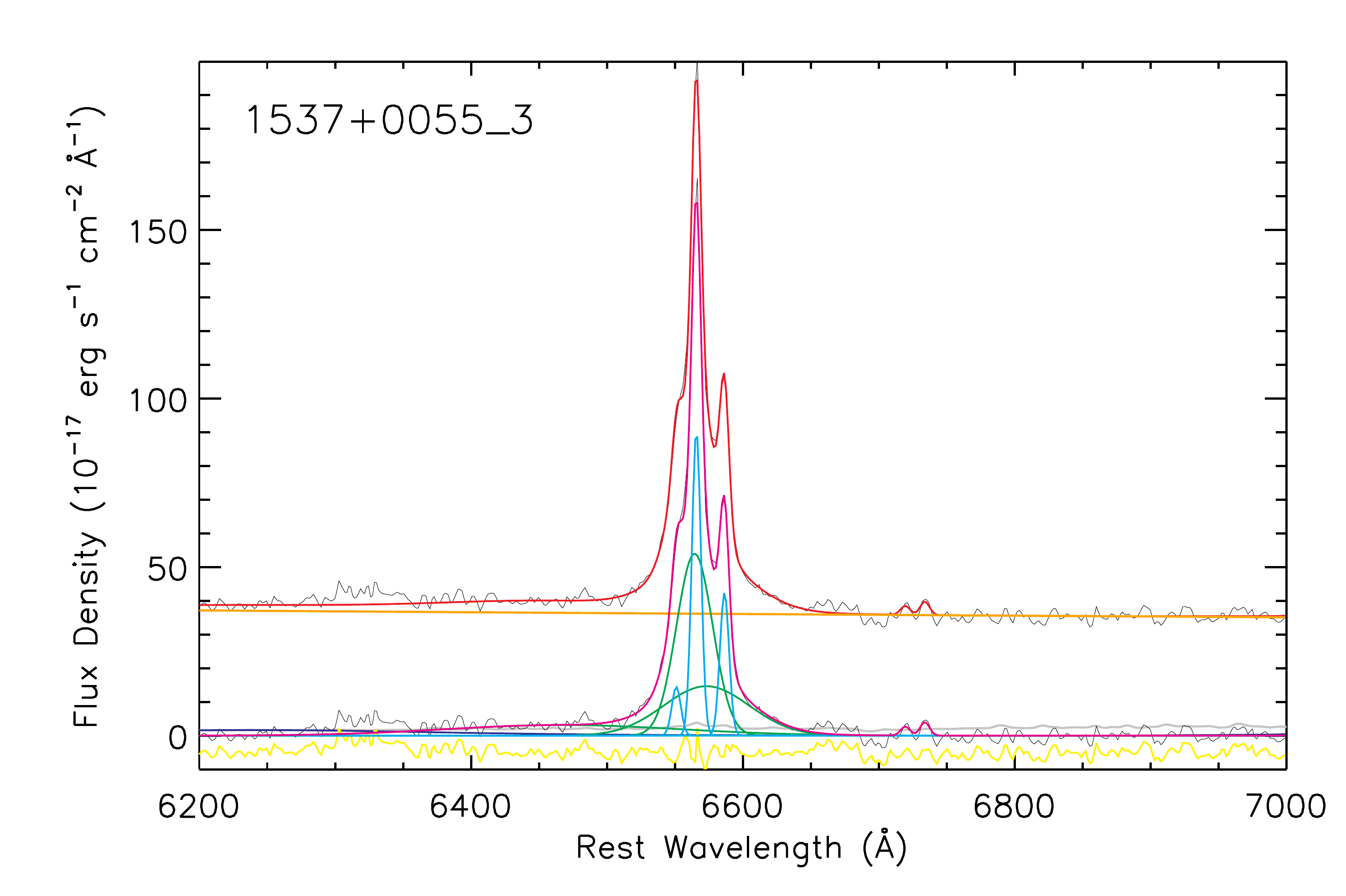}
    \includegraphics[width=80mm]{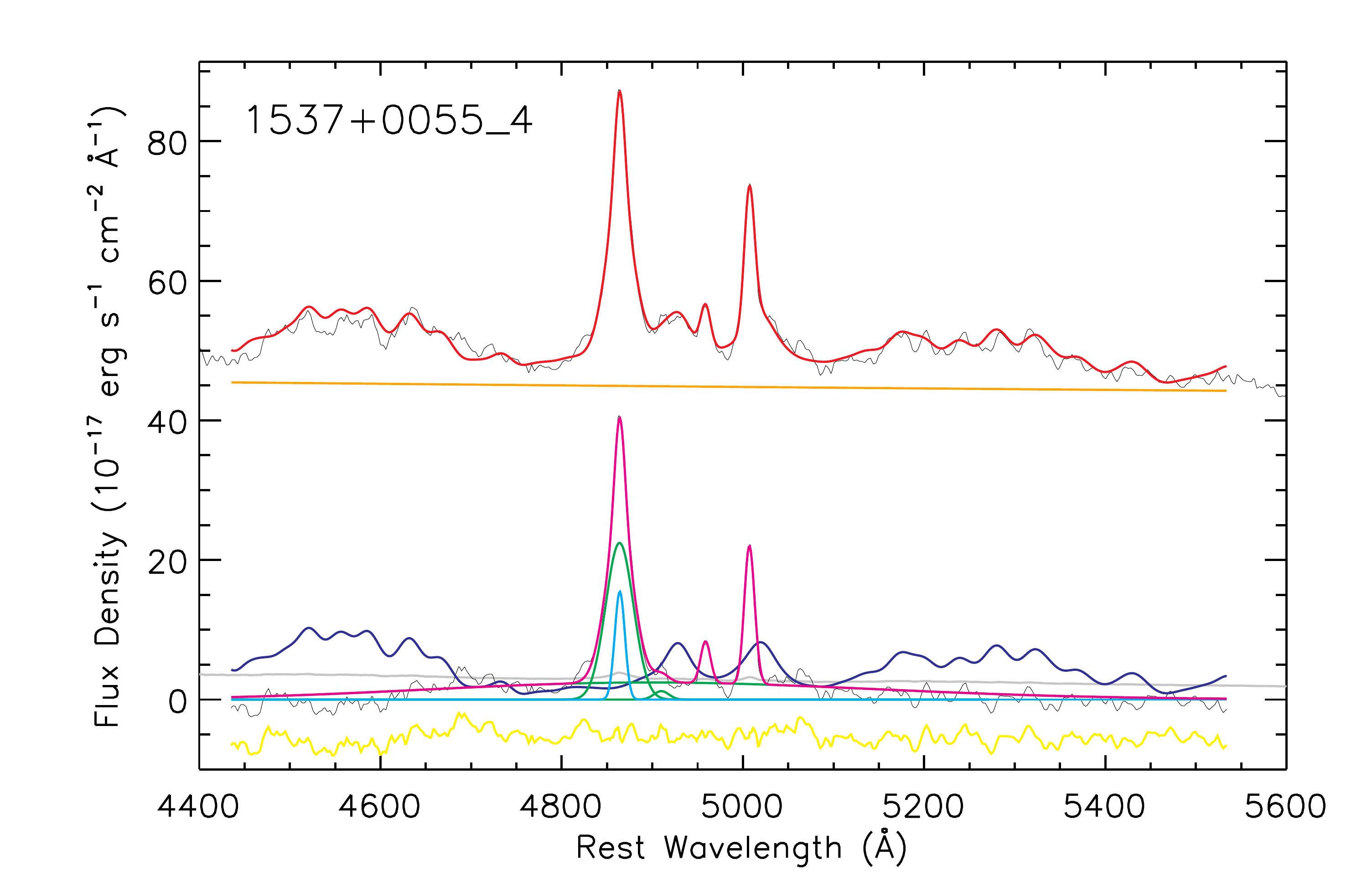}
    \includegraphics[width=80mm]{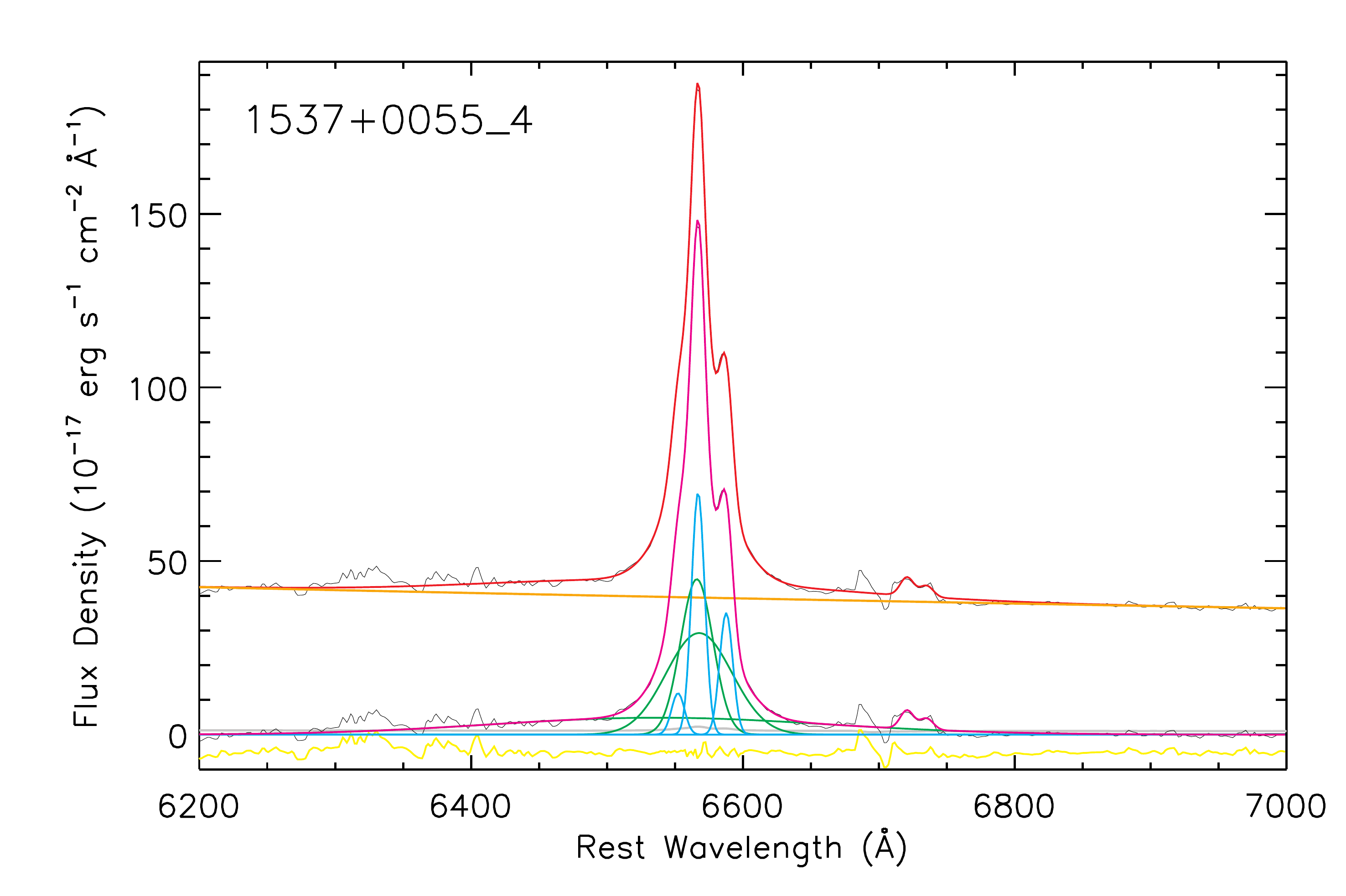}
       \centerline{Figure A1 -- Continued.    } 
\end{figure*}
\clearpage

\begin{figure*}
   \centering
    \includegraphics[width=80mm]{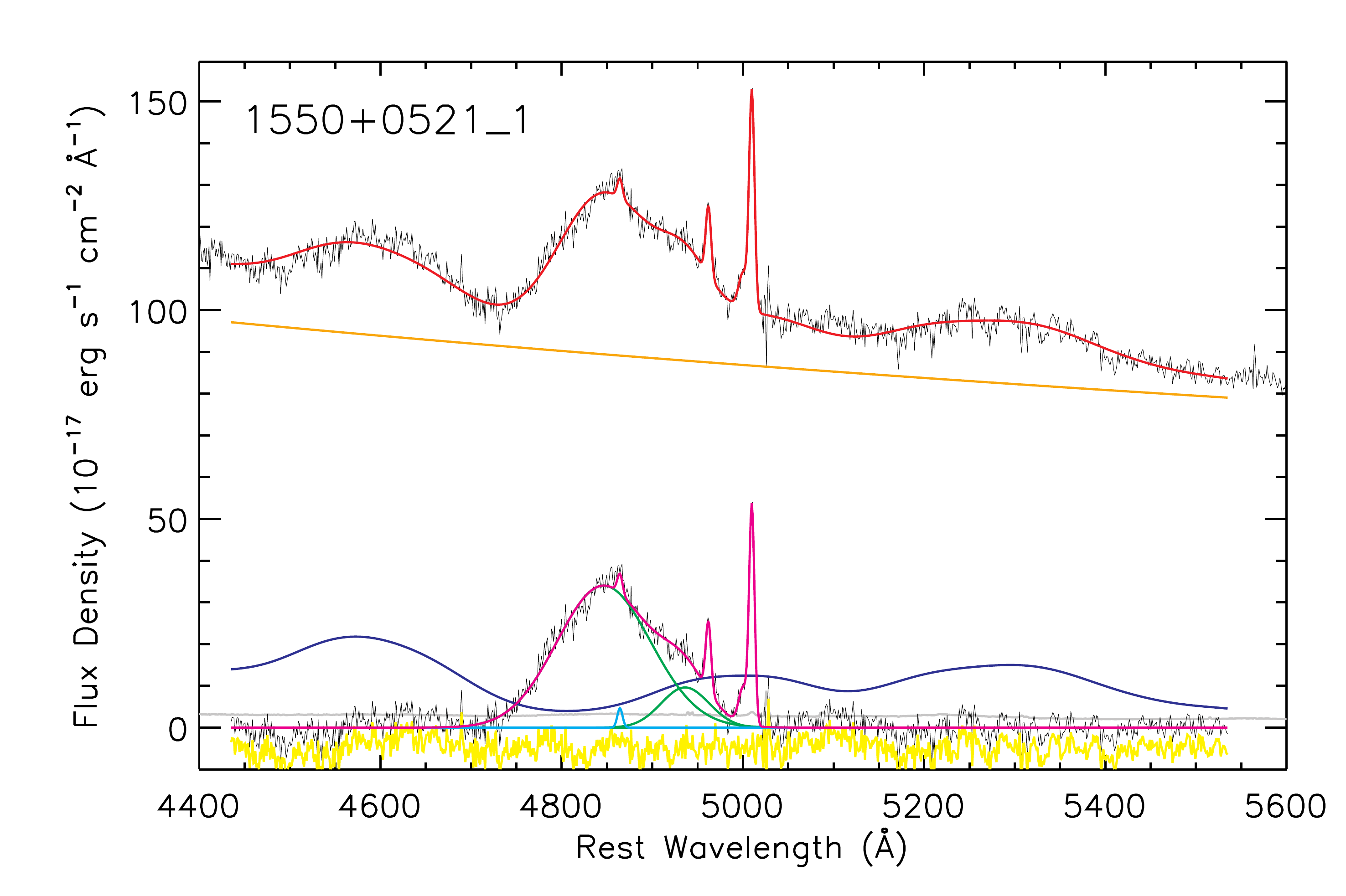}
    \includegraphics[width=80mm]{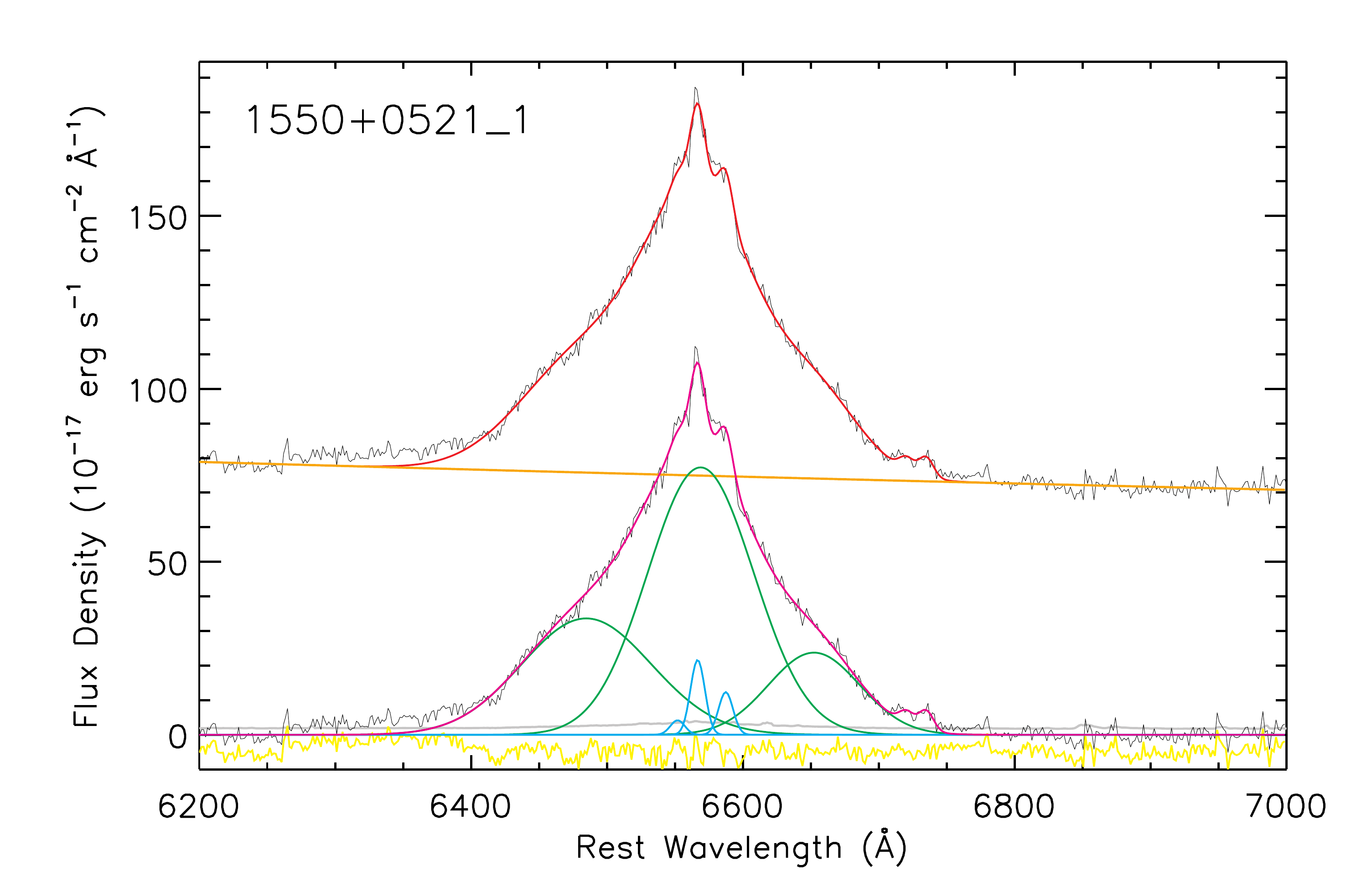}
    \includegraphics[width=80mm]{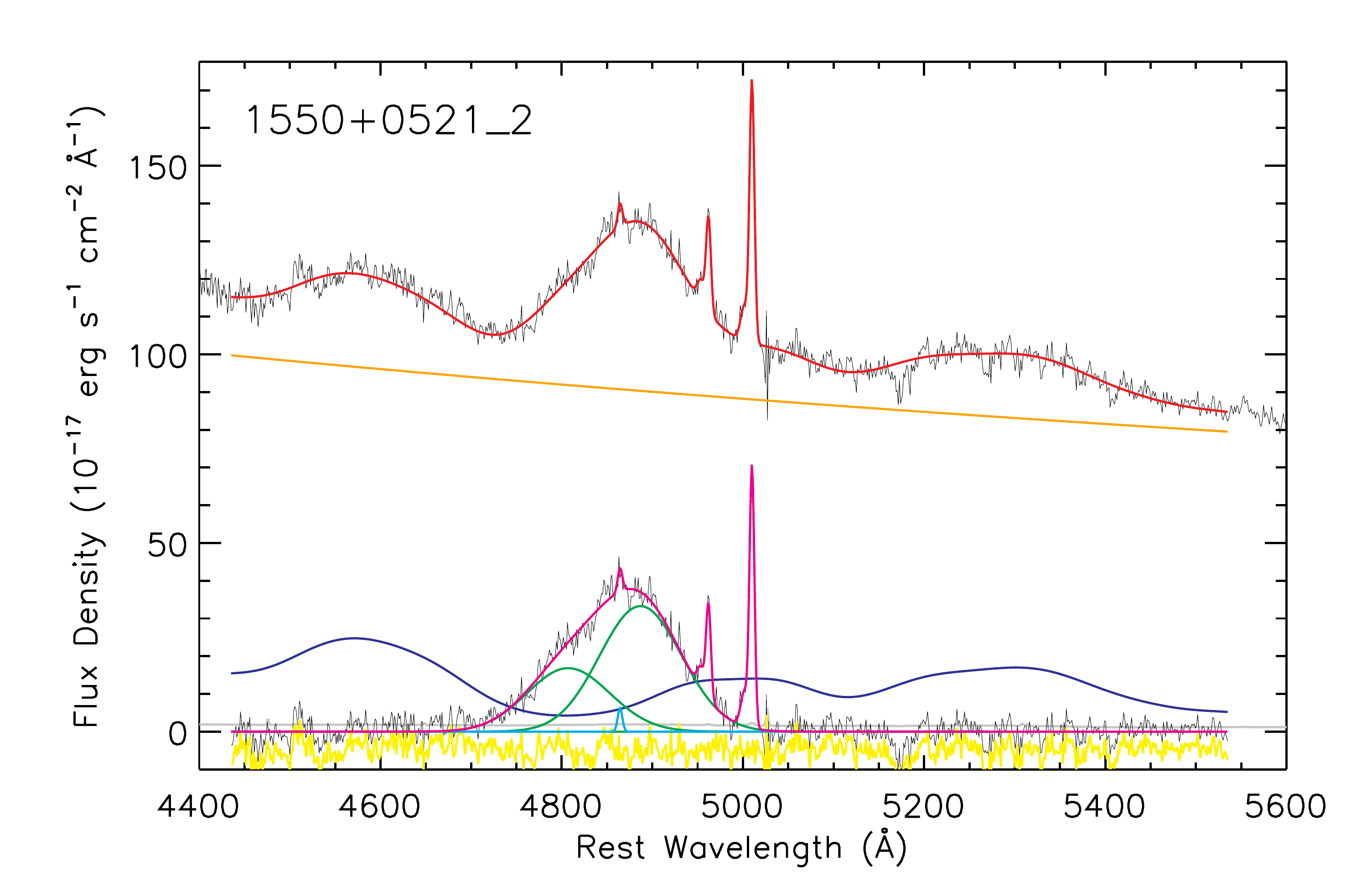}
    \includegraphics[width=80mm]{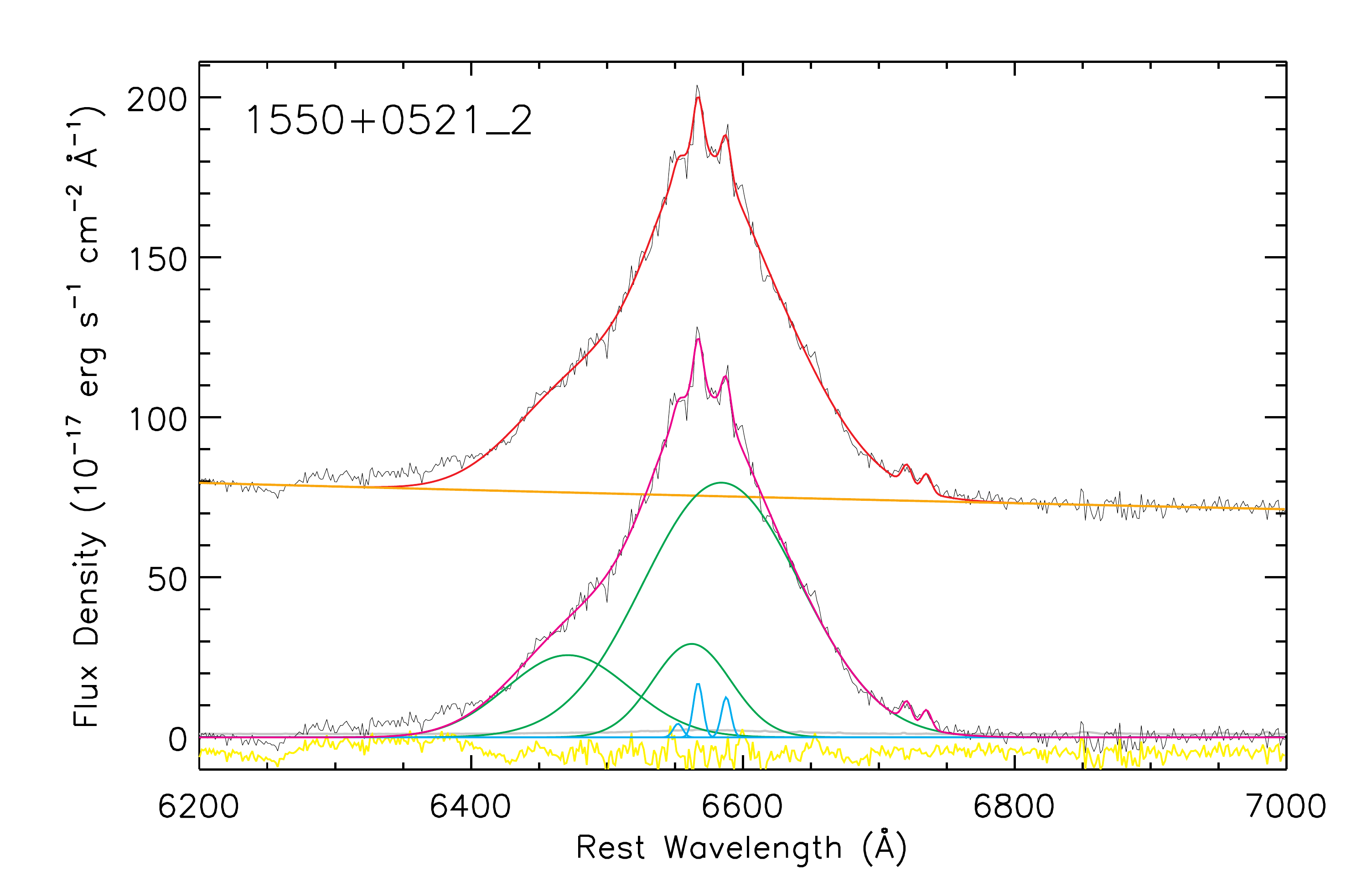}
    \includegraphics[width=80mm]{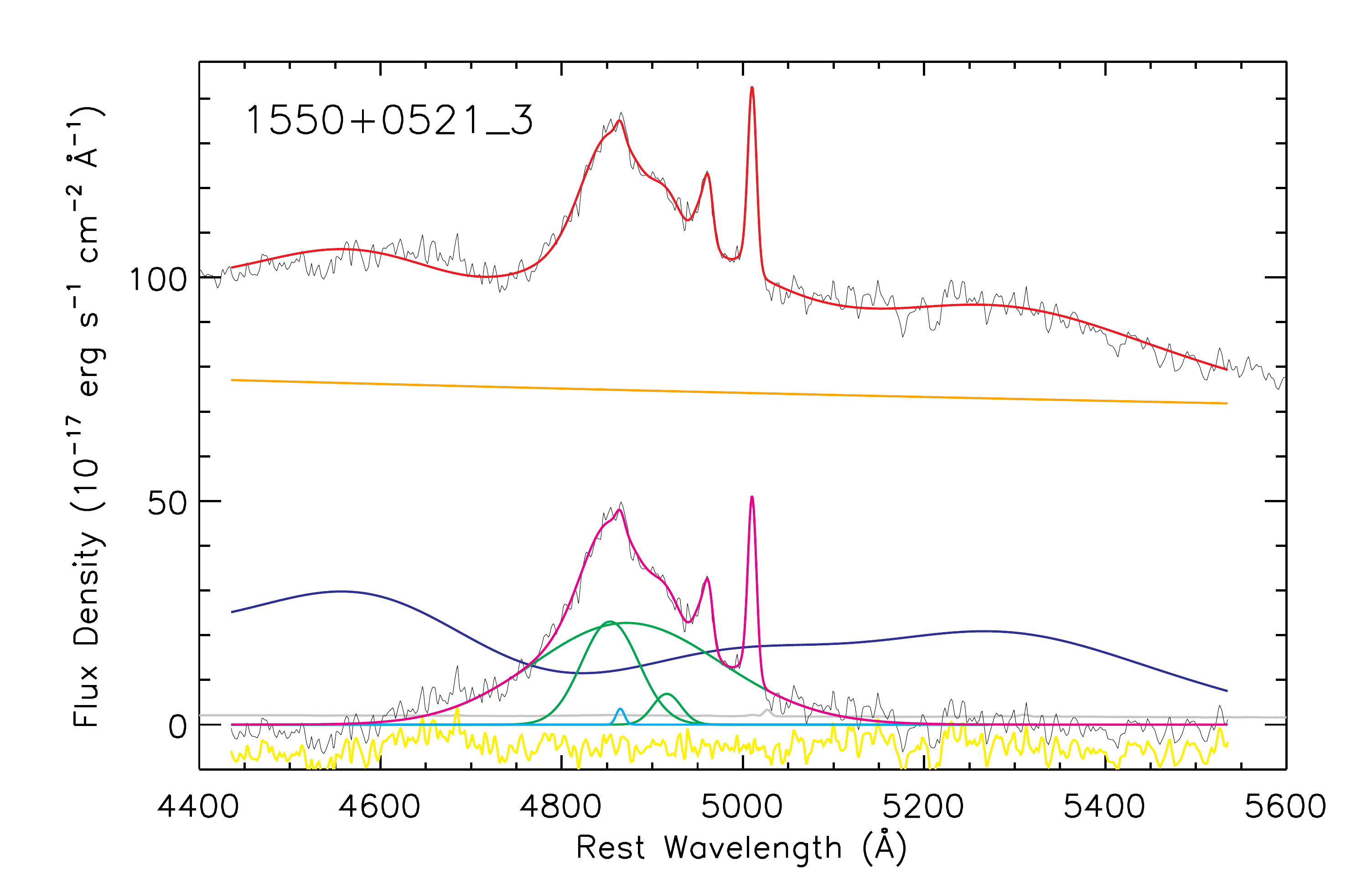}
    \includegraphics[width=80mm]{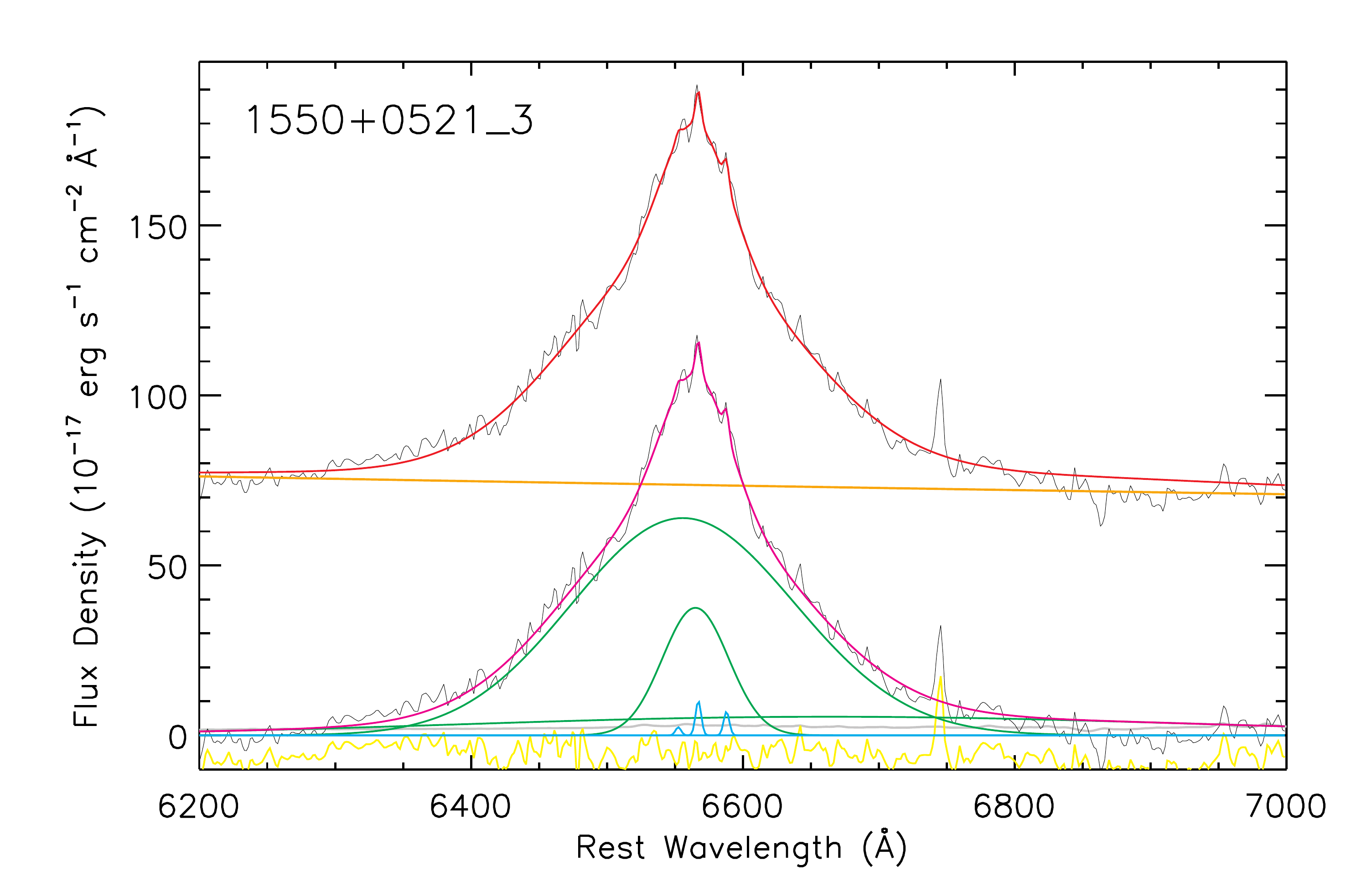}
    \includegraphics[width=80mm]{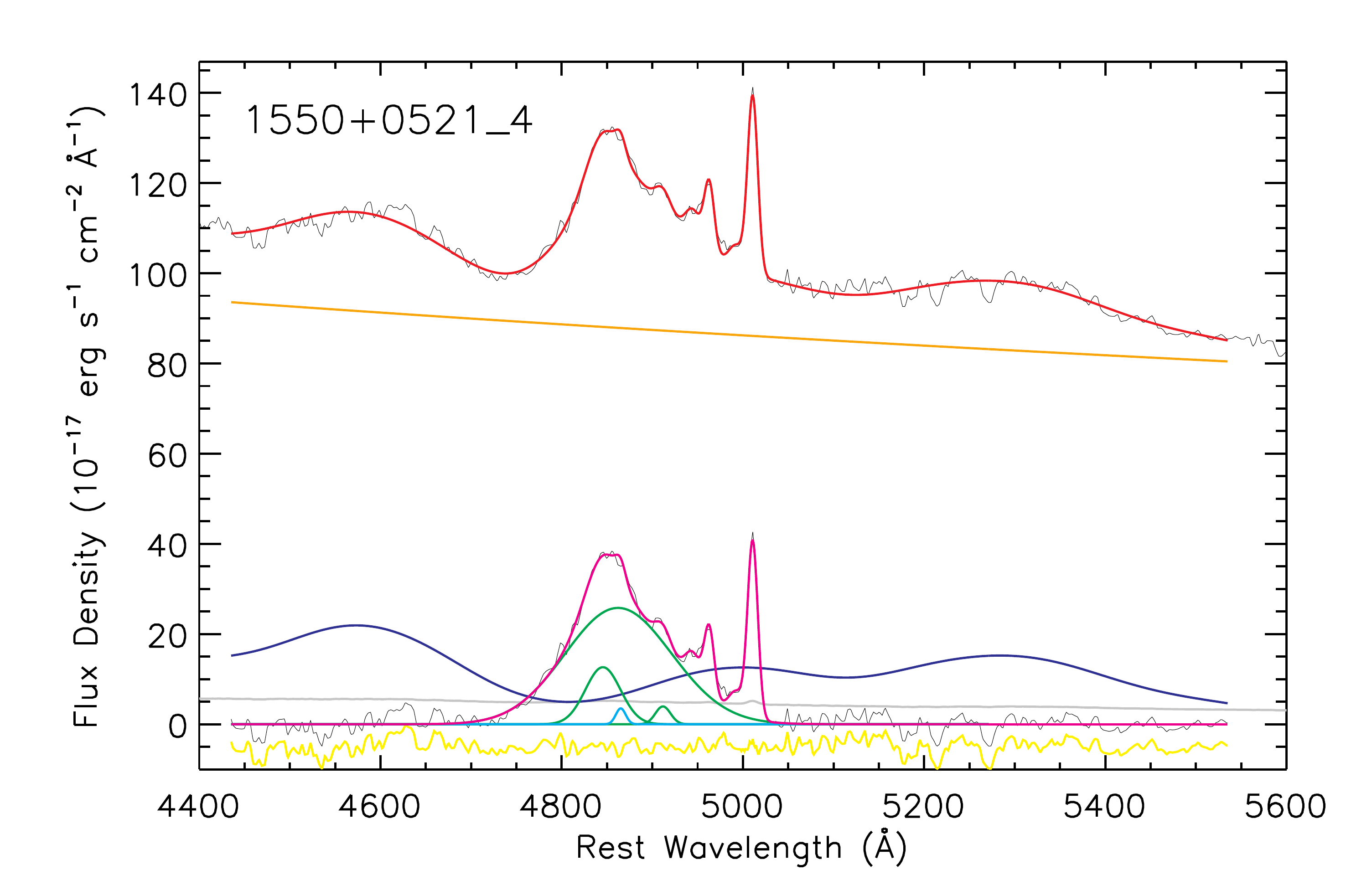}
    \includegraphics[width=80mm]{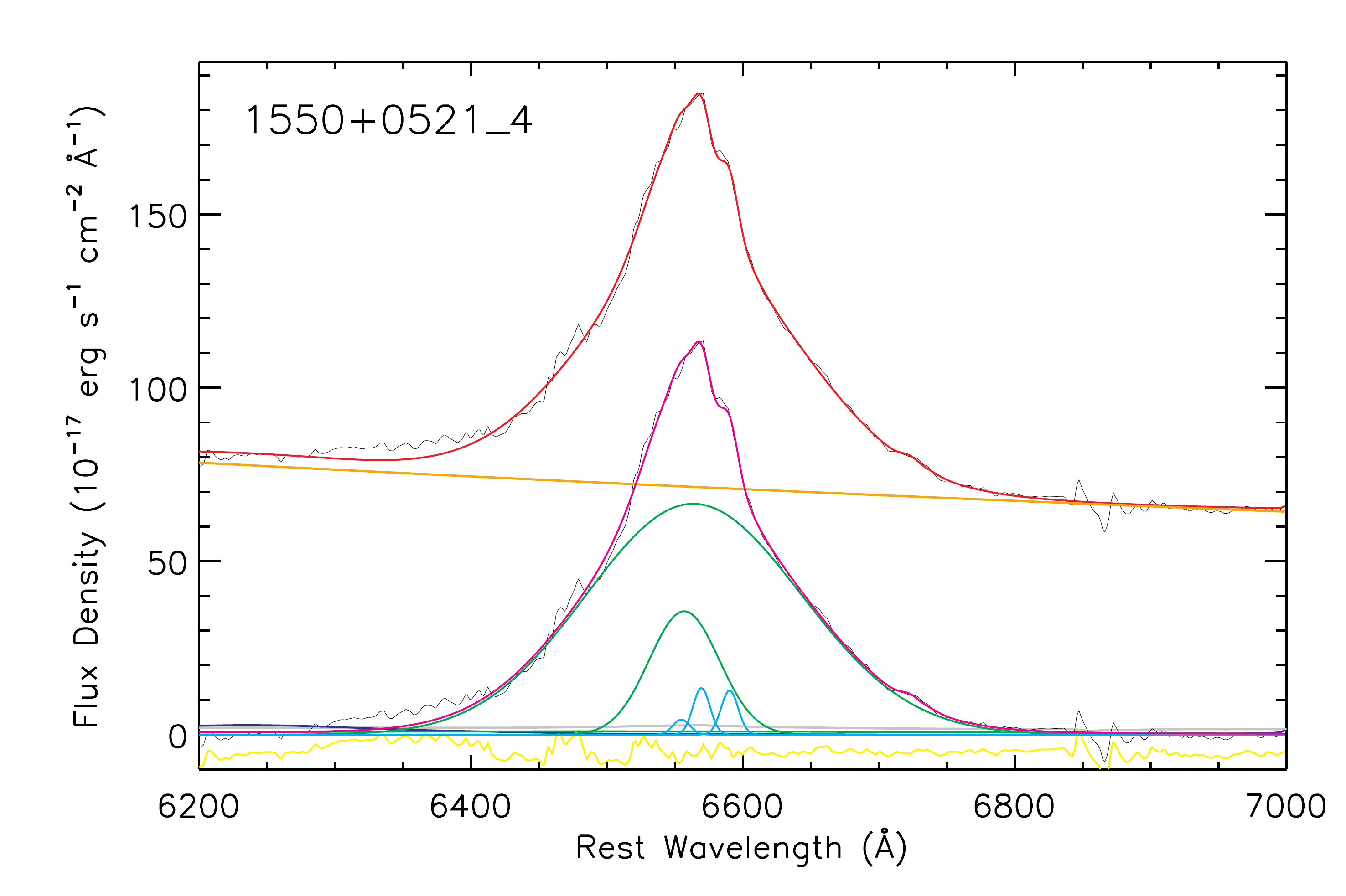}
       \centerline{Figure A1 -- Continued.    }   
\end{figure*}
\clearpage

\begin{figure*}
  \centering
    \includegraphics[width=80mm]{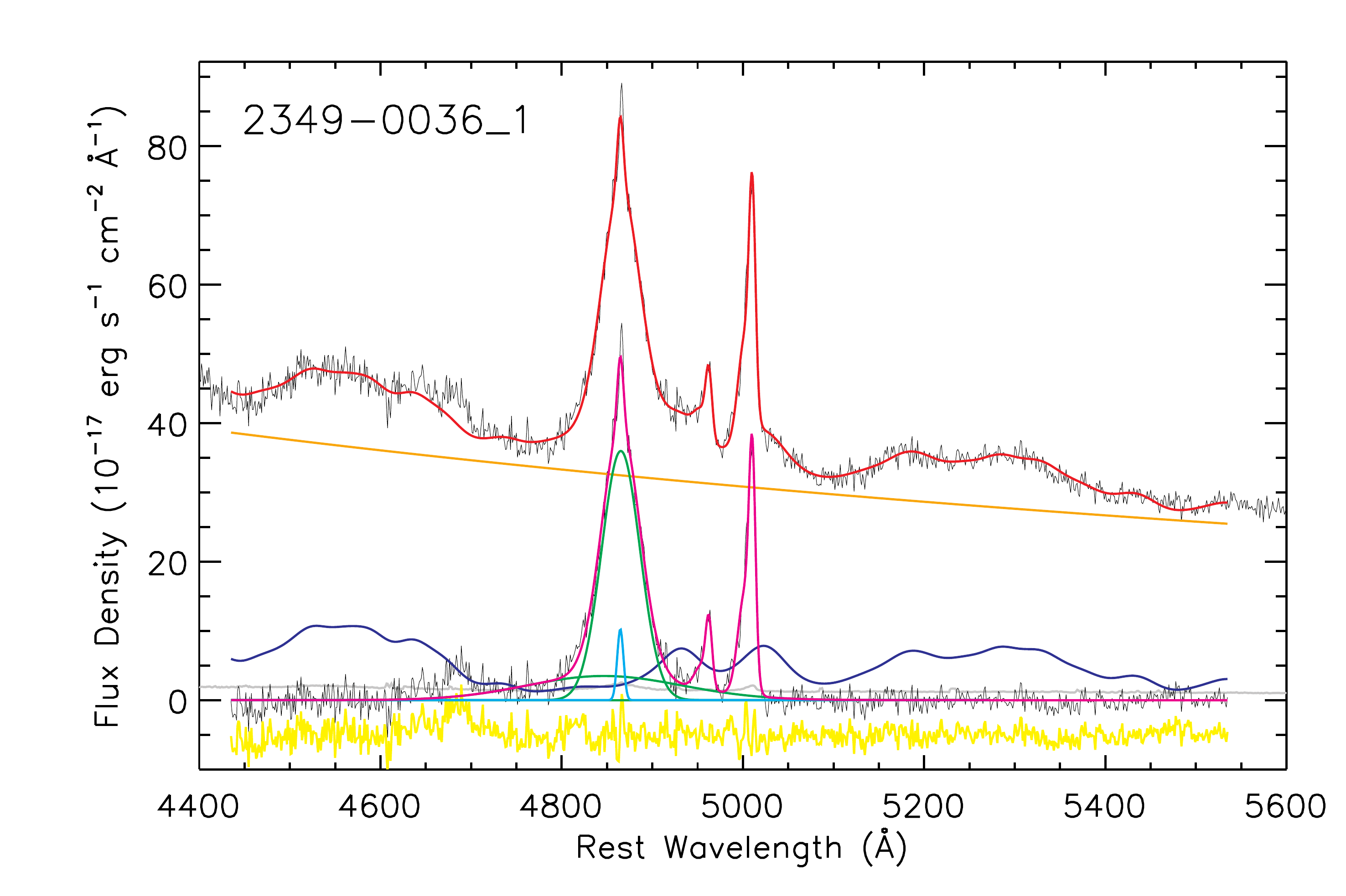}
    \includegraphics[width=80mm]{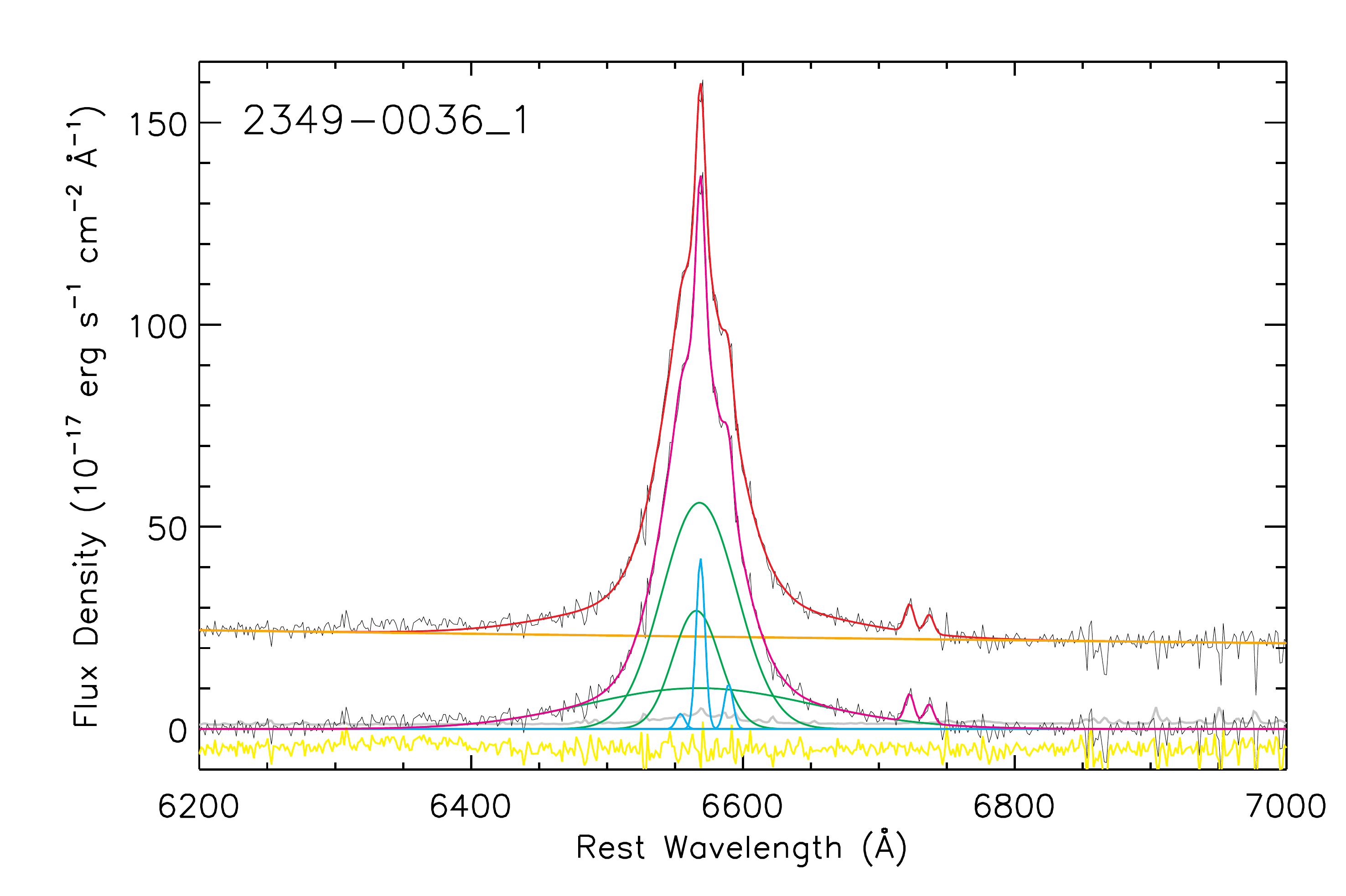}
    \includegraphics[width=80mm]{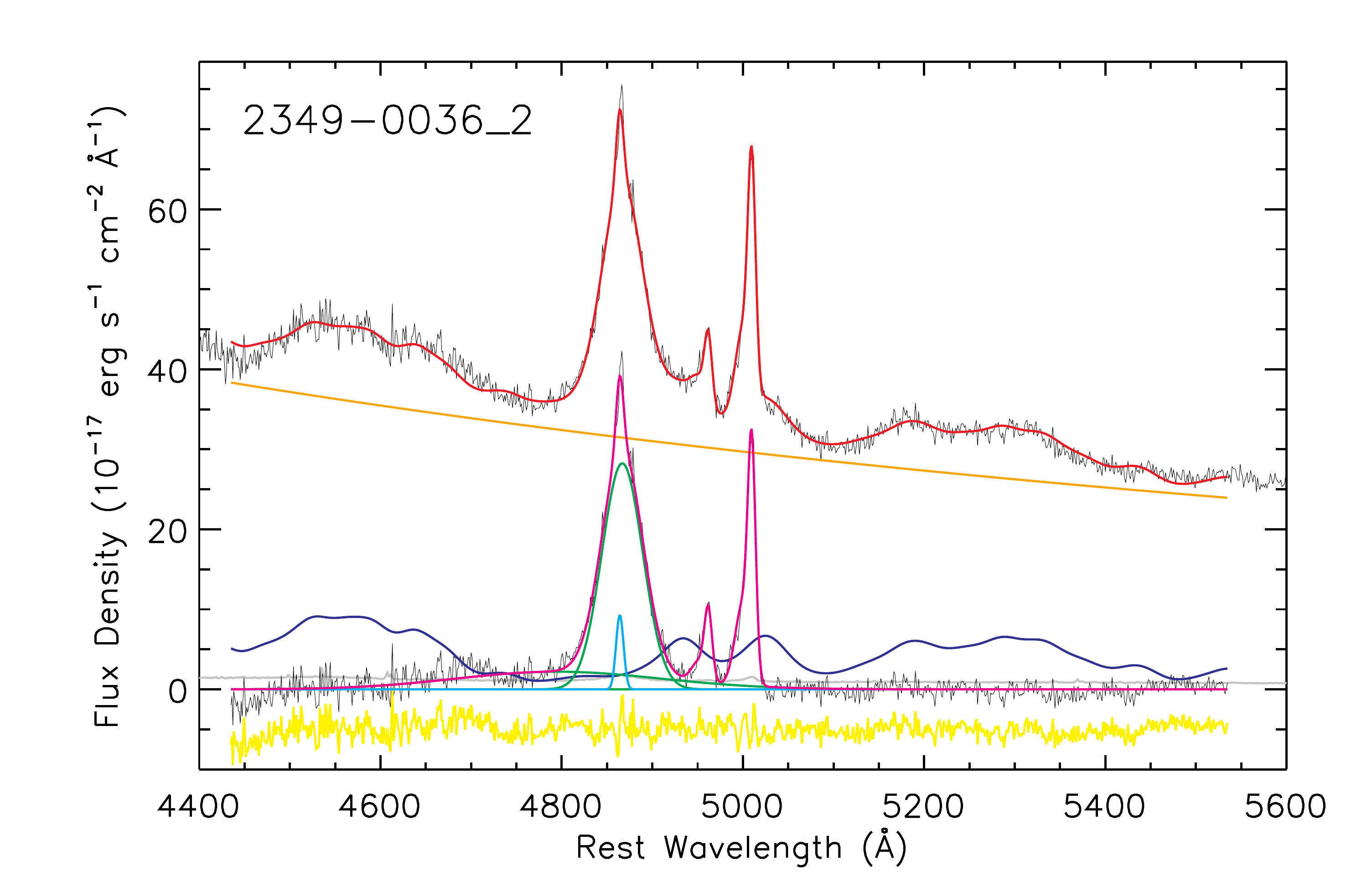}
    \includegraphics[width=80mm]{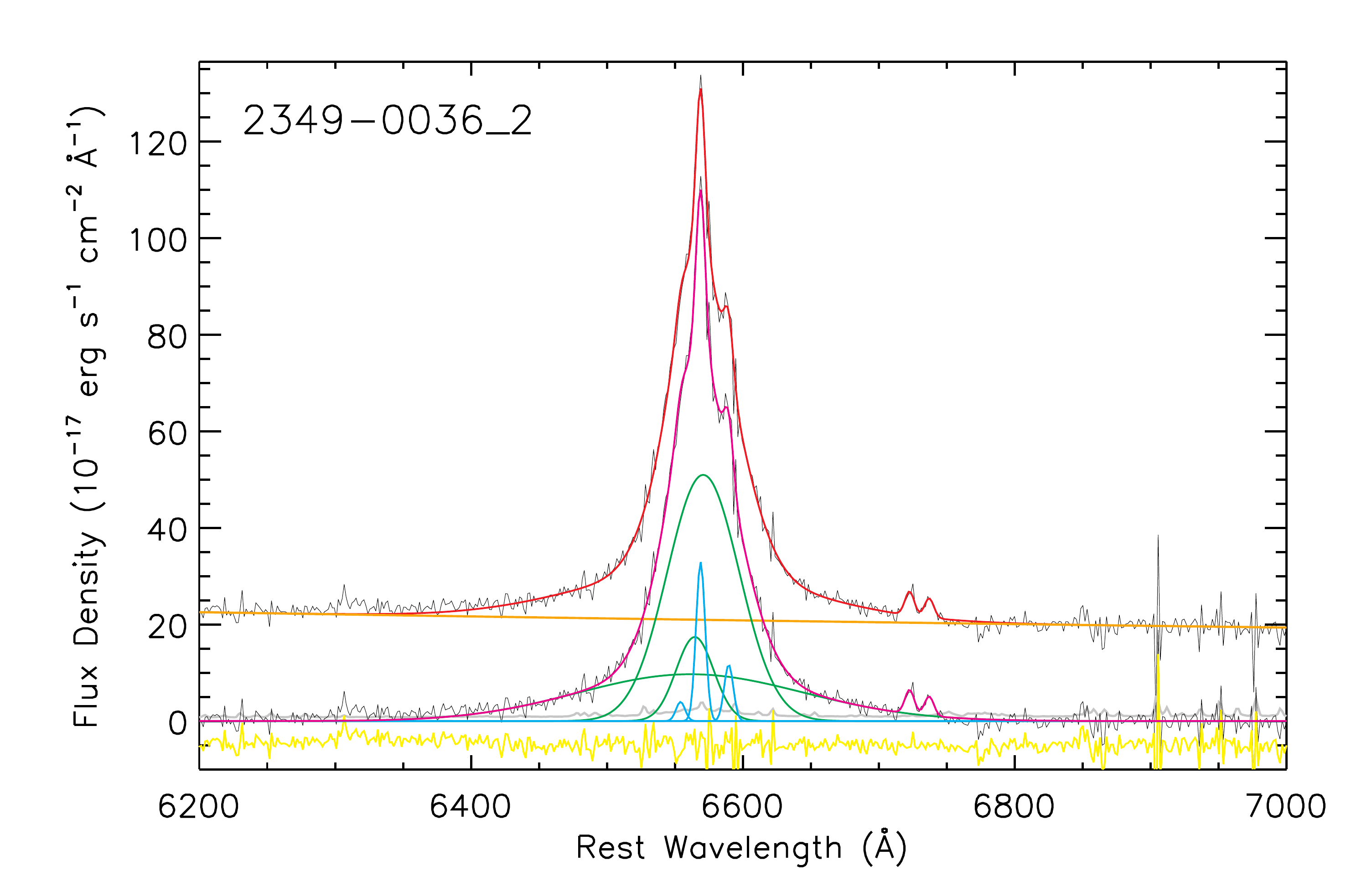}
    \includegraphics[width=80mm]{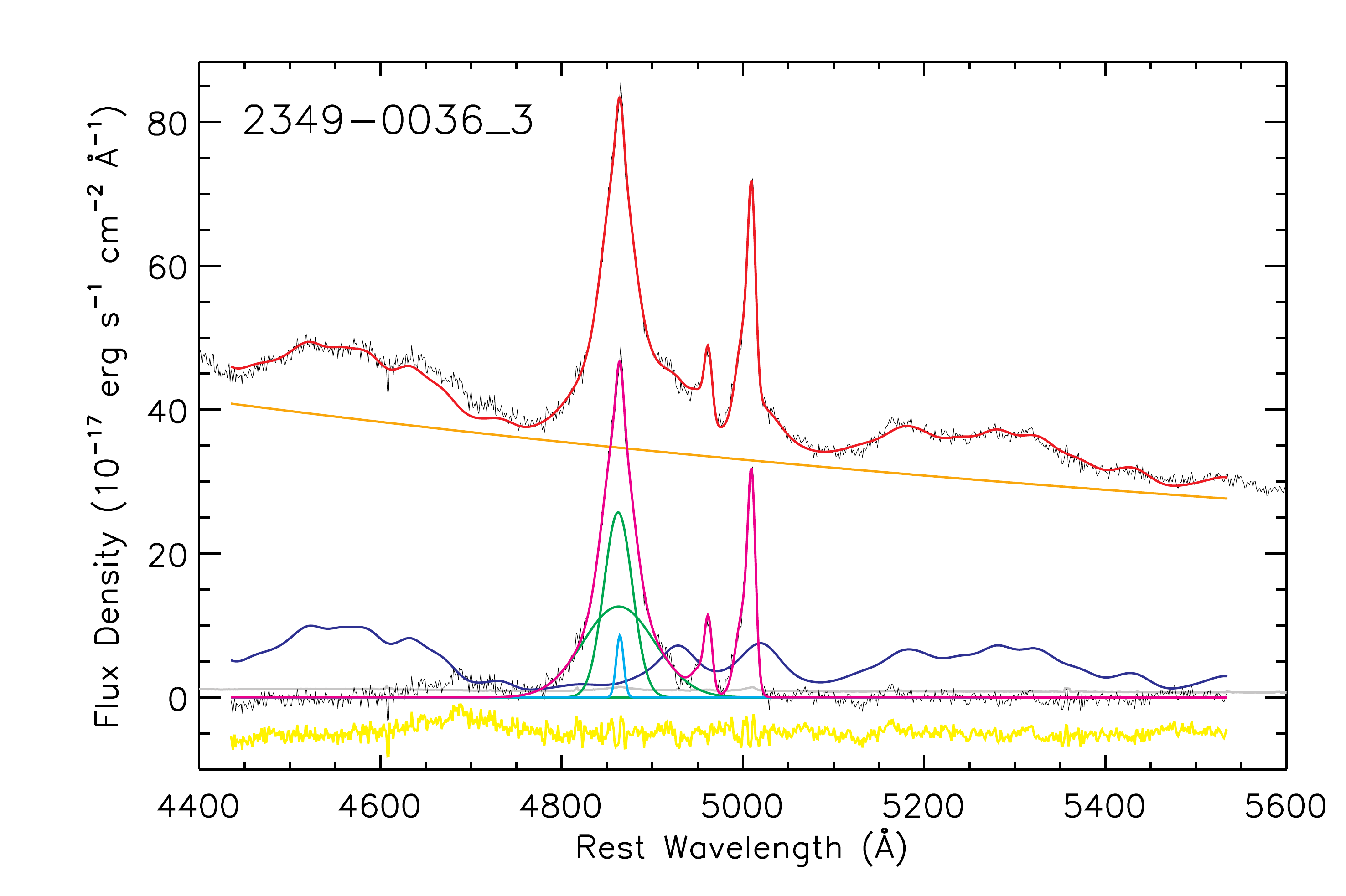}
    \includegraphics[width=80mm]{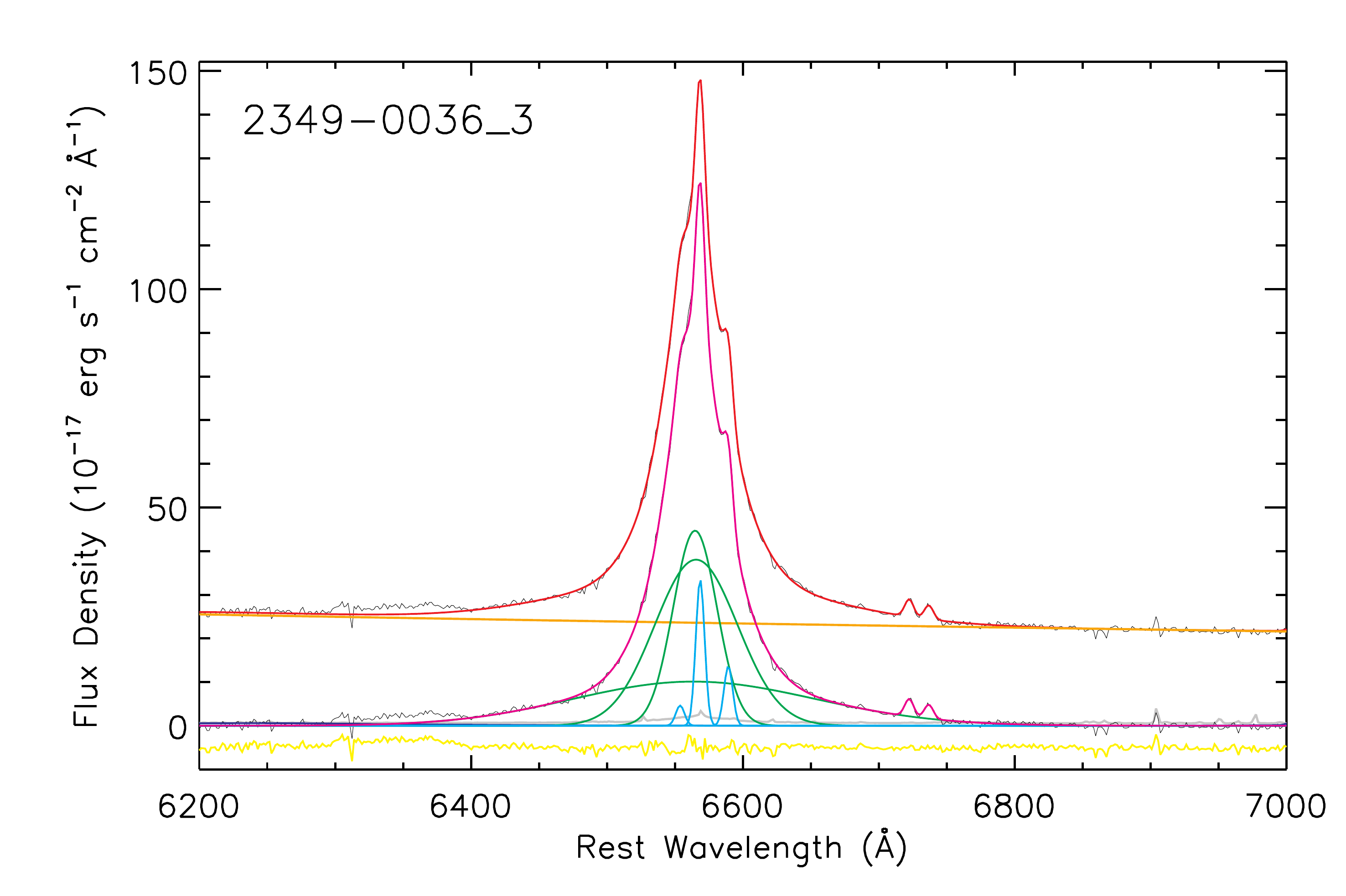}
    \includegraphics[width=80mm]{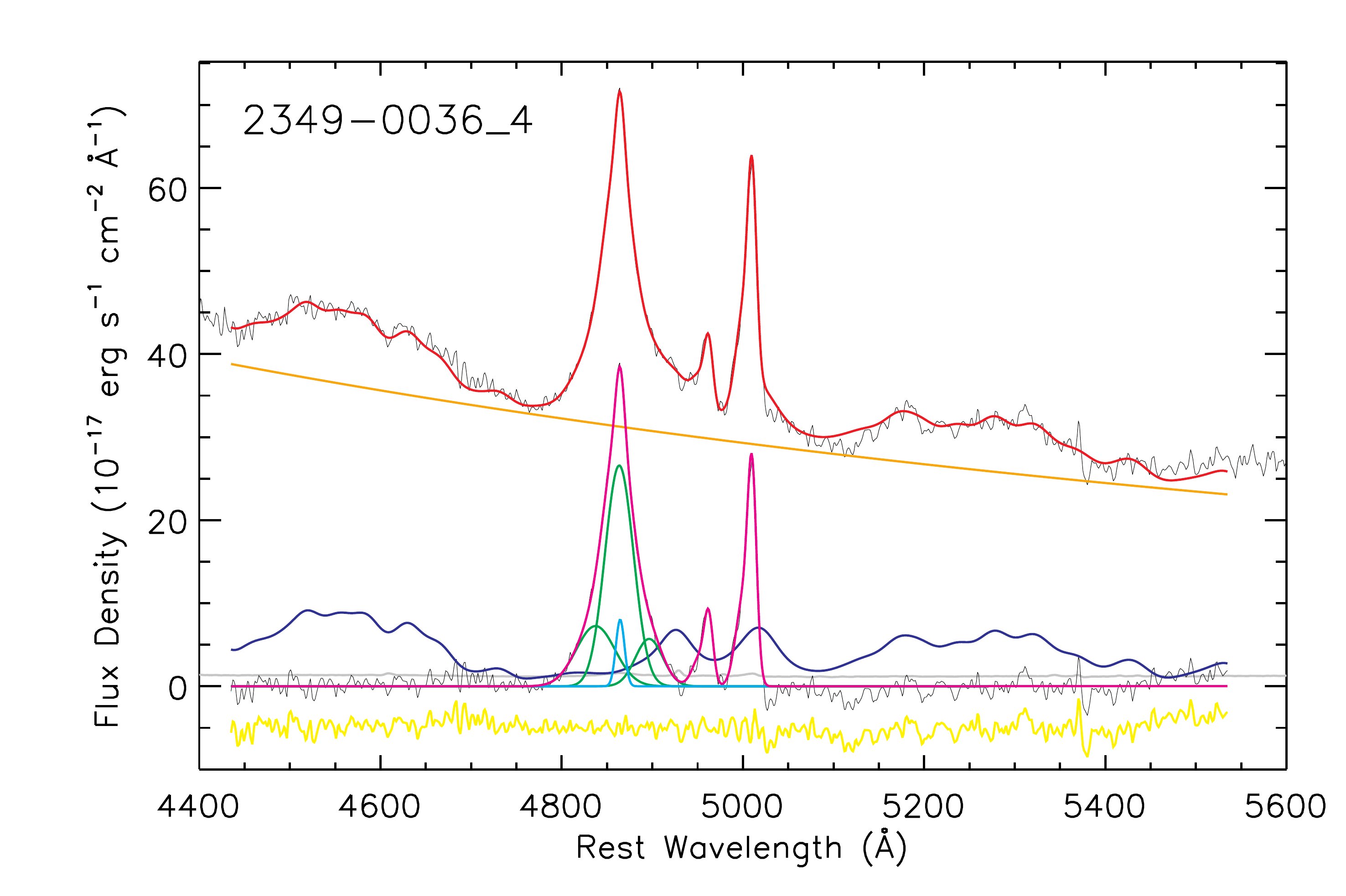}
    \includegraphics[width=80mm]{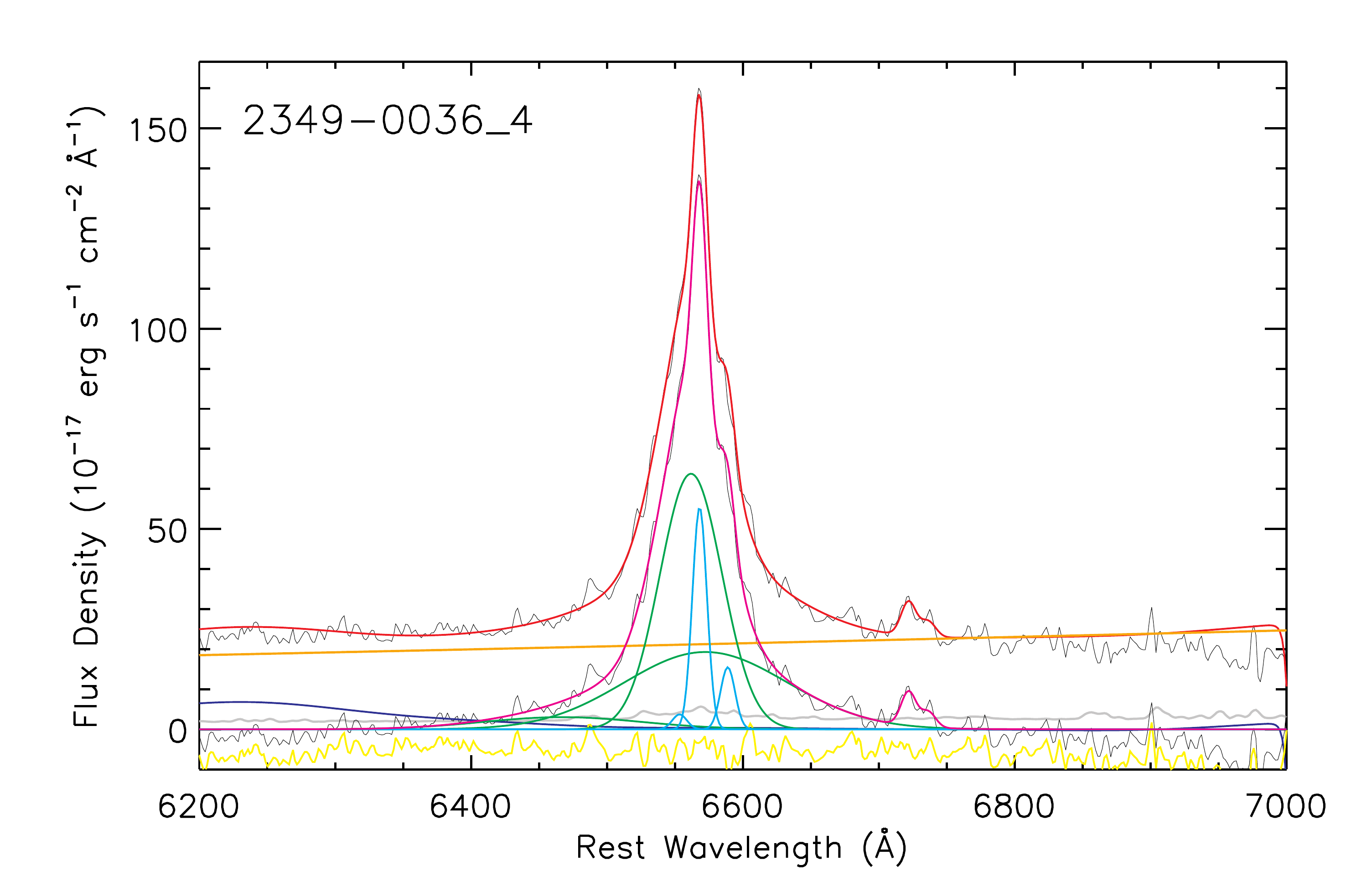}
       \centerline{Figure A1 -- Continued.    }  
\end{figure*}

\clearpage
%\end{comment}

%%%%%%%%%

\section{Cross-correlation analysis results for all targets}\label{appendix:ccf}
All the figures are available online.

%\begin{comment}
\begin{figure*}
  \centering
    \includegraphics[width=80mm]{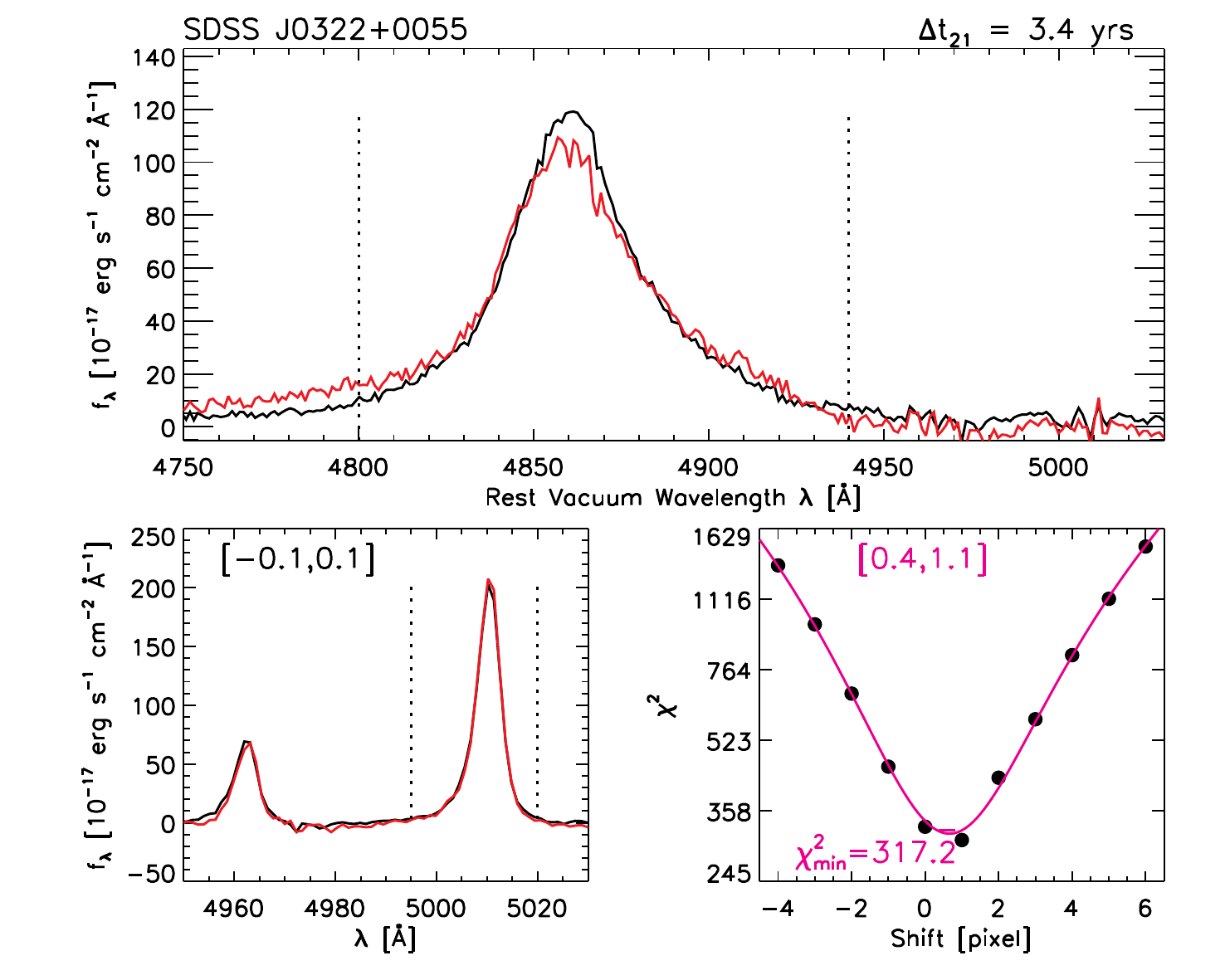}
    \includegraphics[width=80mm]{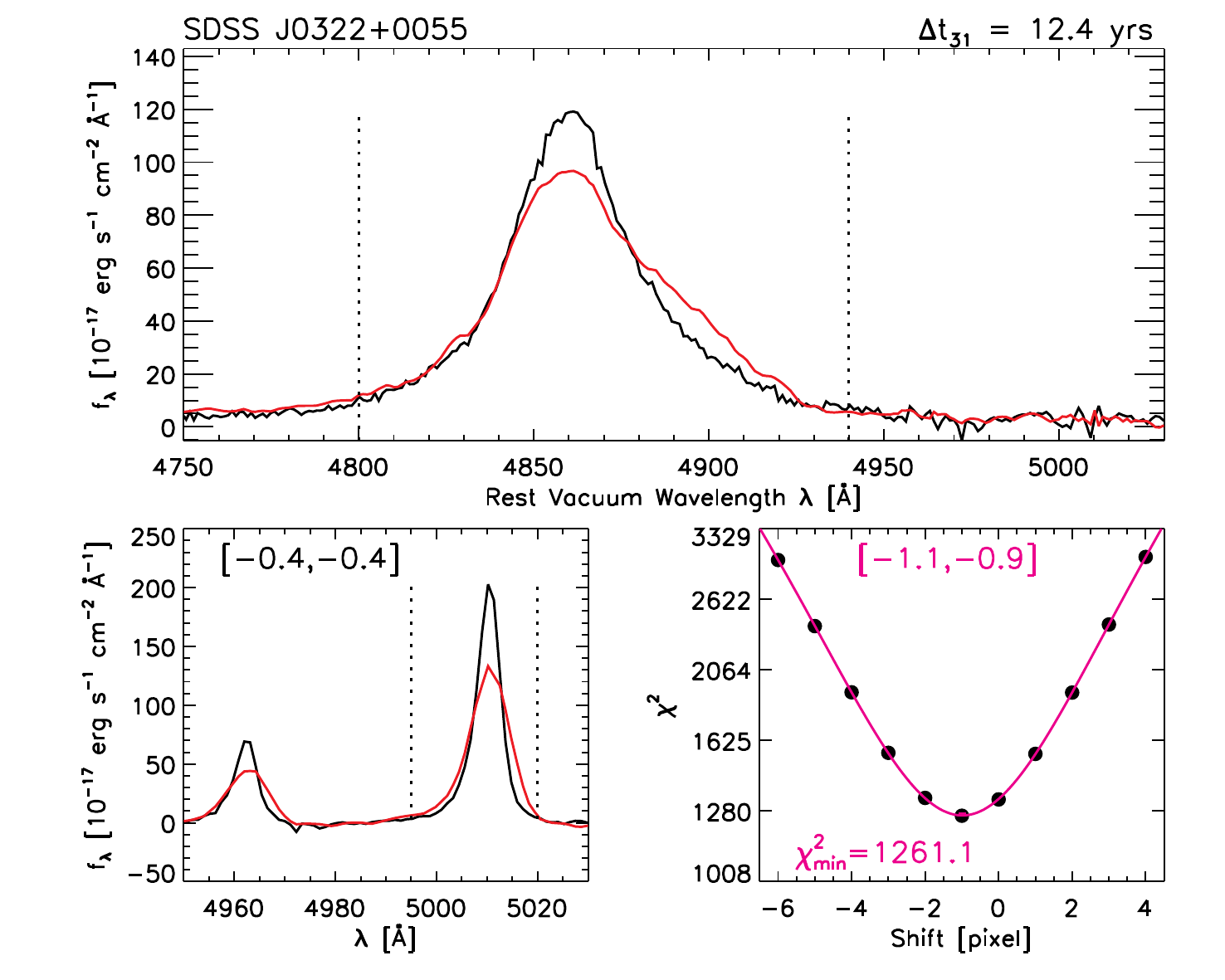}
    \includegraphics[width=80mm]{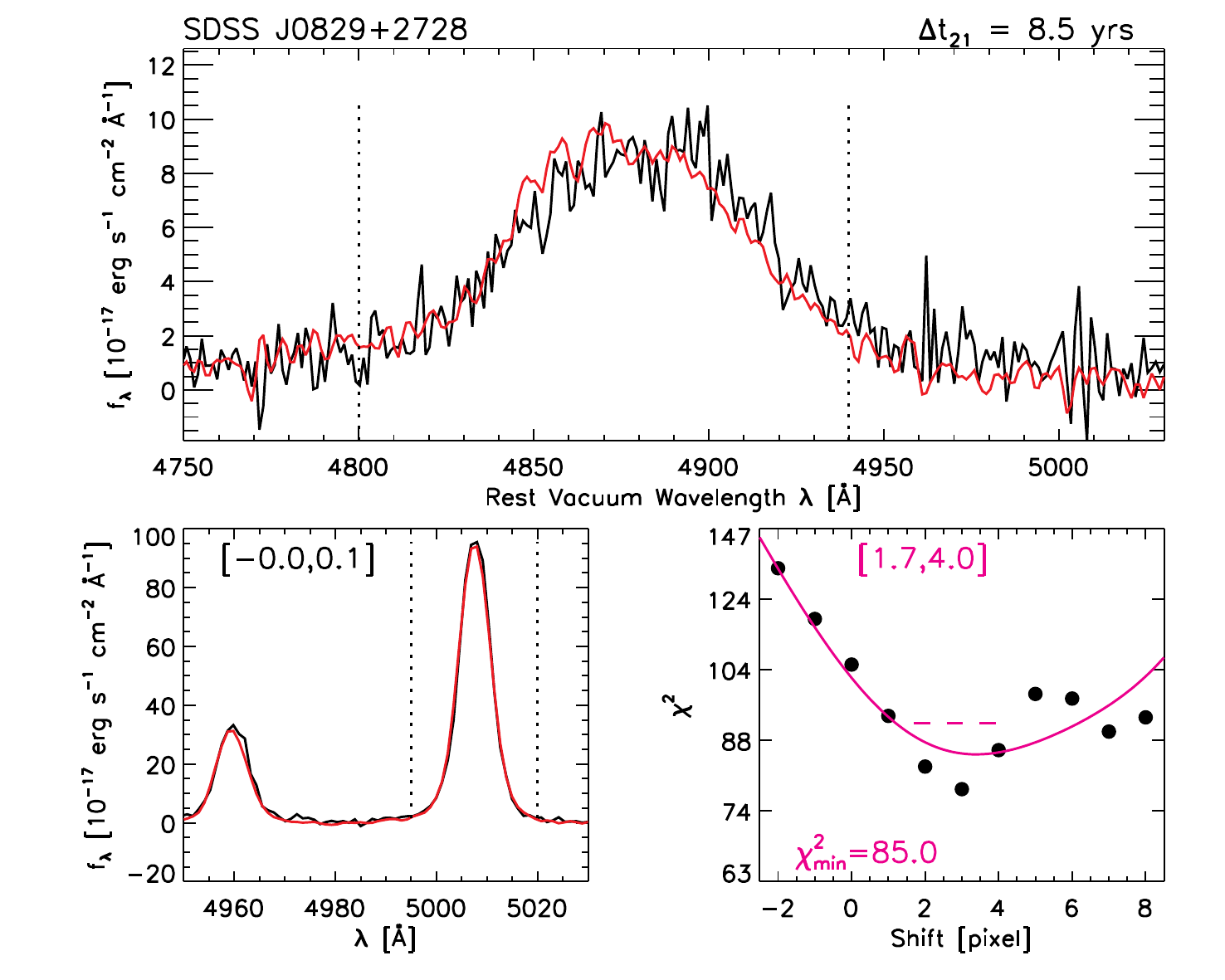}
    \includegraphics[width=80mm]{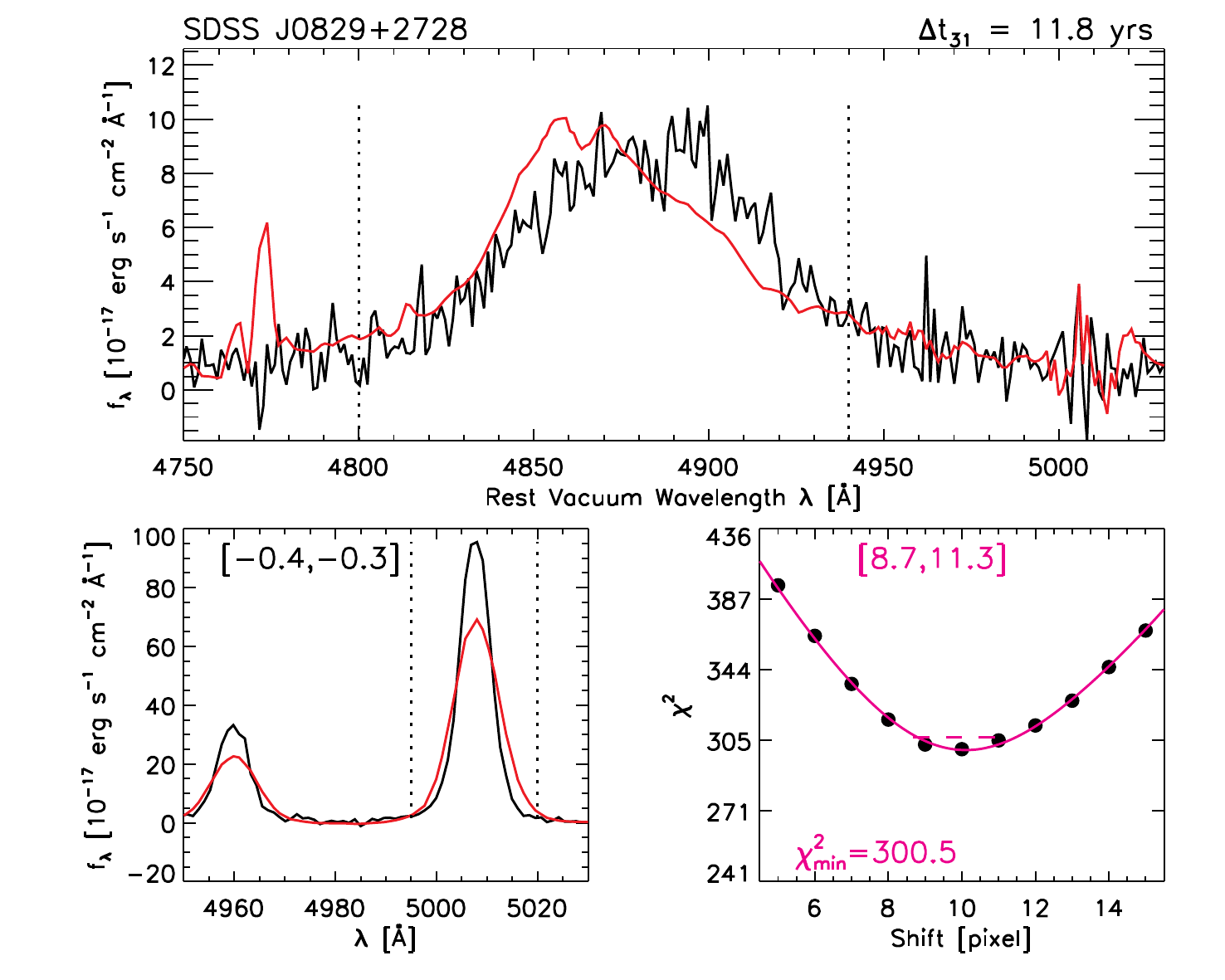}
       \centerline{Figure B1 -- Similar to Figure \ref{fig:ccf_eg}, but for the other targets and/or epochs.}  
\end{figure*}

\begin{figure*}
  \centering
    \includegraphics[width=80mm]{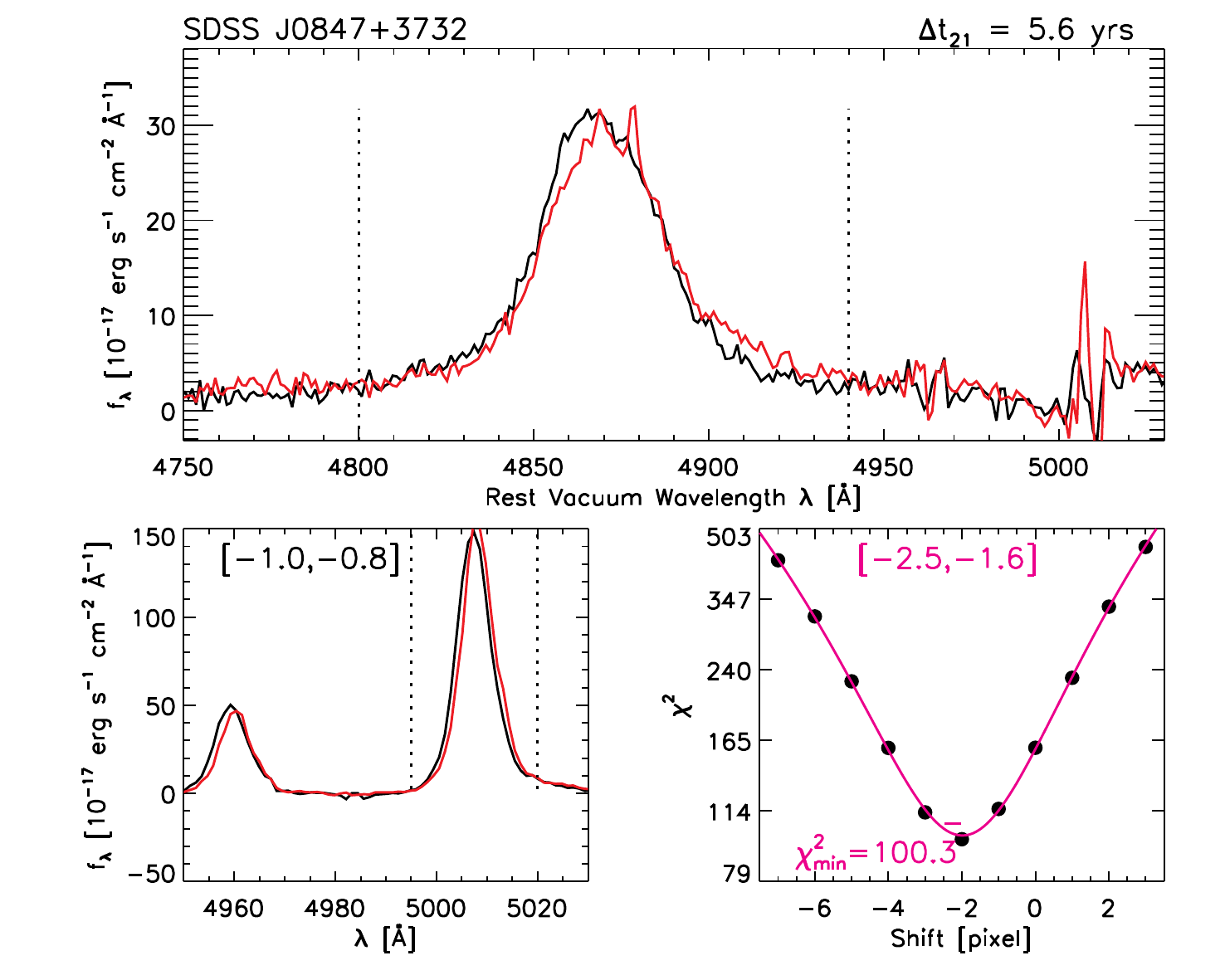}
    \includegraphics[width=80mm]{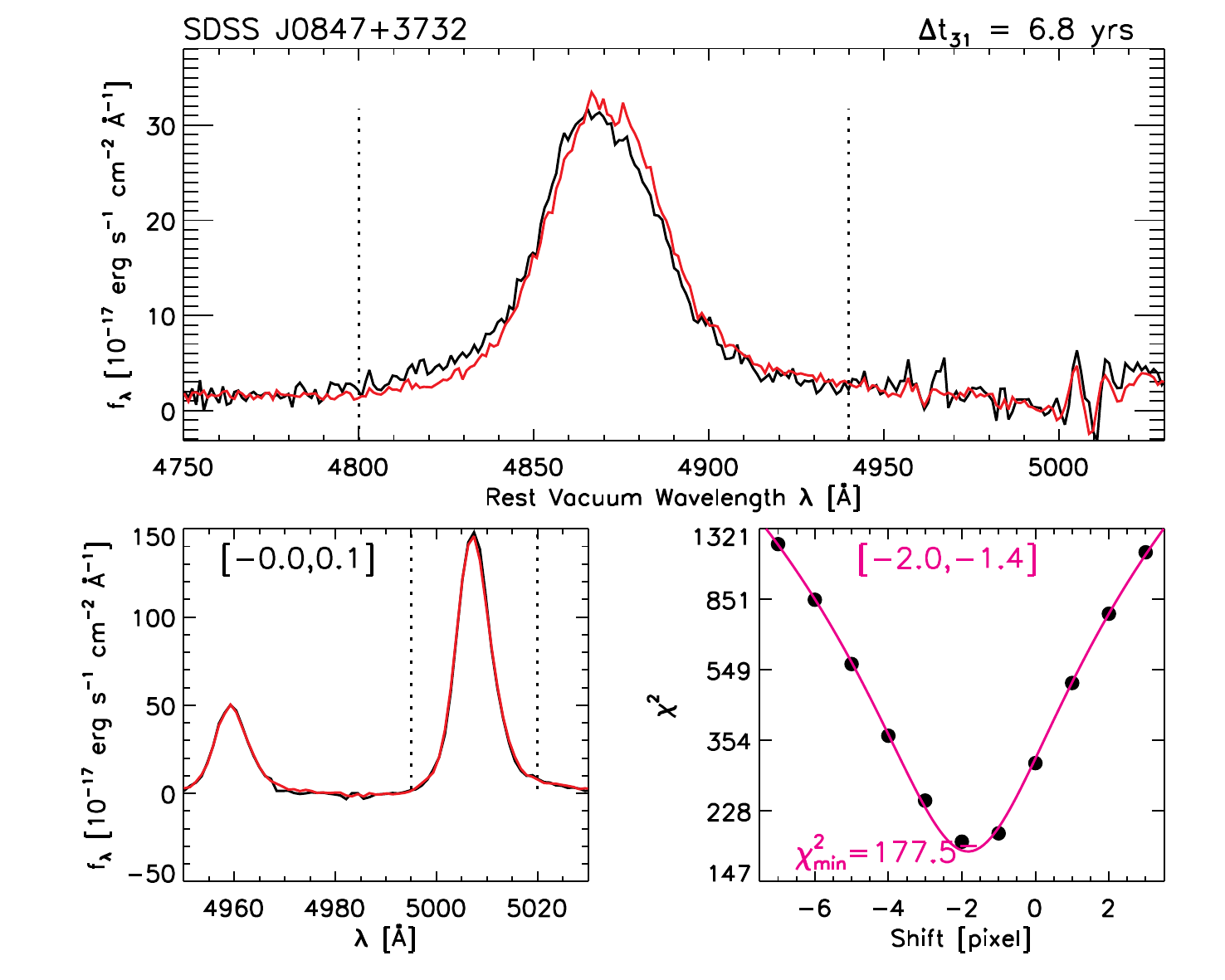}
    \includegraphics[width=80mm]{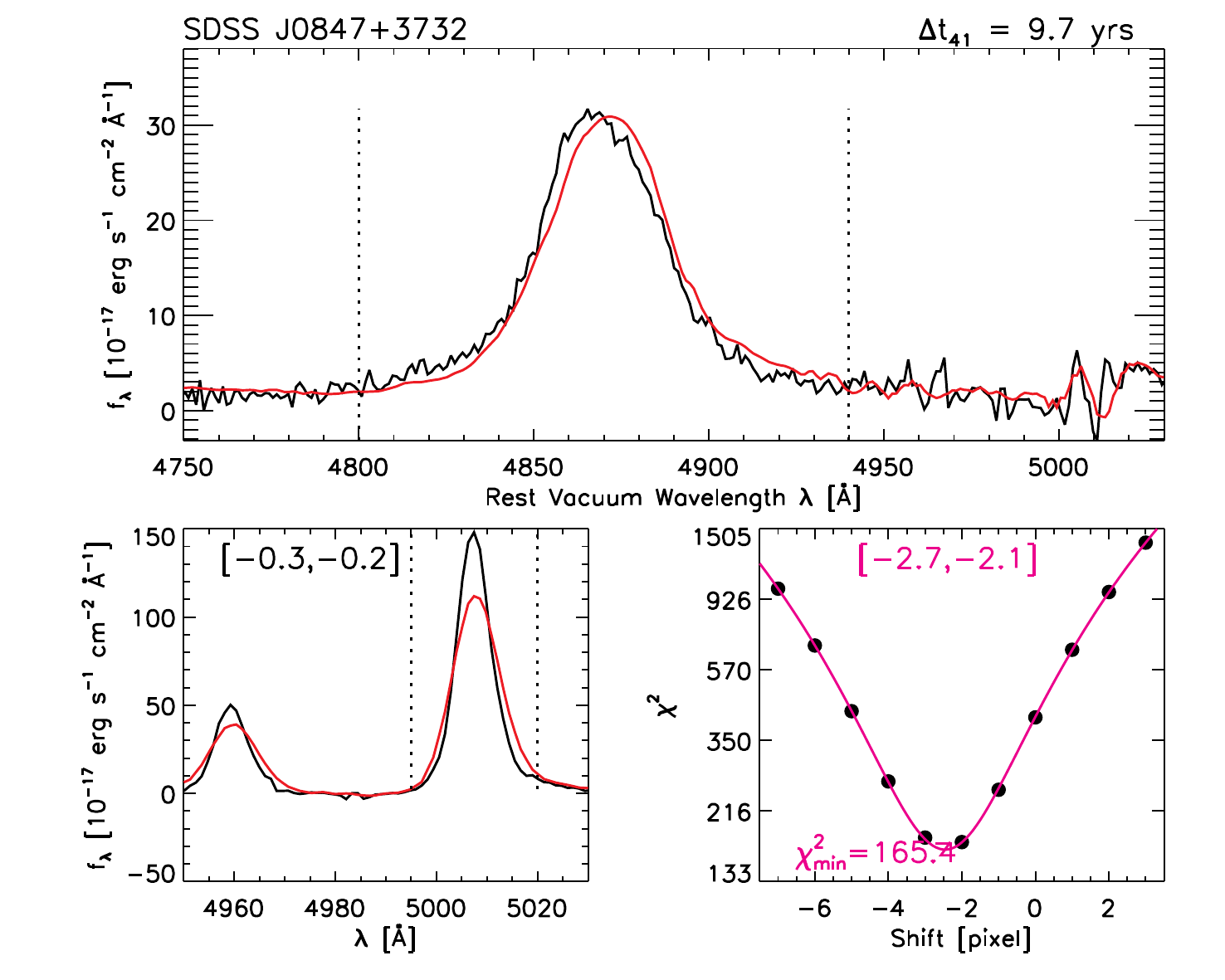}
    \includegraphics[width=80mm]{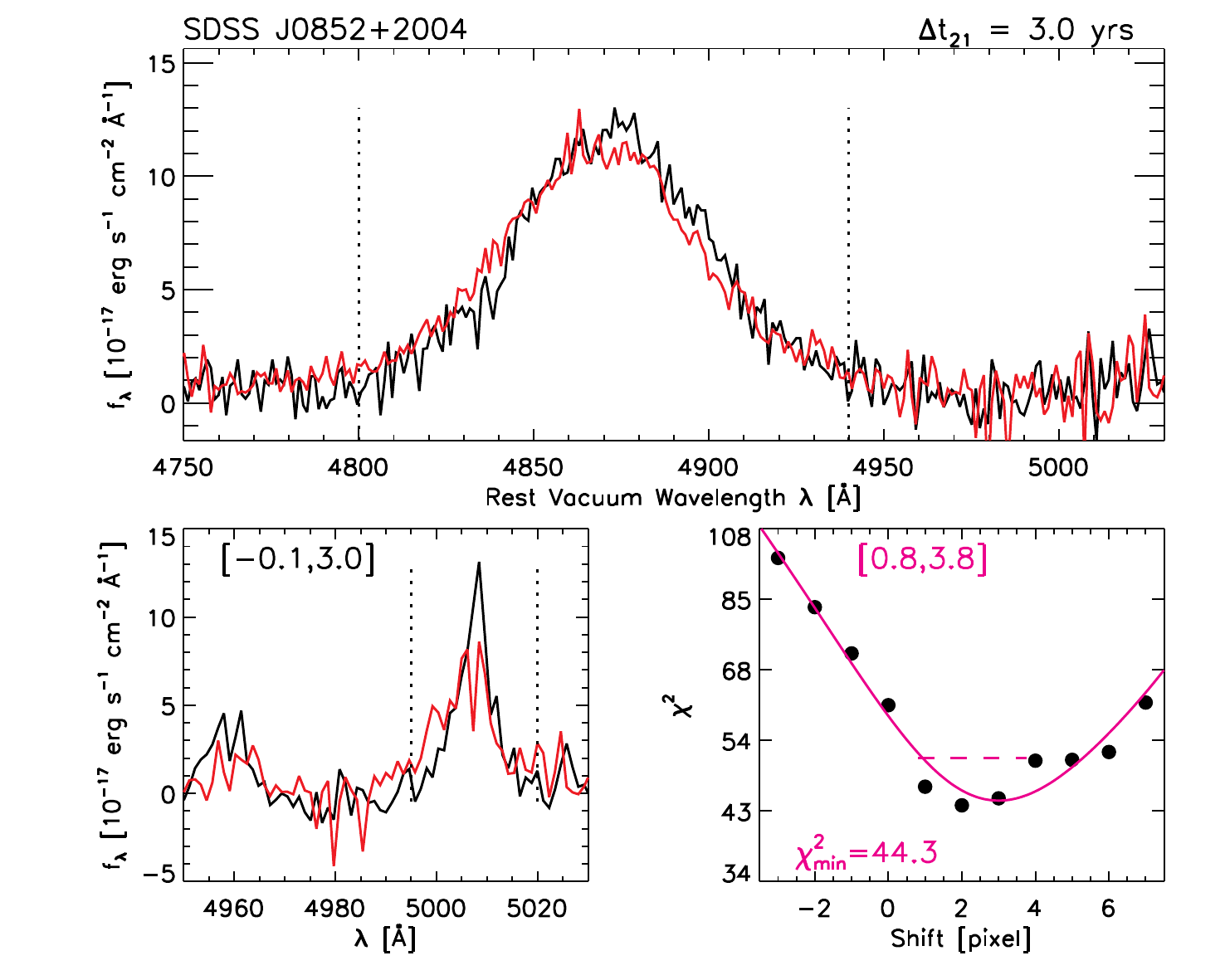}
    \includegraphics[width=80mm]{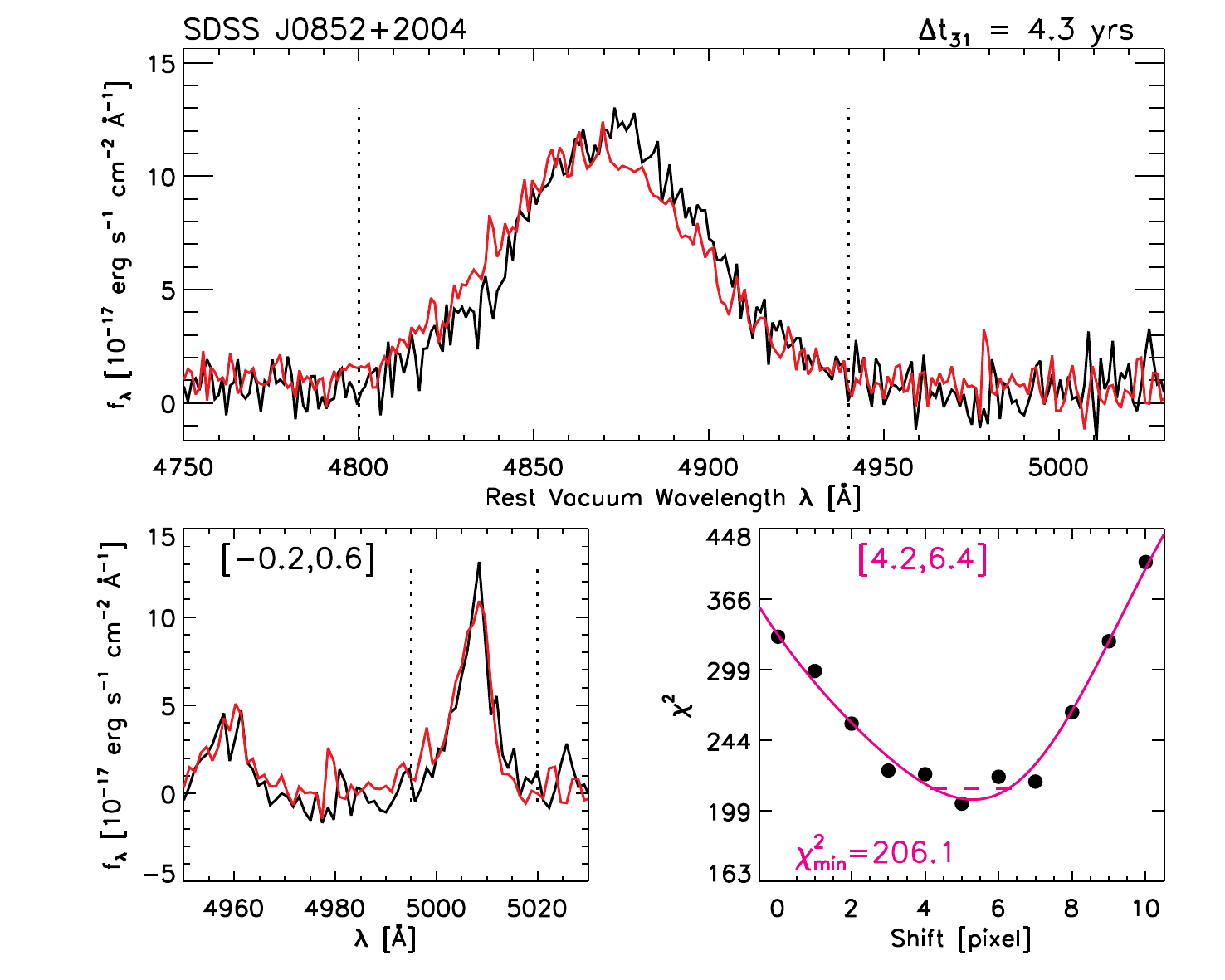}
    \includegraphics[width=80mm]{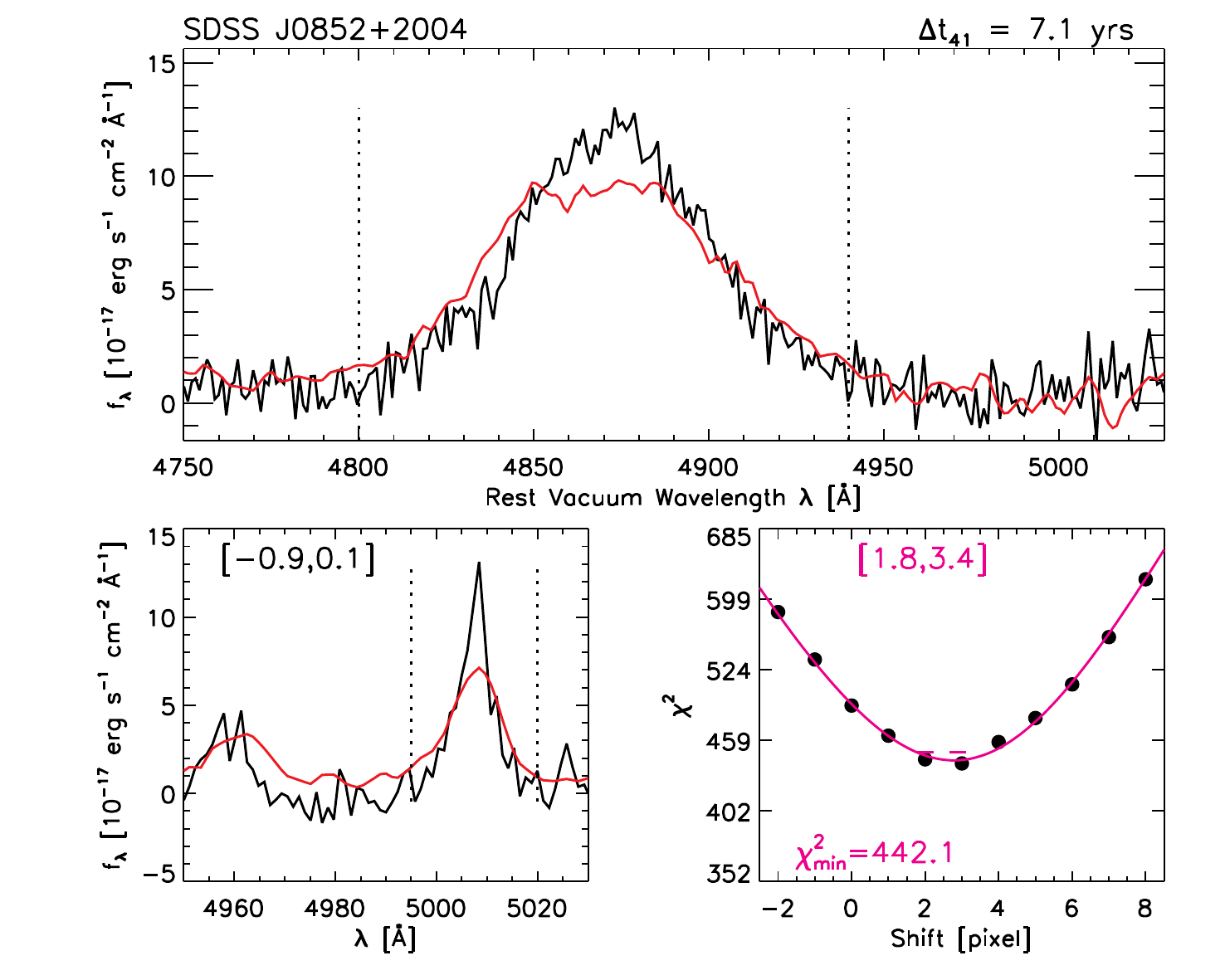}
       \centerline{Figure B1 -- Continued.   }  
\end{figure*}

\begin{figure*}
  \centering
    \includegraphics[width=80mm]{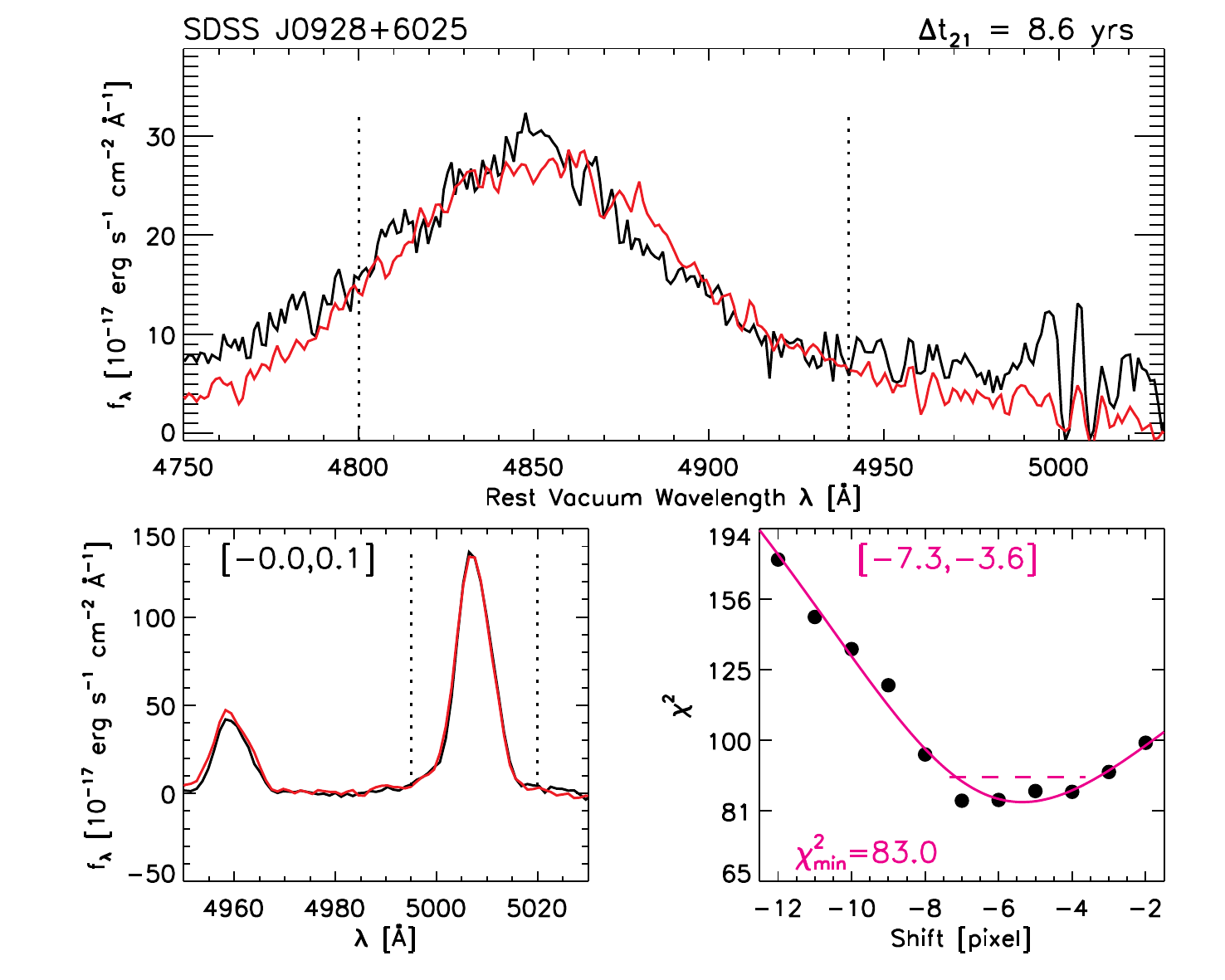}
    \includegraphics[width=80mm]{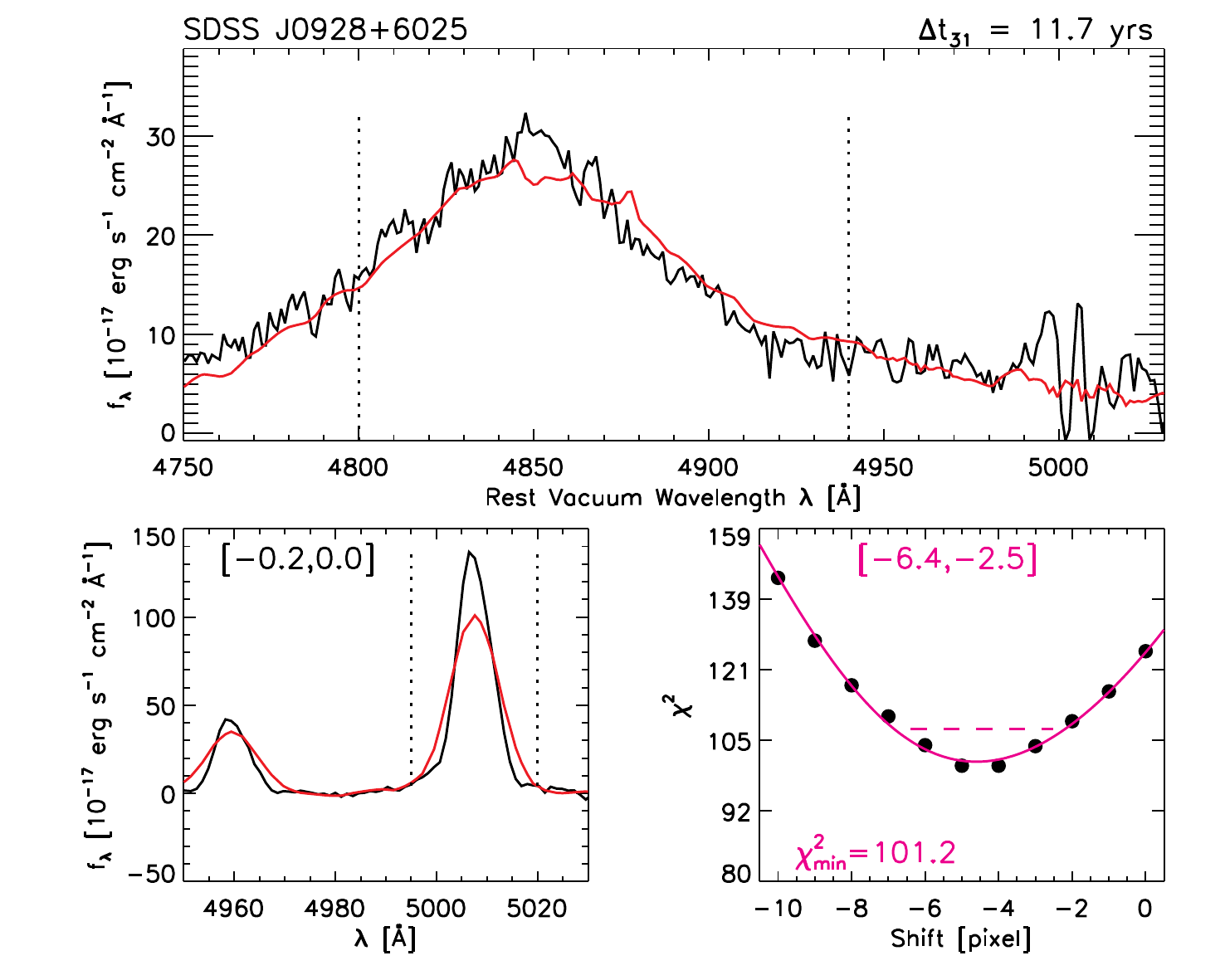}
    \includegraphics[width=80mm]{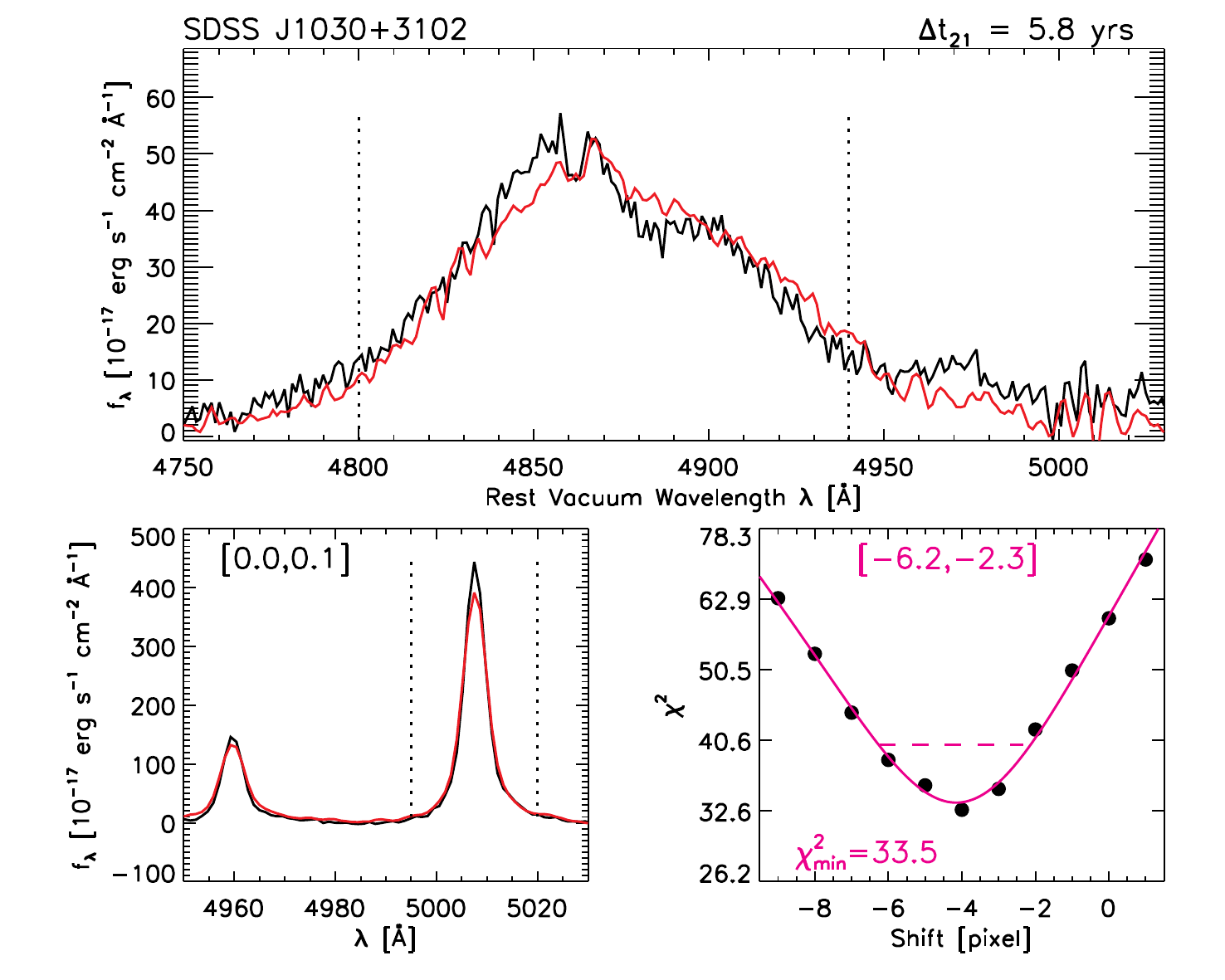}
    \includegraphics[width=80mm]{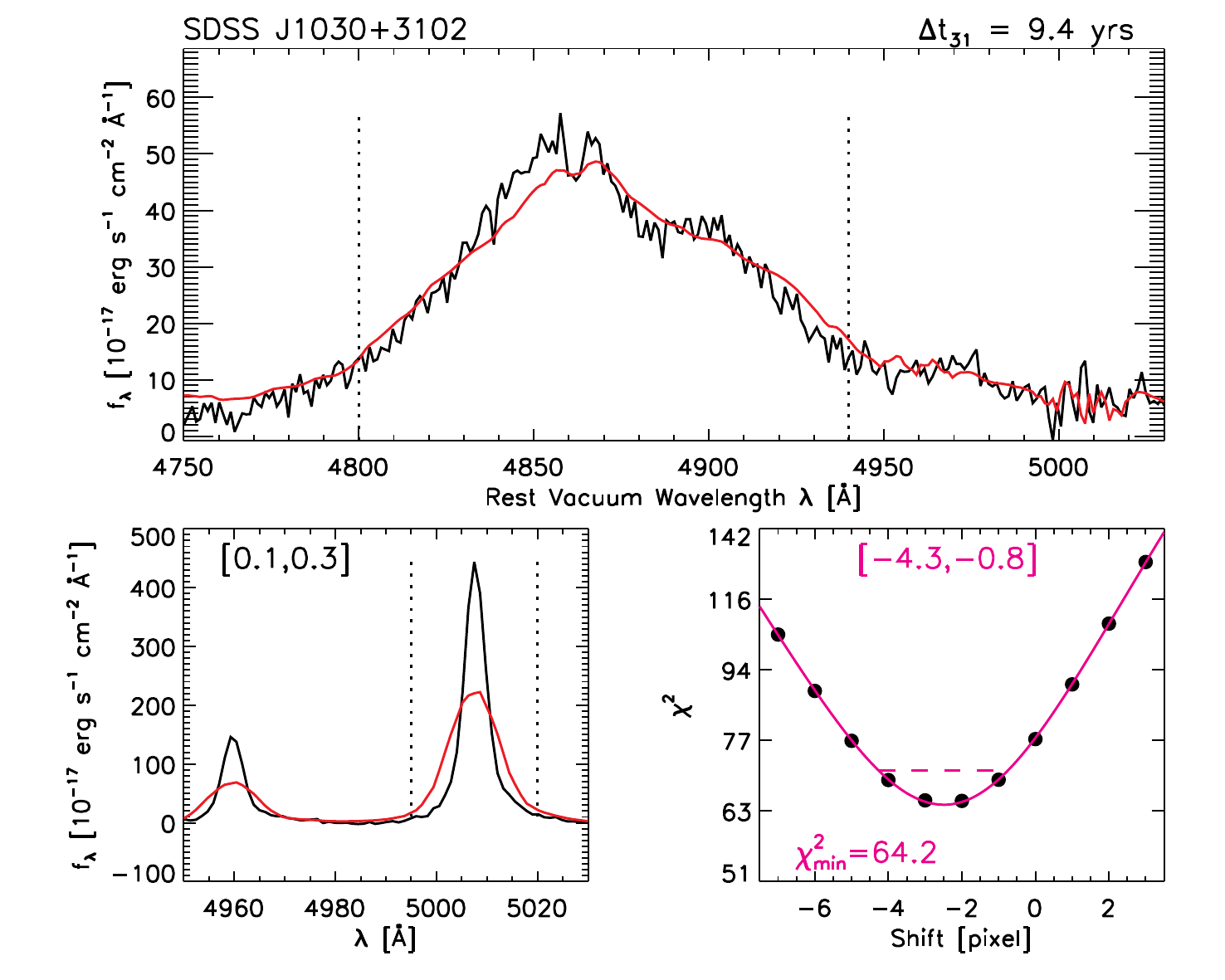}
    \includegraphics[width=80mm]{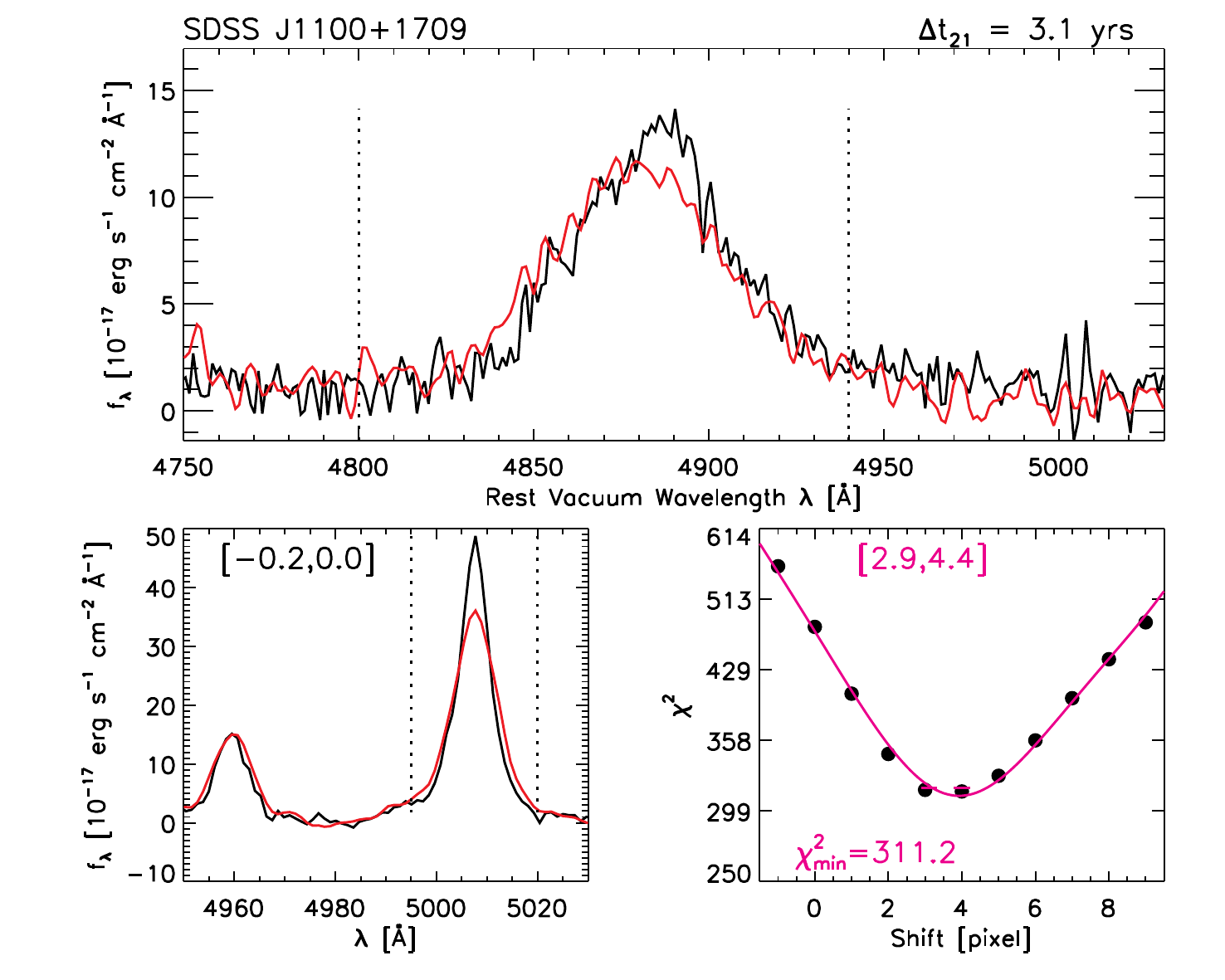}	
    \includegraphics[width=80mm]{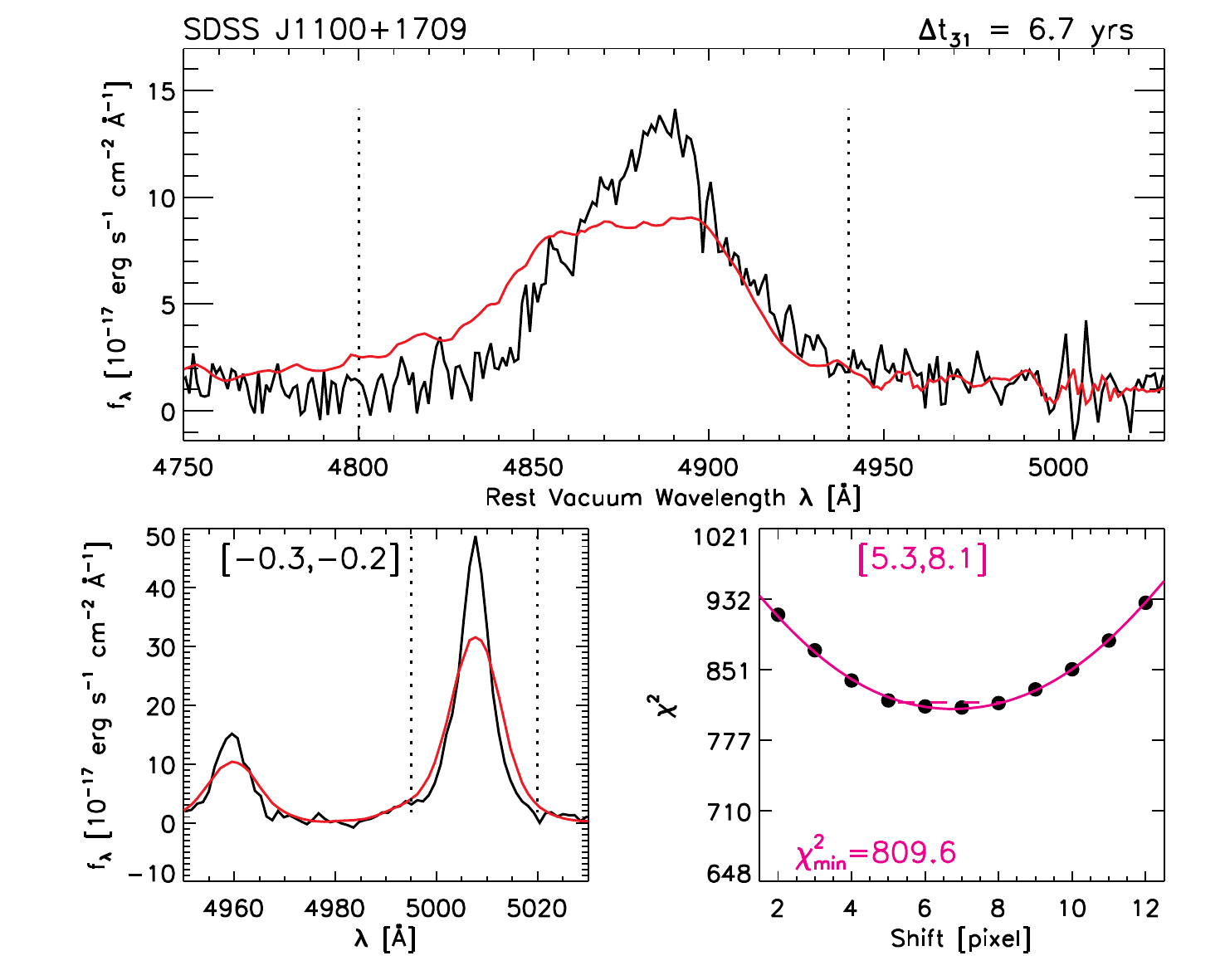}
       \centerline{Figure B1 -- Continued.   }  
\end{figure*}

\begin{figure*}
  \centering
    \includegraphics[width=80mm]{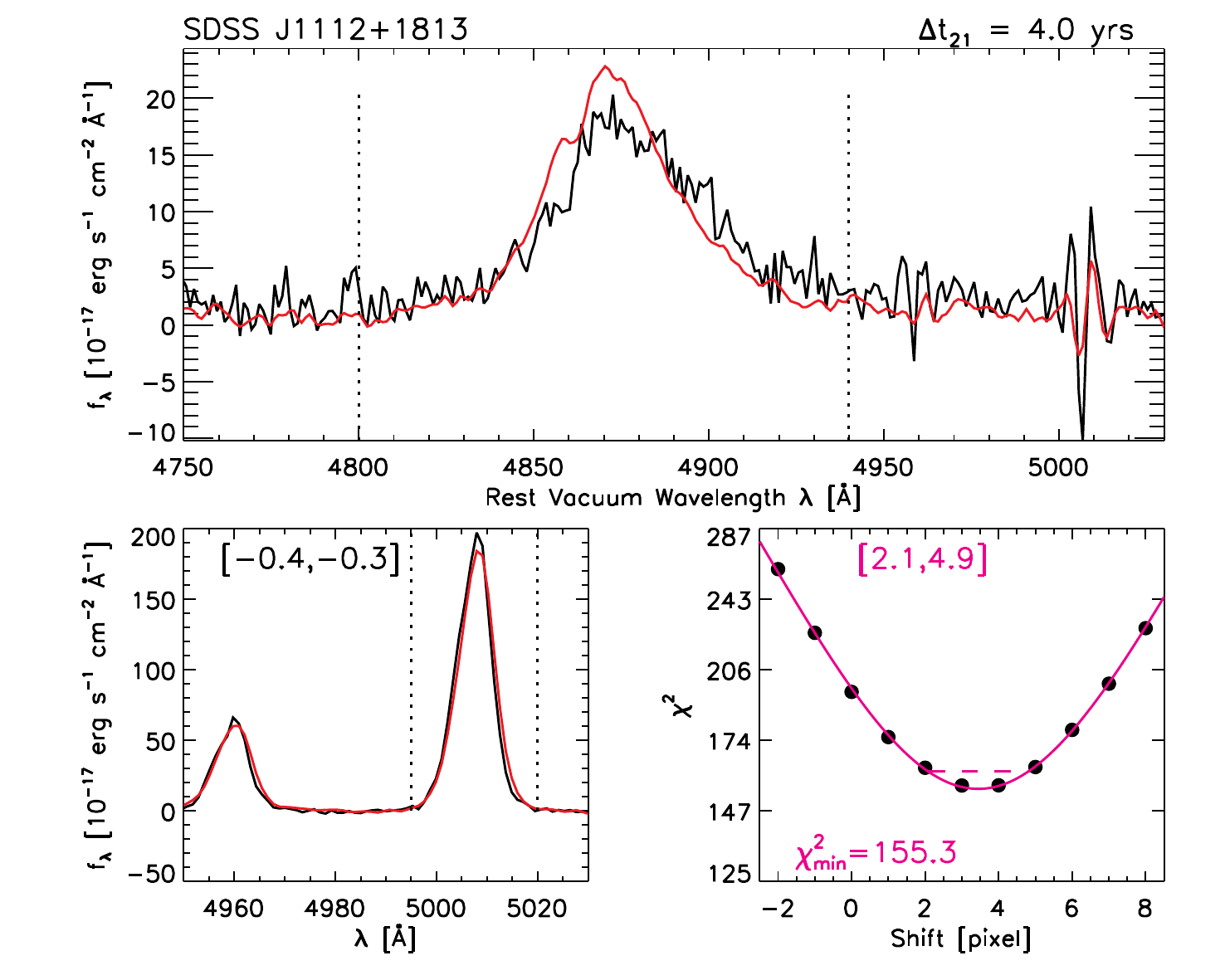}
    \includegraphics[width=80mm]{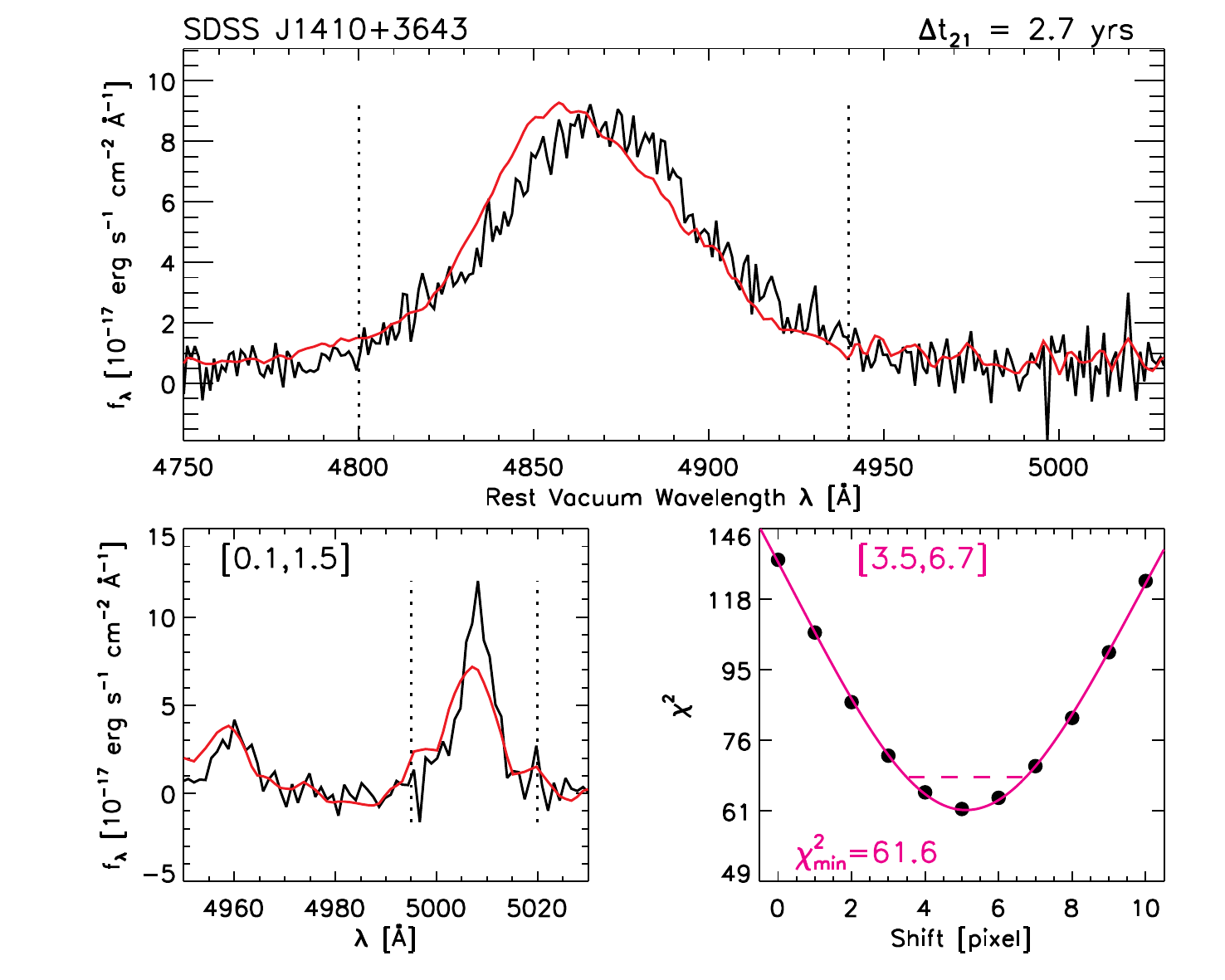}
    \includegraphics[width=80mm]{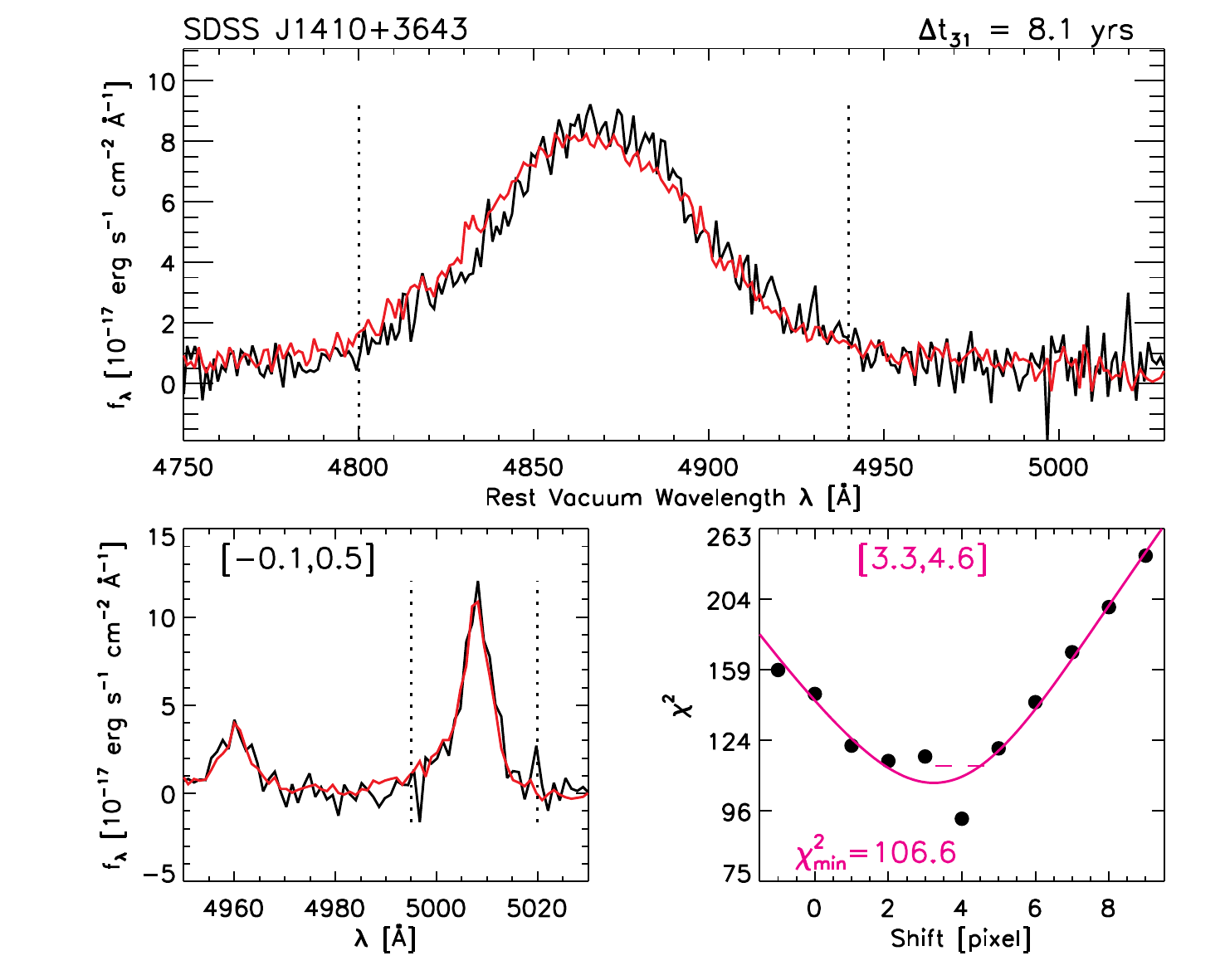}
    \includegraphics[width=80mm]{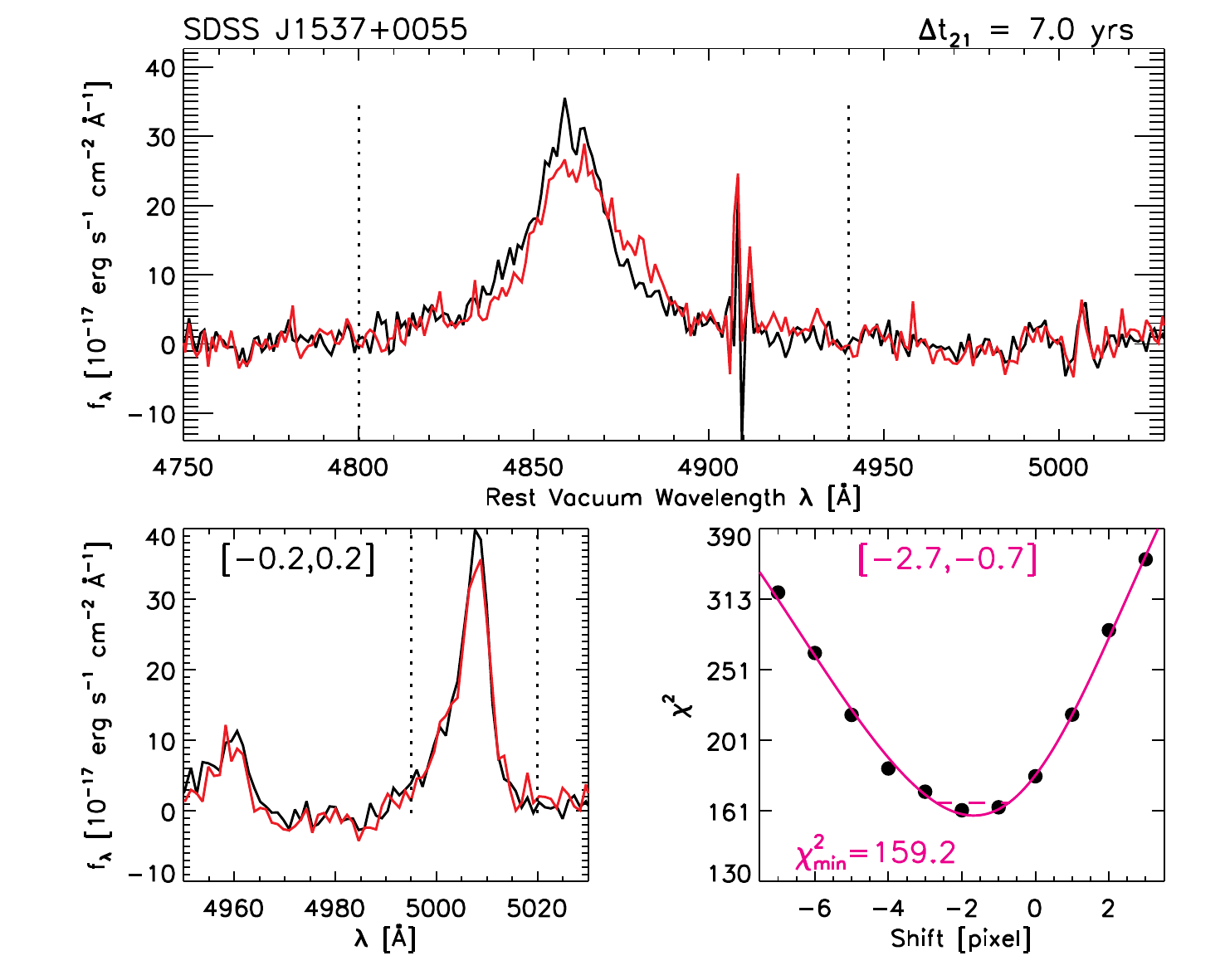}
    \includegraphics[width=80mm]{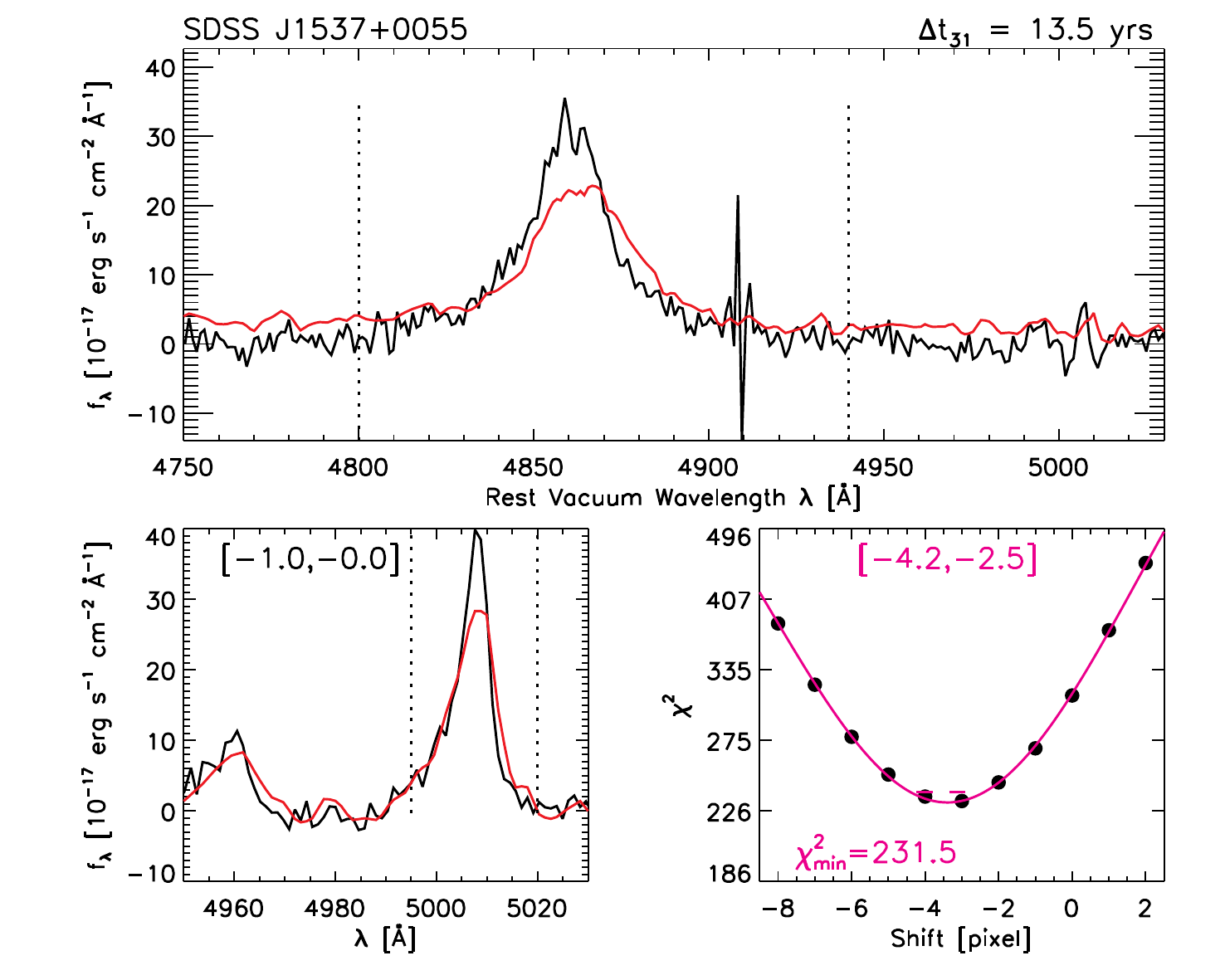}
    \includegraphics[width=80mm]{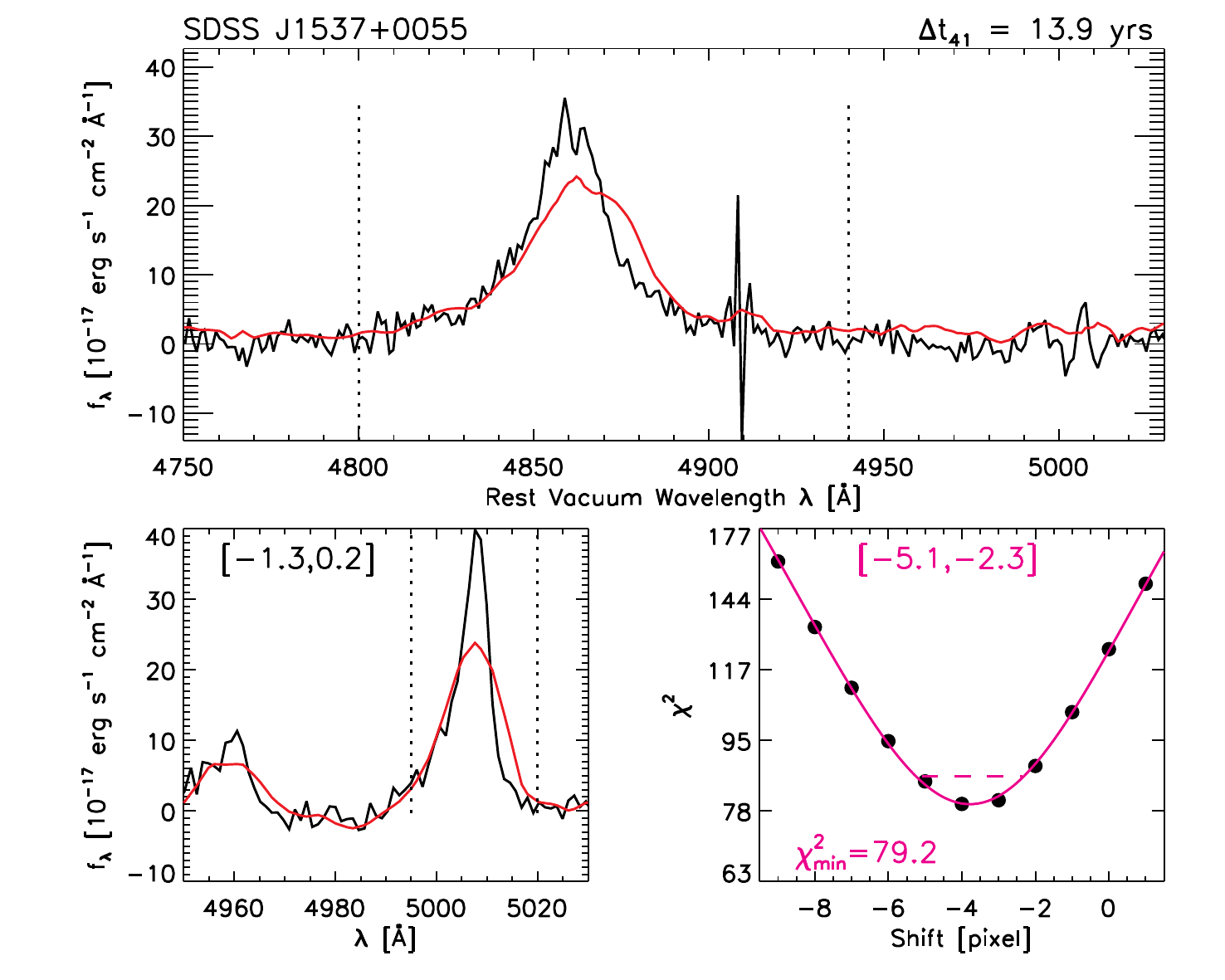}
       \centerline{Figure B1 -- Continued.   }  
\end{figure*}

\clearpage

\begin{figure*}
  \centering
    \includegraphics[width=80mm]{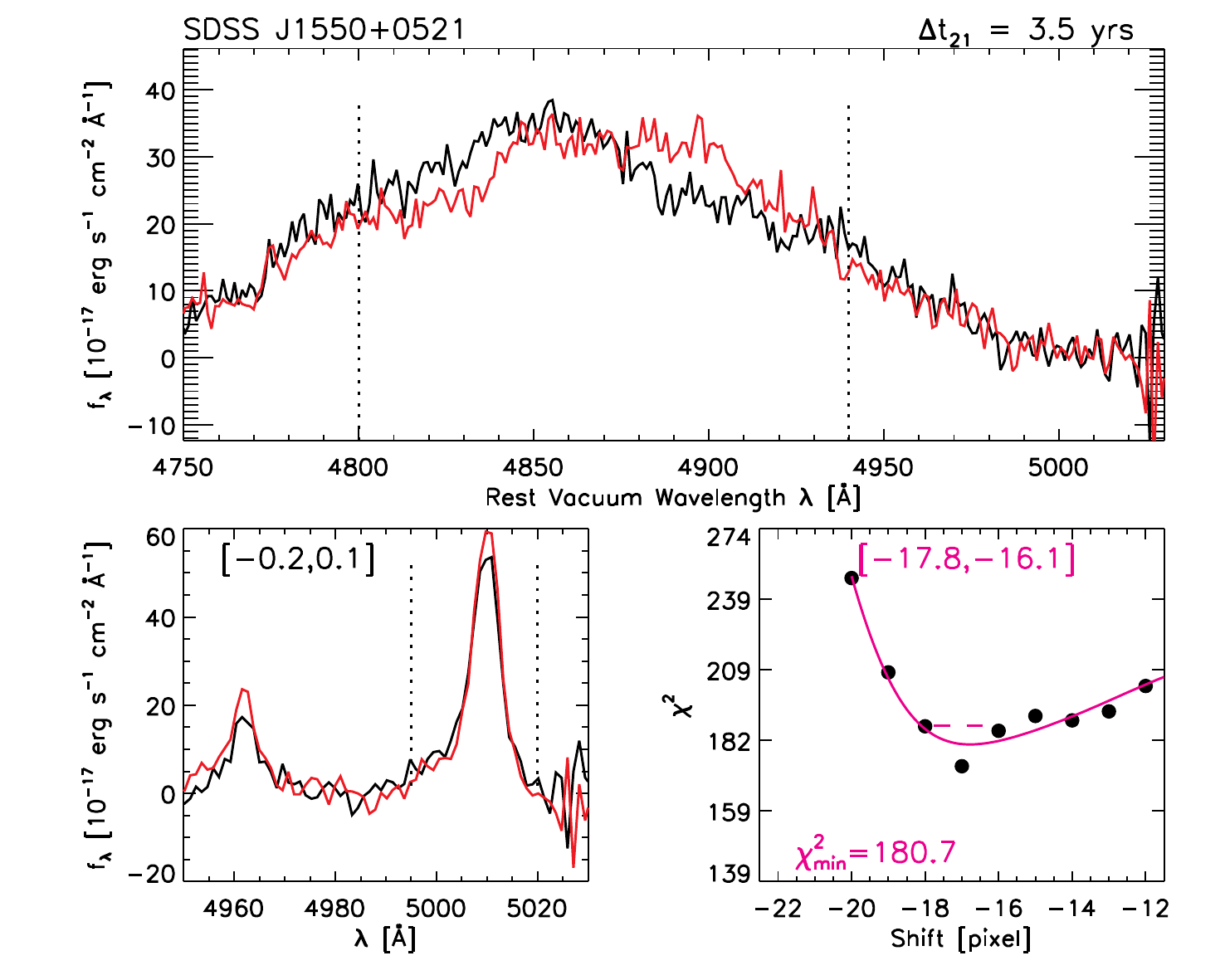}
    \includegraphics[width=80mm]{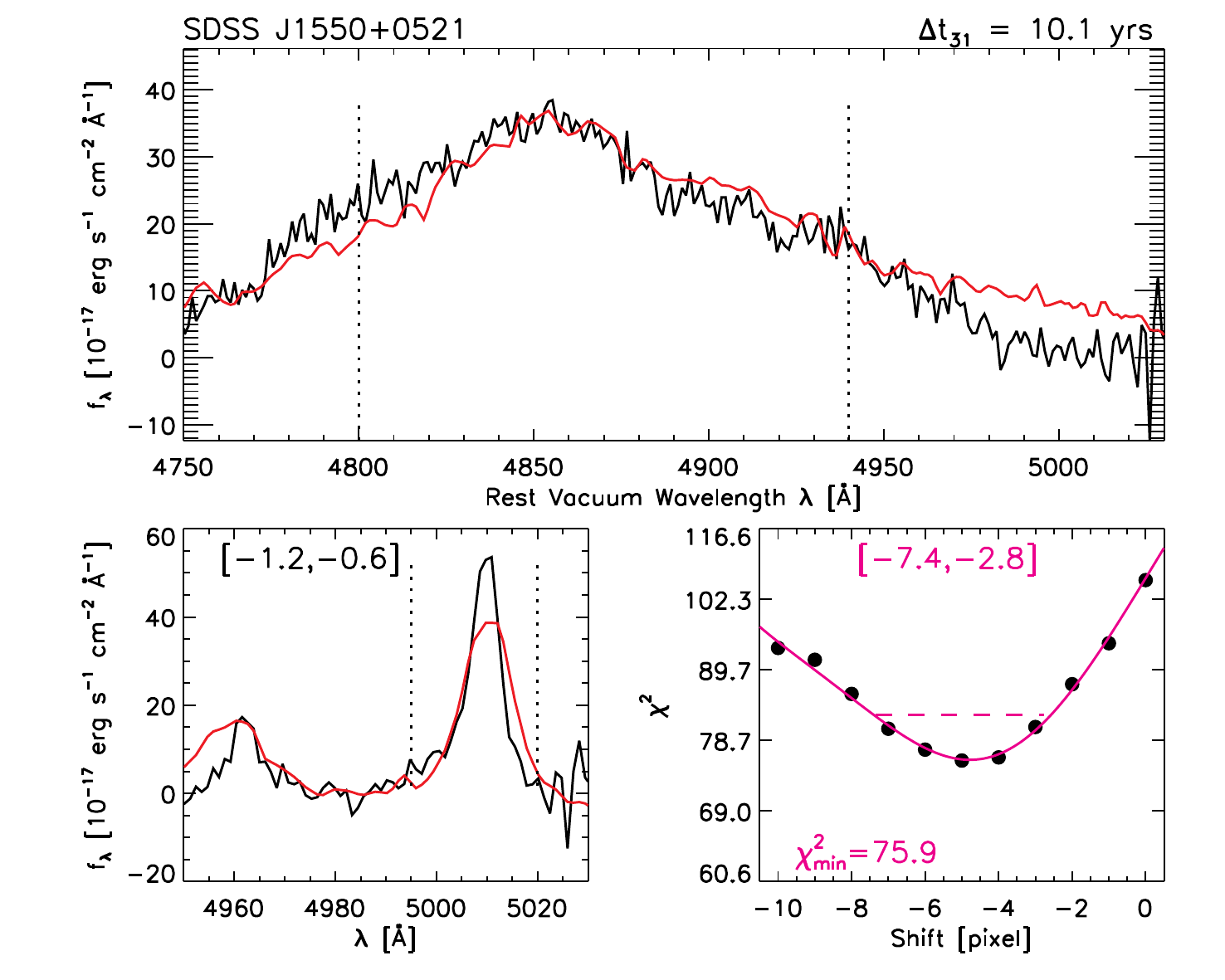}
    \includegraphics[width=80mm]{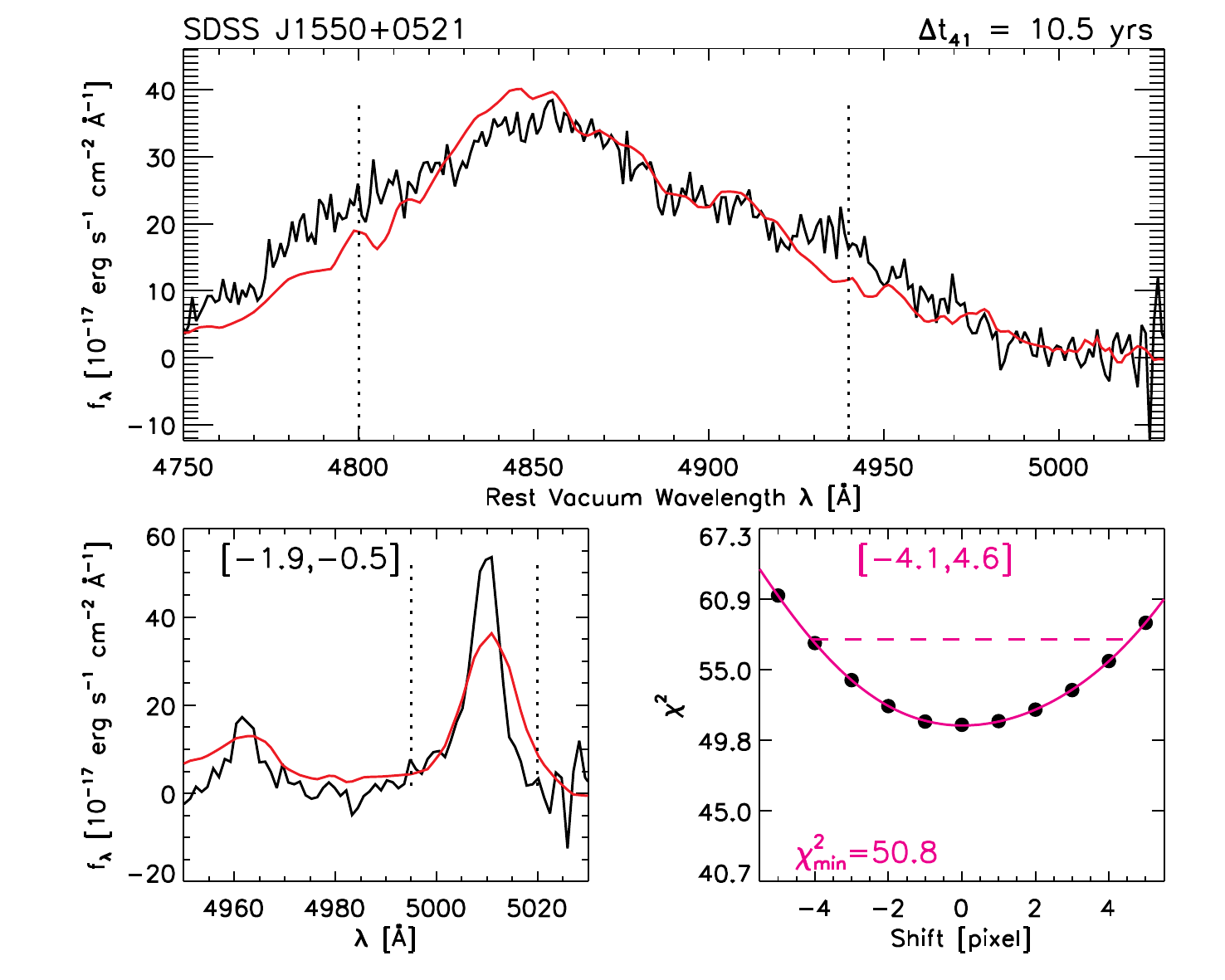}
    \includegraphics[width=80mm]{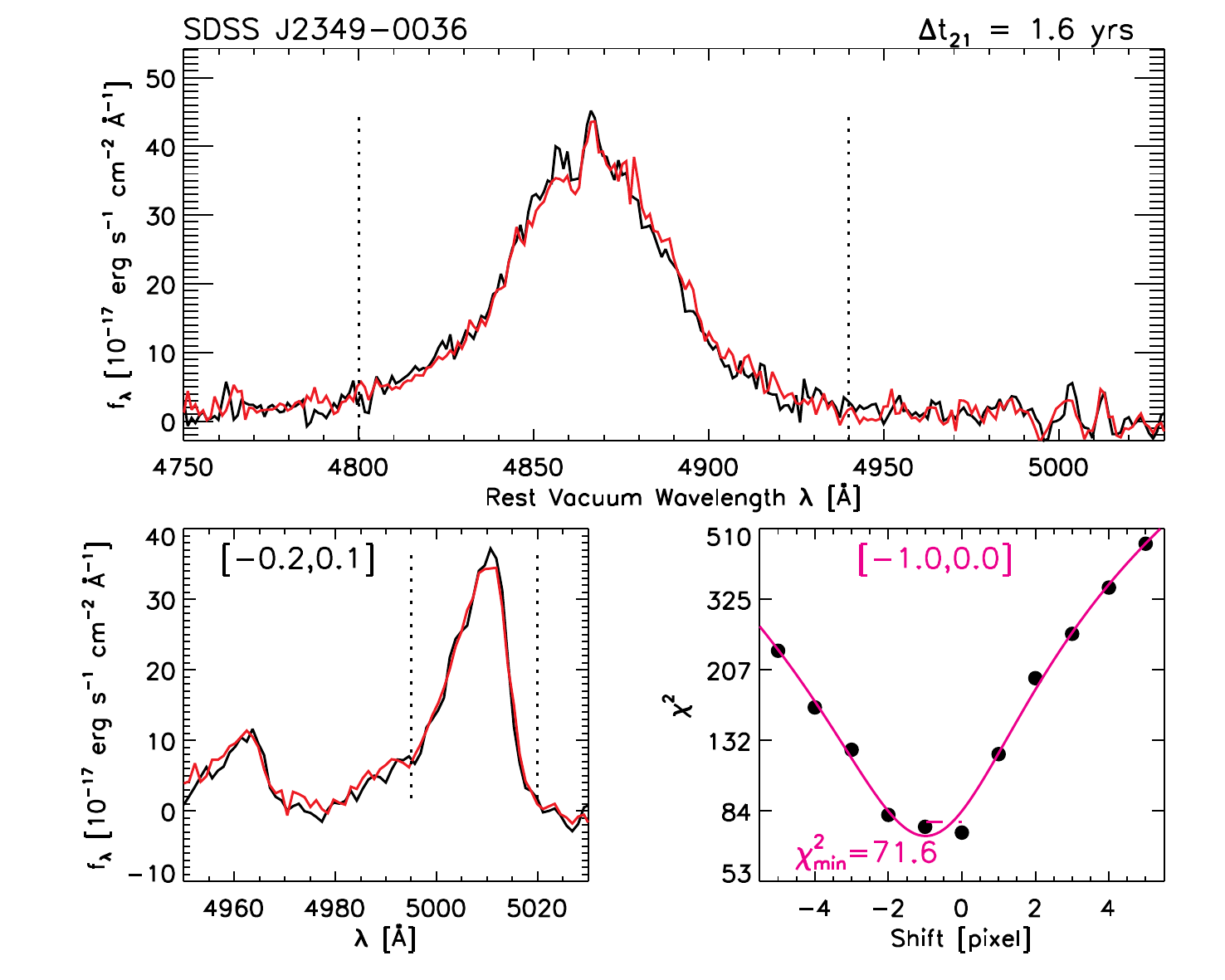}
    \includegraphics[width=80mm]{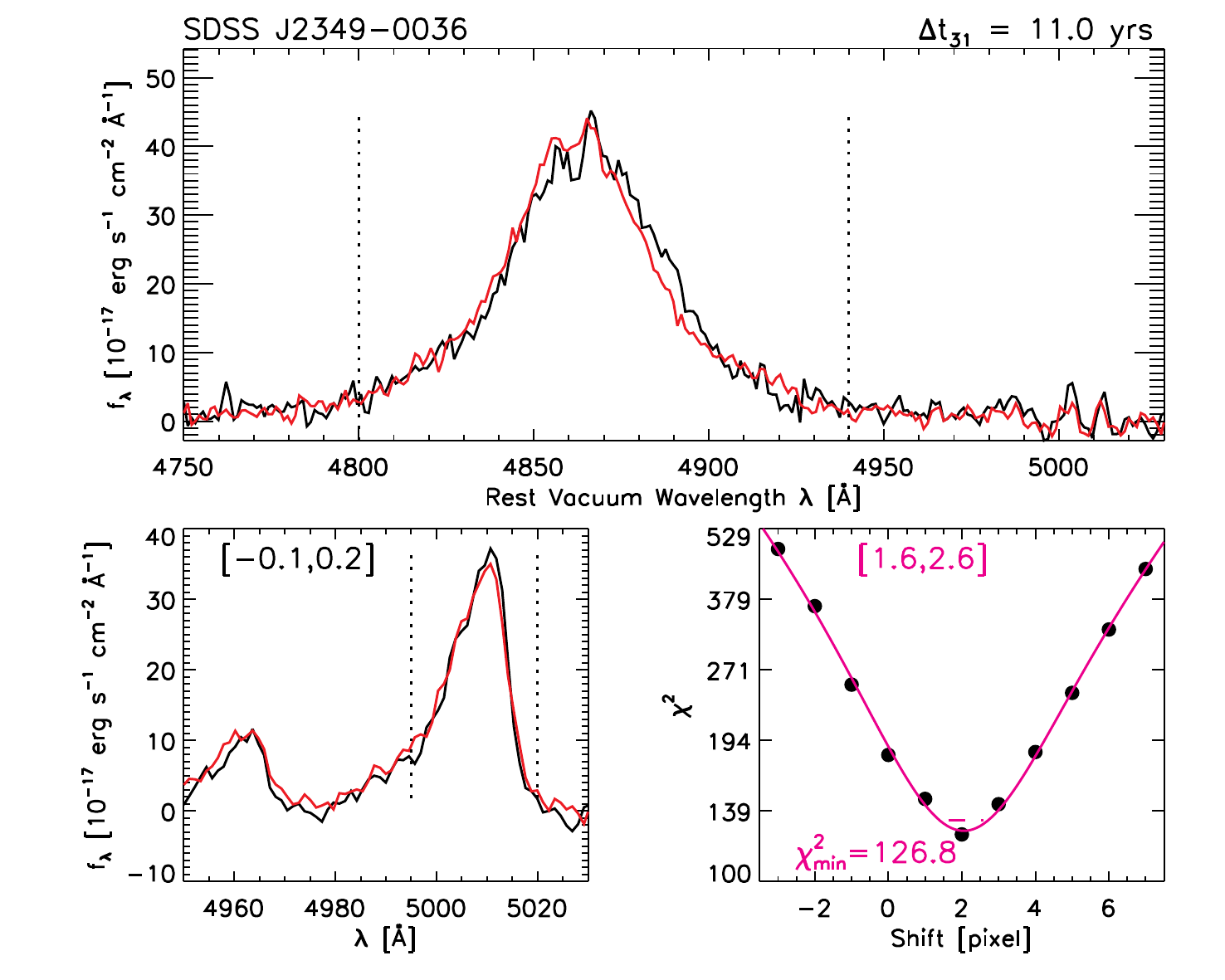}
    \includegraphics[width=80mm]{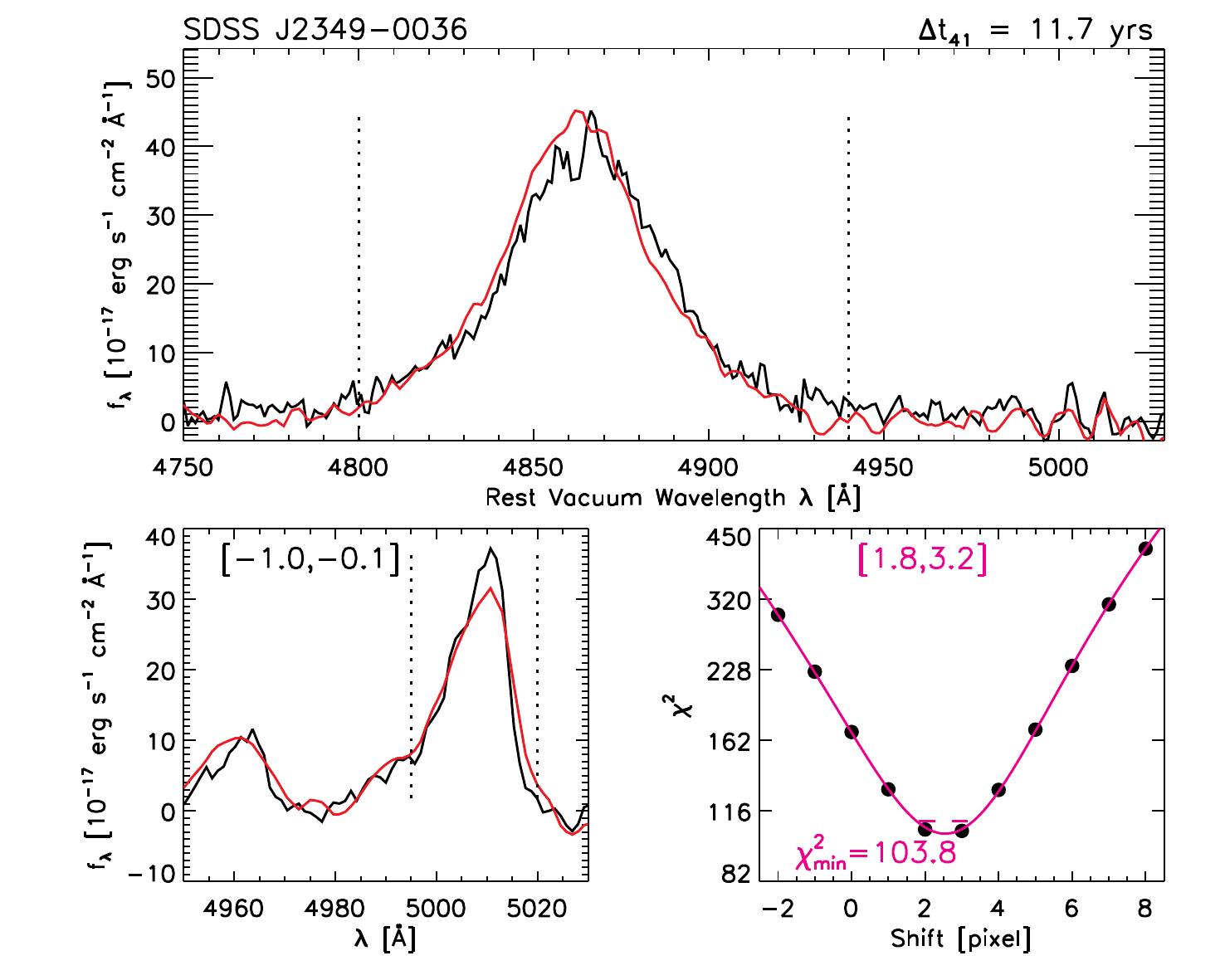}
       \centerline{Figure B1 -- Continued.   }  
\end{figure*}

\clearpage
%\end{comment}

\section{Dependence of the MCMC Results on the Initial Values}\label{appendix:mcmc}

Because our RV measurements only sampled 3 or 4 epochs, the parameter space was not very well constrained. We therefore try a large range of initial values (in particular, for the parameter $P$; the results are relatively insensitive to the other two free parameters, i.e., orbital phase and the initial velocity offset) to make sure that our result is representative of the maximum likelihood from the global posterior distribution. In particular, we loop through different initial values of $P$ spanning the whole range allowed by the prior.  Depending on the initial value, the MCMC chain may be trapped in different local maxima of the loosely constrained parameter space. To avoid running the MCMC chain for too long given our limited computational resources, we first find the local maxima in all the likely converged chains and then choose the global maximum likelihood region in the parameter space according to Equation \ref{eq:likelihood} as our final result. Figure C1 shows an example for the dependence of the MCMC results on the adopted initial values of $P$.  

\begin{figure*}
  \centering
    \includegraphics[width=88mm]{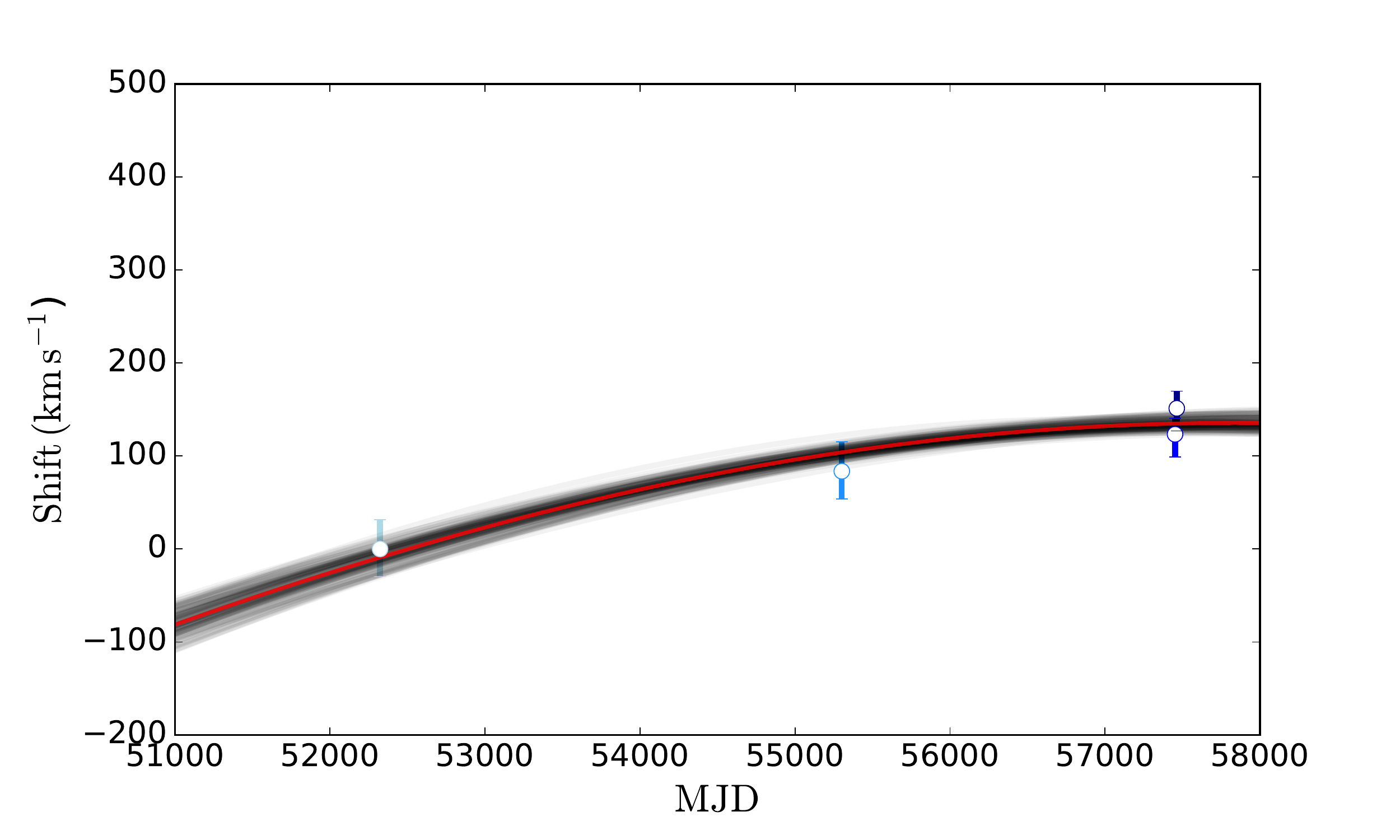}
    \includegraphics[width=88mm]{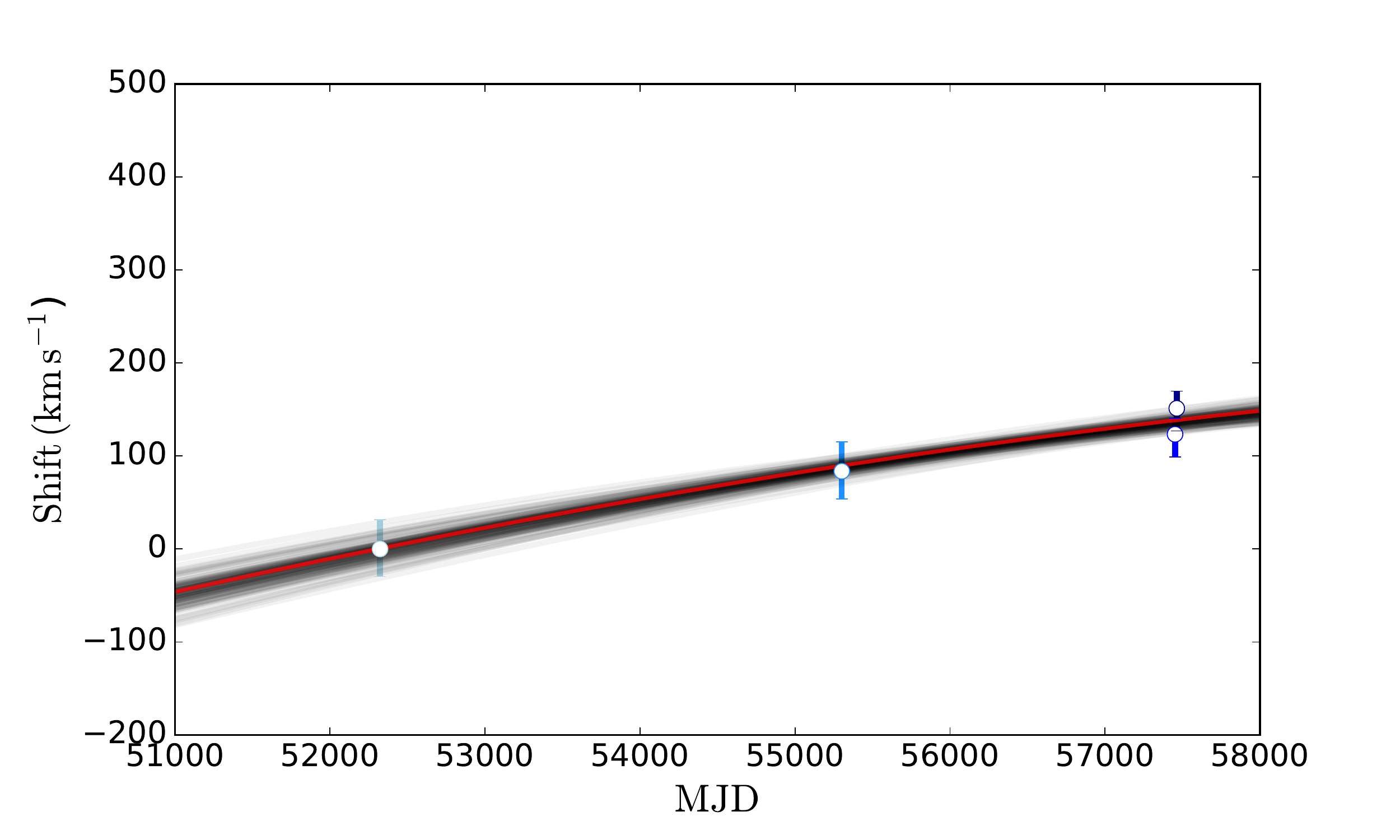}
    \includegraphics[width=88mm]{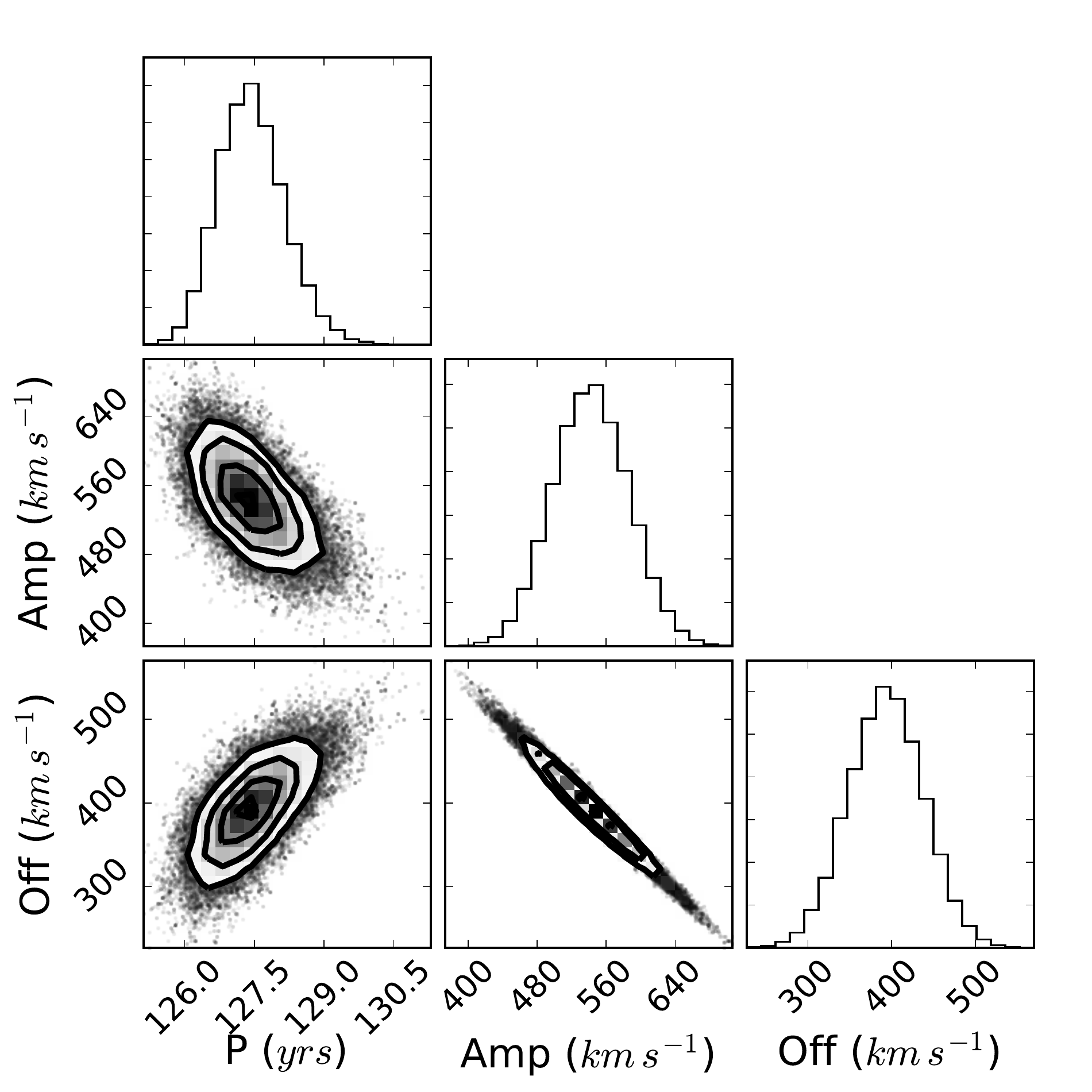}
    \includegraphics[width=88mm]{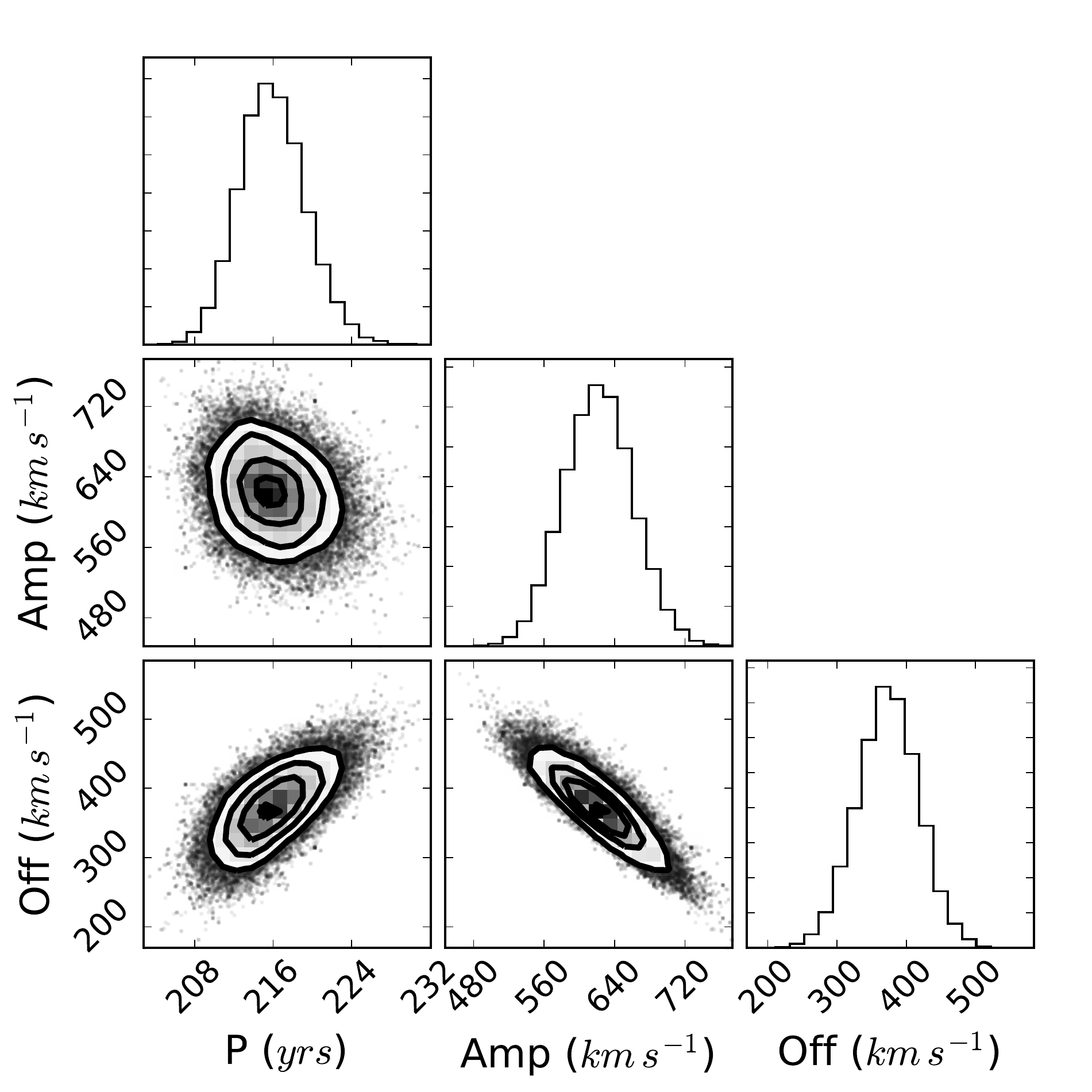}
    \label{fig:mcmc_test}	
    \caption{Dependence of the MCMC results on the initial values of $P$. 
    The two columns show results from two different initial values of $P$ = 150 yr (left) and 330 yr (right). 
    The upper rows of panels exihit the RV data (based on the broad \hb ) and the best-fit model (the red curve) along with the 1$\sigma$ uncertainty (sampled by the grey curves).  
    The lower rows of panels show the corner plots exhibiting the local posterior distribution and the covariance between the parameters. 
    While the model shown in the left column is preferred over the one shown in the right column according to 
    the likelihood function given by Equation \ref{eq:likelihood} at face value, a more extended time baseline is needed to better constrain $P$.  
     } 
\end{figure*}

\section{The effect of broad emission line short-term variability on measuring radial velocity from cross-correlation analysis}\label{appendix:noise}

We discuss how short-term variability ``jitter'' noise may affect the measurement and the modeling of broad emission-line radial velocities from multi-epoch spectra.  

Figure D1 (left panel) shows the effect of short-term variability on our RV shift results using an example of simulated broad \hb\ emission line profiles. We use a Gaussian (with $\sigma$$\sim$2000 \kms\ chosen to match the typical value in our sample) plus white noise for the baseline toy model of a simulated broad \hb\ (the black curve). To mimic the induced variability of short-term ``jitter'' noise, we add a red noise component (modeled with a power-law power spectral density distribution with a spectral index of $-2$ in the wavelength space) convolved with a Gaussian (whose parameters have been adjusted to determine the width and the amplitude of the ``jitter'' noise as described below) to the baseline model of broad \hb . Unlike the bulk RV shifts one would expect from a real BSBH orbital motion, the red noise component primarily accounts for asymmetric changes in the broad \hb\ line profiles, producing more variations in the line core than in the wing component (e.g., Paper I). We have experimented with a range of the width and the amplitude of the Gaussian component that was convolved with the jitter noise component so that the resulting peak-velocity variation distribution matches with the observed value found by the SDSS-RM sample \citep[which can be modeled with a Gaussian centered around zero with a standard deviation of $\sigma_{{\rm peak}}\sim$400 \kms , based on the measurement of 849 broad-line quasars with 32 spectroscopic epochs over six months;][]{Shen2015a,Shen2016a}, as shown in Figure D1 (right panel). We then run the ccf analysis of the jitter-noise-added broad \hb\ line (to simulate follow-up spectra) w.r.t. its original baseline model (to simulate the first-epoch spectrum) to quantify the resulting RV shifts in the same way as we would for the real data. Figure D1 (right panel) also shows the distribution of the ccf-based RV shifts, which can be modeled with a Gaussian centered around zero with a standard deviation of $\sigma_{{\rm CCF}}\sim$130 \kms . We therefore adopt $\sim$130 \kms\ as the typical uncertainty induced by the short-term jitter noise. We have also tested other types of models for the jitter noise (such as the ``see saw'' model studied by \citealt{Runnoe2017} and found a similar result on the ccf-based RV uncertainty.

\begin{figure*}
  \centering
    \includegraphics[width=88mm]{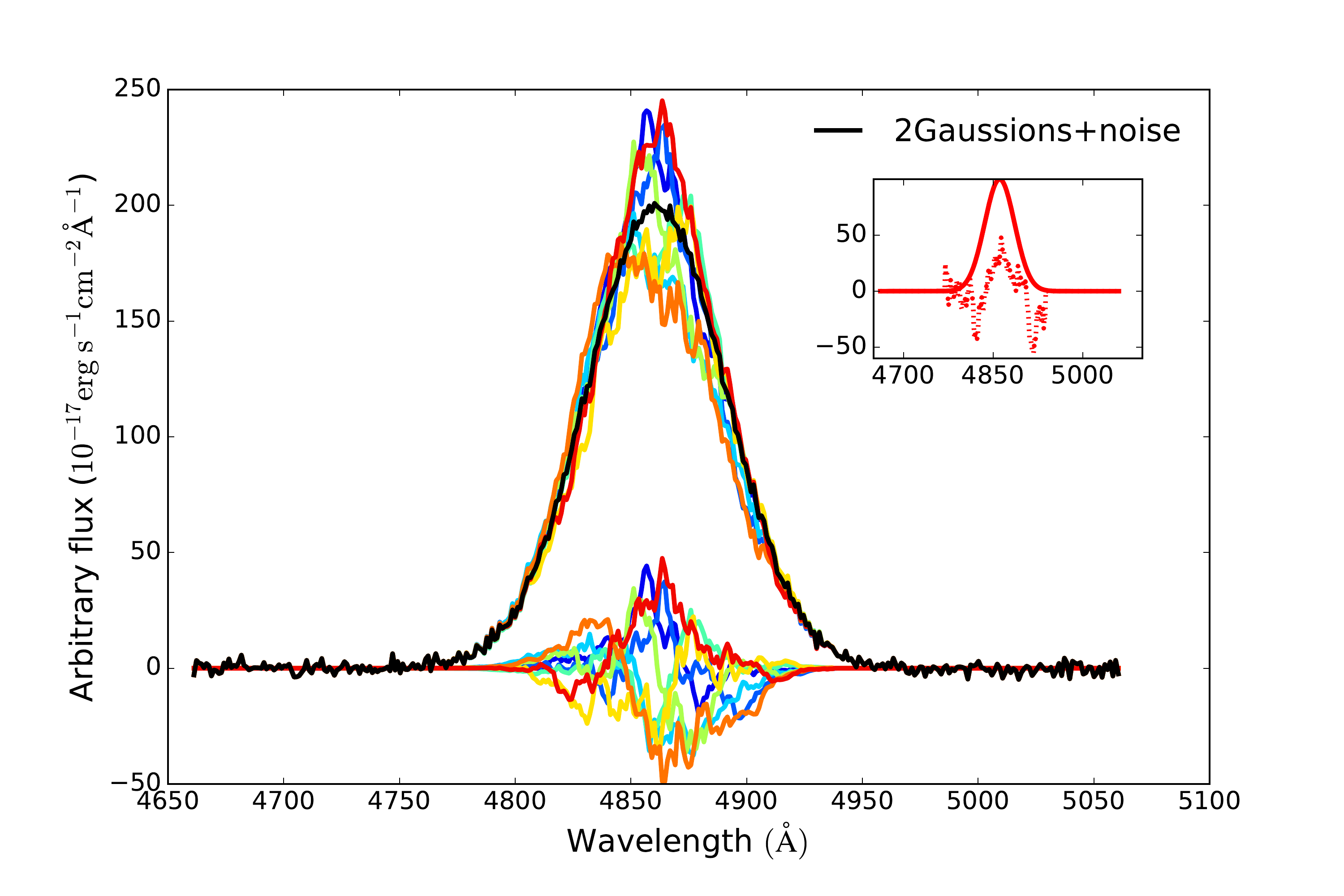}
    \includegraphics[width=88mm]{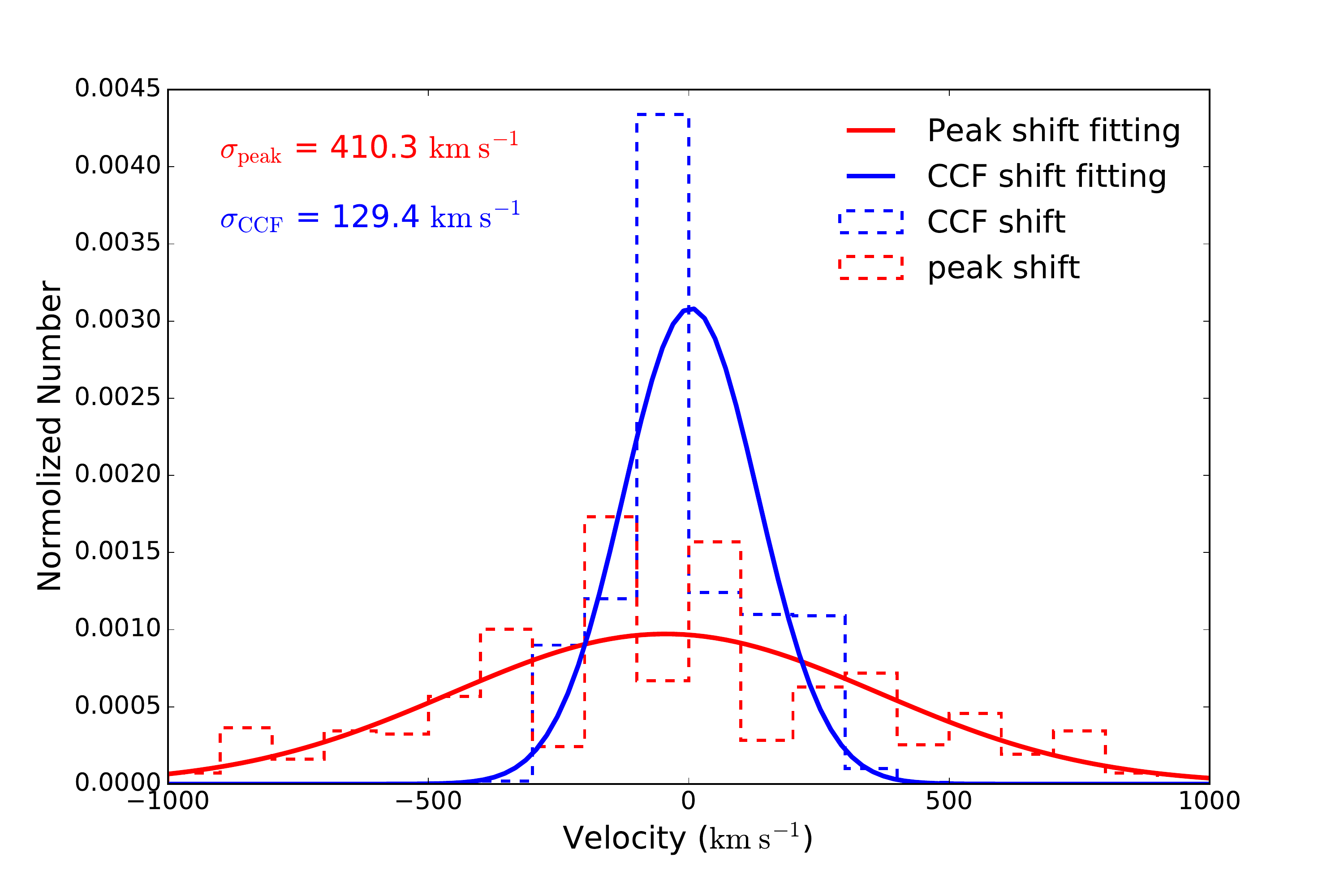}
    \label{fig:jitter_sim}	
    \caption{The effect of short-term variability jitter noise on the measurement of broad \hb\ RV shifts. Left panel: The black curve shows the baseline model of broad \hb\ (to simulate the first-epoch spectrum) consisting of a Gaussian (with a standard deviation of 2000 \kms\ chosen to match the typical value observed in our sample) plus white noise. The colored curves represent six examples of the red-noise-added models to simulate the follow-up spectra to illustrate the effect of short-term variability. The inset panel shows the red-noise component (the dotted curve) before convolving with the Gaussian component (the solid curve), whose width and amplitude have been chosen to produce a resulting peak-velocity shift distribution that matches with that observed by the SDSS-RM sample \citep{Shen2016a} as demonstrated in the right panel. Right panel: Probability distributions of the peak-velocity \citep[red dashed histograms;][]{shen11} and the RV shifts based on the cross-correlation analysis (blue dashed histograms) measured from a sample of 1,000 simulated broad \hb\ profiles induced by the short-term variability jitter noise as shown in the left panel. The solid curves show their best-fit models with a standard deviation of $\sigma_{{\rm peak}}\sim$400 \kms\ and $\sigma_{{\rm CCF}}\sim$130 \kms\ assuming a Gaussian distribution function.      
     } 
\end{figure*}

%%%%%%%%%%%%%%%%%%%%%%%%%%%%%%%%%%%%%%%%%%%%%%%%%%

% Don't change these lines
\bsp	% typesetting comment
\label{lastpage}
\end{document}